\documentclass[12pt]{article}
\pdfoutput=1
\usepackage{amsmath,amssymb,amsthm,mathtools,bm,mathdots,mathrsfs,xfrac}
\usepackage{graphicx,float,array,multirow,multicol,rotfloat,caption,subcaption,booktabs}
\usepackage{enumerate,parskip,adjustbox}
\usepackage[shortlabels]{enumitem}
\usepackage[normalem]{ulem}
\usepackage{xcolor}
\usepackage{pdflscape}
\usepackage{rotating}
\usepackage{empheq}

\usepackage{geometry}
\usepackage{dsfont}
\usepackage{pifont}
\DeclareSymbolFont{largesymbolsA}{U}{jkpexa}{m}{n}
\SetSymbolFont{largesymbolsA}{bold}{U}{jkpexa}{bx}{n}
\DeclareMathSymbol{\varprod}{\mathop}{largesymbolsA}{16}

\usepackage[vcentermath]{youngtab}
\usepackage[boxsize=.8em]{ytableau}
\usepackage[all]{xy}
\usepackage{tikz,tikz-cd}
\usetikzlibrary{arrows,shapes.misc,positioning,decorations.pathmorphing,decorations.markings,decorations.pathreplacing,matrix,patterns,backgrounds,calc}
\usepackage{pgfplots}
\pgfplotsset{compat=1.18}
\usepgfplotslibrary{fillbetween}

\tikzset{
    scale cd/.style={every label/.append style={scale=#1}, cells={nodes={scale=#1}}},
    gauge/.style={rounded rectangle, draw=black!100, thick, minimum size=2mm}, 
    gaugeD/.style={rounded rectangle, draw=black!100,double,thick,minimum size=2mm},  
    empty/.style={rounded rectangle, draw=white!100, thick, minimum size=2mm}, 
    flavor/.style={rectangle, draw=black!100, thick, minimum size=2mm},
    flavorD/.style={rectangle, draw=black!100, double,thick, minimum size=2mm},
    node/.style={circle, thick, draw=black!100,fill=white!100,  minimum size=2mm, inner sep=0pt},
    sqnode/.style={rectangle, thick, draw=black!100,fill=white!100,  minimum size=2mm, inner sep=0pt},
    sonode/.style={circle, thick, draw=black!100,fill=red!100,  minimum size=2mm, inner sep=0pt},
    spnode/.style={circle, thick, draw=black!100,fill=blue!100,  minimum size=2mm, inner sep=0pt},
    fnode/.style={rectangle, thick, draw=black!100,fill=white!100,  minimum size=2mm, inner sep=0pt},
    tnode/.style={rounded rectangle, outer sep=0pt, thick, minimum size=2mm},
    brace/.style={decoration={brace, mirror},decorate},
    ncbar angle/.initial=90,
    ncbar/.style={
        to path=(\tikztostart)
        -- ($(\tikztostart)!#1!\pgfkeysvalueof{/tikz/ncbar angle}:(\tikztotarget)$)
        -- ($(\tikztotarget)!($(\tikztostart)!#1!\pgfkeysvalueof{/tikz/ncbar angle}:(\tikztotarget)$)!\pgfkeysvalueof{/tikz/ncbar angle}:(\tikztostart)$)
        -- (\tikztotarget)
    },
    ncbar/.default=0.5cm,
    square left brace/.style={ncbar=0.2cm},
    square right brace/.style={ncbar=-0.2cm},
    snake it/.style={decorate, decoration={snake, amplitude=.4mm, segment length=2mm, post length=0mm,pre length=0mm}}
}

\usepackage[numbers,sort&compress]{natbib}
\usepackage{csquotes}
\usepackage{hhline}
\usepackage{longtable}
\usepackage{todonotes}
\usepackage{tikz}
\usepackage{tikz-cd}
\graphicspath{{figs/}}

\usepackage[hidelinks,linktocpage]{hyperref}
\usepackage[capitalise,sort]{cleveref}

\hypersetup{
	pdftitle={Discrete Gauging of 5d N=1 En SCFTs},
	pdfauthor={\textcopyright\ Guillermo Arias-Tamargo, Mario De Marco, Craig Lawrie, Shani Nadir Meynet, Alessandro Mininno},
	pdfsubject={hep-th},
	pdfcreator={pdfLaTex},
	pdfproducer={LaTex},
	pdfkeywords={},
	colorlinks=true,
    urlcolor=blue,
    anchorcolor=blue,
    citecolor=purple,
    filecolor=blue,
    linkcolor=orange,
    menucolor=blue,
    linktocpage=true
}

\crefname{figure}{Figure}{Figures}
\crefname{table}{Table}{Tables}
\crefname{definition}{Definition}{Definitions}
\crefname{proposition}{Proposition}{Propositions}
\crefname{claim}{Claim}{Claims}
\crefname{conjecture}{Conjecture}{Conjectures}

\numberwithin{equation}{section}
\allowdisplaybreaks[1]
\tikzset{
  gtp edge/.style={line width=0.85pt,black},
  black point/.style={circle,fill=black,draw=black,minimum size=5.4pt,inner sep=0pt},
  white point/.style={circle,fill=white,draw=black,line width=0.85pt,
                      minimum size=5.4pt,inner sep=0pt},
  dots label/.style={fill=white,inner xsep=1.5pt,inner ysep=0.6pt,
                     font=\footnotesize},
  block brace/.style={decorate,decoration={brace,amplitude=5pt},line width=0.7pt}
}

\makeatletter
\def\myrotate{\ifodd\c@page\else+\fi 90}
\g@addto@macro{\landscape}{\PLS@Rotate{\myrotate}}
\makeatother

\newcommand{\Lpagenumber}{\ifdim\textwidth=\linewidth\else\bgroup
	\dimendef\margin=0 
	\ifodd\value{page}\margin=\oddsidemargin
	\else\margin=\evensidemargin
	\fi
	\raisebox{\dimexpr -3\topmargin-\headheight-\headsep-0.5\linewidth}[0pt][0pt]{%
		\rlap{\hspace{\dimexpr \margin+\textheight+3\footskip}%
			\llap{\rotatebox{90}{\hspace{-4.5cm}\thepage\hfill}}}}%
	\egroup\fi}
\AddToHook{shipout/background}{\Lpagenumber}%

\def\fnote#1#2{\begingroup\def\thefootnote{#1}\footnote{#2}
     \addtocounter{footnote}{-1}\endgroup}

\theoremstyle{plain}
\newtheorem*{thm*}{Theorem}

\DeclareMathOperator{\U}{\mathfrak{u}}
\DeclareMathOperator{\SU}{SU}
\DeclareMathOperator{\SO}{SO}

\DeclareMathOperator{\USp}{USp}

\def\su{\mathfrak{su}}

\def\so{\mathfrak{so}}

\def\usp{\mathfrak{usp}}

\newcommand{\CC}{\mathbb{C}}

\newcommand{\ZZ}{\mathbb{Z}}

\newcommand{\ID}{\mathds{1}}

\newcommand{\coma}{\ensuremath{\,,\quad}}
\newcommand{\fstop}{\ensuremath{\,.}}

\newcommand{\PE}{\text{PE}}
\newcommand{\PL}{\text{PL}}
\newcommand{\HS}{\text{HS}}

\newcommand{\ttiny}[1]{\text{\tiny #1}}
\newcommand{\scalemath}[2]{\scalebox{#1}{$\displaystyle #2$}}
\newcommand{\mailref}[1]{\href{mailto:#1}{\color{black}\nolinkurl{#1}}}


\begin{document}

\begin{titlepage}
\vspace*{-3cm} 
\begin{flushright}
{\tt DESY-26-032}\\
\end{flushright}
\begin{center}
\vspace{1.7cm}
{\LARGE\bfseries Discrete Gauging of 5d \boldmath{$\mathcal{N}=1$} \boldmath{$E_n$} SCFTs}
\vspace{0.4cm}

{\large Guillermo Arias-Tamargo,$^1$ Mario De Marco,$^2$ \\[0.3em] Craig Lawrie,$^3$  Shani Nadir Meynet,$^{4}$\fnote{$\dagger$}{Part of this project was conducted when the author was affiliated to the Mathematics Institute and the Centre for Geometry and Physics, Uppsala University, Box 480, SE-75106 Uppsala, Sweden.} and Alessandro Mininno$^{5}$}\\
\vspace{.6cm}

{$^1$ Abdus Salam Centre for Theoretical Physics, The Blackett Laboratory, \\ 
Imperial College London, Prince Consort Road, London, SW7 2AZ, UK}\\
\vspace{.1cm}
{$^2$ Physique Théorique et Mathématique and International Solvay Institutes, \\
Université Libre de Bruxelles, C.P.~231, 1050 Brussels, Belgium}\\
\vspace{.1cm}
{$^3$ Deutsches Elektronen-Synchrotron DESY,  Notkestr.~85, 22607 Hamburg, Germany}\\
\vspace{.1cm}
{$^4$ Department of Physics and Astronomy, University of Pennsylvania,\\
209 South 33rd Street, Philadelphia, PA 19104, USA}\\
\vspace{.1cm}
{$^5$ Department of Physics, University of Wisconsin--Madison,\\1150 University Avenue, Madison, WI 53706, USA}\par
\vspace{.2cm}

\scalebox{0.8}{\tt \mailref{g.arias-tamargo@imperial.ac.uk}, \mailref{mario.de.marco@ulb.be}, \mailref{craig.lawrie1729@gmail.com},} 
\scalebox{0.8}{\tt \mailref{smeynet@sas.upenn.edu}, \mailref{mininno@physics.wisc.edu}}
\vspace{0.3cm}

\textbf{Abstract}
\end{center}

We study the gauging of discrete symmetries of the 5d rank-one $E_n$ SCFTs that fix the Coulomb branch coordinate but act nontrivially on the Higgs branch chiral ring. Using generalized toric polygons and $(p,q)$-brane webs, we identify symmetries of the moduli space that extend to symmetries of the full SCFT. In many cases they are non-anomalous, and we analyze the moduli space of the gauged theories. We realize the resulting Higgs branch quotients as Coulomb branches of wreathed magnetic quivers and compute their Hilbert series; the flavor algebras extracted from moment maps agree with the fixed-point subalgebras of $\mathfrak{e}_n$. We also lift the discrete actions to the Seiberg--Witten geometry and recover the expected restriction on the mass deformations. For $\mathbb{Z}_2$-wreathings that exchange adjacent gauge nodes, we find that the monopole formula requires a flux-dependent sign and propose a corrected prescription. Finally, we conjecture a higher-rank extension and verify it for the rank-two $E_6/\mathbb{Z}_2$ Higgs branch.

\vspace{1cm}
\vfill 
\end{titlepage}

\tableofcontents
\bigskip\medskip
\hrule
\bigskip\bigskip
\newpage

\section{Introduction}\label{sec:intro}

Over the last two decades, substantial progress has been made towards classifying superconformal field theories (SCFTs) with eight supercharges in four or more dimensions. 
These efforts proceed along two complementary avenues: bottom-up approaches, which constrain putative SCFTs through consistency conditions and the structure of their moduli spaces, and top-down approaches, which construct SCFTs through compactifications of higher-dimensional theories or geometric engineering in string theory. Geometry, therefore, typically plays a crucial role in both approaches, e.g., whether through supersymmetric moduli spaces of vacua or engineering geometries in string theory. We refer to \cite{Akhond:2021xio,Argyres:2022mnu} for recent reviews of these constructions and classification programs. A proposed classification should, in particular, be closed under gauging non-anomalous finite global symmetries; in this paper, we explore the robustness of such classifications via this procedure of discrete gauging. 

Let $\mathcal{T}$ denote an eight-supercharge supersymmetric quantum field theory (SQFT) in dimension $3\leq d\leq6$, equipped with a finite bosonic internal symmetry $\Gamma$.\footnote{The group $\Gamma$ need not be a discrete factor or component group of the global symmetry group. In several of the examples here, it is instead a finite noncentral subgroup of a continuous flavor symmetry and acts nontrivially on the flavor algebra by inner automorphisms.}  Provided the relevant 't Hooft anomaly vanishes, one may gauge $\Gamma$ to obtain a new theory, which we denote by $\widetilde{\mathcal{T}}=\mathcal{T}/\Gamma$. For $d\geq3$, there are no twisted sectors of local operators: a principal $\Gamma$-bundle on the sphere $S^{d-1}$ linking a local insertion is necessarily trivial. Consequently, discrete gauging projects the local operator algebra onto its invariant subalgebra.
Nevertheless, it modifies the spectrum of extended operators and, more generally, the theory on manifolds supporting nontrivial $\Gamma$-bundles. This differs markedly from two dimensions, where nontrivial $\Gamma$-bundles on the circle surrounding a local insertion give rise to twisted sectors of local operators. 
Discrete gauging often produces non-paradigmatic members of the SCFT landscape; that is, theories which violate standard conjectures that were developed outside of the discretely-gauged context. See \cite{Argyres:2017tmj,Bourton:2018jwb,Bourget:2018ond,Argyres:2018wxu} for some evidence of such features.

In this paper, we focus on the 5d $\mathcal{N}=1$ rank-one Seiberg theories \cite{Seiberg:1996bd,Morrison:1996xf}, and we present evidence that discrete gauging produces an abundance of heretofore unbeknownst rank-one SCFTs that evade known geometric classifications.\footnote{A small sample of the recent literature on the geometric constructions of 5d SCFTs is \cite{DeMarco:2026tnc,DeMarco:2026jyj,delzotto20265dscftfixtures,Bourget:2026ono,Collinucci:2025rrh,DeMarco:2023irn,Dramburg:2025tlb,Mu:2023uws,Xie:2022lcm,DelZotto:2022fnw,Cvetic:2022imb,Tian:2021cif,Apruzzi:2019vpe,Apruzzi:2019opn,Apruzzi:2019enx,Arias-Tamargo:2024fjt,Bourget:2023wlb,Bhardwaj:2020gyu,Jefferson:2018irk,Bhardwaj:2019fzv,Collinucci:2026kom,CarrenoBolla:2024fxy,CarrenoBolla:2025rkv,Bonelli:2025juz}.}
Discrete gaugings of 5d SCFTs have previously been studied through M-theory orbifolds and U-folds \cite{Acharya:2021jsp}, S-fold quotients of five-brane webs and their holographic duals \cite{Kim:2021fxx,Apruzzi:2022nax}, and quotients of the Seiberg--Witten (SW) geometries associated with their circle reductions \cite{Furrer:2024zzu}. Discrete gaugings also appear in the study of free or rank-zero sectors engineered by terminal Calabi--Yau singularities \cite{Closset:2020scj,DeMarco:2021try,Collinucci:2021ofd,DeMarco:2022dgh}. The gaugings considered here are of a different type from those of \cite{Acharya:2021jsp,Kim:2021fxx,Apruzzi:2022nax,Furrer:2024zzu}: they leave the rank-one Coulomb branch (CB) coordinate invariant while acting nontrivially on the mass parameters, Higgs branch (HB) chiral ring, and flavor algebra. 

One central object of our study is the sector of $1/2$-BPS operators of $\mathcal{T}$ and $\widetilde{\mathcal{T}}$ that belong to the so-called Higgs branch chiral ring. Associated to any eight-supercharge SQFT there is a hyperk\"ahler space known as the Higgs branch, $\mathcal{H}$, and the chiral ring is $\mathbb{C}[\mathcal{H}]$. Recently, a powerful method to understand the Higgs branch of an eight-supercharge theory, known as the magnetic quiver method, has been developed \cite{Hanany:1996ie, Hanany:1997gh, Ferlito:2017xdq}. In this approach, one aims to find a 3d $\mathcal{N}=4$ Lagrangian SQFT, $\mathcal{T}_M$, such that the Coulomb branch is isomorphic to the Higgs branch of the theory of interest, i.e.:
\begin{equation}
    \mathcal{H}(\mathcal{T}) \cong \mathcal{C}(\mathcal{T}_M)\fstop
\end{equation}
Since $\mathcal{T}_M$ is Lagrangian, one can use techniques like the monopole formula \cite{Cremonesi:2013lqa} to compute the Hilbert series of the Coulomb branch chiral ring, which is identical to that of the Higgs branch chiral ring of $\mathcal{T}$.

Since $\mathcal{T}_M$ is a Lagrangian theory, we can capture that Lagrangian by writing down a quiver. Let us suppose that $\Gamma$ is a finite global symmetry of $\mathcal{T}$, under which the Higgs branch chiral ring of $\mathcal{T}$, as a ring, is invariant, and that it acts on the quiver $\mathcal{T}_M$ via a diagram automorphism. Then, it has been proposed that the magnetic quiver for the Higgs branch of the gauged theory $\widetilde{\mathcal{T}}$, $\widetilde{\mathcal{T}}_M$, which captures the Higgs branch chiral ring after discrete gauging, is given via the $\Gamma$-wreathing of the magnetic quiver $\mathcal{T}_M$:\footnote{Briefly, a wreathing of a quiver gauge theory has a disconnected gauge group defined via a wreath product. Let us consider a Lie group $G$ and a discrete group $\Gamma \subset S_k$. Then, the wreath product of $G$ by $\Gamma$ is defined as
\begin{equation*}
    G\wr \Gamma \equiv \left( \bigtimes_{i=1}^k G_i\right) \rtimes \Gamma\coma
\end{equation*}
where $k$ copies of $G$ are taken as $k$-fold direct products, and the action with $\Gamma$ is defined by the function $\phi\,: \, \Gamma \rightarrow \text{Aut}(G^k)$. For instance, if $\Gamma = S_k$, the $k$ copies of $G$ are permuted by $\Gamma$, with the action given by $\phi(\sigma)(\alpha_1,\ldots \alpha_k)=(\alpha_{\sigma(1)},\ldots, \alpha_{\sigma(k)})$ for $\sigma \in S_k$. The matter spectrum of the quiver similarly has to be well-behaved under the action of $\Gamma$.} 
\begin{equation}
    \widetilde{\mathcal{T}}_M = \mathcal{T}_M \, \wr \, \Gamma\fstop
\end{equation}
In other cases, various pieces of evidence for this conjecture have been put forth in \cite{Hanany:2018cgo,Arias-Tamargo:2019jyh,Bourget:2020bxh,Arias-Tamargo:2022nlf,Hanany:2023uzn,Giacomelli:2024sex,Grimminger:2024mks,Lawrie:2025exx}. For unitary quivers, this is related to the BFN construction as applied to disconnected gauge groups \cite[Prop 2.7(3)]{Braverman:2016wma}.

The existence of a diagram automorphism, $\Gamma$, of a magnetic quiver for the Higgs branch of a higher-dimensional theory, $\mathcal{T}$, is neither a sufficient nor a necessary condition for the existence of a discrete global symmetry of $\mathcal{T}$. It is not sufficient since, while the diagram automorphism indicates that the protected sector of Higgs branch operators, the Higgs branch chiral ring, has a consistent $\Gamma$-action; there is no guarantee that the action extends to the full operator algebra. It is also not a necessary condition, as there is no requirement that all discrete global symmetries of $\mathcal{T}$ act as a diagram automorphism on some magnetic quiver. Indeed, in this paper, we will see multiple magnetic quivers for the same Higgs branch that have diagram automorphisms corresponding to different discrete global symmetries. In particular, a given global symmetry might be visible in only one of these magnetic quivers but not in the others. Furthermore, even if $\Gamma$ is an honest discrete global symmetry of $\mathcal{T}$, it is still necessary to analyze the relevant 't Hooft anomalies before one can confirm the possibility of gauging $\Gamma$.

To circumvent these uncertainties, we turn to geometric engineering. In \cite{Lawrie:2025exx}, it was observed that the magnetic quiver for the Higgs branch of the 6d $\mathcal{N}=(1,0)$ SCFTs known as (rank $N$ Higgsed) conformal matter \cite{DelZotto:2014hpa} admits a $\mathbb{Z}_2$ diagram automorphism. These 6d SCFTs are known to be engineered in F-theory via the atomic construction \cite{Heckman:2015bfa,Heckman:2013pva,Distler:2022yse}, in terms of certain non-compact elliptically-fibered Calabi--Yau (CY) threefolds. The $\mathbb{Z}_2$ diagram automorphism of the magnetic quiver can be seen to arise from a geometric Green--Schwarz automorphism \cite{Apruzzi:2017iqe} of the Calabi--Yau threefold $Y$. Since the entire 6d SCFT should follow from $Y$, a Green--Schwarz automorphism has been argued to give rise to a discrete global symmetry of the SCFT \cite{Apruzzi:2017iqe}. The gauging of this $\mathbb{Z}_2$ was discussed in \cite{Lawrie:2025exx}.

The goal of this work is to tackle the problem with a similar approach, now in the context of 5d SCFTs, which necessarily have eight supercharges. Such theories admit several complementary realizations in geometric engineering; here, we focus primarily on their construction by compactifying M-theory on non-compact, generalized toric Calabi--Yau threefolds. The engineering data can be encoded by a generalized toric polygon and its dual $(p,q)$-five-brane web, from which one may moreover extract a magnetic quiver for the Higgs branch. For many of the examples studied in this paper, we identify the relevant diagram automorphisms of the magnetic quiver with automorphisms of the generalized toric polygon or brane web. Since these act on the complete engineering geometry, rather than only on the geometry of the Higgs branch, they uplift to symmetries of the full five-dimensional SCFTs \cite{DelZotto:2024tae, DeMarco:2025pza}.\footnote{Alternatively to the toric picture, we can consider the ``combined-fiber-diagram'' geometries of \cite{Apruzzi:2019enx,Apruzzi:2019opn,Apruzzi:2019vpe}. These involve a non-flat resolution, followed by a sequence of flop transitions, of the (non-minimal) singularity of the elliptically-fibered Calabi--Yau that engineers the 6d $(1,0)$ parent of the 5d SCFTs of interest. These geometries are depicted explicitly for the rank-one Seiberg theories in \cite{Apruzzi:2019enx}, and we can ask if the quiver automorphism group is a geometric symmetry there.}

The toric geometry also determines a mirror curve, which may be interpreted as the Seiberg--Witten curve governing the Coulomb branch dynamics of the theory compactified on a circle. An automorphism of the generalized toric polygon induces an action on the coordinates and mass parameters of this curve. Whenever this action can be made manifest, we construct the Seiberg--Witten geometry of the discretely-gauged theory. In the rank-one examples considered here, the Coulomb branch coordinate is invariant under the discrete symmetry. The quotient preserves the rank of the theory, while projecting its mass deformations, and the number of invariant mass parameters agrees with the rank of the flavor algebra obtained independently from the Higgs branch.

\subsubsection*{Summary of Hilbert Series Results}

One of the results of this paper is the computation of the Coulomb branch Hilbert series for the wreathed versions of the magnetic quivers for the rank-one Seiberg theories. To help the reader, we have summarized the results of the Hilbert series computations in Tables \ref{tbl:Uwreathing} and \ref{tbl:OSwreathing}. 

\subsubsection*{Structure of the Paper}

The remainder of this paper is organized as follows. In Section \ref{sec:2}, we review the brane web and magnetic quiver descriptions of the Seiberg theories (Section \ref{sec:brane_webs}), determine how affine Dynkin diagram automorphisms act on their flavor algebras (Section \ref{sec:dg}), and discuss the relevant anomaly constraints in Section \ref{sec:tHooftanomalies}. We then review wreathing of magnetic quivers in Section \ref{sec:wreathingMQ} and propose a corrected prescription to compute the wreathing in certain special cases in Section \ref{sec:Z2cuttingedges}. Subsequently, we explain how the same discrete actions are implemented on the associated Seiberg--Witten curves in Section \ref{sec:swsection2}. The bulk of this paper is the application of these methods systematically to all the rank-one $E_n$ theories in Section \ref{sec:GaugeAutoEnTheories}. Next, in Section \ref{sec:higherrank}, we discuss the extension to higher rank, with the rank-two $E_6$ theory as an explicit example. We conclude in Section \ref{sec:disc} with a summary of some interesting future directions. Finally, in Appendix \ref{app:birch}, we go beyond 5d SCFTs and discuss the $\mathbb{Z}_2$-wreathing of the affine $A_{K}$ quivers for all values of $K$. 

\subsubsection*{Conventions and Notation}

\begin{enumerate}
\item In the following, we consider both unitary and orthosymplectic quivers, and we follow the usual language in which \tikz{\node[gauge] {};} represents a gauge node and its label is the rank of the unitary gauge group, while \tikz{\node[flavor] {};} denotes a flavor node and its label denotes the number of fundamentals. When the gauge or flavor node is $\mathfrak{so}$, we will color the node in {\color{red}{red}}, while we will color the $\mathfrak{usp}$ nodes in {\color{blue}{blue}}. The line connecting two nodes represents a hypermultiplet in the bifundamental representation of the two groups. If the edge corresponds to multiple hypermultiplets, we will specify this. A loop attached to a unitary gauge node indicates an adjoint-valued hypermultiplet, whereas a loop attached to a symplectic gauge node refers to an antisymmetric hypermultiplet. If the hypermultiplets have charge $2$, we will denote it via a wiggle line.
\item We adopt the notions of excess number, balance, underbalance, and overbalance as in \cite{Gaiotto:2008ak}.  For a $\U(N)$ gauge group with $N_f$ hypermultiplets transforming in the fundamental representation, the excess number for such a gauge group is defined as
\begin{equation}
\texttt{e}_{\U(N)} = N_f - 2N\fstop
\end{equation}
If, in a given quiver theory, all the excess numbers are $\texttt{e}_{\U(N)} \geq 0$, the theory is said to be \textit{good}, because all the monopole operators are above the unitarity bound. If a gauge group has $\texttt{e}_{\U(N)}=0$, we call that node \textit{balanced}; while if $\texttt{e}_{\U(N)} >0$, the gauge group is said to be overbalanced. If $\texttt{e}_{\U(N)} <0$, the gauge group is said to be underbalanced. In particular, if any of the gauge groups has $\texttt{e}_{\U(N)}=-1$, the theory is called \textit{ugly} because there is a monopole operator saturating the unitarity bound; otherwise, the theory is \textit{bad} because it admits monopole operators below the unitarity bound.\footnote{In some cases, it may happen that monopoles valued in the magnetic lattice of multiple gauge groups render an apparently good theory bad. This is the case for some affine Dynkin-shaped unitary quiver theories \cite{Cremonesi:2014xha}. A similar discussion can be applied to some orthosymplectic magnetic quivers, and recently an adjustment to the balance notion for $\SO$ groups has been proposed in \cite{Lawrie:2024wan}.}  The excess number for $\SO$ and $\USp$ gauge groups with $N_f$ (full) hypermultiplets transforming in the fundamental representation is defined, respectively, as
\begin{equation}
    \texttt{e}_{\so(N)} = N_f-N+1\coma \texttt{e}_{\usp(2N)} = N_f - 2N-1\fstop
\end{equation}
For orthosymplectic nodes, having any gauge group with $\texttt{e}_{\so(N)/\usp(2N)}=-1$ already signals that the theory is \textit{bad}. 
\item The Hilbert series counts the operators belonging to the chiral ring of a given theory $\mathcal{T}$. We will make use of the plethystic exponential (PE) and its inverse, the plethystic logarithm (PL), to extract the generators of the chiral ring and the relations among them. The PE of a multivariate function $f(t_1,\ldots t_k)$ (vanishing at the origin) is defined as 
\begin{equation}
    \PE\left[f(t_1,\ldots, t_k)\right] = \exp\left(\sum_{n=1}^\infty \frac{1}{n}f(t_1^n,\ldots, t_k^n)\right)\coma
\end{equation} 
while the PL is defined as
\begin{equation}
    \PL\left[f(t_1,\ldots,t_k)\right] = \sum_{n=1}^\infty \frac{\mu(n)}{n}\log \left(f(t_1^n,\ldots, t_k^n)\right)\coma
\end{equation}
where $\mu(n)$ is the M\"obius function defined by
\begin{equation}
    \mu(n) = \begin{cases} 1 & n = 1\coma \\ (-1)^k & \text{if $n$ is the product of $k$ distinct primes,}\\ 0 & \text{otherwise.} \end{cases}
\end{equation}
\item We denote the characters of a representation of a Lie algebra $\mathfrak{g}$ as $\chi^{\mathfrak{g}}_{[\lambda_1, \dots, \lambda_n]}(\mathbf{z})$, where the integers $\lambda_i$ are the Dynkin labels. If no algebra $\mathfrak{g}$ is explicitly specified, the $A$ algebra is assumed. We adopt the Slansky convention \cite{Slansky:1981yr} used by the \texttt{LieART} package \cite{Feger:2019tvk} to order the nodes. For the classical algebras $A_n$, $B_n$, $C_n$, and $D_n$, this matches the standard Bourbaki numbering \cite{Bourbaki:2002}. $A_n$ nodes are numbered from $1$ to $n$ from left to right along the chain. For $B_n$ and $C_n$, the chain is numbered from $1$ to $n$, terminating at the short root for $B_n$ and at the long root for $C_n$, while for $D_n$, the chain is numbered from $1$ to $n-2$, and the two spinor nodes are $n-1$ and $n$. Under this convention, the $B_n$ spinor is $[0, \dots, 0, 1]$, the vector is $[1, 0, \dots, 0]$, the rank-$2$ symmetric traceless representation is $[2, 0, \dots, 0]$, and the adjoint is $[0, 1, 0, \dots, 0]$ for $n \ge 3$, which reduces to $[0, 2]$ for $B_2$. For the exceptional $E_n$ algebras, the nodes $1$ through $n-1$ label the longest horizontal chain from left to right. This convention places the shorter leg of the horizontal chain on the left, ensuring the branching node is always at position $3$, and node $n$ labels the vertical node of the short leg attached to node $3$. Consequently, the labels for the adjoint representations of the exceptional algebras shift relative to the Bourbaki convention. The adjoint is $[0, \ldots, 0, 1]$ for $E_6$, $[1, 0, \ldots,0]$ for $E_7$, and $[0,\ldots, 0, 1, 0]$ for $E_8$. Finally, $F_4$ nodes are numbered from $1$ to $4$ along the chain, with nodes $1$ and $2$ being long roots and nodes $3$ and $4$ being short roots; the $F_4$ adjoint is therefore $[1, 0, 0, 0]$ and the fundamental $\mathbf{26}$ is $[0, 0, 0, 1]$. For $G_2$, the short root is labeled $1$ and the long root is $2$, with the Dynkin label $[1, 0]$ for the fundamental $\mathbf{7}$ and $[0, 1]$ for the adjoint representations.
\end{enumerate}

\section{Discrete Gauging in the Moduli Space}
\label{sec:2}

We begin in this section by developing/reviewing the framework used throughout this paper to analyze discrete gauging of the rank-one $E_n$ theories. We first review (generalized) toric polygons, $(p,q)$-brane webs, and the associated magnetic quivers, and then determine how diagram automorphisms of the latter act on the continuous flavor algebra. Because such an automorphism initially defines only an action on the Higgs branch chiral ring, we separately discuss anomaly freedom and, where possible, its lift to a symmetry of the full brane web geometry. We discuss how to implement the corresponding Higgs branch quotient by wreathing the magnetic quiver, including a modified monopole prescription when a $\mathbb{Z}_2$ axis bisects a bifundamental edge, and finally describe how to study the induced restriction of the Seiberg--Witten family to the invariant mass-deformation locus.

\subsection{Recap: Brane Webs and Magnetic Quivers} 
\label{sec:brane_webs}

In this section, we will review how to read the magnetic quiver from the brane web/toric diagram that engineers the theory. A system of $(p,q)$-5-branes and $[p,q]$-7-branes in Type IIB represents the five-dimensional generalization of the Hanany--Witten setup \cite{Hanany:1996ie} and defines the brane webs we discuss in our work. We start from branes extending as shown in Table \ref{tbl:braneweb}.

\begin{table}[t]
    \centering
    \begin{tabular}{c||c|c|c|c|c|c|c|c|c|c}
                        & $x^0$ & $x^1$ & $x^2$ & $x^3$ & $x^4$ & $x^5$ & $x^6$ & $x^7$ & $x^8$ & $x^9$   \\ \hhline{=#=|=|=|=|=|=|=|=|=|=}
     NS5-brane          & $\bullet$ & $\bullet$ & $\bullet$ & $\bullet$ & $\bullet$ & $\bullet$ & $-$ & $-$ & $-$ & $-$  \\
        \hhline{-|-|-|-|-|-|-|-|-|-|-}
     D5-brane           & $\bullet$ & $\bullet$ & $\bullet$ & $\bullet$ & $\bullet$ & $-$ & $\bullet$ & $-$ & $-$ & $-$  \\ \hhline{-|-|-|-|-|-|-|-|-|-|-}
     $(p,q)$-5-brane  & $\bullet$ & $\bullet$ & $\bullet$ & $\bullet$ & $\bullet$ & \multicolumn{2}{c|}{angle} & $-$ & $-$ & $-$ \\\hhline{-|-|-|-|-|-|-|-|-|-|-}
     $[p,q]$-7-brane & $\bullet$ & $\bullet$ & $\bullet$ & $\bullet$ & $\bullet$ & $-$ & $-$ & $\bullet$ & $\bullet$ & $\bullet$ 
    \end{tabular}
    \caption{Brane web setup where $\bullet$ mark the direction spanned by the brane, while $-$ the direction where the branes are pointlike. The $(p,q)$-5-branes lie at angle given by $\arctan(q/p)$.}
    \label{tbl:braneweb}
\end{table}

We use the notation that a D5-brane is a $(1,0)$-5-brane, while an NS5-brane is a $(0,1)$-5-brane. A generic $(p,q)$-5-brane is a bound state of $p$ D5-branes and $q$ NS5-branes. A similar discussion applies to $[p,q]$-7-branes.  Since the only unshared directions that D5-, NS5-, and $(p,q)$-5-branes have are along $(x^5,x^6)$, it is possible to draw the branes on this plane. Supersymmetry fixes the corresponding angle in terms of their charges: D5-branes will be shown horizontally, NS5-branes vertically, while $(p,q)$-5-branes will be at an angle $\alpha=\arctan(q/p)$. Orthogonal to the plane, there will be $[p,q]$-7-branes on which $(p,q)$-5-branes can end. For example, here is the brane web of a single D5-brane, NS5-brane, and $(1,1)$-5-brane ending at the corresponding 7-branes:
\begin{equation}
    \begin{tikzpicture}[baseline=0,font=\footnotesize]
         \node[node, fill=black,label=left:{$[1,0]$}] (D5) at (-1,0) {};
        \node[node, fill=black,label=below:{$[0,1]$}] (NS5) at (0,-1) {};
        \node[node, fill=black,label=above right:{$[1,1]$}] (pq) at (1,1) {};
       \draw[thick] (0,0) -- (D5);
       \draw[thick] (0,0) -- (NS5);
       \draw[thick] (0,0) -- (pq);
       \node at (-3,-1) {
        \begin{tikzpicture}[baseline=0,font=\footnotesize]
        \node[node,draw=gray,label=below left:{\color{gray}{$x^{7,8,9}$}}] at (0,0) {};
        \draw[thick,gray,-Triangle] (0,0) -- node[above,pos=1] {$x^6$} (1,0);
		\draw[thick,gray,-Triangle] (0,0) -- node[left,pos=1] {$x^5$} (0,1);
        \end{tikzpicture}
       };
    \end{tikzpicture}\fstop
\end{equation}
More generally, one considers junctions where multiple $(p_i,q_i)$-5-branes meet, and at each junction, charge conservation requires that 
\begin{equation}
    \sum_i p_i = 0 = \sum_i q_i\fstop
\end{equation}
Finally, in order for the junction of three $(p,q)$-5-branes to be supersymmetric, the branes must satisfy the so-called ``s-rule": namely, that the number of D5-branes that may connect a single D7-brane and a stack of NS5-branes is less than or equal to the number of NS5-branes in the stack. It has been proposed in \cite{DeWolfe:1998bi,Iqbal:1998xb} (see also \cite{DeWolfe:1999hj,DeWolfe:1998eu,DeWolfe:1998pr,vanBeest:2020kou,Bergman:2020myx}) that for irreducible junctions, the intersection of branes $(p_i,q_i)$ obeys the s-rule if
\begin{equation}
\left|\det \left( \begin{array}{cc} p_i & q_i \\ p_j & q_j \end{array}\right)\right| \geq \gcd(p_k,q_k)^2\coma    
\end{equation}
for every permutation $(i,j,k)$ of $(1,2,3)$.

Each brane web describes a 5d $\mathcal{N}=1$ theory. The Coulomb branch of the theory corresponds to faces in the brane web. In the case of parallel D5-branes, the distance separating the branes corresponds to the VEV of the Coulomb branch scalars in the theory. Analogously, the separation between NS5-branes is identified with the mass of the instanton. For generic $(p,q)$-5-brane webs, one can construct more complicated string junctions; however, the rank of the 5d theory will still correspond to the number of compact faces in the web. Note that not all the 5d SCFTs we consider admit an infrared gauge theory description. One example is the $E_0$ theory, with a brane web given by
\begin{equation}
    \begin{tikzpicture}[scale=0.85,baseline=0,font=\footnotesize]
			\node[node, fill=black,label=left:{$[-1,-1]$}] (A) at (-1,-1) {};
			\node[node, fill=black,label=left:{$[-1,2]$}] (B) at (-1,2) {};
            \node[node, fill=black,label=right:{$[2,-1]$}] (C) at (2,-1) {};
           \draw[line width=1pt] (0,0)--(A);
           \draw[line width=1pt] (0,0)--(B);
           \draw[line width=1pt] (0,0)--(C);
           \end{tikzpicture}\coma
\end{equation}
where no parallel D5-branes are present in any phase of the web.

We are also interested in the Higgs branch of the theory, which is necessarily hyperk\"ahler. This is given by consistent brane subwebs obtained when moving along the 7-branes. When 5-branes move along the 7-branes, the hypermultiplets will get a VEV, hence parametrizing the Higgs branch. In Section \ref{sec:intro}, we reviewed the technique of magnetic quivers, which is a powerful approach to study the Higgs branch of non-Lagrangian eight-supercharge SCFTs, $\mathcal{T}$, by finding an appropriate 3d $\mathcal{N}=4$ Lagrangian quiver, $\mathcal{T}_M$, such that
\begin{equation}
    \text{HB}(\mathcal{T}) \cong \text{CB}(\mathcal{T}_M)\fstop
\end{equation}
Given a brane web, we can determine a magnetic quiver of the Higgs branch by following the prescription developed in \cite{Cabrera:2018jxt}, which we are now briefly reviewing:
\begin{enumerate}
    \item First of all, we need to determine the consistent subwebs of a given brane web. These are obtained by looking for subwebs for which the $(p,q)$ charge is conserved at the vertices and that satisfy the s-rule. 
    \item For every such subweb, we associate a $\U(1)$ gauge node. If there are $n$ identical and coincident subwebs, the algebra is enhanced to $\U(n)$.
    \item The number of bifundamental matter multiplets connecting each node associated to two subwebs $S_1$ and $S_2$ is determined by
    \begin{equation}
        \text{E}(S_1,S_2) = \text{SI}(S_1,S_2) + \sum_k X(S_1,S_2,k) - \sum_k Y(S_1,S_2,k)\coma
    \end{equation}
    where $\{k\}$ is the set of all 7-branes in the web, $\text{SI}(S_1,S_2)$ is the \textit{stable intersection number} \cite{Fulton:1994aaa,RichterGebert:2003aaa,Mikhalkin:2006aaa} given by
    \begin{equation}\label{eq:SIcomp}
       \text{SI}(S_1,S_2) = \left|\det \left( \begin{array}{cc} p_1 & q_1 \\ p_2 & q_2 \end{array}\right)\right|\coma
    \end{equation}
    at each point where 5-branes from two subwebs with $(p_1,q_1)$ and $(p_2,q_2)$ charges intersect; and finally, $X(S_1,S_2,k)$ counts the number of pairs of 5-branes belonging to different subwebs that end on opposite sides of a 7-brane, while $Y(S_1,S_2,k)$ counts the number of pairs of 5-branes belonging to different subwebs that end on the same side of a 7-brane. 
\end{enumerate}

Given a toric polytope that geometrically engineers the SCFT in M-theory, the brane web is obtained by taking its dual graph. This means that given an edge in the polytope, it is mapped to $(p,q)$-5-branes perpendicular to the edge, in a number that depends on the length of the edge. In this work, we consider only rank-one $E_n$ theories; thus, in the following, we review how to derive the magnetic quivers for these theories from their brane webs. 

Let us consider an explicit example, and we will construct the magnetic quivers for the $E_1$ and $E_3$ theories. The brane web for $E_1$ can be derived from its toric polytope: 
\begin{equation}\label{eq:fromTDtoPQWeb}
   \begin{tikzpicture}[scale=1,baseline=0]
   \node (EL) {$\Longrightarrow$};
  \begin{scope}[xshift=-4cm]
  			\draw[thick,gray,-Triangle] (-1.5,0) -- node[above,pos=1] {$x$} (1.5,0);
			\draw[thick,gray,-Triangle] (0,-1.5) -- node[left,pos=1] {$y$} (0,1.5);
      	\node[node, fill=black] (O) at (0,0) {};
			\node[node, fill=black] (A) at (-1,0) {};
			\node[node, fill=black] (B) at (0,-1) {};
            \node[node, fill=black] (C) at (1,0) {};
             \node[node, fill=black] (D) at (0,1) {};
           \draw[line width=1pt] (A)--(B)--(C) -- (D) -- (A);
           \draw[line width=1pt,red] (-1.5,1.5) -- (-0.5,0.5) -- (-0.5,-0.5) -- (0.5,-0.5) -- (0.5,0.5) -- (1.5,1.5);
           \draw[line width=1pt,red] (-1.5,-1.5)  --(-0.5,-0.5);
           \draw[line width=1pt,red] (1.5,-1.5)  --(0.5,-0.5);
           \draw[line width=1pt,red] (-0.5,0.5)  --(0.5,0.5);
  \end{scope}
        \begin{scope}[xshift=4cm,font=\footnotesize]
            \node[node, fill=black,label=left:{$[-1,1]$}] (A) at (-1,1) {};
			\node[node, fill=black,label=right:{$[1,-1]$}] (B) at (1,-1) {};
            \node[node, fill=black,label=right:{$[1,1]$}] (C) at (1,1) {};
            \node[node, fill=black,label=left:{$[-1,-1]$}] (D) at (-1,-1) {};
           \draw[line width=1pt] (0,0)--(A);
           \draw[line width=1pt] (0,0)--(B);
           \draw[line width=1pt] (0,0)--(C);
           \draw[line width=1pt] (0,0)--(D);
        \end{scope}
        \end{tikzpicture} \,.
\end{equation}
Since there are only two segments, we associate each of them with a $\U(1)$ node, and it is then sufficient to compute the stable intersection between them using equation \eqref{eq:SIcomp}, i.e.,
\begin{equation}
   \text{SI}(S_1,S_2) =  \left|\det \left( \begin{array}{cc} -1 & 1 \\ 1 & 1 \end{array}\right)\right| = 2\fstop
\end{equation}
Since there are no instances of different subwebs ending in the same 7-brane, the $X$ and $Y$ contributions to the number of edges vanish. Hence, the magnetic quiver for the $E_1$ theory is 
\begin{equation}\label{eqn:dora}
    \begin{tikzpicture}[baseline=0,font=\footnotesize]
        \node[node, label=below:{$1$}] (A1) {};
        \node[node, label=below:{$1$}] (A2) [right=8mm of A1] {};
        \draw[double distance=3pt] (A1) -- (A2);
        \node (Z2) [right=6mm of A2] {$/\U(1)\,,$};
    \end{tikzpicture}
\end{equation}
where $/\U(1)$ means that we need to decouple the overall $\U(1)$ associated to the center-of-mass of this unframed unitary quiver.

More interesting is how to obtain the magnetic quivers for the two components of the Higgs branch of the $E_3$ theory. From the toric diagram of $E_3$, we can extract the brane web:
\begin{equation}
   \begin{tikzpicture}[scale=1,baseline=0]
   \node (EL) {$\Longrightarrow$};
  \begin{scope}[xshift=-4cm]
  			\draw[thick,gray,-Triangle] (-1.5,0) -- node[above,pos=1] {$x$} (1.5,0);
			\draw[thick,gray,-Triangle] (0,-1.5) -- node[left,pos=1] {$y$} (0,1.5);
			\node[node, fill=black] (O) at (0,0) {};
			\node[node, fill=black] (A) at (-1,0) {};
			\node[node, fill=black] (B) at (0,-1) {};
            \node[node, fill=black] (C) at (1,0) {};
             \node[node, fill=black] (D) at (0,1) {};
             \node[node, fill=black] (E) at (1,1) {};
             \node[node, fill=black] (F) at (-1,-1) {};
           \draw[line width=1pt] (A)--(F) -- (B)--(C) -- (E) -- (D) -- (A);
  \end{scope}
        \begin{scope}[xshift=4cm,font=\footnotesize]
            \node[node, fill=black,label=left:{$[-1,1]$}] (A) at (-1,1) {};
			\node[node, fill=black,label=right:{$[1,-1]$}] (B) at (1,-1) {};
            \node[node, fill=black,label=above:{$[0,1]$}] (C) at (0,1) {};
            \node[node, fill=black,label=right:{$[1,0]$}] (E) at (1,0) {};
            \node[node, fill=black,label=left:{$[-1,0]$}] (F) at (-1,0) {};
            \node[node, fill=black,label=below:{$[0,-1]$}] (G) at (0,-1) {};
           \draw[line width=1pt] (0,0)--(A);
           \draw[line width=1pt] (0,0)--(B);
           \draw[line width=1pt] (0,0)--(C);
           \draw[line width=1pt] (0,0)--(E);
           \draw[line width=1pt] (0,0)--(F);
           \draw[line width=1pt] (0,0)--(G);
        \end{scope}
        \end{tikzpicture} \,.
\end{equation}
There are two different consistent decompositions of this web into various subwebs, which implies that the Higgs branch is a union of two cones, each of which is given by a different magnetic quiver. We distinguish the two decompositions by colors:
\begin{equation}
    \begin{tikzpicture}[scale=1,baseline=0]
   \node (EL) {$\cup$};
  \begin{scope}[xshift=-4cm,font=\footnotesize]
      	\node[node, fill=black,label=left:{$[-1,1]$}] (A) at (-1,1) {};
			\node[node, fill=black,label=right:{$[1,-1]$}] (B) at (1,-1) {};
            \node[node, fill=black,label=above:{$[0,1]$}] (C) at (0,1) {};
            \node[node, fill=black,label=right:{$[1,0]$}] (E) at (1,0) {};
            \node[node, fill=black,label=left:{$[-1,0]$}] (F) at (-1,0) {};
            \node[node, fill=black,label=below:{$[0,-1]$}] (G) at (0,-1) {};
           \draw[line width=1pt,red] (0,0)--(A);
           \draw[line width=1pt,red] (0,0)--(B);
           \draw[line width=1pt,green] (0,0)--(C);
           \draw[line width=1pt,blue] (0,0)--(E);
           \draw[line width=1pt,blue] (0,0)--(F);
           \draw[line width=1pt,green] (0,0)--(G);
  \end{scope}
        \begin{scope}[xshift=4cm,font=\footnotesize]
         \node[node, fill=black,label=left:{$[-1,1]$}] (A) at (-1,1) {};
			\node[node, fill=black,label=right:{$[1,-1]$}] (B) at (1,-1) {};
            \node[node, fill=black,label=above:{$[0,1]$}] (C) at (0,1) {};
            \node[node, fill=black,label=right:{$[1,0]$}] (E) at (1,0) {};
            \node[node, fill=black,label=left:{$[-1,0]$}] (F) at (-1,0) {};
            \node[node, fill=black,label=below:{$[0,-1]$}] (G) at (0,-1) {};
           \draw[line width=1pt,red] (0,0)--(A);
           \draw[line width=1pt,blue] (0,0)--(B);
           \draw[line width=1pt,blue] (0,0)--(C);
           \draw[line width=1pt,red] (0,0)--(E);
           \draw[line width=1pt,blue] (0,0)--(F);
           \draw[line width=1pt,red] (0,0)--(G);
        \end{scope}
        \end{tikzpicture} \,.
\end{equation}
To each subweb, we associate a $\U(1)$ algebra. To determine the bifundamental matter, we need to compute the stable intersection for any pair of $(p,q)$-5-branes in the web. By computing the stable intersection for all the subwebs, we always get a contribution $1$. This means that the subweb on the left leads to a magnetic quiver given by
\begin{equation}
    \begin{tikzpicture}[baseline=0,font=\footnotesize]
        \node (A0) at (0,0) {};
        \node[node, label=above:{$1$}] (A1) [above=4mm of A0] {};
        \node[node, label=below:{$1$}] (A2) [below left=6mm of A0] {};
        \node[node, label=below:{$1$}] (A3) [below right=6mm of A0] {};
        \draw (A1) -- (A2) -- (A3) -- (A1);
        \node (Z2) [above right=5mm of A3] {$/\U(1)\,.$};
    \end{tikzpicture}
\end{equation}
On the other hand, for the subweb on the right, we need to sum over all stable intersections, leading to a contribution to the bifundamental matter equal to 2, and the magnetic quiver is 
\begin{equation}
     \begin{tikzpicture}[baseline=0,font=\footnotesize]
        \node[node, label=below:{$1$}] (A1) {};
        \node[node, label=below:{$1$}] (A2) [right=8mm of A1] {};
        \draw[double distance=3pt] (A1) -- (A2);
        \node (Z2) [right=6mm of A2] {$/\U(1)\,,$};
    \end{tikzpicture}
\end{equation}
One can compute the magnetic quivers for all the $E_n$ theories analogously, and we report them in Table \ref{tbl:Endata}.

    \renewcommand{\arraystretch}{1.6}
    \begingroup
    \setlength{\tabcolsep}{3pt}
 \begin{longtable}{c|c|c|c}
         \label{tbl:Endata}   Theory  & GTP & Brane Web   & Magnetic Quiver  \\
            \hhline{=|=|=|=}
  \endfirsthead
   Theory  & GTP & Brane Web   & Magnetic Quiver  \\\hhline{=|=|=|=} 
  \endhead
     $E_0$  & \begin{tikzpicture}[scale=0.6,baseline=-10]
			\node[node, fill=black] (O) at (0,0) {};
			\node[node, fill=black] (A) at (-1,0) {};
			\node[node, fill=black] (B) at (0,-1) {};
            \node[node, fill=black] (C) at (1,1) {};
           \draw[line width=1pt] (A)--(B)--(C) -- (A);
           \end{tikzpicture}  & \begin{tikzpicture}[scale=0.6,baseline=0]
			\node[node, fill=black] (A) at (-1,-1) {};
			\node[node, fill=black] (B) at (-1,2) {};
            \node[node, fill=black] (C) at (2,-1) {};
           \draw[line width=1pt] (0,0)--(A);
           \draw[line width=1pt] (0,0)--(B);
           \draw[line width=1pt] (0,0)--(C);
           \end{tikzpicture}  & $\emptyset$   \\
      \hline                                  
      $\widetilde{E}_1$  & \begin{tikzpicture}[scale=0.6,baseline=0]
			\node[node, fill=black] (O) at (0,0) {};
			\node[node, fill=black] (A) at (-1,0) {};
			\node[node, fill=black] (B) at (0,-1) {};
            \node[node, fill=black] (C) at (1,1) {};
             \node[node, fill=black] (D) at (0,1) {};
           \draw[line width=1pt] (A)--(B)--(C) -- (D) -- (A);
           \end{tikzpicture}  & 
           \begin{tikzpicture}[scale=0.6,baseline=0]
			\node[node, fill=black] (A) at (-1,1) {};
			\node[node, fill=black] (B) at (0,1) {};
            \node[node, fill=black] (C) at (2,-1) {};
            \node[node, fill=black] (D) at (-1,-1) {};
           \draw[line width=1pt] (0,0)--(A);
           \draw[line width=1pt] (0,0)--(B);
           \draw[line width=1pt] (0,0)--(C);
           \draw[line width=1pt] (0,0)--(D);
           \end{tikzpicture}
           & $\emptyset$   \\
      \hline                 
       $E_1$  & \begin{tikzpicture}[scale=0.6,baseline=0]
			\node[node, fill=black] (O) at (0,0) {};
			\node[node, fill=black] (A) at (-1,0) {};
			\node[node, fill=black] (B) at (0,-1) {};
            \node[node, fill=black] (C) at (1,0) {};
             \node[node, fill=black] (D) at (0,1) {};
           \draw[line width=1pt] (A)--(B)--(C) -- (D) -- (A);
           \end{tikzpicture} & \begin{tikzpicture}[scale=0.6,baseline=0]
			\node[node, fill=black] (A) at (-1,1) {};
			\node[node, fill=black] (B) at (1,-1) {};
            \node[node, fill=black] (C) at (1,1) {};
            \node[node, fill=black] (D) at (-1,-1) {};
           \draw[line width=1pt] (0,0)--(A);
           \draw[line width=1pt] (0,0)--(B);
           \draw[line width=1pt] (0,0)--(C);
           \draw[line width=1pt] (0,0)--(D);
           \end{tikzpicture} &   \begin{tikzpicture}[baseline=-5,font=\footnotesize]
        \node[node, label=below:{$1$}] (A1) {};
        \node[node, label=below:{$1$}] (A2) [right=6mm of A1] {};
        \draw[double distance=3pt] (A1) -- (A2);
    \end{tikzpicture}  \\
      \hline                 
       $E_2$  & \begin{tikzpicture}[scale=0.6,baseline=0]
			\node[node, fill=black] (O) at (0,0) {};
			\node[node, fill=black] (A) at (-1,0) {};
			\node[node, fill=black] (B) at (0,-1) {};
            \node[node, fill=black] (C) at (1,0) {};
             \node[node, fill=black] (D) at (0,1) {};
             \node[node, fill=black] (E) at (1,1) {};
           \draw[line width=1pt] (A)--(B)--(C) -- (E) -- (D) -- (A);
           \end{tikzpicture} & \begin{tikzpicture}[scale=0.6,baseline=0]
			\node[node, fill=black] (A) at (-1,1) {};
			\node[node, fill=black] (B) at (1,-1) {};
            \node[node, fill=black] (C) at (0,1) {};
            \node[node, fill=black] (D) at (-1,-1) {};
            \node[node, fill=black] (E) at (1,0) {};
           \draw[line width=1pt] (0,0)--(A);
           \draw[line width=1pt] (0,0)--(B);
           \draw[line width=1pt] (0,0)--(C);
           \draw[line width=1pt] (0,0)--(D);
           \draw[line width=1pt] (0,0)--(E);
           \end{tikzpicture} &  \begin{tikzpicture}[baseline=0,font=\footnotesize]
        \node[node, label=below:{$1$}] (A1) {};
        \node[node, label=below:{$1$}] (A2) [right=6mm of A1] {};
        \draw[double distance=3pt] (A1) -- (A2);
    \end{tikzpicture}  \\
      \hline                 
       $E_3$  & \begin{tikzpicture}[scale=0.6,baseline=0]
			\node[node, fill=black] (O) at (0,0) {};
			\node[node, fill=black] (A) at (-1,0) {};
			\node[node, fill=black] (B) at (0,-1) {};
            \node[node, fill=black] (C) at (1,0) {};
             \node[node, fill=black] (D) at (0,1) {};
             \node[node, fill=black] (E) at (1,1) {};
             \node[node, fill=black] (F) at (-1,-1) {};
           \draw[line width=1pt] (A)--(F) -- (B)--(C) -- (E) -- (D) -- (A);
           \end{tikzpicture} & \begin{tikzpicture}[scale=0.6,baseline=0]
			\node[node, fill=black] (A) at (-1,1) {};
			\node[node, fill=black] (B) at (1,-1) {};
            \node[node, fill=black] (C) at (0,1) {};
            \node[node, fill=black] (E) at (1,0) {};
            \node[node, fill=black] (F) at (-1,0) {};
            \node[node, fill=black] (G) at (0,-1) {};
           \draw[line width=1pt] (0,0)--(A);
           \draw[line width=1pt] (0,0)--(B);
           \draw[line width=1pt] (0,0)--(C);
           \draw[line width=1pt] (0,0)--(E);
           \draw[line width=1pt] (0,0)--(F);
           \draw[line width=1pt] (0,0)--(G);
           \end{tikzpicture} &   \begin{tikzpicture}[baseline=0,font=\footnotesize]
        \node (A0) at (0,0) {};
        \node[node, label=above:{$1$}] (A1) [above=3mm of A0] {};
        \node[node, label=below:{$1$}] (A2) [below left=3.464mm of A0] {};
        \node[node, label=below:{$1$}] (A3) [below right=3.464mm of A0] {};
        \draw (A1) -- (A2) -- (A3) -- (A1);
    \end{tikzpicture} $\quad\cup\quad$  \begin{tikzpicture}[baseline=0,font=\footnotesize]
        \node[node, label=below:{$1$}] (A1) {};
        \node[node, label=below:{$1$}] (A2) [right=6mm of A1] {};
        \draw[double distance=3pt] (A1) -- (A2);
    \end{tikzpicture} \\
      \hline                 
       $E_4$  & \begin{tikzpicture}[scale=0.6,baseline=0]
			\node[node, fill=black] (O) at (0,0) {};
			\node[node, fill=black] (A) at (-1,0) {};
			\node[node, fill=black] (B) at (0,-1) {};
            \node[node, fill=black] (C) at (1,0) {};
             \node[node, fill=black] (D) at (0,1) {};
             \node[node, fill=black] (E) at (1,1) {};
             \node[node, fill=black] (F) at (-1,-1) {};
              \node[node, fill=black] (G) at (1,-1) {};
           \draw[line width=1pt] (A)--(F) -- (B) -- (G) --(C) -- (E) -- (D) -- (A);
           \end{tikzpicture} & \begin{tikzpicture}[scale=0.6,baseline=-10]
			\coordinate (O) at (0,0);
            \coordinate (O2) at (-0.07,0.07);
    \node[node, fill=black] (TL) at (-1,1) {};   
    \node[node, fill=black] (T)  at (0.07,1) {};    
    \node[node, fill=black] (L)  at (-1,-0.07) {};   
    \node[node, fill=black] (R)  at (1,0) {};    
    \node[node, fill=black] (B)  at (0,-1) {};   
    \node[node, fill=black] (RR) at (2,0) {};    
    \node[node, fill=black] (BB) at (0,-2) {};   
    \draw[line width=1pt] (TL) -- (O2);
    \draw[line width=1pt] (O2) -- (1,0.07);
    \draw[line width=1pt] (O2) -- (-0.07,-1);
    \draw[line width=1pt] (0.07,1) -- (0.07,-1);
    \draw[line width=1pt] (-1,-0.07) -- (1,-0.07);
    \draw[line width=1pt] (R) -- (RR);
    \draw[line width=1pt] (B) -- (BB);
           \end{tikzpicture} & \begin{tikzpicture}[baseline=0,font=\footnotesize]
        \def\radius{7mm};
        \node[node, label=above:{$1$}] (A1) at (90:\radius) {}; 
        \node[node, label=above:{$1$}] (A2) at (162:\radius) {}; 
        \node[node, label=below:{$1$}] (A3) at (234:\radius) {};
        \node[node, label=below:{$1$}] (A4) at (306:\radius) {};
        \node[node, label=above:{$1$}] (A5) at (18:\radius) {};
        \draw (A1) -- (A2) -- (A3) -- (A4) -- (A5) -- (A1);
    \end{tikzpicture}    \\
      \hline                 
       $E_5$  & \begin{tikzpicture}[scale=0.6,baseline=0]
			\node[node, fill=black] (O) at (0,0) {};
			\node[node, fill=black] (A) at (-1,0) {};
			\node[node, fill=black] (B) at (0,-1) {};
            \node[node, fill=black] (C) at (1,0) {};
             \node[node, fill=black] (D) at (0,1) {};
             \node[node, fill=black] (E) at (1,1) {};
             \node[node, fill=black] (F) at (-1,-1) {};
              \node[node, fill=black] (G) at (1,-1) {};
               \node[node, fill=black] (H) at (-1,1) {};
           \draw[line width=1pt] (A)--(F) -- (B) -- (G) --(C) -- (E) -- (D) -- (H) -- (A);
           \end{tikzpicture}  & 
           \begin{tikzpicture}[scale=0.6,baseline=0]
    \node[node, fill=black] (LL) at (-2,0) {};   
    \node[node, fill=black] (L)  at (-1,0) {};    
    \node[node, fill=black] (T)  at (0,1) {};   
    \node[node, fill=black] (TT)  at (0,2) {};   
    \node[node, fill=black] (R)  at (1,0) {};    
    \node[node, fill=black] (RR) at (2,0) {};    
    \node[node, fill=black] (B)  at (0,-1) {};   
    \node[node, fill=black] (BB) at (0,-2) {};   
    \draw[line width=1pt] (LL) -- (L);
    \draw[line width=1pt] (TT) -- (T);
    \draw[line width=1pt] (RR) -- (R);
    \draw[line width=1pt] (BB) -- (B);
    \draw[line width=1pt] (-0.07,1) -- (-0.07,-1);
    \draw[line width=1pt] (0.07,1) -- (0.07,-1);
    \draw[line width=1pt] (-1,0.07) -- (1,0.07);
    \draw[line width=1pt] (-1,-0.07) -- (1,-0.07);
           \end{tikzpicture}
           & \begin{tikzpicture}[baseline=0,font=\footnotesize]
        \node[node, label=below:{$2$}] (A2) {};
        \node[node, label=below:{$2$}] (A3) [right=6mm of A2] {};
        \node[node, label=left:{$1$}] (Al1) [above left=6mm of A2] {};
        \node[node, label=left:{$1$}] (Al2) [below left =6mm of A2]{};
        \node[node, label=right:{$1$}] (Ar1) [above right=6mm of A3] {};
        \node[node, label=right:{$1$}] (Ar2) [below right =6mm of A3]{};
        \draw (Al1) -- (A2) -- (A3) -- (Ar1);
        \draw   (Al2) -- (A2) ;
        \draw   (Ar2) -- (A3) ;
    \end{tikzpicture}   \\
      \hline                 
       $E_6$  & \begin{tikzpicture}[scale=0.6,baseline=20]
			\node[node, fill=black] (O) at (0,0) {};
			\node[node, fill=black] (A) at (1,0) {};
            \node[node, fill=black] (B) at (2,0) {};
            \node[node, fill=black] (C) at (3,0) {};
            \node[node, fill=black] (D) at (2,1) {};
            \node[node, fill=black] (E) at (1,2) {};
            \node[node, fill=black] (F) at (0,3) {};
            \node[node, fill=black] (G) at (0,2) {};
            \node[node, fill=black] (H) at (0,1) {};
            \node[node, fill=black] (I) at (1,1) {};
           \draw[line width=1pt] (O)--(C) -- (F) -- (O);
           \end{tikzpicture} & \begin{tikzpicture}[scale=0.6,baseline=0]
    \node[node, fill=black] (LLL) at (-3,0) {};   
    \node[node, fill=black] (LL) at (-2,0) {};   
    \node[node, fill=black] (L)  at (-1,0) {};    
    \node[node, fill=black] (T)  at (1,1) {};   
    \node[node, fill=black] (TT)  at (2,2) {};   
    \node[node, fill=black] (TTT)  at (3,3) {};   
    \node[node, fill=black] (B)  at (0,-1) {};   
    \node[node, fill=black] (BB) at (0,-2) {};   
    \node[node, fill=black] (BBB) at (0,-3) {};   
    \draw[line width=1pt] (LLL) -- (LL);
    \draw[line width=1pt] (TTT) -- (TT);
    \draw[line width=1pt] (BBB) -- (BB);
    \draw[line width=1pt] (-2,0.07) -- (-1,0.07);
    \draw[line width=1pt] (-2,-0.07) -- (-1,-0.07);
    \draw[line width=1pt] (0.07,-2) -- (0.07,-1);
    \draw[line width=1pt] (-0.07,-2) -- (-0.07,-1);
    \draw[line width=1pt] (1.93,2)  -- (1,1.07);
    \draw[line width=1pt] (2.07,2) -- (1,0.93);
    \draw[line width=0.6pt] (L) -- (0,0) -- (T);
    \draw[line width=0.6pt] (B) -- (0,0);
    \draw[line width=0.6pt] (-1,0.07) -- (0,0.07) -- (1,1.07);
    \draw[line width=0.6pt] (0.07,-1) -- (0.07,0) -- (1,0.93);
    \draw[line width=0.6pt] (-1,-0.07) -- (-0.07,-0.07) -- (-0.07,-1);
           \end{tikzpicture}
           &  \begin{tikzpicture}[baseline=10,font=\footnotesize]
        \node[node, label=below:{$3$}] (A2) {};
        \node[node, label=below:{$2$}] (Al1) [left=6mm of A2] {};
        \node[node, label=below:{$2$}] (Ar1) [right =6mm of A2]{};
        \node[node, label=below:{$1$}] (Al2) [left=6mm of Al1] {};
        \node[node, label=below:{$1$}] (Ar2) [right =6mm of Ar1]{};
        \node[node, label=right:{$2$}] (A3) [above=4mm of A2] {};
        \node[node, label=right:{$1$}] (A3a) [above=4mm of A3] {};
        \draw (A2) -- (A3) -- (A3a);
        \draw   (Al2) -- (Al1) -- (A2) -- (Ar1) -- (Ar2) ;
    \end{tikzpicture}  \\
      \hline                 
       $E_7$  & \begin{tikzpicture}[scale=0.6,baseline=30]
			\node[node, fill=black] (O) at (0,0) {};
			\node[node, fill=black] (A) at (1,0) {};
            \node[node, fill=black] (B) at (2,0) {};
            \node[node, fill=black] (C) at (3,0) {};
            \node[node, fill=black] (D) at (4,0) {};
            \node[node, fill=black] (F) at (2,2) {};
            \node[node, fill=black] (H) at (0,4) {};
            \node[node, fill=black] (I) at (0,3) {};
            \node[node, fill=black] (L) at (0,2) {};
            \node[node, fill=black] (M) at (0,1) {};
            \node[node, fill=black] (N) at (1,1) {};
            \node[node, fill=black] (P) at (2,1) {};
            \node[node, fill=black] (Q) at (1,2) {};
           \draw[line width=1pt] (O)--(D) -- (H) -- (O);
           \node[node, fill=white] (E) at (3,1) {};
           \node[node, fill=white] (G) at (1,3) {};
           \end{tikzpicture} & \cite[Figure 10]{Benini:2009gi} &  \begin{tikzpicture}[baseline=0,font=\footnotesize]
        \node[node, label=below:{$4$}] (A2) {};
        \node[node, label=below:{$3$}] (Al1) [left=6mm of A2] {};
        \node[node, label=below:{$3$}] (Ar1) [right =6mm of A2]{};
        \node[node, label=below:{$2$}] (Al2) [left=6mm of Al1] {};
        \node[node, label=below:{$2$}] (Ar2) [right =6mm of Ar1]{};
        \node[node, label=below:{$1$}] (Al3) [left=6mm of Al2] {};
        \node[node, label=below:{$1$}] (Ar3) [right =6mm of Ar2]{};
        \node[node, label=right:{$2$}] (A3) [above=4mm of A2] {};
        \draw (A2) -- (A3);
        \draw  (Al3) -- (Al2) -- (Al1) -- (A2) -- (Ar1) -- (Ar2) -- (Ar3) ;
    \end{tikzpicture}  \\
      \hline                 
       $E_8$  & \begin{tikzpicture}[scale=0.6,baseline=40,font=\footnotesize]
			\node[node, fill=black] (O) at (0,0) {};
			\node[node, fill=black] (A) at (1,0) {};
            \node[node, fill=black] (B) at (2,0) {};
            \node[node, fill=black] (C) at (3,0) {};
            \node[node, fill=black] (D) at (4,0) {};
            \node[node, fill=black] (F) at (5,0) {};
            \node[node, fill=black] (H) at (6,0) {};
            \node[node, fill=black] (I) at (4,2) {};
            \node[node, fill=black] (L) at (2,4) {};
            \node[node, fill=black] (M) at (0,6) {};
            \node[node, fill=black] (N) at (0,3) {};
           \draw[line width=1pt] (O)--(H) -- (M) -- (O);
           \node[node, fill=white] at (5,1) {};
           \node[node, fill=white] at (3,3) {};
           \node[node, fill=white] at (1,5) {};
           \node[node, fill=white] at (0,5) {};
           \node[node, fill=white] at (0,4) {};
           \node[node, fill=white] at (0,2) {};
           \node[node, fill=white] at (0,1) {};
           \node[node, fill=white] at (1,1) {};
           \node[node, fill=black] at (1,2) {};
           \node[node, fill=white] at (1,3) {};
           \node[node, fill=white] at (1,4) {};
           \node[node, fill=black] at (2,1) {};
           \node[node, fill=black] at (2,2) {};
           \node[node, fill=black] at (2,3) {};
           \node[node, fill=black] at (3,1) {};
           \node[node, fill=black] at (3,2) {};
           \node[node, fill=black] at (4,1) {};
           \end{tikzpicture} & \cite[Figure 12]{Benini:2009gi} &  
   \begin{tikzpicture}[baseline=0,font=\footnotesize]
        \node[node, label=below:{$6$}] (A2) {};
        \node[node, label=below:{$5$}] (Al1) [left=6mm of A2] {};
        \node[node, label=below:{$4$}] (Ar1) [right =6mm of A2]{};
        \node[node, label=below:{$4$}] (Al2) [left=6mm of Al1] {};
        \node[node, label=below:{$2$}] (Ar2) [right =6mm of Ar1]{};
        \node[node, label=below:{$3$}] (Al3) [left=6mm of Al2] {};
        \node[node, label=below:{$2$}] (Al4) [left =6mm of Al3]{};
        \node[node, label=below:{$1$}] (Al5) [left =6mm of Al4]{};
        \node[node, label=right:{$3$}] (A3) [above=4mm of A2] {};
        \draw (A2) -- (A3);
        \draw  (Al5) -- (Al4) -- (Al3) -- (Al2) -- (Al1) -- (A2) -- (Ar1) -- (Ar2);
    \end{tikzpicture}
\\           
     \caption{We show generalized toric polytope (GTP), brane web, and corresponding magnetic quiver for the (reduced) Higgs branch of the rank-one $E_n$ theories.}
    \label{tab:magquiverspqwebs}
\end{longtable}
\endgroup

\subsection{Flavor Symmetry Automorphisms and Discrete Gauging}\label{sec:dg}

Let us suppose that an eight-supercharge theory has a continuous semisimple flavor
symmetry algebra, $\mathfrak{g}$, and, in addition, a finite global
symmetry, $H$. The induced action of $H$ on the flavor algebra is
specified by a homomorphism
\begin{equation}
    \rho:\, H \rightarrow \operatorname{Aut}(\mathfrak{g})\fstop
\end{equation}
For our purposes, $\rho$ is defined only up to simultaneous conjugation
in $\operatorname{Aut}(\mathfrak{g})$.\footnote{That is, the
homomorphisms $\rho$ and $\rho'$ are regarded as equivalent if there
exists a fixed $a\in\operatorname{Aut}(\mathfrak{g})$ such that
$\rho'(h)=a\rho(h)a^{-1}$ for every $h\in H$.}
We assume for the purposes of this discussion that $H$ is
non-anomalous and can be gauged.
The existence of the flavor symmetry $\mathfrak{g}$ implies the
existence of moment-map operators transforming in the adjoint
representation of $\mathfrak{g}$. Upon gauging $H$, the non-invariant
components of these moment maps are projected out of the local operator
algebra. Consequently, the continuous flavor algebra inherited from
$\mathfrak{g}$ is the fixed-point subalgebra
\begin{equation}\label{eqn:gHdef}
    \mathfrak{g}^{\rho(H)}
    \,=\,
    \left\{\,
        X\in\mathfrak{g}
        \,\middle|\,
        \rho(h)X=X
        \quad\text{for every }\,h\in H\,
    \right\}\fstop
\end{equation}
To determine this fixed-point subalgebra, we can define the averaging projector
\begin{equation}\label{eqn:projector}
    P_H = \frac{1}{|H|} \sum_{h \in H} \rho(h)\coma
\end{equation}
which is a linear projector $\mathfrak{g} \rightarrow \mathfrak{g}$ whose image is precisely the fixed-point subalgebra:
\begin{equation}
    \mathfrak{g}^{\rho(H)} = \operatorname{Im}(P_H)\fstop
\end{equation}
The Lie algebra structure on $\mathfrak{g}^{\rho(H)}$ is obtained by restricting the bracket of $\mathfrak{g}$ to this invariant subspace. Thus, one may compute $\mathfrak{g}^{\rho(H)}$ by representing the action of $\rho(H)$ on $\mathfrak{g}$, determining the image of $P_H$, and identifying the resulting Lie algebra from its restricted bracket. For cyclic groups generated by a finite-order automorphism, the same fixed-point algebra can often be determined more efficiently using the corresponding untwisted or twisted Kac diagram \cite{MR1104219,MR2674853}.

Briefly, Kac diagrams classify homomorphisms
$\ZZ_N \to\operatorname{Aut}(\mathfrak g)$ up to conjugacy. After fixing an outer class of order $\ell$, one assigns labels $s_i$ to the nodes of the corresponding ($\ell$-twisted) affine Dynkin diagram $X^{(\ell)}$. The untwisted diagram $X^{(1)}$ describes inner automorphisms, while nontrivial outer classes are described by the twisted affine diagrams. If $a_i$ are the affine marks\footnote{We recall that the affine marks $a_i$ for $i > 0$ are the coefficients of the expansion of the highest root $\theta$ on the simple roots: $\theta = \sum_{i=1}^{\text{rank}(\mathfrak g)}a_i\alpha_i$, while $a_0 \equiv 1$.} of $X^{(\ell)}$, a (not necessarily faithful) $\ZZ_N$ homomorphism is classified by non-negative integers $s_i$ (called \emph{Kac labels}) such that:
\begin{align}\label{eq:Kac_label_def}
  N=\ell\sum_i a_i s_i  \,,
\end{align}
and the (semisimple part of the) fixed-point subalgebra is read from the zero-labeled subdiagram. In the absence of accidental symmetry enhancement, this is the (semisimple part of the) continuous flavor algebra of the discretely-gauged theory.

We now specialize to a particular class of discrete symmetries
arising from automorphisms of affine Dynkin diagrams. Let $\widetilde{\mathcal D}_{\mathfrak g}$ denote the untwisted affine
Dynkin diagram associated to $\mathfrak{g}$, and denote the group of automorphisms of this affine Dynkin diagram as $\operatorname{Aut}(\widetilde{\mathcal D}_{\mathfrak g})$. In \cite{MR3000483}, it was shown that, up to simultaneous
conjugation in $\operatorname{Aut}(\mathfrak g)$, there exists a homomorphism
\begin{equation}\label{eqn:BIGHOMO}
    \widehat\rho:\,
    \operatorname{Aut}(\widetilde{\mathcal D}_{\mathfrak g})
    \rightarrow
    \operatorname{Aut}(\mathfrak g)\fstop
\end{equation}
See \cite{MR3000483} for the explicit homomorphisms. Thus, every subgroup $H\subseteq \operatorname{Aut}(\widetilde{\mathcal D}_{\mathfrak g})$ determines
an action on the finite-dimensional flavor algebra through the above homomorphism restricted to the subgroup:
\begin{equation}
    \rho_H
    =
    \left.\widehat\rho\,\right|_H:\,
    H\rightarrow\operatorname{Aut}(\mathfrak g)\fstop
\end{equation}
Often we refer to $H$, the subgroup of the automorphism group of the quiver, as being either outer or inner, depending on whether its image under the homomorphism $\rho_H$ has non-trivial embedding in the outer automorphism group of $\mathfrak{g}$, or not, respectively. 
Supposing that such an $H$ is a global symmetry of the SCFT in question, we can discuss its gauging (modulo questions of anomalies, which will be discussed in the next subsection). We note that the freedom of simultaneous conjugation in the definition of $\widehat{\rho}$ does not change the isomorphism class of the fixed-point algebra, $\mathfrak{g}^{\rho_H(H)}$, only the representative embedding, and thus is irrelevant for the surviving global symmetry of the discretely-gauged theory.

\subsubsection{Example: the $E_6$ Theory}

As an example, we consider the rank-one $E_6$ SCFT. The unitary magnetic quiver for the Higgs branch takes the form of the affine $E_6$ Dynkin diagram as in Table \ref{tab:magquiverspqwebs}.
The automorphisms of this quiver are the automorphisms of the untwisted affine $\mathfrak{e}_6$ Dynkin diagram:
\begin{equation}
    \operatorname{Aut}(\widetilde{\mathcal D}_{\mathfrak{e}_6}) \cong S_3 \cong \mathbb{Z}_3 \rtimes \mathbb{Z}_2\fstop
\end{equation}
Up to conjugacy, the nontrivial subgroups of
$S_3$ are
\begin{equation}
    \mathbb Z_2\coma \mathbb Z_3\coma S_3\fstop
\end{equation}
We have depicted the actions of the $\mathbb{Z}_2$ and $\mathbb{Z}_3$ on the quiver in equations \eqref{eq:E6-wrZ2} and \eqref{eq:E6wrZ3-quiver}, respectively. Under the homomorphism $\widehat\rho$, the normal subgroup
$\mathbb Z_3\triangleleft S_3$ acts by an inner automorphism of
$\mathfrak e_6$, whereas each of the three elements of order two acts
by an outer automorphism. Since the three $\mathbb Z_2$ subgroups are
conjugate in $S_3$, they determine conjugate actions on
$\mathfrak e_6$ and hence isomorphic fixed-point subalgebras.

We briefly summarize the procedure to compute the fixed-point subalgebras directly here; the technical details of the calculation are in \cite{MR3000483}, to which we also refer for any not-explicitly-defined terminology below. Write
\begin{equation}
S_3=\langle r,s\mid r^3=s^2=1,\ srs=r^{-1}\rangle\coma
\end{equation}
where $r$ is an affine rotation and $s$ is a reflection. In \cite[Table 19($3_b$)]{MR3000483}, it was shown that the affine rotation induces an automorphism $\widehat\rho(r)$ on $\mathfrak{e}_6$, which has the untwisted Kac diagram with label one on the three terminal nodes, i.e.,
\begin{equation}
    \begin{tikzpicture}[baseline=0,font=\footnotesize]
        \node[node, label=below:{$0$}] (A2) {};
        \node[node, label=below:{$0$}] (Al1) [left=6mm of A2] {};
        \node[node, label=below:{$1$}] (Al2) [left=6mm of Al1] {};
        \node[node, label=below:{$0$}] (Ar1) [right=6mm of A2] {};
        \node[node, label=below:{$1$}] (Ar2) [right=6mm of Ar1] {};
        \node[node, label=right:{$0$}] (A3) [above=4mm of A2] {};
        \node[node, label=right:{$1$}] (A3a) [above=4mm of A3] {};
        \draw (A2) -- (A3) -- (A3a);
        \draw (Al2) -- (Al1) -- (A2) -- (Ar1) -- (Ar2);
    \end{tikzpicture} \,.
\end{equation}

The zero-labeled subdiagram is the Dynkin diagram of $\mathfrak{so}(8)$, and hence we have determined the semisimple part of the fixed-point algebra. Moreover, because $\widehat{\rho}(r)$ is a finite-order, Abelian inner automorphism, it is conjugate to a toral automorphism and therefore fixes a Cartan subalgebra pointwise; thus, the fixed-point subalgebra has rank six. It follows that
\begin{equation}\label{eqn:E6Z3}
\mathfrak e_6^{\rho_{\mathbb{Z}_3}(\mathbb{Z}_3)}
=
\mathfrak{so}(8)\oplus\mathfrak u(1)^{\oplus2}\fstop
\end{equation}
For $s$, we may choose the reflection fixing the affine node. Its restriction to the finite $\mathfrak{e}_6$ diagram is the nontrivial diagram involution; that is, an outer automorphism. Thus, we must consider Kac diagrams for the twisted affine Dynkin diagram of $\mathfrak{e}_6$; there are two Kac diagrams capturing $\mathbb{Z}_2$ embeddings, and the relevant one has only the affine vertex nonzero, and hence the zero-labeled subdiagram, which is the same as the folding of $\mathfrak{e}_6$, is $\mathfrak{f}_4$:\footnote{The other $\mathbb{Z}_2$-homomorphism into $\operatorname{Aut}(\mathfrak{e}_6)$, with nontrivial support in the outer-automorphism subgroup, has fixed-point subalgebra $\mathfrak{usp}(8)$. This embedded $\ZZ_2$ is not relevant for our discussion, as it is not induced by an automorphism of the affine $\mathfrak{e}_6$ Dynkin diagram under equation \eqref{eqn:BIGHOMO}.}
\begin{equation}\label{eqn:E6Z2}
\mathfrak e_6^{\rho_{\mathbb{Z}_2}(\mathbb{Z}_2)}
=
\mathfrak f_4\fstop
\end{equation}
There is no single Kac diagram associated to the non-cyclic group $S_3$. Instead, one takes the simultaneous fixed points. Indeed, the projector, as in equation \eqref{eqn:projector}, becomes
\begin{equation}
P_{S_3}
=
\frac{1}{2}\bigl(1+\widehat\rho(s)\bigr)P_{\langle r\rangle} \qquad \Rightarrow
\qquad
\mathfrak e_6^{\widehat\rho(S_3)}
=
\left(
\mathfrak e_6^{\langle\widehat\rho(r)\rangle}
\right)^{\langle\widehat\rho(s)\rangle}\coma
\end{equation}
since $s$ normalizes the subgroup generated by $r$. For the compatible representatives determined by $\widehat\rho$, the $\mathfrak{so}(8)$ factor above is contained in the $\mathfrak{f}_4$ fixed by $s$, while $s$ acts without invariant vectors on the $\mathfrak{u}(1)^{\oplus 2}$. Thus, only the $\mathfrak{so}(8)$ factor is invariant:
\begin{equation}\label{eqn:E6S3}
\mathfrak e_6^{\rho_{S_3}(S_3)}
=
\mathfrak{so}(8)\fstop
\end{equation}

In this example, these computations can be repackaged a posteriori in terms of a decomposition of $\mathfrak{e}_6$ under the commuting actions of $\mathfrak{so}(8)$ and $S_3$. Since $S_3$ fixes the $\mathfrak{so}(8)$ subalgebra pointwise, its action on $\mathfrak{e}_6$ commutes with the restricted adjoint action of $\mathfrak{so}(8)$. Explicitly,
\begin{equation}
    \widehat{\rho}(h) \circ \operatorname{ad}_X = \operatorname{ad}_X \circ\, \widehat{\rho}(h) \qquad \text{ for all } \quad X \in \mathfrak{so}(8) \,,\,\, h \in S_3\fstop
\end{equation}
Consequently, the adjoint representation of $\mathfrak{e}_6$ carries commuting actions of
$\mathfrak{so}(8)$ and $S_3$, under which it decomposes as
\begin{equation}
\label{eq:branchinge6sec2}
    \bm{78} \rightarrow (\bm{28}, \bm{1}) \oplus (\bm{8_v}, \bm{2}) \oplus (\bm{8_s}, \bm{2}) \oplus (\bm{8_c}, \bm{2}) \oplus 2(\bm{1}, \bm{1_\mathrm{sgn}})\fstop
\end{equation}
Here, $\bm{2}$ denotes the
two-dimensional irreducible representation of $S_3$, and
$\bm{1_\mathrm{sgn}}$ denotes the one-dimensional sign representation. Under the $\ZZ_3$ subgroup, the $\bm{2}$ contains no invariant
vector, whereas the sign representation restricts trivially; therefore, all eight-dimensional representations of $\mathfrak{so}(8)$ are projected out in the discrete gauging, and the last summand of equation \eqref{eq:branchinge6sec2} contributes the Abelian factor in equation \eqref{eqn:E6Z3}. Similarly, a $\mathbb{Z}_2$ reflection in $S_3$ has a one-dimensional invariant subspace in
$\bm{2}$ and acts as $-1$ on $\bm{1_\mathrm{sgn}}$.\footnote{That is, considering such invariant subspaces, we retain $(\bm{28}, \bm{1}) \oplus (\bm{8_v}, \bm{1}) \oplus (\bm{8_s}, \bm{1}) \oplus (\bm{8_c}, \bm{1})$ as the adjoint of the fixed-point subalgebra, which is exactly the branching of $\mathfrak f_4$ to its $\mathfrak{so}(8)$ subalgebra.}
Therefore, by recombining the invariant moment-maps into the adjoint representations of enhanced algebras, we reproduce the fixed-point subalgebras in equations \eqref{eqn:E6Z3}, \eqref{eqn:E6Z2}, and \eqref{eqn:E6S3}.

\subsection{Comments on 't Hooft anomalies}
\label{sec:tHooftanomalies}

In this section, we discuss the possible existence of 't Hooft anomalies of the rank-one $E_n$ theories involving the discrete symmetries that we want to gauge. We study both the pure and mixed-gravitational anomalies that could obstruct gauging, as well as mixed anomalies with other internal symmetries, which may lead to other symmetry structures in the discretely-gauged theory.

We consider a $d$-dimensional spin-QFT with a finite internal symmetry, $H$, which is both bosonic and independent of $(-1)^F$. The pure-$H$ anomalies and mixed-$H$-gravitational anomalies are encoded by a $(d+1)$-dimensional invertible field theory in the presence of a background $H$-bundle. For the finite internal symmetries considered here, the relevant six-dimensional spin-bordism groups are finite, and the corresponding anomalies are classified by their Pontryagin duals:
\begin{equation}
    \operatorname{Hom}\left(\Omega_{6}^\text{Spin}(BH), \text{U}(1)\right) \,.
\end{equation}
That is, a nontrivial anomaly corresponds to a nontrivial $\text{U}(1)$-valued spin-bordism invariant \cite{Freed:2016rqq,Davighi:2022icj,Yonekura:2016wuc,Garcia-Etxebarria:2018ajm}. The following spin-bordism groups have been computed explicitly in \cite{Davighi:2022icj}
\begin{equation}
\begin{aligned}
    \Omega_6^\text{Spin}(BH) &= 0 \quad & &\text{for} \quad H = \ZZ_2, \ZZ_3, \ZZ_4, \ZZ_5, S_3, D_{10} \,, \\
    \Omega_6^\text{Spin}(BH) &= \mathbb{Z}_2 \quad & &\text{for} \quad H =  \ZZ_2 \times \ZZ_2, D_{8} \,.
\end{aligned}
\end{equation}
These are all the groups $H$ that arise from the automorphisms of the magnetic quivers of the $E_n$ SCFTs.\footnote{For the $E_3$ SCFT, the moduli space is a union of two hyperk\"ahler cones, and the two factors of $S_3\times\mathbb Z_2$ act nontrivially on different cones. We assume that the anomaly class contains no term mixing these two factors so that anomaly freedom may be checked factor by factor.} Therefore, under the assumptions stated here, the only gaugings that we consider that \emph{may} have a pure-discrete or mixed-discrete-gravitational anomaly are the $\ZZ_2 \times \ZZ_2$ or $D_8$ gaugings of the $E_5$ SCFT, while for all other cases, there are no anomaly obstructions to the discrete gauging.

Further, we briefly discuss the existence of mixed anomalies between the discrete symmetry that we gauge and other internal symmetries. In particular, we would like to check that the discrete symmetry does not have a mixed anomaly with either supersymmetry or the R-symmetry, which is a sufficient condition to guarantee that the gauged theory has the same amount of SUSY. In what follows, we show explicitly that the mixed anomalies with the $\SU(2)_R$ R-symmetry discovered in \cite{Sacchi:2023omn} involve discrete internal symmetries that differ from the ones we consider in this work. The anomaly of \cite{Sacchi:2023omn} has the following inflow action
\begin{equation}\label{eq:anomaly_inflow}
S_{\mathrm{anom}} = \frac{4\pi i}{3} \int_{M_6} w_2(\mathcal{G}/\ZZ_n) \left( c_2(R) \bmod 3 \right)\coma
\end{equation}
where $\mathcal{G}/\ZZ_n$ is the global symmetry of the theories under consideration, and the Stiefel--Whitney class $w_2 \in H^2(M_6, \ZZ_n)$ represents the topological obstruction to lifting the $\mathcal{G}/\ZZ_n$-bundle to a $\mathcal{G}$-bundle over the spacetime $M_6$.\footnote{Sometimes the name `Stiefel--Whitney class' is reserved for the $\ZZ_2$-valued class that controls the obstruction to lifting SO bundles to Spin bundles; the more general version we consider here is instead called the `generalized Stiefel--Whitney class' or `Brauer class'.} This class can be written as the pullback $w_2 = f^* \widetilde{w}_2$ of the universal class $\widetilde{w}_2 \in H^2(B(\mathcal{G}/\ZZ_n),\ZZ_n)$ along the map $f: M_6 \to B(\mathcal{G}/\ZZ_n)$; this allows for the separation of the information of the particular configuration of the background gauge field bundle, which may depend on the spacetime $M_6$, and which is contained in the map $f$; from a universal part $\widetilde{w}_2$.

We want to gauge a discrete group $H\subset \mathcal{G}/\ZZ_n$. A sufficient condition to show there is no mixed anomaly involving $H$ is to show that the following pullback trivializes in cohomology:
\begin{equation}
\varphi^* \widetilde{w}_2(\mathcal{G}/\ZZ_n) \in H^2(BH, \ZZ_n)\,,
\end{equation}
where $\varphi: BH \to B(\mathcal{G}/\ZZ_n)$ is the embedding. As explained in Section \ref{sec:dg}, $\varphi$ is specified by a set of Kac labels $\{s_i\}$. From now on, we focus on the case in which $H$ is a cyclic group $\ZZ_N$, that is the only case in which the quiver automorphisms we want to gauge may have a mixed anomaly, according to \cite{Sacchi:2023omn}.  

Recall the definition of the Stiefel--Whitney classes: on triple overlaps, the transition functions of the background gauge bundle must satisfy a cocycle condition. For discrete groups, the background gauge fields are locally flat, and the transition functions are simply group elements. Consider such a transition function $g_{ij}\in H$, which satisfies $g_{ij} g_{jk} g_{ki}=1$; and consider the map $\eta:\mathcal{G}/\ZZ_n\to\mathcal{G}$ that lifts the transition functions to the covering group (a choice of $\eta$ is guaranteed to exist as a set-theoretic section). If we are able to prove that, as a group element, $\eta(g)$ still has order $N$ for $g$ any generator of $H$; then it follows that $\eta$ is a well defined group homomorphism from $\ZZ_N$ to $\ZZ_N$, and for any choice of open covering, $\eta(g_{ij})\eta(g_{jk})\eta(g_{ki})=\eta(g_{ij}g_{jk}g_{ki})=\eta(1)=1$. Consequently, the Stiefel--Whitney class vanishes; that is, $\varphi^* \widetilde{w}_2=0$.

Any element of the Cartan torus can be written in terms of the fundamental coweights $\omega_i^\vee$. To proceed, we write the group element $g\in \ZZ_N$ as
\begin{align}\label{eq:group_element_general}
    g=\exp \left(\frac{2\pi i}{N}\sum_i s_i \omega^\vee_i\right)\,,
\end{align}
where the Kac labels $s_i$ are non-negative integers, as a requirement that $g$ has order $N$, and they satisfy the constraint in equation \eqref{eq:Kac_label_def}. For the anomaly to vanish, we need to show that $g^N$ is the identity in $\mathcal{G}$, as opposed to a non-trivial element in $\ZZ_n$. There are two rank-one theories discussed in \cite{Sacchi:2023omn} that have a mixed anomaly with the R-symmetry that is relevant for us: $E_3$ and $E_6$. Let us consider them case by case.

\subsubsection{The $E_3$ Theory}

The $E_3$ theory possesses a flavor algebra $\mathfrak{e}_3 \simeq \su(3) \times \su(2)$. Following \cite{Apruzzi:2021vcu}, the global form of this symmetry is $E_3/\ZZ_6 \cong \text{PSU}(3) \times \SO(3)$, which admits potential mixed anomalies for discrete subgroups of its $\mathrm{PSU}(3)$ factor. Because the obstruction class $w_2(\mathrm{PSU}(3))$ has order three, its pullback to any $\ZZ_2$ subgroup lands in $H^2(B\ZZ_2, \ZZ_3) = \ZZ_{\gcd(2,3)} = 0$, trivializing the anomaly immediately for all the $\ZZ_2$-gaugings. For $\ZZ_3$ and $S_3$ discrete gaugings, however, the potential anomaly is determined by the $\ZZ_3$ subgroup that we now evaluate.

As explained above, Kac labels classify the order-three automorphisms of the $\su(3)$ component via its affine Dynkin diagram. We number the nodes of the affine $A_2$ diagram as $0, 1, 2$. Each node has a corresponding affine mark $a_i$, which in this case are all equal to $1$:
\begin{equation}
    \begin{tikzpicture}[baseline=0,font=\footnotesize]
        \node[node, label=below:{$1$}] (A1) at (0,0) {};
        \node[node, label=below:{$1$}] (A2) at (1,0) {};
        \node[node, label=above:{$1$}] (A0) at (0.5,0.866) {};
        \draw (A1) -- (A2) -- (A0) -- (A1);
    \end{tikzpicture} \,.
\end{equation}
A valid order-three subgroup requires $\sum_{i=0}^2 a_i s_i = 3$. The quiver automorphism symmetry we are interested in gauging is the one corresponding to Kac labels $s_0 = s_1 = s_2 = 1$. Having labeled the affine node by $0$, the nodes of the $A_2$ diagram are numbered $1$ and $2$ from left to right. Below, we highlight in {\color{red}{red}} the activated nodes:
\begin{equation}
    \begin{tikzpicture}[baseline=0,font=\footnotesize]
        \node[sonode, label=below:{$1$}] (A1) at (0,0) {};
        \node[sonode, label=below:{$2$}] (A2) at (1,0) {};
        \node[sonode, label=above:{$0$}] (A0) at (0.5,0.866) {};
        \draw (A1) -- (A2) -- (A0) -- (A1);
    \end{tikzpicture} \,.
\end{equation}
The group generator of the discrete symmetry comes from equation \eqref{eq:group_element_general},
\begin{equation}
    g = \exp_{\text{PSU}(3)} \left[ \frac{2\pi i}{3} (\omega_1^\vee + \omega_2^\vee) \right] \fstop 
\end{equation}
Consider its lift to the simply-connected $\SU(3)$ as 
\begin{equation} 
\eta(g) = \exp_{\SU(3)} \left[ \frac{2\pi i}{3} (\omega_1^\vee + \omega_2^\vee) \right] \fstop 
\end{equation} 
The question is whether $\eta(g)^3=1$ or $\eta(g)^3\in\ZZ_3=Z(\SU(3))$. Recall that for a Lie group, the center of the group coincides with the quotient of the coweight lattice by the coroot lattice, which is usually denoted $P^\vee/Q^\vee$. For $A_2$, the sum of the two fundamental coweights coincides with a coroot, $\omega_1^\vee + \omega_2^\vee = \alpha_1^\vee + \alpha_2^\vee \in Q^\vee$. Consequently, the cube of the lifted generator is the identity, $\eta(g)^3 = 1$, the Stiefel--Whitney class is trivial, and the mixed anomaly vanishes.

While we are not considering gauging the following subgroup, 
in contrast and for elucidative purposes, we could consider the Kac diagrams $s_0 = 2, s_1 = 1$ and $s_0 = 2, s_2 = 1$. These Kac labels delete fewer nodes, leaving an invariant Lie algebra $\mathfrak{su}(2) \oplus \mathfrak{u}(1)$. More importantly, for the $s_0=2, s_1=1$ embedding, the cube of the lifted generator evaluates to the generator of the center $\mathcal{Z}(\SU(3)) \simeq \ZZ_3$:\footnote{In general, it is possible to show that the quotient $P^\vee/Q^\vee$ is generated by the so-called `minuscule coweights', namely the coweights $\omega^\vee_i$ that correspond to nodes with $a_i=1$.}
\begin{equation}
    \eta(g)^3 = \exp_{\SU(3)} (2\pi i \omega_1^\vee)\neq 1\fstop
\end{equation}
Because the subgroup does not lift to a $\ZZ_3$, these alternative embeddings evaluate to nonzero obstruction classes, and hence have a mixed anomaly with the R-symmetry. The outer-conjugate $s_0 = 2, s_2 = 1$ behaves identically, with $\eta(g)^3 = \exp(-2\pi i \omega_1^\vee)$. 

\subsubsection{The $E_6$ Theory}

Next, consider the $E_6$ theory, whose global symmetry is $E_6/\ZZ_3$. Similarly, as for the $E_3$ theory, the fact that the obstruction class takes values in $\ZZ_3$ guarantees that any $\ZZ_2$ subgroup does not suffer from this mixed anomaly. To discuss its $\ZZ_3$ subgroups, consider the affine $E_6$ Dynkin diagram. We number the nodes of the affine $E_6$ Dynkin diagram following the Slansky convention \cite{Slansky:1981yr}, where the central node is $3$, the left and right arms are $(1,2)$ and $(4,5)$, and the affine node $0$ attaches to the short top arm $6$. The integers assigned to each node represent the affine marks $a_i$:
\begin{equation}\begin{tikzpicture}[baseline=0,font=\footnotesize]
\node[node, label=below:{$3$}] (A2) {};
\node[node, label=below:{$2$}] (Al1) [left=6mm of A2] {};
\node[node, label=below:{$1$}] (Al2) [left=6mm of Al1] {};
\node[node, label=below:{$2$}] (Ar1) [right=6mm of A2]{};
\node[node, label=below:{$1$}] (Ar2) [right=6mm of Ar1]{};
\node[node, label=right:{$2$}] (A3) [above=4mm of A2] {};
\node[node, label=right:{$1$}] (A3a) [above=4mm of A3] {};
\draw (A2) -- (A3) -- (A3a);\draw (Al2) -- (Al1) -- (A2) -- (Ar1) -- (Ar2);
\end{tikzpicture} \,.
\end{equation}
An order-three automorphism ($N=3$) requires a set of non-negative integers $s_i$ such that $\sum_{i=0}^6 a_i s_i = 3$. 
As discussed around equation \eqref{eqn:E6Z3}, the automorphism $\widehat\rho(r)$ we gauge corresponds to the Kac diagram with $s_0 = s_1 = s_5 = 1$, yielding the invariant Lie algebra $\mathfrak{so}(8) \oplus \mathfrak{u}(1)^2$,
\begin{equation}\label{eq:E6_Kac_Z3}
\begin{tikzpicture}[baseline=0,font=\footnotesize]
\node[node, label=below:{$3$}] (A2) {};
\node[node, label=below:{$2$}] (Al1) [left=6mm of A2] {};
\node[sonode, label=below:{$1$}] (Al2) [left=6mm of Al1] {};
\node[node, label=below:{$4$}] (Ar1) [right=6mm of A2]{};
\node[sonode, label=below:{$5$}] (Ar2) [right=6mm of Ar1]{};
\node[node, label=right:{$6$}] (A3) [above=4mm of A2] {};
\node[sonode, label=right:{$0$}] (A3a) [above=4mm of A3] {};
\draw (A2) -- (A3) -- (A3a);
\draw (Al2) -- (Al1) -- (A2) -- (Ar1) -- (Ar2) ;
\end{tikzpicture} \,.
\end{equation}
The generator for this discrete subgroup is
\begin{equation}
g = \exp_{E_6/\ZZ_3} \left[ \frac{2\pi i}{3} (\omega^\vee_1 + \omega^\vee_5) \right]\fstop
\end{equation}
As in the $E_3$ case, the center $\mathcal{Z}(E_6) \simeq \ZZ_3$ is generated by the two minuscule coweights $\omega^\vee_1$ and $\omega^\vee_5$, but $\omega_1^\vee+\omega_5^\vee\in Q^\vee$; thus, the lift to $E_6$ still has order 3, and the anomaly vanishes. By contrast, embeddings such as $s_0=2$ and $s_1=1$ will have a non-trivial obstruction, and the corresponding $\ZZ_3$ subgroups will have a non-vanishing anomaly.

\subsection{Wreathing of Magnetic Quivers}
\label{sec:wreathingMQ}

Let us first review the definition of the Coulomb branch Hilbert series of a quiver. Consider a 3d $\mathcal{N}=4$ quiver gauge theory $\mathcal{T}_M$ with (reductive) gauge group $G$ connected by a set of edges associated with bifundamental hypermultiplets. The unrefined Coulomb branch Hilbert series is given as \cite{Cremonesi:2013lqa}:
\begin{equation}\label{eq:HSunrefinedCB}
    \HS_{\mathcal{T}_M}(t)=\frac{1}{|W|}\sum_{\mathbf{m}}\sum_{\gamma\in W(\mathbf{m})}\frac{t^{2\Delta(\mathbf{m})}}{\det(\ID-t^2\gamma)}\coma
\end{equation}
where $W$ is the Weyl group of $G$, and $\Delta(\mathbf{m})$ is the dimension of the monopole operator with magnetic flux $\mathbf{m}$, generically given by \cite{Gaiotto:2008ak,Cremonesi:2013lqa}
\begin{equation}
    \Delta(\mathbf{m}) = -\sum_{\alpha \in \Delta_+}|\alpha(\mathbf{m})|+\frac{1}{2}\sum_{i=1}^{N_f}\sum_{\rho_i\in\mathcal{R}_i}|\rho_i(\mathbf{m})|\fstop
\end{equation}
Here, $\alpha\in\Delta_+$ are the positive roots of the gauge group $G$, $\rho_i$ are the weights of the irreducible matter field representation $\mathcal{R}_i$ under the gauge group, and $N_f$ is the total number of irreducible representations under which the matter transforms. Respectively, they represent the vector multiplet and hypermultiplet contributions to the dimension of the monopole operators. The summation of the magnetic fluxes depends on the gauge group $G$; for a quiver with only unitary gauge nodes, the summation is over $\mathbf{m}\in \ZZ^r$, with $r$ being the rank of the gauge group, while if the quiver contains orthosymplectic nodes, then $\mathbf{m}\in \ZZ^r\cup \left(\ZZ+\frac{1}{2}\right)^r$ \cite{Bourget:2020xdz}. In this work, we will consider only certain representations for unitary and orthosymplectic groups, which we list in Table \ref{tab:conformaldimensions}, together with the corresponding contributions to the conformal dimension of the monopole operators \cite{Cremonesi:2013lqa,Bourget:2020xdz}. The prefactor
\begin{equation}
    P_G(t,\mathbf{m}) = \frac{1}{|W|}\sum_{\gamma\in W(\mathbf{m})}\frac{1}{\det(\ID-t^2\gamma)}\coma
\end{equation}
is a classical contribution that counts the gauge invariant operators of the residual gauge group unbroken by the magnetic flux $\mathbf{m}$. If the gauge group $G$ is not simply connected, there is a nontrivial topological symmetry under which monopole operators may be charged. One can then refine the Hilbert series by introducing fugacities $\mathbf{z}$ valued in the topological symmetry group, with topological charge $J(\mathbf{m})$ for a monopole operator with charge $\mathbf{m}$ \cite{Cremonesi:2013lqa}. The Hilbert series of the Coulomb branch \eqref{eq:HSunrefinedCB} can be refined as
\begin{equation}\label{eq:HSrefinedCB}
    \HS_{\mathcal{T}_M}(t,\mathbf{z})=\frac{1}{|W|}\sum_{\mathbf{m}}\sum_{\gamma\in W(\mathbf{m})}\frac{t^{2\Delta(\mathbf{m})}\mathbf{z}^{J(\mathbf{m})}}{\det(\ID-t^2\gamma)}\fstop
\end{equation}

\begin{table}[!htp]
    \centering
    \begin{subtable}[t]{\textwidth}
    \centering
    \renewcommand*{\arraystretch}{2}
        \begin{tabular}{c|c}
       Group  &  $\displaystyle -\sum_{\alpha \in \Delta_+}|\alpha(\mathbf{m})|$ \\\hhline{=|=}
       $\text{U}(N)_\mathbf{m}$ & $\displaystyle -\sum_{i<j}^N|m_i-m_j|$\\\hline
       $\SO(2N)_\mathbf{m}$ & $\displaystyle -\sum_{i<j}^N\left(|m_i+m_j|+|m_i-m_j|\right)$\\\hline
       $\USp(2N)_\mathbf{m}$ & $\displaystyle -\sum_{i<j}^N\left(|m_i+m_j|+|m_i-m_j|\right)-2\sum_{i=1}^N|m_i|$
    \end{tabular}
    \caption{Vector multiplet contribution.}
    \end{subtable}\\[2.2em]
    \begin{subtable}[t]{\textwidth}
    \centering
    \renewcommand*{\arraystretch}{2}
    \begin{tabular}{c|c}
       Representation & $\displaystyle \frac{1}{2}\sum_{i=1}^{N_f}\sum_{\rho_i\in\mathcal{R}_i}|\rho_i(\mathbf{m})|$\\\hhline{=|=}
       Bifundamental of $\text{U}(N)_\mathbf{m}\times \text{U}(M)_\mathbf{n}$ & $\displaystyle \frac{1}{2}\sum_{i=1}^N\sum_{j=1}^M|n_i-m_j|$\\\hline
       Bifundamental of $\SO(2N)_\mathbf{m}\times \USp(2M)_\mathbf{n}$ & $\displaystyle \frac{1}{2}\sum_{i=1}^N\sum_{j=1}^M\left(|n_i+m_j|+|n_i-m_j|\right)$\\\hline
       Bifundamental of $\SO(2N)_\mathbf{m}\times \text{U}(M)_\mathbf{n}$ & $\displaystyle \frac{1}{2}\sum_{i=1}^N\sum_{j=1}^M\left(|n_i+m_j|+|n_i-m_j|\right)$\\\hline
       Bifundamental of $\USp(2N)_\mathbf{m}\times \text{U}(M)_\mathbf{n}$ & $\displaystyle \frac{1}{2}\sum_{i=1}^N\sum_{j=1}^M\left(|n_i+m_j|+|n_i-m_j|\right)$\\\hline
       Adjoint of $\text{U}(N)_\mathbf{m}$ & $\displaystyle \sum_{i<j}^N|m_i-m_j|$\\\hline
       Antisymmetric $\Lambda^2$ of $\USp(2N)_\mathbf{m}$ & $\displaystyle \sum_{i<j}^N\left(|m_i+m_j|+|m_i-m_j|\right)$\\\hline
       $\text{U}(N)_\mathbf{m}$ with $\ell$ charge 2 hypermultiplets & $\displaystyle \ell\sum_{i=1}^N|m_i|$
    \end{tabular}
    \caption{Hypermultiplet contribution.}
    \end{subtable}
    \caption{Contributions to the conformal dimension $\Delta(\mathbf{m})$ in the monopole formula \cite{Cremonesi:2013lqa,Bourget:2020xdz}, for magnetic fluxes $\mathbf{m}=(m_1,\ldots,m_N)$ and $\mathbf{n}=(n_1,\ldots,n_M)$. The subscripts on the groups denote the magnetic fluxes in the lattice associated to that group.}
    \label{tab:conformaldimensions}
\end{table}

The action of the $\Gamma$-wreathing is implemented at the level of the Coulomb branch Hilbert series of the 3d $\mathcal{N}=4$ SCFT by taking the wreath product of the Weyl group $W$ of the gauge group of the quiver with $\Gamma$. As explained in \cite{Bourget:2020bxh}, a quiver has a diagram automorphism by a finite group $\Gamma$ if this group leaves $\Delta(\mathbf{m})$ invariant. Then, it is possible to consider the wreathing by $\Gamma$ of the quiver, obtaining the wreathed quiver which we call $\widetilde{\mathcal{T}}_M$. The general expression for the Hilbert series for the Coulomb branch of $\widetilde{\mathcal{T}}_M$ is 
\begin{equation}
    \HS_{\widetilde{\mathcal{T}}_M}(t,\widetilde{\mathbf{z}})=\frac{1}{|W_\Gamma|}\sum_{\widetilde{\mathbf{m}}}\sum_{\gamma\in W_\Gamma(\widetilde{\mathbf{m}})}\frac{t^{2\Delta(\widetilde{\mathbf{m}})}\widetilde{\mathbf{z}}^{J(\widetilde{\mathbf{m}})}}{\det(\ID-t^2\gamma)}\coma
\end{equation}
where we have denoted $W_\Gamma = W\wr \Gamma$. The fugacities $\widetilde{\mathbf{z}}$ are those associated to each $\U(1)$ factor of the wreathed group. Moreover, we have modified the summation from $\mathbf{m}$ to $\widetilde{\mathbf{m}}$ because, in general, the summation on the fluxes and the prefactors may differ after the wreathing; the reason is that the wreathing by $\Gamma$ acts only on a subset of the gauge nodes of the quiver. In practice, one needs to determine all the matrices in $W_\Gamma$ that leave a given choice of fluxes invariant. This way of computing the Coulomb branch Hilbert series for wreathed unitary quivers has been explored in \cite{Bourget:2020bxh,Arias-Tamargo:2021ppf,Giacomelli:2024sex,Grimminger:2024mks} and, more recently, in \cite{Lawrie:2025exx} applied to orthosymplectic quivers. Alternatively, it was noted in \cite{Grimminger:2024mks} that the wreathing of 3d $\mathcal{N}=4$ SCFTs can be implemented at the level of the superconformal index by summing over the elements of $\Gamma$ and rescaling the fugacities of the contributions involved in the wreathing. In particular, for $\ZZ_2$-wreathing, \cite[Appendix A]{Grimminger:2024mks} provides a detailed prescription of how to implement the wreathing on unitary quivers. We notice that the same prescription applies directly to the Coulomb branch limit of the superconformal index, such that in the following, we will also use this method to compute the wreathing, either to confirm our findings or because it is more convenient for the considered wreathing. However, as we will explain in Section \ref{sec:Z2cuttingedges}, whenever the axis of symmetry of the $\ZZ_2$-wreathing cuts a bifundamental edge, the prescription introduced in \cite[Appendix A]{Grimminger:2024mks} must be modified. This is a novel result of this work and is necessary for consistency with the correct flavor symmetries of the theories after the discrete gauging.\footnote{Let us stress that the two prescriptions described above to determine the wreathed Hilbert series are precisely analogous and equivalent. Therefore, both require this modification to obtain the correct result. However, we find that modifying the prescription introduced in \cite[Appendix A]{Grimminger:2024mks} leads to a more direct way to compute the Hilbert series and to implement the modification.}

Below, we will also consider the wreathings of the orthosymplectic realizations of the $E_n$ theories as proposed in \cite{Bourget:2020xdz}, despite their lack of a direct interpretation in terms of toric diagrams or Seiberg--Witten curves. The only possible wreathings allowed for these quivers are $\ZZ_2$, and they can be computed as introduced in \cite{Lawrie:2025exx}. Analogously, as mentioned above, the prescription in  \cite[Appendix A]{Grimminger:2024mks} extends naturally to orthosymplectic quivers, so it can be applied to obtain the same result after the wreathing.\footnote{There are no orthosymplectic quivers that we will consider for which the axis of symmetry cuts a bifundamental edge in half. However, we believe that a generalized modification of the prescription that we describe in Section \ref{sec:Z2cuttingedges} should hold, although a check of this expectation is beyond the scope of this work.} Interestingly, the resulting flavor symmetries do not always correspond to a $\ZZ_2$-wreathing as computed via the unitary quiver. To find the flavor symmetry more precisely, we refine the Hilbert series of the orthosymplectic quivers by a fugacity $\omega$ that accounts for a residual $\ZZ_2^{[0]}$ zero-form symmetry in the 3d quiver. This fugacity can also be seen as distinguishing the contributions coming from $\mathbf{m}\in \ZZ^r$ and $\mathbf{m}\in \left(\ZZ+\frac{1}{2}\right)^r$ for the following reason. All these quivers have an overall $\ZZ_2$ symmetry that acts trivially on the matter fields, and it can be considered to be, e.g., the subgroup of the center of the central node of the quiver that is not screened by the bifundamental fields. This gives rise to a $\ZZ_2^{[1]}$ one-form symmetry which can be gauged or not. If the one-form symmetry is gauged, as we consider here,\footnote{The gauging of such one-form symmetry has been observed in \cite{Cremonesi:2014vla} to lead to the correct 3d mirror theory of class $\mathcal{S}$ theories \cite{Gaiotto:2009we,Gaiotto:2009gz} of type $D$. At the level of the Hilbert series, it corresponds to a sum over odd-half-integer fluxes.} the resulting 3d Coulomb branch of the theory has a $\ZZ_2^{[0]}$ zero-form symmetry, and we associate it with the fugacity $\omega$. The presence of $\omega$ is important to understand the patterns of branching rules that lead to the flavor symmetry of the theory after wreathing.

\subsubsection{Example: the \texorpdfstring{$E_6$}{E6} Theory}
\label{sec:E6th}

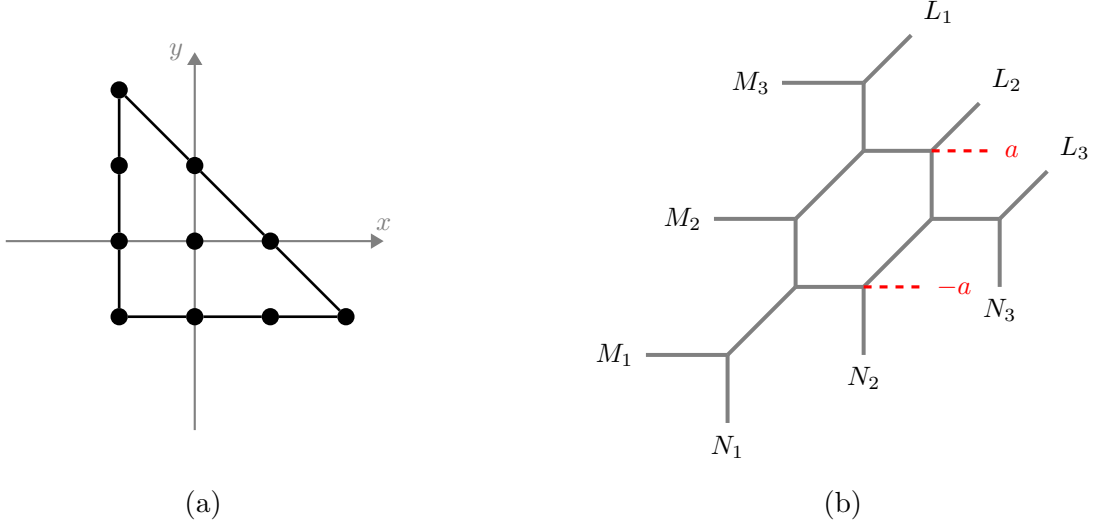
\begin{figure}[t]
\centering
\begin{subfigure}[b]{0.49\textwidth}
\centering
    \begin{tikzpicture}[scale=1,font=\footnotesize,baseline=0]
            \path (0, -3) -- (0, 4.0);
            
			\draw[thick,gray,-Triangle] (-2.5,0) -- node[above,pos=1] {$x$} (2.5,0);
			\draw[thick,gray,-Triangle] (0,-2.5) -- node[left,pos=1] {$y$} (0,2.5);
			
			\node[node, fill=black] (O) at (0,0) {};
			\node[node, fill=black] (A) at (-1,-1) {};
			\node[node, fill=black] (B) at (0,-1) {};
			\node[node, fill=black] (C) at (1,-1) {};
            \node[node, fill=black] (D) at (2,-1) {};
            \node[node, fill=black] (E) at (1,0) {};
            \node[node, fill=black] (F) at (0,1) {};
            \node[node, fill=black] (G) at (-1,2) {};
            \node[node, fill=black] (H) at (-1,1) {};
            \node[node, fill=black] (I) at (-1,0) {};
           \draw[line width=1pt] (A) -- (B) -- (C) -- (D) -- (E) -- (F) -- (G) -- (H) -- (I) -- (A);
    \end{tikzpicture}
\caption{}\label{fig:E6toricdiagram}
\end{subfigure}\hfill
\begin{subfigure}[b]{0.49\textwidth}
\centering
\begin{tikzpicture}[font=\footnotesize,baseline=0, scale=0.9,
    brane/.style={gray,line width=1.5pt},
    midarrow/.style={
      postaction={decorate},
      decoration={markings,
        mark=at position .55 with {
          \arrow[black]{Stealth[length=2.2mm,width=2.2mm]}
        }
      }
    }
]


\coordinate (J1) at (-1,-1);
\coordinate (J2) at ( 0, 0);
\coordinate (J3) at ( 0, 1);

\coordinate (J4) at ( 1, 2);
\coordinate (J5) at ( 1, 3);

\coordinate (J6) at ( 2, 2);
\coordinate (J7) at ( 2, 1);

\coordinate (J8) at ( 3, 1);

\coordinate (J9) at ( 1, 0);


\draw[brane] (J1) -- (J2);

\draw[brane] (J2) -- (J3);

\draw[brane] (J2) -- (J9);

\draw[brane] (J3) -- (J4);

\draw[brane] (J4) -- (J5);

\draw[brane] (J4) -- (J6);

\draw[brane] (J6) -- (J7);

\draw[brane] (J9) -- (J7);

\draw[brane] (J7) -- (J8);

\draw[brane] (J1) -- ++(-1.2,0)
    node[left,black] {$M_1$};

\draw[brane] (J3) -- ++(-1.2,0)
    node[left,black] {$M_2$};

\draw[brane] (J5) -- ++(-1.2,0)
    node[left,black] {$M_3$};

\draw[brane] (J1) -- ++(0,-1.0)
    node[below,black] {$N_1$};

\draw[brane] (J9) -- ++(0,-1.0)
    node[below,black] {$N_2$};

\draw[brane] (J8) -- ++(0,-1.0)
    node[below,black] {$N_3$};

\draw[brane] (J5) -- ++(0.7,0.7)
    node[above right,black] {$L_1$};

\draw[brane] (J6) -- ++(0.7,0.7)
    node[above right,black] {$L_2$};

\draw[brane] (J8) -- ++(0.7,0.7)
    node[above right,black] {$L_3$};
\draw[red, dashed, line width=1.2pt] (J6) -- ++(0.9,0) node[right,red] {$a$}; \draw[red, dashed, line width=1.2pt] (J9) -- ++(0.9,0) node[right,red] {$-a$};
\end{tikzpicture}
\caption{}
\end{subfigure}
\caption{
Toric diagram and corresponding $(p,q)$-web for the $E_6$ theory.
}
\label{fig:e6pwweb}
\end{figure}

The rank-one $E_6$ theory can be realized using the $(p,q)$-web and the toric diagram in Figure \ref{fig:e6pwweb}.
The $E_6$ theory admits a unitary magnetic quiver describing the Higgs branch of the theory, which is
\begin{equation}\label{eq:unitaryE6}
    \begin{tikzpicture}[baseline=0,font=\footnotesize]
        \node[node, label=below:{$3$}] (A2) {};
        \node[node, label=below:{$2$}] (Al1) [left=6mm of A2] {};
        \node[node, label=below:{$2$}] (Ar1) [right =6mm of A2]{};
        \node[node, label=below:{$1$}] (Al2) [left=6mm of Al1] {};
        \node[node, label=below:{$1$}] (Ar2) [right =6mm of Ar1]{};
        \node[node, label=right:{$2$}] (A3) [above=4mm of A2] {};
        \node[node, label=right:{$1$}] (A3a) [above=4mm of A3] {};
        \draw (A2) -- (A3) -- (A3a);
        \draw   (Al2) -- (Al1) -- (A2) -- (Ar1) -- (Ar2) ;
        \node (Z2) [right=6mm of Ar2] {$/\U(1)\fstop$};
    \end{tikzpicture}
\end{equation}
Let us compute the refined Hilbert series for this quiver. We introduce seven fugacities, $w_i$, with $i = 1,\cdots, 7$, where $w_1,w_3,w_5$ are associated to the three $\U(1)$s, $w_2,w_4,w_6$ are the three $\U(2)$s, and $w_7$ is for the $\U(3)$. We decouple the overall $\U(1)$ by summing over fluxes of $\U(3)$ such that $m_1\leq m_2\leq m_3 = 0$ and impose 
\begin{equation}
    w_7^3 = \frac{1}{w_1w_2^2w_3w_4^2w_5w_6^2}\fstop
\end{equation}
It is possible to compute the Hilbert series in full totality, summing all the magnetic fluxes of the whole quiver. A faster yet analogous method is achieved by noting that the quiver is given by the gauging of the $\SU(3)$ of three copies of $T_{[1,1,1]}[\SU(3)]$ tails,
\begin{equation}\label{eq:HSE6formal}
\scalebox{0.88}{$\displaystyle
\begin{aligned}
    \text{HS}_{E_6}(t,\mathbf{w}) = &\,(1-t^2)\sum_{-\infty \leq m_1\leq m_2\leq m_3 = 0}^\infty P_{\U(3)}(t,\mathbf{m}) w_7^{m_1+m_2+m_3}t^{2\Delta_\text{res.}}\times \\
    &\text{HS}_{T_{[1,1,1]}[\SU(3)]}(t,w_1,w_2,\mathbf{m})\text{HS}_{T_{[1,1,1]}[\SU(3)]}(t,w_3,w_4,\mathbf{m})\text{HS}_{T_{[1,1,1]}[\SU(3)]}(t,w_5,w_6,\mathbf{m})\coma    
\end{aligned}$}
\end{equation}
where, for example, in the first tail, we have
\begin{equation}\label{eq:HSTSU3tail}
    \text{HS}_{T_{[1,1,1]}[\SU(3)]}(t,w_1,w_2,\mathbf{m}) = \frac{1}{1-t^2}\sum_{-\infty\leq n_1}^\infty \sum_{-\infty\leq s_1\leq s_2}^\infty P_{\U(2)}(t,\mathbf{s})w_1^{n_1}w_2^{s_1+s_2}t^{2\Delta_{T[\SU(3)]}}\coma
\end{equation}
with 
\begin{equation}
    \begin{split}
        \Delta_{T[\SU(3)]} = -| s_1-s_2|+\frac{1}{2}&\left(| m_1-s_1| +| m_1-s_2| +| m_2-s_1| +| m_2-s_2|\right.\\
        &\left.+| m_3-s_1| +| m_3-s_2| +| n_1-s_1| +| n_1-s_2| \right)
    \end{split}
\end{equation}
while
\begin{equation}
    \Delta_\text{res.} = -|m_1-m_2| -|m_1-m_3| -| m_2-m_3|\fstop
\end{equation} 
Note that in equation \eqref{eq:HSTSU3tail} there is a dependence on the magnetic fluxes $\mathbf{m}$ that has not been summed over yet, and they function as background magnetic charges associated with the flavor group $\U(3)$, gauged only when equation \eqref{eq:HSE6formal} is considered. Computationally, it is then more convenient to first compute the Hilbert series of the single tails and perform the gauging of the central node at the end. This method will play an important role when performing wreathing of the quivers because of the method developed in \cite[Appendix A]{Grimminger:2024mks} that we exemplify anon. Once the Hilbert series is computed,  it is instructive, first, to rewrite the fugacities in terms of the characters of $\su(3)^{\oplus 3}$, i.e., defining
\begin{equation}\label{eq:E6toSU33}
    w_1\to x_1x_2\coma w_2\to \frac{x_1}{x_2^2}\coma w_3\to y_1 y_2\coma w_4\to \frac{y_1}{y_2^2}\coma w_5\to z_1 z_2\coma w_6\to \frac{z_1}{z_2^2}\fstop
\end{equation}
The Hilbert series is given by
\begin{equation}\label{eq:E6-HSwSU3chr}
\scalebox{1}{$\displaystyle
\begin{aligned}
     \HS_{E_{6}}(t,\mathbf{x},\mathbf{y},\mathbf{z}) = \PE&
     \left[\left(\chi^A_{[1,1]}(\mathbf{x})+\chi^A_{[1,1]}(\mathbf{y})+\chi^A_{[1,1]}(\mathbf{z})+\chi^A_{[1,0]}(\mathbf{x})\chi^A_{[1,0]}(\mathbf{y})\chi^A_{[1,0]}(\mathbf{z})+\right.\right.\\
    & \left.\left.+
    \chi^A_{[0,1]}(\mathbf{x})\chi^A_{[0,1]}(\mathbf{y})\chi^A_{[0,1]}(\mathbf{z})\right)t^2+\mathcal{O}(t^{4})\right]\fstop
\end{aligned}
$}
\end{equation}
However, we can observe that there is an enhanced global symmetry in accordance with the expected $\mathfrak{e}_6$ algebra, given by the following decomposition:
\begin{equation}
    \begin{split}
      \mathfrak{e}_6 & \rightarrow \su(3)\oplus \su(3)\oplus \su(3)\coma\\
      \mathbf{78} & \rightarrow (\mathbf{3}, \mathbf{3}, \mathbf{3}) \oplus (\overline{\mathbf{3}}, \overline{\mathbf{3}}, \overline{\mathbf{3}}) \oplus (\mathbf{8}, \mathbf{1}, \mathbf{1}) \oplus (\mathbf{1}, \mathbf{8}, \mathbf{1}) \oplus (\mathbf{1}, \mathbf{1}, \mathbf{8})\fstop
    \end{split}
\end{equation}
In this way, the Hilbert series has a manifest $\mathfrak{e}_6$ global symmetry, realized by redefining the fugacities as
\begin{equation}
\renewcommand{\arraystretch}{1.7}
\begin{array}{rclrclrcl}
 w_1 & \to & \dfrac{x_4 x_6}{x_1x_3}\coma & w_2 & \to & \dfrac{x_1 x_6}{x_4} \coma & w_3 & \to & \dfrac{x_1 x_4}{x_3 x_5}\coma\\ 
 w_4 & \to & \dfrac{x_5}{x_4^2}\coma & w_5 & \to & \dfrac{x_2}{x_1x_4}\coma & w_6 & \to & \dfrac{x_1x_3}{x_2^2}\fstop
\end{array}
\end{equation}
The Hilbert series is given by
\begin{equation}\label{eq:E6-HS}
\scalebox{1}{$\displaystyle
\begin{aligned}
     \HS_{E_{6}}(t,\mathbf{x}) = \PE\left[\chi^{\mathfrak{e}_6}_{[0,0,0,0,0,1]}(\mathbf{x})t^2+\mathcal{O}(t^{4})\right]\coma
\end{aligned}
$}
\end{equation}
or, unrefined,
\begin{equation}
\begin{split}
    \text{HS}_{E_6}(t,1) =&\, \frac{1+56 t^2+945 t^4+6776 t^6+23815 t^8+43989 t^{10}+43989 t^{12}+
    \cdots+t^{22}}{(1-t^{2})^{22}} \\
    & = \PE\left[78 t^2-651 t^4+12376 t^6+\mathcal{O}\left(t^{8}\right)\right]\fstop
\end{split}
\end{equation}
The coefficients packaged in the $\cdots$ are fixed as the numerator of the Hilbert series is palindromic.
The quiver in equation \eqref{eq:unitaryE6} has an $S_3$ diagram-automorphism group, which acts via permutation of the three identical legs. In the following, we consider the putative Higgs branch obtained via discrete gauging of the $\ZZ_2$ subgroup of $S_3$. In Section \ref{sec:E56th_more}, we further consider the $\ZZ_3$ and $S_3$ discrete gauging.

\subsubsection*{The \texorpdfstring{$\ZZ_2$}{Z2}-Wreathing of the Rank-one \texorpdfstring{$E_6$}{E6} Theory}

We first perform the $\ZZ_2$-wreathing of equation \eqref{eq:unitaryE6} by exchanging two of the legs, i.e.,
\begin{equation}\label{eq:E6-wrZ2}
    \begin{tikzpicture}[baseline=0,font=\footnotesize]
        \node[node, label=below right:{$3$}] (A2) {};
        \node[node, label=below:{$2$}] (Al1) [left=6mm of A2] {};
        \node[node, label=below:{$2$}] (Ar1) [right =6mm of A2]{};
        \node[node, label=below:{$1$}] (Al2) [left=6mm of Al1] {};
        \node[node, label=below:{$1$}] (Ar2) [right =6mm of Ar1]{};
        \node[node, label=right:{$2$}] (A3) [above=4mm of A2] {};
        \node[node, label=right:{$1$}] (A3a) [above=4mm of A3] {};
        \draw (A2) -- (A3) -- (A3a);
        \draw   (Al2) -- (Al1) -- (A2) -- (Ar1) -- (Ar2) ;
        \draw[Triangle-Triangle,red] ([yshift=-5mm]Al1.south) to[bend right=50] node[below,pos=0.5] {$\wr \ZZ_2$} ([yshift=-5mm]Ar1.south);
        \node (Z2) [right=6mm of Ar2] {$/\U(1)\,.$};
        \draw[densely dashed, red,thin] ([yshift=-3mm]A2.south) -- ([yshift=3mm]A3a.north);
    \end{tikzpicture} 
\end{equation}
There are two possible ways to compute the wreathing: one is the prescription given in the original papers that introduced the wreathing procedure \cite{Bourget:2020bxh}, and it results in a generalization of the cases considered in \cite{Arias-Tamargo:2021ppf,Giacomelli:2024sex,Lawrie:2025exx}. This method considers the various contributions to the Hilbert series that are consistent with the wreathing, computing the modified version of the prefactor associated to the chiral in the vector multiplet, as explained above. The other method is the one discovered in \cite{Giacomelli:2024sex} and further explored in \cite{Grimminger:2024mks,Grimminger:2024doq}. In fact, precisely for $\ZZ_2$-wreathing, \cite[Appendix A]{Grimminger:2024mks} lists a set of rules that we can use to compute the Hilbert series for equation \eqref{eq:E6-wrZ2}. The way in which the wreathing, as in equation \eqref{eq:E6-wrZ2}, can be computed is by noting that the wreathing is only acting on two of the three $T_{[1,1,1]}[\SU(3)]$ tails we used to compute the Hilbert series in equation \eqref{eq:E6-HS}. The action of the wreathing is to first identify the fugacities of the two $T_{[1,1,1]}[\SU(3)]$ tails, i.e.,
\begin{equation}
    w_3 = w_1 \coma w_4 = w_2 \fstop
\end{equation}
Then, following \cite[Appendix A]{Grimminger:2024mks}, the wreathed Hilbert series is given by
\begin{equation}
\scalebox{0.88}{$\displaystyle
\begin{aligned}
     \text{HS}_{E_6\wr \ZZ_2}(t,\mathbf{w}) = &\,(1-t^2)\sum_{-\infty\leq m_1\leq m_2\leq m_3 = 0}^\infty P_{\U(3)}(t,\mathbf{m}) w_7^{m_1+m_2+m_3}t^{2\Delta_\text{res.}}\text{HS}_{T_{[1,1,1]}[\SU(3)]}(t,w_5,w_6,\mathbf{m})\times \\
     &\frac{1}{2}\left(\text{HS}_{T_{[1,1,1]}[\SU(3)]}(t,w_1,w_2,\mathbf{m})^2+\text{HS}_{T_{[1,1,1]}[\SU(3)]}(t^2,w_1^2,w_2^2,\mathbf{m})\right)\fstop
\end{aligned}$}
\end{equation}
This prescription was derived by analyzing the wreathing of the 3d superconformal index and then taking the Coulomb branch limit of the index. A straightforward redefinition of the fugacities makes manifest an $\mathfrak{su}(3)^{\oplus 2}$ global symmetry, and we find
\begin{equation}\label{eq:E6Z2-HS}
\begin{split}
     \HS_{E_{6}\wr \ZZ_2}(t,\mathbf{x},\mathbf{z}) = &\, \PE\left[\left(\chi^A_{[1,1]}(\mathbf{x})+\chi^A_{[1,1]}(\mathbf{z})+\chi^A_{[2,0]}(\mathbf{x})\chi^A_{[0,1]}(\mathbf{z})\right.\right.\\
     &\qquad\left.\left.+\chi^A_{[0,2]}(\mathbf{x})\chi^A_{[1,0]}(\mathbf{z})\right)t^2 -t^4+\mathcal{O}(t^{6})\right]\coma
\end{split}
\end{equation}
or unrefined,
\begin{equation}
\begin{split}    \HS_{E_{6}\wr \ZZ_2}(t,1,1) &= \frac{\left(\begin{aligned}        &1+30 t^2+464 t^4+3383 t^6+11946 t^8 \\        &\quad +21967 t^{10}+21967 t^{12}+\cdots+t^{22}    \end{aligned}\right)}{(1-t^2)^{22}} \\      &= \PE\left[52t^2-t^4-1547t^6 +\mathcal{O}(t^{8})\right]\fstop \end{split}
\end{equation}
By studying the $t^2$ term in the refined Hilbert series, we can see that the global symmetry is enhanced to $\mathfrak{f}_4$, as can be observed from the branching rule
\begin{equation}
    \begin{split}
        \mathfrak{f}_4 &\rightarrow \su(3)\oplus \su(3)\\
        \mathbf{52} &\rightarrow (\mathbf{8},\mathbf{1})\oplus (\mathbf{1},\mathbf{8}) \oplus (\mathbf{6},\overline{\mathbf{3}})\oplus (\overline{\mathbf{6}},\mathbf{3})\fstop
    \end{split}
\end{equation}
In fact, one can directly express the $w_i$ fugacities in terms of $\mathfrak{f}_4$ characters by defining, e.g.,
\begin{equation}
    w_1 \rightarrow \frac{y_2}{y_1y_3}\coma w_2 \rightarrow \frac{y_1y_3}{y_2y_4}\coma w_5 \rightarrow \frac{y_3^2}{y_1y_4^2}\coma w_6\rightarrow \frac{y_1y_2}{y_3^2}\coma
\end{equation}
resulting in the following Hilbert series
\begin{equation}\label{eq:E6Z2-HS-f4}
\scalebox{1}{$\displaystyle
    \HS_{E_{6}\wr \ZZ_2}(t,\mathbf{y}) = \PE\left[\chi^{\mathfrak{f}_4}_{[1,0,0,0]}(\mathbf{y})t^2 -t^4 -\left(\chi^{\mathfrak{f}_4}_{[0,1,0,0]}(\mathbf{y})+\chi^{\mathfrak{f}_4}_{[0,0,1,0]}(\mathbf{y})\right)t^6+\mathcal{O}(t^{8})\right]\fstop$}
\end{equation}
In fact, this Hilbert series is nothing other than the Hilbert series of the closure of the $\widetilde{A}_1$ nilpotent orbit of $\mathfrak{f}_4$ \cite{Hanany:2017ooe}.\footnote{We use the standard Bala--Carter \cite{bala1976classesI,bala1976classesII} labeling for the nilpotent orbits of the exceptional Lie algebras.}

\subsection{\texorpdfstring{$\ZZ_2$}{Z2}-wreathing Cutting Quiver Edges}
\label{sec:Z2cuttingedges}

The computation of wreathings of magnetic quivers that involve the exchange of two nodes connected by a bifundamental requires a subtle modification of the wreathing technique. To exemplify the need for this modification of the prescription, we consider the Coulomb branch relations associated to the closure of the minimal nilpotent orbit of $\mathfrak{su}(3)$, $\mathcal{O}_\text{\tiny min}^{\mathfrak{su}(3)}$, whose magnetic quiver is
\begin{equation}
    \begin{tikzpicture}[baseline=0,font=\footnotesize]
        \node (A0) at (0,0) {};
        \node[node, label=above:{$1$}] (A1) [above=4mm of A0] {};
        \node[node, label=below:{$1$}] (A2) [below left=6mm of A0] {};
        \node[node, label=below:{$1$}] (A3) [below right=6mm of A0] {};
        \draw (A1) -- (A2) -- (A3) -- (A1);
        \node (Z2) [above right=5mm of A3] {$/\U(1)$};
    \end{tikzpicture}\coma
\end{equation}
which we later study in detail in Section \ref{sec:E3th}. The fundamental monopole operators are labeled by the magnetic charges of the two nodes at the base of the triangle, $v_{(m_1, m_2, 0)}$, while we decouple the node at the top. Denote $\varphi_1$ and $\varphi_2$ as the Casimirs of the lower two nodes. 

Among all the Coulomb branch relations (see, e.g., \cite[(5.27)]{Hanany:2023uzn}), let us study the two that are of particular interest when considering the $\ZZ_2$-wreathing; these are
\begin{equation}\label{eq:CBrel}
    \begin{split}
    v_{(1,0,0)}v_{(0,1,0)} &= v_{(1,1,0)}(\varphi_1 - \varphi_2)\coma\\
    v_{(-1,0,0)}v_{(0,-1,0)} &= v_{(-1,-1,0)}(\varphi_1 - \varphi_2)\fstop
    \end{split}
\end{equation}
If we consider the $\ZZ_2$-wreathing exchanging the two lower nodes of the quiver, the involution $\sigma$ acts on the scalar fields and monopoles as
\begin{equation}
\begin{split}
     \sigma(\varphi_1) = \varphi_2\,, &\quad \sigma(\varphi_2) = \varphi_1 \coma \\
        \sigma(v_{(1,0,0)}) = v_{(0,1,0)}\,, &\quad \sigma(v_{(0,1,0)}) = v_{(1,0,0)} \fstop
\end{split}
\end{equation}
Hence, the RHS of the relations in equation \eqref{eq:CBrel} would naively pick up a minus sign, unless accompanied by the action:
\begin{equation}
    \sigma(v_{(\pm 1, \pm 1, 0)}) = - v_{(\pm 1, \pm 1, 0)} \,.
\end{equation}
Taking this into account, in order to project onto the $\ZZ_2$-invariant subspace of the Coulomb branch, we propose a modification to the wreathing procedure introduced in \cite[(A.20)-(A.25)]{Grimminger:2024mks},\footnote{A. M. is extremely grateful to W. Harding for his patience and continuous correspondence on this topic.} by adding a phase factor 
\begin{equation}
    \sum \epsilon_\sigma(m) t^{2\Delta}\coma
\end{equation}
to the Hilbert series contribution obtained when the bifundamental is cut by the axis of symmetry of the $\ZZ_2$-wreathing. The phase factor is 
\begin{equation}
    \epsilon_\sigma(m) = (-1)^{\sum_{i=1}^N m_{i}}\coma
\end{equation}
where $m_{i}$ are the magnetic fluxes of the $\U(N)$ gauge group. In this way, the Hilbert series corresponds to that of the actual Coulomb branch moduli space, invariant under the gauging of the $\ZZ_2$ symmetry.\footnote{This modification is somewhat reminiscent of the modification introduced to the superconformal index in the presence of a parity anomaly (see, e.g., \cite{Garavaglia:2025cgz}). It would be interesting to clarify any possible connections.}  Moreover, if the edges cut by the axis of symmetries are multiple, there will be a factor $\epsilon_\sigma(m)$ for each edge. In Sections \ref{sec:E3th} and \ref{sec:E5th}, we will use this new prescription extensively to reproduce the expected flavor symmetries obtained when gauging discrete groups containing $\ZZ_2$ elements cutting a bifundamental edge.

\subsection{Discrete Gauging: UV and SW Curve Analysis}
\label{sec:swsection2}

We have seen in the previous sections that, at the level of the Higgs branch of a 5d SCFT, one can implement a discrete gauging by wreathing the corresponding magnetic quiver. It is, in principle, not obvious how and if the group action of the wreathing procedure corresponds to a symmetry of the full theory. The goal of this section is to show how to extract this information and use it to obtain the Seiberg--Witten curve of the discretely gauged theory. The key ingredient is the relation between the brane web that engineers the 5d SCFT in Type IIB and the mirror of the CY that engineers the same SCFT in M-theory. 

In Section \ref{subsec:action_on_curve}, we discuss how to lift the wreathing action to the $(p,q)$-web, showing the existence of a symmetry of the SCFT associated with the wreathing. In Section \ref{sec:swcurvediscrgauging}, we show how to use this lift to obtain the SW curves of the discretely-gauged theories. Finally, in Section \ref{sec:warmupcurve}, we show our prescription at work in the example of the $E_6$ theory, complementing the analysis of Section \ref{sec:E6th} with the SW curve analysis.

\subsubsection{Lifting of the Wreathing Action to the \texorpdfstring{$(p,q)$}{(p,q)}-web}
\label{subsec:action_on_curve}

The duality chain we wish to exploit requires placing the Type IIB setup introduced in Section \ref{sec:brane_webs} on $\mathbb{R}^{1,7}\times T^2$, with the branes extended along one cycle of the torus and transverse to the other. In particular, this means that we are considering the 5d SCFT compactified on a circle; in other words, a 4d KK theory. We can then T-dualize along the two different directions.

T-duality along the transverse $S^1$, followed by the M-theory uplift, transforms the $(p,q)$-web 5-branes into KK monopoles \cite{Leung:1997tw} reconstructing the toric Calabi--Yau threefold. If instead we T-dualize along the other $S^1$ present in the Type IIB setup, we find D4-branes and NS5-branes in the Type IIA dual. One can also uplift these to M-theory, producing a recombined M5-brane \cite{Witten:1997sc}. This second setup is more convenient to make contact with the Seiberg--Witten curve of the 4d KK-theory \cite{Aharony:1997bh} (see also \cite{Magureanu:2023rrg} for a nice review). The tension of the M2-branes suspended on the recombined M5 naturally gives rise to the Seiberg--Witten differential \cite{Witten:1997sc}, and the geometry that the M5 wraps gives rise to the SW curve. Determining what curve corresponds to each brane web is also straightforward \cite{Aharony:1997bh}, and it amounts to finding the local mirror of the toric Calabi--Yau  \cite{Hori:2000kt}. Starting from the toric diagram, assign to each point with lattice coordinates $(a,b)$ a monomial $x^a y^b$. Then the local mirror is
\begin{equation}
    \begin{dcases}
        w = P(x,y)= \sum c_{a,b}\, x^ay^b\coma\\
        w=uv\coma
    \end{dcases}
\end{equation}
where the sum is over the points in the toric diagram, $u,v,w\in\mathbb{C}$ and $x,y\in\mathbb{C}^*$. In what follows, we will focus on the fiber
\begin{align}
    P(x,y)=0\coma
\end{align}
which we will refer to as the \emph{mirror curve} and denote by $\Sigma$. All the relevant information for our purposes is contained in this equation.

The constants $c_{a,b}$ are associated with the complex structure moduli of the mirror CY. Importantly, they are not all independent; one can exploit rescalings of the coordinates $(x,y)$ to note that
\begin{align}\label{eq:miror_equation}
    P(x,y)=0\quad\Leftrightarrow\quad \lambda_1 P(\lambda_2 x,\lambda_3 y) = 0\fstop
\end{align}
Namely, we can use this freedom to set three of the constants equal to 1. The remaining ones correspond to K\"ahler moduli of the toric Calabi--Yau and extended Coulomb branch parameters of the 4d KK-theory. More precisely, one still needs to take into account the redundancy given by the Weyl group, $W$, of the continuous global symmetry: the space of inequivalent mass deformations is $T_{\CC} \, / \, W$ where $T_{\CC}$ is the complexified Cartan torus. The surviving $c_{a,b}$ are coordinates on $T_\CC$ and are thus subject to equivalence under $W$. This can be clearly seen in the Weierstrass form of the curve, where the $T_{\mathbb C}$-coordinates re-organize in $W$-invariant combinations. For our purposes, this means that we take the Seiberg--Witten geometry to be a fibration over $\CC \times T_{\CC} / W$.\footnote{Sometimes it can be useful to consider the SW geometry as a fibration over $\mathbb{C}\times T_{\mathbb C}$ instead; as some properties, such as how the Weyl group acts on the BPS spectrum of the theory, are encoded in that structure \cite{Argyres:2015gha}.}

\begin{figure}[!htp]
\centering
\begin{tikzpicture}[font=\footnotesize,baseline=0]
\begin{axis}[
    xlabel={$\ln|x|$},
    ylabel={$\ln|y|$},
    xmin=-4, xmax=4,
    ymin=-5, ymax=6,
    axis lines=center,
    domain=-4:4,
    width=10cm,
    height=8cm,
    axis line style={-},
    tick label style={font=\footnotesize},
    label style={font=\footnotesize},
    scale=0.9,
]

\addplot[name path=upper, blue!80!black, thick, samples=100] {ln(exp(2*x) + 4.7048*exp(x) + 1)};
\addplot[name path=lower, blue!80!black, thick, samples=501, restrict y to domain=-5:10] {ln(abs(exp(2*x) - 4.7048*exp(x) + 1))};

\addplot [blue!10] fill between [of=upper and lower];

\draw[red!80, dashed, thick] (-1.5,-5) -- (-1.5,0); 
\draw[red!80, dashed, thick] (1.5,-5) -- (1.5,3); 
\draw[red!80, dashed, thick] (-1.5,0) -- (1.5,3); 
\draw[red!80, dashed, thick] (-4,0) -- (-1.5,0); 
\draw[red!80, dashed, thick] (1.5, 3) -- (3, 6); 

\node[below left] at (-1.5,-4) {$\ln|x_1|$};
\node[below right] at (1.5,-4) {$\ln|x_2|$};
\fill[black] (-1.5,0) circle (1.5pt);
\fill[black] (1.5,3) circle (1.5pt);

\end{axis}
\end{tikzpicture}
\caption{Example of amoeba projection constructed using $P(x,y) = (x-x_1)(x-x_2) + y = 0$.}
\label{fig:amoeba_comparison}
\end{figure}
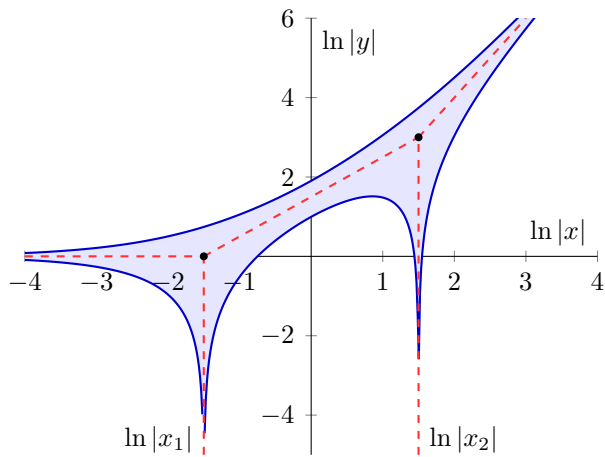

The key for our purposes is that there is an intimate relation between the brane web and the mirror curve \cite{Feng:2005gw,Kim:2014nqa,Arias-Tamargo:2024fjt}. One way to make this connection explicit is via the so-called \emph{amoeba projection} \cite{Feng:2005gw},
\begin{align}\label{eq:amoeba}
    \mathcal{A}_\Sigma = \{(\log |x|,\log|y|)\,, \text{ s.t. }\, P(x,y)=0\}\subset \mathbb{R}^2\fstop
\end{align}
Without loss of generality, assume that the mirror curve is written in the following form (one can always make use of SL$(2,\mathbb{Z})$ transformations to ensure that this is the case):
\begin{align}\label{eq:mirror_curve_choice}
    P(x,y) = \prod_i (x-x_i) + yP'(x,y)=0\coma
\end{align}
with only positive powers of $y$ appearing in $P'(x,y)$, and the roots $x_i$ are combinations of the constants $c_{a,b}$ in equation \eqref{eq:miror_equation}. In the limit $y\to 0$, the amoeba projection in equation \eqref{eq:amoeba} develops asymptotes in the vertical direction; and from equation \eqref{eq:mirror_curve_choice}, we see that they are located at points $\log|x_i|$ on the horizontal axis. These ``spikes'' of the amoeba correspond precisely to the external legs of the brane web in Type IIB, and the roots $x_i$ appearing in the polynomial correspond to the positions of the said external legs. We show an example of the amoeba projection in Figure \ref{fig:amoeba_comparison}. 

This connection allows us to translate the wreathing action from the magnetic quiver to the SW curve. The outcome of the discussion above is that, via the mapping from the brane web to the mirror curve, each external leg of the brane web is mapped to a pair of monomials in $P(x,y)$. 
The wreathing action on the magnetic quiver can be uplifted to an action on the subwebs of the brane web, and thus to an action on the monomials that appear in the Seiberg--Witten curve. Thus, the wreathing action can be related to an action on the coordinates $x$ and $y$. 

\subsubsection{SW curve of the discretely gauged theory}
\label{sec:swcurvediscrgauging}

In this section, we will provide the prescription to obtain the SW curve of the discretely gauged theory that we will use in Section \ref{sec:GaugeAutoEnTheories} to reproduce the results of the various wreathings.

In general, the discrete gauging of a zero-form symmetry broken on the CB produces a flat orbifold of it as the new CB \cite{Argyres:2016yzz,Arias-Tamargo:2023duo}. Stated differently, the CB special K\"ahler metric after the gauging can be obtained as an orbifold of the one before the quotient. The SW geometry has to be modified accordingly to encode the (mild) orbifold singularities \cite{Argyres:2016yzz,Arias-Tamargo:2023duo}. The symmetries that we are studying are, however, preserved on the CB. Consequently, the IR gauge coupling $\tau(u;m)$ and the special K\"ahler metric on the CB do not change. In particular, this means that since $\tau(u;m)$ uniquely fixes the holomorphic class of the fibral curve, the birational class of the SW geometry has to be the same as the theory before the quotient.\footnote{Let us notice that our analysis of the curve ignores the global structure of the theory that can be seen via more refined presentations of the SW curve \cite{Magureanu:2023rrg}; however, this has no effect on our analysis.  Our analysis hence just fixes the Jacobian of the SW geometry, or equivalently its birational class \cite{Donagi:1995cf,Magureanu:2023rrg}.} Such a birational class is uniquely labeled by the Jacobian fibration of the SW solution, which, in the case of rank-one theories, is obtained by passing to the Weierstrass presentation of the mirror curve:\footnote{This can be systematically achieved through a sequence of standard birational transformations, specifically, rescaling the polynomial to an algebraic curve, completing the square to obtain a hyperelliptic form, and applying a fractional linear (M\"obius) transformation.} 
\begin{equation}
\label{eq:weiform}
    Y^2 = 4X^3 - g_2(y_i, u) X - g_3(y_i, u) \,,
\end{equation}
with $Y,X$ being rational functions of the coordinates $(x,y)$, and the $y_i$ parametrizing the Cartan torus of the flavor symmetry. 

Since we need to preserve the K\"ahler metric on the CB, and hence its Weierstrass form in equation \eqref{eq:weiform}, \textit{the Weierstrass form of the SW geometry after the discrete gauging is the same as it was before the discrete gauging, with at most some restrictions on the space of mass parameters.} Hence, providing the SW curve after the discrete gauging is equivalent to finding the necessary restrictions (if any) to be imposed on the mass parameters. In the remaining part of this section, we provide the prescription to find such constraints.  
 
Choosing $y_i$ to be the fugacities used in \cite{Mitev:2014jza}, the Weierstrass form in equation \eqref{eq:weiform} reveals a fundamental property: the Weierstrass coefficients $g_2$ and $g_3$ are expressed in terms of characters of the UV flavor symmetry group \cite{Closset:2021lhd} and are therefore  invariant under the action of the Weyl group (and, more generally, under any inner automorphism of the flavor group).\footnote{In particular, in Section \ref{sec:GaugeAutoEnTheories} we will give the explicit action, as in equation \eqref{eq:actiongeneric}, of the various symmetry transformations. In some particular cases, the group composition relations satisfied by the action on $(x,y)$ will not mimic the inner-automorphism action on the $y_i$. However, since we are free to choose any inner automorphism action on the masses without changing the physics, it is irrelevant how that inner automorphism commutes (or not) with the automorphism induced by the diagram symmetry of the quiver.} Consequently, even if a given mirror curve $P(x,y)$ is not manifestly invariant under a Weyl transformation, its birational equivalence class, and thus the underlying physics, is preserved under such transformations. This also holds true for outer automorphisms if we restrict the fugacities to those of a non-simply laced subalgebra of the flavor algebra. This means that in both cases we are allowed to balance the action on $(x,y)$ given in Section \ref{subsec:action_on_curve} with an inner or an outer automorphism of the flavor algebra, leading to the following prescription to compute the mass constraints.

Consider the action
\begin{equation}
\label{eq:actiongeneric}
     (x,y) \to (x',y'), \qquad (y_i) \to (y_i') \,,
\end{equation}
with $y_i \to y_i'$ being either an inner or outer automorphism, and with $(x,y) \to (x',y')$ dictated by the procedure of Section \ref{subsec:action_on_curve}.\footnote{We note that $x'$ (similarly for $y'$) will always be of the form $\alpha(y_i)x^{n_1}y^{n_2}$, with $\alpha(y_i)$ a prefactor depending only on flavor fugacities.}
Then, impose that 
\begin{equation}
\label{eq:paramsfixingcondition}
    P(x,y;y_i) = k(y_i') P(x',y';y_i')
\end{equation}
for some functions $k(y_i')$ of the fugacities. 
We will see in Section \ref{sec:GaugeAutoEnTheories} that if the automorphism group is abelian and inner, equation \eqref{eq:paramsfixingcondition} does not impose any constraint on the $y_i$, and the rank of the flavor symmetry is preserved. On the other hand,  if the automorphism group is outer, equation \eqref{eq:paramsfixingcondition} can be satisfied only by imposing specific conditions on the $y_i$, which are exactly those that select the Cartan torus of the non-simply laced subalgebra  predicted by the magnetic quiver analysis inside the Cartan torus of the original flavor symmetry. 

Let us conclude with three comments. First, the Cartan torus is the same for both the mirror curve written in terms of $x,y$ and for its Weierstrass form. Hence, if we compute the Weierstrass form of the curve again after imposing the aforementioned constraints for the curve written in terms of $x,y$, we find the same expression that we would obtain by imposing the constraints directly at the level of the original Weierstrass curve.

Secondly, to infer the flavor algebra (and not just its rank) from the curve of  a (4d) SCFT \cite{Argyres:2015ffa,Argyres:2016xua}, one assigns spurionic weights $\omega_i$ to the deformation coefficients of the curve by exploiting the scaling symmetry, and then looks for a Lie group whose generators of $\mathfrak h/W$, $\mathfrak h$ the Cartan algebra, are polynomials of degree $\omega_i$ when written in terms of $\mathfrak h$ coordinates. Given the proposed Seiberg--Witten curves for our discretely-gauged SCFTs, a 5d analogue of this method may be useful to predict consistent continuous flavor algebras.

Finally, after the gauging of either an inner or an outer automorphism, the mirror curve written in terms of $x,y$ is no longer gauge invariant, as it contains fugacities that are not fixed by the discretely gauged symmetry. Hence, it has to be thought of as a gauge-covariant object, whose gauge-invariant physical information is encoded in its Jacobian fibration, or equivalently in its Weierstrass form. This obstructs an immediate interpretation in Type IIA for the SW geometry of the discretely gauged theory, which cannot be regarded as the mirror of the original toric Calabi--Yau.

\subsubsection{Example: the \texorpdfstring{$E_6$}{E6} Theory}\label{sec:warmupcurve}

Let us continue our warm-up by considering again the example of the $E_6$ theory, whose toric diagram is given in Figure \ref{fig:E6toricdiagram}. The curve is
\begin{equation}\label{eq:SW_E6_first}
    \begin{split}
    P(x,y) = &\,u + c_{(1,0)} x + c_{(0,1)} y +  c_{(-1,0)} \frac{1}{x} +  c_{(0,-1)} \frac{1}{y} \\
    &+ c_{(-1,-1)} \frac{1}{xy} + c_{(-1,1)} \frac{y}{x} + c_{(1,-1)} \frac{x}{y} + c_{(2,-1)} \frac{x^2}{y} + c_{(-1,2)} \frac{y^2}{x}\fstop 
    \end{split}
\end{equation}

In this section, we want to reproduce the wreathing of Section \ref{sec:E6th} at the level of the SW curve. Following \cite{Kim:2014nqa}, let us first
use the rescaling  freedom to fix the curve as follows:\footnote{Our curve is obtained from that of \cite{Kim:2014nqa} by calling $t_{there} \to x, w_{there} \to y$ and by shifting the origin by $(-1,-1)$. Equivalently, we divide the curve of \cite{Kim:2014nqa} by $\frac{1}{xy}$.} 
\begin{equation}
\label{eq:e6curvekimyagi}
\begin{alignedat}{5}
  & \frac{y^{2}}{x} & 
  & -\left(\sum_{i=1}^{3} M_i\right)\frac{y}{x} & 
  & +\left(\sum_{i=1}^{3} L_i\right)y & 
  & +\left(\sum_{i=1}^{3} M_i^{-1}\right)\frac{1}{x} & 
  & +U -\frac{1}{xy} \\
  {}+{} & \frac{x^{2}}{y} & 
  & -\left(\sum_{i=1}^{3} N_i\right)\frac{x}{y} & 
  & +\left(\sum_{i=1}^{3} L_i^{-1}\right)x & 
  & +\left(\sum_{i=1}^{3} N_i^{-1}\right)\frac{1}{y} & 
  & =0\coma
\end{alignedat}
\end{equation}
with 
\begin{equation}
\label{eq:determinantconstraint}
   M_1 M_2 M_3 = N_1 N_2 N_3 = L_1 L_2 L_3  = 1\fstop
\end{equation}
The $L_i$ (respectively, the $N_i$ and the $M_i$) are the fugacities of the three factors of the maximal regular subalgebra 
\begin{equation}
    \mathfrak{su}(3)_{L} \oplus\mathfrak{su}(3)_{M} \oplus \mathfrak {su}(3)_N \subset E_6\coma
\end{equation}
with equation \eqref{eq:determinantconstraint} being the requirement that the determinant of the element of the Cartan torus is one. 

Let's consider the $\mathbb Z_2$ wreathing of equation \eqref{eq:E6-wrZ2}; whose Hilbert series was given in equation \eqref{eq:E6Z2-HS-f4}. The action on the nodes of the magnetic quiver corresponds to exchanging the horizontal and vertical sides of the toric diagram of Figure \ref{fig:E6toricdiagram}. This can be achieved by acting on the $(x,y)$ coordinates as
\begin{align}
\label{eq:coords_action_E6_example}
    (x,y)\to(y,x)\fstop
\end{align}
From the discussion in Section \ref{sec:dg}, we know that this $\mathbb{Z}_2$ acts as an outer automorphism of the flavor algebra. Mass parameters transform in the adjoint representation of the flavor symmetry. Because the outer automorphism permutes the simple roots of the flavor Lie algebra, it induces a simultaneous permutation of the associated mass parameters $M_i$, $N_i$, and $L_i$ so as to leave the Seiberg--Witten curve invariant. The (outer) action on them is\footnote{This is the $\mathbb{Z}_2$ outer-automorphism induced by the affine $\mathfrak{e}_6$ Dynkin diagram automorphism, composed with non-physically-relevant inner automorphisms.} 
\begin{equation}
\label{eq:outerz2flavorpart}
  (M_i,N_i,L_i)  \to    (M_i^{-1},N_i^{-1},L_i^{-1})\fstop
\end{equation}
One sees immediately that equation \eqref{eq:coords_action_E6_example} together with equation \eqref{eq:outerz2flavorpart} does not, in general, preserve the mirror curve in equation \eqref{eq:e6curvekimyagi}. For the symmetry to be preserved, we need to impose
\begin{equation}
\label{eq:fugacitiesidoutere6}
    M_i = N_i^{-1}\fstop
\end{equation}
Since the number of independent constants counts the rank of the global symmetry, we see that this condition reduces said rank from six to four. Furthermore, note that by applying equation \eqref{eq:outerz2flavorpart} to a solution of equation \eqref{eq:fugacitiesidoutere6}, we obtain another solution of equation \eqref{eq:fugacitiesidoutere6}. Namely, equation \eqref{eq:fugacitiesidoutere6} defines a subalgebra of the flavor group that is invariant under the outer automorphism; hence, from general Lie algebra theory, it is isomorphic to either $\mathfrak{f}_4$ or $\mathfrak{usp}(8)$. This analysis confirms that the rank of the resulting algebra is insufficient to distinguish between the two options; however, it is consistent with the reduction of the flavor algebra as
\begin{equation}
    \mathfrak{e}_6 \to \mathfrak{f}_4\coma 
 \end{equation}
confirming the wreathed magnetic quiver computation.

\section{Discrete Gauging in the Rank-one \texorpdfstring{\boldmath{$E_n$}}{En} Theories}
 \label{sec:GaugeAutoEnTheories}

Now that we have introduced the necessary material, we turn to the analysis of the proposed discrete gauging of the remaining rank-one Seiberg theories. For each rank-one $E_n$ SCFT, we follow this procedure. 

\begin{itemize}
\item We begin with known magnetic quivers for the Higgs branch of the SCFT and find the finite group of diagram-automorphisms of these quivers.\footnote{As we will see, there may be multiple magnetic quivers for the Higgs branch, and the Higgs branch can also be the union of the Coulomb branches of a collection of magnetic quivers.} For each magnetic quiver $\mathcal{T}_M$, this produces a group $\operatorname{Aut}(\mathcal{T}_M)$. As discussed in Section \ref{sec:dg}, there exists a homomorphism
\begin{equation}
    \rho\,:\,\,\operatorname{Aut}(\mathcal{T}_M) \rightarrow \operatorname{Aut}(\mathfrak{e}_n) \,,
\end{equation}
into the automorphism group of the (finite) $\mathfrak{e}_n$ flavor symmetry. When $\mathcal{T}_M$ is the unitary magnetic quiver in the shape of the affine Dynkin diagram of $\mathfrak{e}_n$, the homomorphism $\rho$ is known, and we determine the fixed-point subalgebra, $\mathfrak{e}_n^{\rho|_H(H)}$, defined in equation \eqref{eqn:gHdef}, of the automorphism $\rho|_H$ restricted to each subgroup $H \subseteq \operatorname{Aut}(\mathcal{T}_M)$. This $\mathfrak{e}_n^{\rho|_H(H)}$ would be the continuous flavor symmetry of the putative discretely-gauged SCFT, which we refer to as $E_n \, / \,  H$. 

\item Secondly, using the wreathed monopole formula, we compute the Coulomb branch Hilbert series of the wreathed magnetic quiver, $\operatorname{HS}_{E_n \wr H}(t, \bm{x})$, where we introduced the shorthand of referring to the wreathed magnetic quiver as $E_n \wr H$. For the unitary magnetic quivers, we see that the coefficient of the $t^2$ term of $\operatorname{HS}_{E_n \wr H}(t, \bm{x})$ is consistent with an $\mathfrak{e}_n^{\rho|_H(H)}$ Coulomb branch flavor symmetry in all cases. For $\mathcal{T}_M$ orthosymplectic, we observe the nontrivial fact that there exists a homomorphism $H \rightarrow \operatorname{Aut}(\mathfrak{e}_n)$ whose fixed point algebra agrees with the Coulomb branch Hilbert series.
\item Thirdly, when we can construct a Seiberg--Witten curve, describing the Coulomb branch of the $E_n$ SCFT, which admits a manifest $H$-action, we restrict the SW family to the invariant mass locus, and find that the number of mass parameters in the restricted curve reproduces the rank of $\mathfrak{e}_n^{\rho|_H(H)}$.
\end{itemize}
We have summarized the results for the wreathing of the unitary and orthosymplectic magnetic quivers in Tables \ref{tbl:Uwreathing} and \ref{tbl:OSwreathing}, respectively. 

 \begin{table}[p]
 \setlength{\tabcolsep}{4pt}
 \renewcommand{\arraystretch}{1}
 \centering
 \begin{tabular}{c|c|c|c|c}
         Theory  & Quiv.~Aut. & Wr.   & Hilbert Series & Flavor Symm.\\
        \hhline{=|=|=|=|=}
      $E_0$  & $\ID$  & $\ID$  & \eqref{eq:E0HS}  
      &  $\emptyset$ \\
      \hline                                  
      $\widetilde{E}_1$\rule{0pt}{2.8ex}  & $\ID$  & $\ID$  & \eqref{eq:tE1HS}  
      &   $\mathfrak{u}(1)$   \\
      \hline                                  
      \multirow{2}{*}{$E_1$} & \multirow{2}{*}{$\ZZ_2$}       & $\ID$  & \eqref{eq:E1HS-uni}  
      &   $\su(2)$   \\
                             &   & $\ZZ_2$    & \eqref{eq:E1wreathed}  
                             &  $\mathfrak{u}(1)$  \\
      \hline                  
      \multirow{2}{*}{$E_2$}  & \multirow{2}{*}{$\ZZ_2$}       & $\ID$        & \eqref{eq:E2HS}   
      &   $\su(2)\oplus \mathfrak{u}(1)$   \\
                              &    & $\ZZ_2$   & \eqref{eq:E2Z2HS}  
                              &  $\mathfrak{u}(1)^{\oplus 2}$    \\
       \hline
      \multirow{8}{*}{$E_3$}  & \multirow{8}{*}{$S_3 \times \ZZ_2^{(2)}$} & $\ID$   & \eqref{eq:E3HS-uni} & $\su(3)\oplus \su(2)$\\
                                 &    & $\ZZ_2^{(1)}$       & \eqref{eq:O3Z2E1-HS} & $\so(3)\oplus \su(2)$\\
                                   &    & $\ZZ_2^{(2)}$       & \eqref{eq:O3E1Z2-HS} & $\su(3)\oplus \mathfrak{u}(1)$\\
                            &    & $\ZZ_3$ & \eqref{eq:O3Z3E1-HS} & $\mathfrak{u}(1)^{\oplus 2}\oplus \su(2)$\\
                           &    & $\ZZ_2^{DT}$ or $\ZZ_2^{(1)} \times \ZZ_2^{(2)}$ & \eqref{eq:O3Z2E1Z2-HS} & $\so(3)\oplus \mathfrak{u}(1)$\\
                           &    & $\ZZ_3 \times \ZZ_2^{(2)}$ & \eqref{eq:O3Z3E1Z2-HS} & $\mathfrak{u}(1)^{\oplus 3}$ \\
                           &    & $S_3$       & \eqref{eq:O3S3E1-HS} & $\su(2)$\\
                          &   & $S_3^{DT}$ or $S_3 \times \ZZ_2^{(2)}$ & \eqref{eq:O3S3E1Z2-HS} & $\mathfrak{u}(1)$ \\
      \hline
      \multirow{4}{*}{$E_4$}   & \multirow{4}{*}{$D_{10}$}      & $\ID$  & \eqref{eq:E4-HS} & $\su(5)$\\
                               &                                & $\ZZ_2$  & \eqref{eq:E4Z2-HS} & $\so(5)$\\
                               &                                & $\ZZ_5$ & \eqref{eq:E4Z5-HS} & $\mathfrak{u}(1)^{\oplus 4}$\\
                               &                                & $D_{10}$ & \eqref{eq:E4D5-HS} & $\emptyset$ \\
      \hline
       \multirow{8}{*}{$E_5$}   & \multirow{8}{*}{$D_8$} & $\ID$ & \eqref{eq:E5-HS} & $\so(10)$\\
                 &    & $\ZZ_2^{(1)}$ & \eqref{eq:E5Z21-HS} & $\so(9)$\\
                 &      & $\ZZ_2^{DT}$ & \eqref{eq:E5DT-HS} & $\so(8)\oplus \mathfrak{u}(1)$\\
                  &  & $\ZZ_2^{(3)}$  & \eqref{eq:E5Z23-HS} & $\so(5)\oplus \so(5)$\\
                       &      & $\ZZ_4$ & \eqref{eq:E5Z4-HS} & $\so(4)\oplus \so(4)\oplus \mathfrak{u}(1)$\\
                         &  & $\ZZ_2^{(1)} \times \ZZ_2^{(2)}$ & \eqref{eq:E5wrZ2U2adjoints} & $\so(8)$\\
               &      & $\ZZ_2^{DT} \times \ZZ_2^{(3)}$ & \eqref{eq:E5DtZ23-HS} & $\so(4)\oplus \so(4)$\\
   & & $D_8$ & \eqref{eq:E5D8-HS} & $\so(4)\oplus \so(4)$\\
        \hline
        \multirow{4}{*}{$E_6$}   & \multirow{4}{*}{$S_3$}     & $\ID$  & \eqref{eq:E6-HS} & $\mathfrak{e}_6$\\
                                 &       & $\mathbb{Z}_2$ & \eqref{eq:E6Z2-HS} & $\mathfrak{f}_4$\\
                                 &     & $\mathbb{Z}_3$  & \eqref{eq:E6Z3-HS} & $\so(8)\oplus \mathfrak{u}(1)^{\oplus 2}$\\
                                 &       & $S_3$ & \eqref{eq:E6S3-HS} & $\so(8)$\\
         \hline
        \multirow{2}{*}{$E_7$}   & \multirow{2}{*}{$\mathbb{Z}_2$} & $\ID$  & \eqref{eq:E7-HS} & $\mathfrak{e}_7$\\
                                 &                                 & $\mathbb{Z}_2$  & \eqref{eq:E7Z2-HS} & $\mathfrak{e}_6\oplus \mathfrak{u}(1)$\\
        \hline
        \multirow{1}{*}{$E_8$}    & $\ID$       & $\ID$ & \eqref{eq:E8HS-unitary} & $\mathfrak{e}_8$\\
\end{tabular}
\caption{\normalsize The (unitary) $E_n$ wreathings, their Hilbert series and flavor symmetries.}\label{tbl:Uwreathing}
\end{table}

\begin{table}[p]
\setlength{\tabcolsep}{3pt}
\renewcommand{\arraystretch}{2.3}
\resizebox{\textwidth}{!}{
 \begin{tabular}{c|c|c|c|c}   
         Th.  & Quiver & Wr.   & HS & Flavor Symm.\\
            \hhline{=|=|=|=|=}
      \multirow{2}{*}{$E_4$}   & \multirow{2}{*}{ \begin{tikzpicture}[baseline=0,font=\footnotesize]
        \node[node, label=below:{$2$},fill=blue] (A2) {};
        \node[node, label=below:{$2$},fill=red] (Al) [left=6mm of A2] {};
        \node[node, label=below:{$2$},fill=red] (Ar) [right =6mm of A2]{};
        \node[node, label=right:{$1$}] (A3) [above=4mm of A2] {};
        \node[flavor, label=right:{$1$}] (A7) [above=6mm of A3] {};
        \draw (A2.north) -- (A3.south);
        \draw[snake it] (A3.north) -- (A7.south);
        \draw (Al.east) -- (A2.west);
        \draw (A2.east) -- (Ar.west);
    \end{tikzpicture}} & $\ID$    & \eqref{eq:HSE4-ortho}   & $\su(5)$\\
                               &                   & $\ZZ_2$  & \eqref{eq:HSE4wrZ2-ortho}   & $\mathfrak{so}(6)\oplus \mathfrak{u}(1)$\\
      \hline
       \multirow{2}{*}{$E_5$}   & \multirow{2}{*}{ \begin{tikzpicture}[baseline=0,font=\footnotesize]
        \node[node, label=below:{$4$},fill=red] (A2) {};
        \node[node, label=below:{$2$},fill=blue] (Al1) [left=6mm of A2] {};
        \node[node, label=below:{$2$},fill=blue] (Ar1) [right =6mm of A2]{};
        \node[node, label=below:{$2$},fill=red] (Al2) [left=6mm of Al1] {};
        \node[node, label=below:{$2$},fill=red] (Ar2) [right =6mm of Ar1]{};
        \node[node, label=right:{$1$}] (A3) [above=4mm of A2] {};
        \draw (A2.north) -- (A3.south);
        \draw (Al2) -- (Al1) -- (A2) -- (Ar1) -- (Ar2);
    \end{tikzpicture}} & $\ID$    & \eqref{eq:HSE5-ortho}   & $\so(10)$\\
                               &                   & $\ZZ_2$  & \eqref{eq:HSE5wrZ2-ortho}   & $\so(8)\oplus\mathfrak{u}(1)$\\
      \hline
      \multirow{2}{*}{$E_6$}   & \multirow{2}{*}{\begin{tikzpicture}[baseline=0,font=\footnotesize]
        \node[node, label=below:{$4$},fill=blue] (A2) {};
        \node[node, label=below:{$4$},fill=red] (Al1) [left=6mm of A2] {};
        \node[node, label=below:{$4$},fill=red] (Ar1) [right =6mm of A2]{};
        \node[node, label=below:{$2$},fill=blue] (Al2) [left=6mm of Al1] {};
        \node[node, label=below:{$2$},fill=blue] (Ar2) [right =6mm of Ar1]{};
        \node[node, label=below:{$2$},fill=red] (Al3) [left=6mm of Al2] {};
        \node[node, label=below:{$2$},fill=red] (Ar3) [right =6mm of Ar2]{};
        \node[node, label=right:{$2$},fill=red] (A3) [above=4mm of A2] {};
        \draw (A2.north) -- (A3.south);
        \draw (Al3) -- (Al2) -- (Al1) -- (A2) -- (Ar1) -- (Ar2) -- (Ar3);
    \end{tikzpicture}} & $\ID$    & \eqref{eq:HSE6-ortho}   & $\mathfrak{e}_6$\\
                               &                   & $\ZZ_2$  & \eqref{eq:HSE6wrZ2-ortho}   & $\so(10)\oplus \mathfrak{u}(1)$\\
      \hline
      \multirow{2}{*}{$E_7$}   & \multirow{2}{*}{\begin{tikzpicture}[baseline=0,font=\footnotesize]
        \node[node, label=below:{$6$},fill=red] (A2) {};
        \node[node, label=below:{$4$},fill=blue] (Al1) [left=6mm of A2] {};
        \node[node, label=below:{$4$},fill=blue] (Ar1) [right =6mm of A2]{};
        \node[node, label=below:{$4$},fill=red] (Al2) [left=6mm of Al1] {};
        \node[node, label=below:{$4$},fill=red] (Ar2) [right =6mm of Ar1]{};
        \node[node, label=below:{$2$},fill=blue] (Al3) [left=6mm of Al2] {};
        \node[node, label=below:{$2$},fill=blue] (Ar3) [right =6mm of Ar2]{};
        \node[node, label=below:{$2$},fill=red] (Al4) [left=6mm of Al3] {};
        \node[node, label=below:{$2$},fill=red] (Ar4) [right =6mm of Ar3]{};
        \node[node, label=right:{$2$},fill=blue] (A3) [above=4mm of A2] {};
        \node[node, label=right:{$1$}] (A3a) [above=4mm of A3] {};
        \draw (A2) -- (A3) -- (A3a);
        \draw (Al4) -- (Al3) -- (Al2) -- (Al1) -- (A2) -- (Ar1) -- (Ar2) -- (Ar3) -- (Ar4);
    \end{tikzpicture}} & $\ID$    & \eqref{eq:HSE7-ortho}   & $\mathfrak{e}_7$\\
                               &                   & $\ZZ_2$  & \eqref{eq:HSE7wrZ2-ortho}   & $\mathfrak{e}_6\oplus\mathfrak{u}(1)$\\
      \hline
      \multirow{2}{*}{$E_8$}   & \multirow{2}{*}{\begin{tikzpicture}[baseline=0,font=\footnotesize]
        \node[node, label=below:{$8$},fill=red] (A2) {};
        \node[node, label=below:{$6$},fill=blue] (Al1) [left=6mm of A2] {};
        \node[node, label=below:{$6$},fill=blue] (Ar1) [right =6mm of A2]{};
        \node[node, label=below:{$6$},fill=red] (Al2) [left=6mm of Al1] {};
        \node[node, label=below:{$6$},fill=red] (Ar2) [right =6mm of Ar1]{};
        \node[node, label=below:{$4$},fill=blue] (Al3) [left=6mm of Al2] {};
        \node[node, label=below:{$4$},fill=blue] (Ar3) [right =6mm of Ar2]{};
        \node[node, label=below:{$4$},fill=red] (Al4) [left=6mm of Al3] {};
        \node[node, label=below:{$4$},fill=red] (Ar4) [right =6mm of Ar3]{};
        \node[node, label=below:{$2$},fill=blue] (Al5) [left=6mm of Al4] {};
        \node[node, label=below:{$2$},fill=blue] (Ar5) [right =6mm of Ar4]{};
        \node[node, label=below:{$2$},fill=red] (Al6) [left=6mm of Al5] {};
        \node[node, label=below:{$2$},fill=red] (Ar6) [right =6mm of Ar5]{};
        \node[node, label=right:{$2$},fill=blue] (A3) [above=4mm of A2] {};
        \draw (A2) -- (A3);
        \draw (Al6) -- (Al5) -- (Al4) -- (Al3) -- (Al2) -- (Al1) -- (A2) -- (Ar1) -- (Ar2) -- (Ar3) -- (Ar4) -- (Ar5) -- (Ar6);
    \end{tikzpicture}} & $\ID$    & \eqref{eq:HSE8-ortho}   & $\mathfrak{e}_8$\\
                               &                   & $\ZZ_2$  & \eqref{eq:HSE8wrZ2-ortho}   & $\mathfrak{e}_7\oplus \su(2)$
     \end{tabular}
     }
     \caption{\normalsize The (orthosymplectic) $E_n$ wreathings, their Hilbert series and flavor symmetries.}\label{tbl:OSwreathing}
\end{table}

\subsection{The Rank-one \texorpdfstring{$E_0$}{E0} Theory}
\label{sec:E0th}

We begin with the $E_0$ theory, for which the Higgs branch is, in fact, trivial. Being a point, the Hilbert series is just 
\begin{equation}\label{eq:E0HS}
    \HS_{E_0}(t) = 1 \fstop
\end{equation}
We do not need to consider this theory any further; we include it here only for completeness.

\subsection{The Rank-one  \texorpdfstring{$\widetilde{E}_1$}{tE1} Theory}
\label{sec:tE1th}

The Higgs branch of $\widetilde{E}_1$ is given by the gaugino bilinear $S$, which is nilpotent $S^2 = 0$. The space generated by this operator is $\operatorname{Spec} \mathbb{C}[S] / (S^2)$, whereas the classical Higgs branch is trivial. The Hilbert series for $\widetilde{E}_1$ can be computed directly:
\begin{equation}\label{eq:tE1HS}
    \HS_{\widetilde{E}_1}(t) = 1+ t^2\coma
\end{equation}
where the $t^2$ term captures the operator associated to $S$. This is consistent with \cite{Cremonesi:2015lsa,Giacomelli:2024sex}. As for the $E_0$ theory, there are no discrete symmetries that we consider gauging.

\subsection{The Rank-one \texorpdfstring{$E_1$}{E1} Theory}
\label{sec:E1th}

We now turn to the first example of a rank-one Seiberg theory for which there exists a putative discrete gauging of a zero-form symmetry. The $E_1$ theory is obtained by considering the toric diagram in equation \eqref{eq:fromTDtoPQWeb}. The theory has a nontrivial Higgs branch which is $\mathbb{C}^2 / \mathbb{Z}_2$. Let us proceed by presenting the magnetic quiver and the mirror curve of this theory, describing their symmetries.

\paragraph{Discrete Symmetries from Magnetic Quivers for the Higgs Branch:} As discussed around equation \eqref{eqn:dora}, there exists an unframed unitary quiver for a 3d $\mathcal{N}=4$ theory that has $\CC^2/\ZZ_2$ as its Coulomb branch. This is simply
\begin{equation}\label{eq:unitaryE1}
    \begin{tikzpicture}[baseline=0,font=\footnotesize]
        \node[node, label=below:{$1$}] (A1) {};
        \node[node, label=below:{$1$}] (A2) [right=8mm of A1] {};
        \draw[double distance=3pt] (A1) -- (A2);
        \node (Z2) [right=6mm of A2] {$/\U(1)$};
    \end{tikzpicture}\coma
\end{equation}
where, again, we denote by $/ \U(1)$ the decoupling of the center-of-mass $\U(1)$. 
The group of quiver automorphisms of this unitary quiver is simply a $\mathbb{Z}_2$ that is generated by the exchange of the two nodes of the diagram:
\begin{equation}\label{eqn:t}
    t \,: \,\,\begin{tikzpicture}[baseline=-8,font=\footnotesize]
        \node[node, label=below:{$1$}] (A1) {};
        \node[node, label=below:{$1$}] (A2) [right=8mm of A1] {};
        \draw[double distance=3pt] (A1) -- (A2);
        \node (Z2) [right=6mm of A2] {$/\U(1)$};
        \draw[Triangle-Triangle,red] ([yshift=-5mm]A1.south) to[bend right=60] 
        ([yshift=-5mm]A2.south);
    \end{tikzpicture}\fstop
\end{equation}
We have labeled the generator of this $\mathbb{Z}_2$ as $t$ for future reference. This 
automorphism of the affine $\mathfrak{su}(2)$ Dynkin diagram induces a nontrivial automorphism on the finite $\mathfrak{su}(2)$ algebra via the homomorphism in equation \eqref{eqn:BIGHOMO}; this inner automorphism swaps the roots:
\begin{equation}
    \alpha \,\, \longleftrightarrow \,\, - \alpha\fstop
\end{equation}
Thus, the fixed-point subalgebra under the action of this automorphism is 
\begin{equation}
    \mathfrak{su}(2)^{\mathbb{Z}_2} = \mathfrak{u}(1)\fstop
\end{equation}
This is the expected continuous flavor symmetry of the putative rank-one theory obtained by discretely-gauging this $\mathbb{Z}_2$ symmetry. 

\paragraph{Mirror Curve:}
From the toric diagram in equation \eqref{eq:fromTDtoPQWeb}, one obtains the following mirror curve:
\begin{align}
\label{eq:curvee1revisited}
    P_{E_1}(x,y)= \lambda^{1/4}u + \sqrt{\lambda}\left(\frac{1}{x}+x\right)+y+\frac{1}{y}\coma
\end{align}
with the following large-volume identifications: 
\begin{equation}
    \lambda = \frac{Q_{B}}{Q_F}\coma  u = \frac{\lambda^{1/4}}{\sqrt{Q_F}}\coma 
\end{equation}
with $Q_F, Q_B$ the fibral and the base K\"ahler volumes of the local $\mathbb P^1 \times \mathbb P^1$. 

We note that our choice of the CB parameter differs from that of \cite{Magureanu:2023rrg} by a factor of $\lambda^{1/4}$, and matches that of \cite{Mitev:2014jza},\footnote{More precisely, $u$ here is $\tilde{A}^{-1}$ of \cite{Mitev:2014jza}.} producing a CB parameter that is invariant under the Weyl symmetry of the $E_1$ flavor group \cite{Mitev:2014jza}. The reason we are allowed to do so is that the CB coordinate is defined up to shifts by background masses $u \to u f(\lambda,m)$ \cite{Mitev:2014jza,Closset:2021lhd}, as the effective CB metric $\partial_{u} a^D / \partial_u a$   is not affected by it.

\subsubsection{The Higgs Branch of the Rank-one \texorpdfstring{$E_1$}{E1} Theory}

We first recall the features of the Higgs branch of the rank-one $E_1$ theory, and then, in the subsequent subsections, we discuss how the Higgs branch changes if we discretely gauge the aforementioned $\mathbb Z_{2}$ zero-form symmetry. 

As we have said, the Higgs branch of the $E_1$ theory is $\mathbb{C}^2/\mathbb{Z}_2$, which is isomorphic to the reduced moduli space of one $\SU(2)$ instanton. The Hilbert series for $\CC^2/\ZZ_2$ can be computed explicitly, e.g., with \texttt{Macaulay2} \cite{M2}, obtaining the closed form expression
\begin{equation}\label{eq:HSC2Z2}
    \HS_{\CC^2/\ZZ_2}(t,z) = \frac{1-t^4}{(1-t^2)(1-z^2t^2)(1-z^{-2}t^2)}\coma
\end{equation}
where $z$ denotes the fugacity for the $\SU(2)$ isometry of $\CC^2/\ZZ_2$. 

We now explain how to recover the Hilbert series for $\mathbb{C}^2/\mathbb{Z}_2$ as in equation \eqref{eq:HSC2Z2} from the magnetic quiver for the Higgs branch in equation \eqref{eq:unitaryE1}. The general expression for the Hilbert series of this quiver is 
\begin{equation}
    \HS_{E_{1}}(t,z) = \frac{1}{1-t^2}\sum_{(m_1,m_2)\in \Gamma_{\widetilde{G}}} t^{2|m_1-m_2|}z^{2(m_1-m_2)}\coma
\end{equation}
where $\Gamma_{\widetilde{G}}$ is the lattice of magnetic weights for the naive gauge algebra $\widetilde{G}=\U(1)\times \U(1)$, while $z$ is the fugacity of the topological $\U(1)$ symmetry under which the magnetic monopoles are charged.\footnote{We have already defined it as $z^2$ to obtain directly the result in equation \eqref{eq:HSC2Z2} without the need of rescaling the fugacity.} The computation of the Hilbert series for such a quiver is obtained by summing over the magnetic weights $(m_1,m_2)\in \Gamma_{\widetilde{G}}$; we need to choose the constraints on the $m_i$ in order to parametrize such a lattice. There are several ways to parametrize $\Gamma_{\widetilde{G}}$, but for future convenience, we would like to choose a parametrization such that the $\ZZ_2$ symmetry of the quiver is manifest after the decoupling of the overall $\U(1)$. This means that we would like to choose to decouple the diagonal $\U(1)$ of the two gauge $\U(1)$s. As recently explained in \cite[Appendix A]{Grimminger:2024doq}, the lattice $\Gamma_{\widetilde{G}}$ can be summed over by imposing either $m_1$ or $m_2 = 0$, or by requiring that $m_1+m_2 \in \{0,1\}$. This latter option guarantees that the conformal dimension of the monopole operators is still constant, and the whole lattice is summed over ``democratically'' between the two $\U(1)$s.  
The monopole formula then gives
\begin{equation}\label{eq:E1HS-uni}
  \begin{aligned}
    \HS_{E_{1}}(t,z) &= \frac{1}{1-t^2}\sum_{m_1\geq -\infty}^\infty\sum_{\underset{m_1+m_2\in \{0,1\}}{m_2\geq -\infty}}^\infty t^{2|m_1-m_2|}z^{2(m_1-m_2)} \\
    & = \frac{1-t^4}{(1-t^2)(1-z^2t^2)(1-z^{-2}t^2)}
    \,\,=\,\, \PE\left[\left(1+z^2+z^{-2}\right)t^2-t^4\right]\,.
    \end{aligned}
\end{equation}
As expected, this is the Hilbert series of $\mathbb{C}^2/\mathbb{Z}_2$ as in equation \eqref{eq:HSC2Z2}.

\subsubsection{The \texorpdfstring{$\ZZ_2$}{Z2} Discrete Gauging of the Rank-one \texorpdfstring{$E_1$}{E1} Theory}

In this section, we will analyze the discrete gauging of the $\mathbb Z_{2}$ symmetry associated with the automorphism in equation \eqref{eqn:t} of the affine $A_{1}$ Dynkin diagram. We first review the effect on the 5d Higgs branch by computing the wreathing of the associated magnetic quiver, and then we analyze the discrete gauging from the perspective of the mirror curve.

\paragraph{Wreathing the Magnetic Quiver:} The computation of the $\ZZ_2$-wreathing of the affine $E_1$ quiver has already been computed in \cite[(5.5)$_{g=2}$]{Hanany:2023uzn}, and the resulting Hilbert series is
\begin{equation}\label{eq:E1wreathed}
    \HS_{E_{1}\wr \ZZ_2}(t) = \frac{1-t^8}{(1-t^2)(1-t^4)^2} = \PE[t^2+2t^4-t^8]\coma
\end{equation}
which is precisely the Hilbert series of the $A_3$ singularity $\CC^2/\ZZ_4$. Geometrically, this is consistent since $\mathbb{C}^2/\mathbb{Z}_2$ is known to admit a single $\mathbb{Z}_2$ hyperk\"ahler quotient, which leads to $\mathbb{C}^2/\mathbb{Z}_4$.

It is worth noting that wreathing the quiver via the automorphism in equation \eqref{eqn:t} should be performed using the modified prescription introduced in Section \ref{sec:Z2cuttingedges}, since the axis of symmetry cuts through the two bifundamental edges of the magnetic quiver. However, precisely because there are two edges, the phase that we have to add in the Hilbert series cancels.

\paragraph{Curve Analysis:}

The $E_1$ theory is sufficiently constrained such that there are distinct and inequivalent actions on the toric diagram that, for what concerns the prescription given in Section \ref{subsec:action_on_curve}, are consistent with the $\ZZ_2$ symmetry of the magnetic quiver we have considered wreathing. We present the two possible scenarios here, and we postpone for future work to determine which is the appropriate one.

First, we consider the action 
\begin{equation}\label{eqn:E1take1}
    (x, y) \rightarrow (x^{-1}, y) \,,
\end{equation}
on the brane web in Table \ref{tab:magquiverspqwebs}, the two subwebs are swapped, thereby reproducing the action on the nodes of the magnetic quiver. We can see that the action in equation \eqref{eqn:E1take1} transforms the mirror curve in equation \eqref{eq:curvee1revisited} as follows:
\begin{equation}
    P_{E_1}(x, y) \quad \xrightarrow{\,\,\eqref{eqn:E1take1}\,\,} \quad P_{E_1}(x^{-1}, y) \,\,=\,\, P_{E_1}(x, y) \,.
\end{equation}
That is, the Seiberg--Witten curve is invariant on-the-nose; therefore, the mass parameter space does not restrict after discrete gauging. This is consistent with the fact that the mass parameter space before the discrete gauging is
\begin{equation}
    T_{\CC} / W \cong \mathbb{C}^* \,,
\end{equation}
and after discrete gauging it is simply the Cartan torus of the predicted $\mathfrak{u}(1)$ global symmetry, which is also $\mathbb{C}^*$.

Alternatively, the $E_1$ theory has a $\ZZ_4$ particle-instanton duality \cite{Mitev:2014jza}, which comes from the symmetry of the toric diagram under the action
\begin{equation}\label{eqn:E1take2}
    (x,y) \rightarrow (y^{-1}, x) \,.
\end{equation}
This transformation also swaps the two subwebs of the brane web and thus reproduces the action on the nodes of the magnetic quiver; however, the magnetic quiver does not see the full $\mathbb{Z}_4$ symmetry. If we compose the geometric transformation in equation \eqref{eqn:E1take2} with the transformation of the mass parameter
\begin{equation}
    \lambda \rightarrow \lambda^{-1} \,,
\end{equation}
treating $\lambda$ as a coordinate of $T_\CC$, the Seiberg--Witten curve given by the zero-locus of equation \eqref{eq:curvee1revisited} is invariant. Thus, we see that this is also consistent with the wreathing result.

\subsection{The Rank-one \texorpdfstring{$E_2$}{E2} Theory}
\label{sec:E2th}

\begin{figure}[t]
\centering
\begin{subfigure}[b]{0.49\textwidth}
\centering
    \begin{tikzpicture}[font=\footnotesize,baseline=0]
  \path (0, 1.6) -- (0, -1.6);
  \draw[thick,gray,-Triangle]
      (-1.5,0) -- node[above,pos=1] {$x$} (1.5,0);
  \draw[thick,gray,-Triangle]
      (0,-1.5) -- node[left,pos=1] {$y$} (0,1.5);

  \node[node,fill=black] (O2) at ( 0, 0) {};
  \node[node,fill=black] (A2) at (-1, 0) {};
  \node[node,fill=black] (B2) at ( 0, 1) {};
  \node[node,fill=black] (C2) at ( 1, 1) {};
  \node[node,fill=black] (D2) at ( 1, 0) {};
  \node[node,fill=black] (E2) at ( 0,-1) {};

  \draw[line width=1pt] (A2) -- (B2) -- (C2) -- (D2) -- (E2) -- (A2);
\end{tikzpicture}
\caption{}
\end{subfigure}\hfill
\begin{subfigure}[b]{0.49\textwidth}
\centering
   \begin{tikzpicture}[font=\footnotesize,baseline=0,
  scale=1,
  brane/.style={gray,line width=1.25pt},
  midarrow/.style={
    postaction={decorate},
    decoration={markings,
      mark=at position .55 with {\arrow[black]{Stealth[length=2.2mm,width=2.2mm]}}
    }
  },
  sublab/.style={font=\large,gray},
]
  \def\w{1.15}
  \def\h{0.55}
  \def\leg{0.70}
  \def\legE{0.65}

  \path (0, 1.6) -- (0, -1.6);

  \coordinate (B) at (0, \h);
  \coordinate (C) at (0, -\h);
  \coordinate (A) at (-\w, \h);
  \coordinate (D) at (-\w, -\h);

  \coordinate (E) at ($(B) + (\legE, \legE)$);

  \draw[brane] (B) -- (A);
  \draw[brane] (A) -- (D);
  \draw[brane] (D) -- (C);
  \draw[brane] (C) -- (B);

  \draw[brane] ($(A) + (-\leg, \leg)$) -- (A);
  \draw[brane] (D) -- ++(-\leg, -\leg);
  \draw[brane] ($(C) + (\leg, -\leg)$) -- (C);

  \draw[brane] (B) -- (E);
  \draw[brane] (E) -- ++(0, 0.75);
  \draw[brane] (E) -- ++(0.9, 0) node[right, blue] {$m_1$};

  \draw[red,dashed,line width=.9pt]
      ($(A) + (-0.10, -0.03)$) -- ++(-1.0, 0) node[left] {$-a$};
  \draw[red,dashed,line width=.9pt]
      ($(D) + (-0.10,  0.00)$) -- ++(-1.0, 0) node[left] {$a$};
\end{tikzpicture}
\caption{}
\end{subfigure}
\caption{
Toric diagram and corresponding $(p,q)$-web for the $E_2$ theory; the letter $a$ denotes the Coulomb branch parameter of the theory.
}
\label{fig:e2pwwebsec3}
\end{figure}
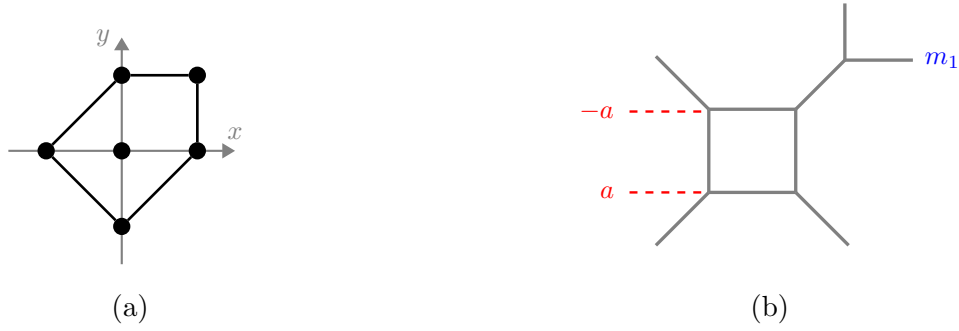

We next consider the rank-one $E_2$ theory. We have depicted the toric diagram and the $(p,q)$-web again in Figure \ref{fig:e2pwwebsec3} for convenience. As we have noted in Table \ref{tab:magquiverspqwebs}, the Higgs branch of $E_2$ is the union of the Higgs branches of the $E_1$ and $\widetilde{E}_1$ theories. However, the unitary quivers arise from brane configurations that are not sensitive to the $\widetilde{E}_1$ part of the moduli space and thus coincide with that of the $E_1$ theory. 

In this case, the $(p,q)$-web, and hence the mirror curve, possesses a $\mathbb Z_{2}$ symmetry that swaps the $x$ and $y$ axes. However, this does not correspond to the $\mathbb{Z}_2$ symmetry of the Higgs branch of the $E_1$ theory we discussed in Section \ref{sec:E1th}. This is not necessarily surprising: the same 5d SCFT can be described by many $(p,q)$-webs, and the $\mathbb Z_{2}$ symmetry that one reads from the magnetic quiver(s) might be evident only in a specific brane web.\footnote{In particular, one might consider webs with many five-branes ending on the same seven-branes, which correspond to the so-called \textit{white dots} in generalized toric geometry \cite{Benini:2009gi}.} Since the $\mathbb{Z}_2$-automorphism of the magnetic quiver is not manifest in the particular $(p,q)$-web or toric diagram of Figure \ref{fig:e2pwwebsec3}, in what follows, we will not consider the curve analysis for the discrete gauging of this $\mathbb Z_{2}$ symmetry.

\subsubsection{The Higgs Branch of the Rank-one \texorpdfstring{$E_2$}{E2} Theory}

Since the Higgs branch of the $E_2$ theory is the union of the Higgs branches of the $E_1$ and $\widetilde{E}_1$ theories, the Hilbert series is determined directly from the union via
\begin{equation}\label{eq:E2HS}
\begin{split}
    \HS_{E_2}(t,z) = & \HS_{E_1}(t,z) + \HS_{\tilde{E}_1}(t) - 1 = \frac{1+2 t^2-t^4 \left(z^2+z^{-2}\right)+t^6}{(1-z^{2}t^2)(1-z^{-2}t^2)} \\
    = & \PE\left[t^2 \left(\chi^A_{[2]}(z)+1\right)-t^4 \left(\chi^A _{[2]}(z)+2\right)+t^6 \left(2 \chi^A_{[2]}(z)+1\right)+\mathcal{O}(t^8)\right]\fstop
\end{split}
\end{equation}
We can see that the flavor symmetry is $\mathfrak{e}_2 = \mathfrak{su}(2) \oplus \mathfrak{u}(1)$, as expected. For convenience, we can also write the unrefined Hilbert series, which is:
\begin{equation}
    \HS_{E_2}(t) = \frac{1+2t^2-2t^4+t^6}{\left(1-t^2\right)^2} =   \PE[4 t^2-5 t^4+7 t^6+\mathcal{O}\left(t^8\right)]\fstop
\end{equation}

\subsubsection{The \texorpdfstring{$\ZZ_2$}{Z2} Discrete Gauging of the Rank-one \texorpdfstring{$E_2$}{E2} Theory}

Since, as discussed for $E_1$, the $\mathbb{C}^2/\mathbb{Z}_2$ magnetic quiver for the $E_1$ sector of the $E_2$ Higgs branch admits a $\mathbb{Z}_2$ wreathing, we can equally perform this wreathing for the $E_2$ theory. The Hilbert series for the putative $\mathbb{Z}_2$-discretely-gauged $E_2$ theory, $E_2 \,/\, \mathbb{Z}_2$, is given by the sum
\begin{equation}\label{eq:E2Z2HS}
\begin{split}
    \HS_{E_2\wr \ZZ_2}(t) =& \HS_{E_1\wr \ZZ_2}(t) + \HS_{\widetilde{E}_1}(t) - 1 = \frac{1+t^2-t^6+t^8}{\left(1-t^2\right)(1-t^4)}  \\
    = &\PE\left[2 t^2- t^6+\mathcal{O}\left(t^8\right)\right]\fstop
\end{split}
\end{equation}
This is just a reflection of the fact that after wreathing, the Higgs branch is the union of the $\mathbb{C}^2/\mathbb{Z}_4$ moduli space of $E_1 \,/\,\ZZ_2$ and the Higgs branch of $\widetilde{E}_1$.

\subsection{The Rank-one \texorpdfstring{$E_3$}{E3} Theory}
\label{sec:E3th}

\begin{figure}[t]
\centering
\begin{subfigure}[b]{0.49\textwidth}
\centering
    \begin{tikzpicture}[scale=1,baseline=0]
    \path (0, 1.95) -- (0, -1.95);
    		\draw[thick,gray,-Triangle] (-1.5,0) -- node[above,pos=1] {$x$} (1.5,0);
			\draw[thick,gray,-Triangle] (0,-1.5) -- node[left,pos=1] {$y$} (0,1.5);
			\node[node, fill=black] (O) at (0,0) {};
			\node[node, fill=black] (A) at (-1,0) {};
			\node[node, fill=black] (B) at (0,-1) {};
            \node[node, fill=black] (C) at (1,0) {};
             \node[node, fill=black] (D) at (0,1) {};
             \node[node, fill=black] (E) at (1,1) {};
             \node[node, fill=black] (F) at (-1,-1) {};
           \draw[line width=1pt] (A)--(F) -- (B)--(C) -- (E) -- (D) -- (A);
           \end{tikzpicture}
\caption{}\label{fig:e3toricdiagram}
\end{subfigure}\hfill
\begin{subfigure}[b]{0.49\textwidth}
\centering
   \begin{tikzpicture}[font=\footnotesize,baseline=0,
  scale=1,
  brane/.style={gray,line width=1.25pt},
  midarrow/.style={
    postaction={decorate},
    decoration={markings,
      mark=at position .55 with {\arrow[black]{Stealth[length=2.2mm,width=2.2mm]}}
    }
  },
  sublab/.style={font=\large,gray}
]
  \def\w{1.15}
  \def\h{0.55}
  \def\leg{0.70}
  \def\legE{0.65}

  \path (0, 1.95) -- (0, -1.95);

  \coordinate (B) at (0, \h);
  \coordinate (C) at (0, -\h);
  \coordinate (A) at (-\w, \h);
  \coordinate (D) at (-\w, -\h);

  \coordinate (E) at ($(B) + (\legE, \legE)$);
  \coordinate (F) at ($(D) + (-\legE, -\legE)$);

  \draw[brane] (B) -- (A);
  \draw[brane] (A) -- (D);
  \draw[brane] (D) -- (C);
  \draw[brane] (C) -- (B);

  \draw[brane] ($(A) + (-\leg, \leg)$) -- (A);

  \draw[brane] (C) -- ++(\leg, -\leg);

  \draw[brane] (B) -- (E);
  \draw[brane] (E) -- ++(0, 0.75);
  \draw[brane] (E) -- ++(0.9, 0) node[right, blue] {$m_1$};

  \draw[brane] (D) -- (F);
  \draw[brane] (F) -- ++(0, -0.75);
  \draw[brane] (F) -- ++(-0.9, 0) node[left, blue] {$m_2$};

  \draw[red,dashed,line width=.9pt]
      ($(A) + (-0.10, -0.03)$) -- ++(-1.0, 0) node[left] {$-a$};
  \draw[red,dashed,line width=.9pt]
      ($(D) + (-0.10,  0.00)$) -- ++(-1.0, 0) node[left] {$a$};
\end{tikzpicture}
\caption{}
\end{subfigure}
\caption{
Toric diagram and corresponding $(p,q)$-web for the $E_3$ theory (following Figure~10 of \cite{Mitev:2014jza}), with the letter $a$ denoting the Coulomb branch parameter of the theory.
}
\label{fig:e3pwweb}
\end{figure}
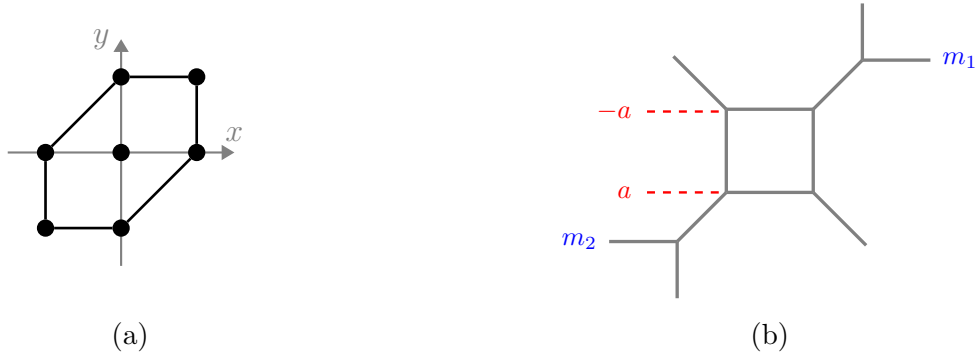

We now consider the myriad discrete gaugings of the rank-one $E_3$ SCFT, whose toric diagram and brane web are reproduced in Figure \ref{fig:e3pwweb} for convenience.

\paragraph{Discrete Symmetries from Magnetic Quivers for the Higgs Branch:} Similarly to the $E_2$ theory, the rank-one $E_3$ SCFT also has a Higgs branch which is a union of two hyperk\"ahler cones. One is $\CC^2 / \ZZ_2$, which was studied as the Higgs branch of the $E_1$ theory in Section \ref{sec:E1th}, and the other is the closure of the minimal nilpotent orbit of $\mathfrak{su}(3)$: $\mathcal{O}_\text{\tiny min}^{\mathfrak{su}(3)}$. The latter can be described via the Coulomb branch of the following magnetic quiver:
\begin{equation}\label{eq:unitaryE3}
    \begin{tikzpicture}[baseline=0,font=\footnotesize]
        \node (A0) at (0,0) {};
        \node[node, label=above:{$1$}] (A1) [above=4mm of A0] {};
        \node[node, label=below:{$1$}] (A2) [below left=6mm of A0] {};
        \node[node, label=below:{$1$}] (A3) [below right=6mm of A0] {};
        \draw (A1) -- (A2) -- (A3) -- (A1);
        \node (Z2) [above right=5mm of A3] {$/\U(1)$};
    \end{tikzpicture}\coma
\end{equation}
where, again, we must recall the decoupling of the center-of-mass $\U(1)$. The automorphism group of this quiver is the $S_3$ generated by the following rotation and reflection:
\begin{equation}\label{eqn:rs}
  r\,:\,\, \begin{tikzpicture}[baseline=0,font=\footnotesize]
        \node (A0) at (0,0) {};
        \node[node, label=right:{$1$}] (A1) [above=4mm of A0] {};
        \node[node, label=below:{$1$}] (A2) [below left=6mm of A0] {};
        \node[node, label=below:{$1$}] (A3) [below right=6mm of A0] {};
        \draw (A1) -- (A2) -- (A3) -- (A1);
        \node (Z2) [above right=5mm of A3] {$/\U(1)$}; 
        \draw[-Triangle,red] ([yshift=-4mm]A2.south) to[bend right=35] node[below,pos=0.5] {} ([yshift=-4mm]A3.south);
        \draw[-Triangle,red] ([yshift=1mm]A3.east) to[bend right=35] node[right,pos=0.5] {} ([yshift=-2mm,xshift=4mm]A1.east);
        \draw[-Triangle,red] ([yshift=0mm,xshift=-1mm]A1.west) to[bend right=35] node[right,pos=0.5] {} ([yshift=1mm]A2.west);
    \end{tikzpicture} \,, \qquad\qquad s\,:\,\,
  \begin{tikzpicture}[baseline=0,font=\footnotesize]
        \node (A0) at (0,0) {};
        \node[node, label=right:{$1$}] (A1) [above=4mm of A0] {};
        \node[node, label=below:{$1$}] (A2) [below left=6mm of A0] {};
        \node[node, label=below:{$1$}] (A3) [below right=6mm of A0] {};
        \draw (A1) -- (A2) -- (A3) -- (A1);
        \draw[Triangle-Triangle,red] ([yshift=-4mm]A2.south) to[bend right=50] node[below,pos=0.5] {} ([yshift=-4mm]A3.south);
        \node (Z2) [above right=5mm of A3] {$/\U(1)$};
        \draw[densely dashed, red,thin]  ([yshift=3mm]A1.north) -- ([yshift=-9mm]A0.south);
    \end{tikzpicture}
    \fstop
\end{equation}
Therefore, the automorphism group for the unitary magnetic quivers for the Higgs branch of the $E_3$ theory is simply the product of the automorphism groups of the two factors:
\begin{equation}
\label{eq:symsaffineE3}
\operatorname{Aut}(\widetilde{\mathcal{D}}_{\mathfrak{e}_3}) = S_3 \times \mathbb{Z}_2\coma
\end{equation}
generated by $r$ and $s$ from equation \eqref{eqn:rs} and $t$ from equation \eqref{eqn:t}. It is then algorithmic, following Section \ref{sec:dg}, to determine the image of each subgroup of $\operatorname{Aut}\left(\widetilde{\mathcal{D}}_{\mathfrak{e}_3}\right)$ inside the automorphism group of $\mathfrak{su}(3) \oplus \mathfrak{su}(2)$, and thus to determine the corresponding fixed-point subalgebras. We have listed all (conjugacy classes of) such subgroups, $H$, and their fixed-point subalgebras in Table \ref{tbl:E3dg}. We will go through each subgroup in turn, determine the $H$-wreathed Coulomb branch Hilbert series, and show that the Coulomb branch flavor symmetry reproduces the expectation from $\mathfrak{g}^H$ for the flavor symmetry after discrete gauging.

\begin{table}[t]
    \centering
    \begin{tabular}{c|c|c}
        $H$ & Generators & $\mathfrak{g}^H$ \\\hhline{=|=|=}
        $1$ & $1$ & $\mathfrak{su}(3) \oplus \mathfrak{su}(2)$ \\
        $\mathbb{Z}_2^{(1)}$ & $s$ & $\mathfrak{su}(2) \oplus \mathfrak{su}(2)$ \\
        $\mathbb{Z}_2^{(2)}$ & $t$ & $\mathfrak{su}(3) \oplus \mathfrak{u}(1)$ \\
        $\mathbb{Z}_2^{DT}$ & $st$ & $\mathfrak{su}(2) \oplus \mathfrak{u}(1)$ \\
        $\mathbb{Z}_3$ & $r$ & $\mathfrak{u}(1) \oplus \mathfrak{u}(1) \oplus \mathfrak{su}(2)$ \\
        $\mathbb{Z}_2^{(1)} \times \mathbb{Z}_2^{(2)}$ & $s, t$ & $\mathfrak{su}(2) \oplus \mathfrak{u}(1)$ \\
        $\mathbb{Z}_3 \times \mathbb{Z}_2^{(2)}$ & $r, t$ & $\mathfrak{u}(1) \oplus \mathfrak{u}(1) \oplus \mathfrak{u}(1)$ \\
        $S_3$ & $r, s$ & $\mathfrak{su}(2)$ \\
        $S_3 \times \mathbb{Z}_2^{(2)}$ & $r, s, t$ & $\mathfrak{u}(1)$ \\
        $S_3^{DT}$ & $r, st$ & $\mathfrak{u}(1)$ 
    \end{tabular}
    \caption{For each subgroup $H$ of $\operatorname{Aut}(\widetilde{\mathcal{D}}_{\mathfrak{e}_3})$, we list the fixed-point subalgebra of $\mathfrak{su}(3) \oplus \mathfrak{su}(2)$, $\mathfrak{g}^H$, as defined in equation \eqref{eqn:gHdef}. We have listed the generators of the subgroup in terms of the generators of the automorphisms of the quivers in equations \eqref{eqn:t} and \eqref{eqn:rs}; we have distinguished subgroups which are identical as groups via superscripts.}
    \label{tbl:E3dg}
\end{table}

\paragraph{Mirror Curve:}

The toric diagram appears in Figure \ref{fig:e3toricdiagram}. The mirror curve before the discrete gauging is
\begin{align}\label{eq:SW_E3_first}
    P(x,y)=u+c_{(1,0)} x+c_{(1,1)}xy+c_{(0,1)}y+c_{(-1,0)}\frac{1}{x}+c_{(-1,-1)}\frac{1}{xy}+c_{(0,-1)}\frac{1}{y}\fstop
\end{align}
We note that the symmetries of the toric diagram reproduce equation \eqref{eq:symsaffineE3} exactly and give a UV realization of the zero-form symmetry associated to them by leveraging the dictionary developed in Section \ref{sec:2}.
We can use the rescaling freedom of the curve to put it in the following form \cite{Magureanu:2023rrg}: 
\begin{equation}
\frac{\sqrt{\lambda}}{t}\left(1+\frac{M_2}{w}\right)
+\frac{1}{w}+w-U
+t\sqrt{\lambda}\left(1+M_1 w\right)=0\coma
\end{equation}
it is convenient to express the coefficients in terms of the fugacities of the $E_3$ flavor symmetries: 
\begin{equation}
    y_1 y_2 y_3 = 1\coma  m \in \mathbb C^* \,,
\end{equation}
as 
\begin{equation}
    M_1  = -m \sqrt{\frac{y_2}{y_3}}\coma  M_2  = -m^{-1} \sqrt{\frac{y_2}{y_3}}\coma  \lambda = \frac{y_1}{y_2}\coma
\end{equation}
and to multiply all the curves by $\sqrt{y_3 y_1^{-1}}$, we get 
\begin{equation}\label{eq:e3curve_param}
    P_{E_3}\equiv -m x y-\frac{1}{m x y}+u
   \sqrt{y_3}+\left(x+\frac{1}{x}\right)
   \sqrt{\frac{y_3}{y_2}}+\left(y+\frac{1}{y}\right)
   \sqrt{\frac{y_3}{y_1}}\coma
\end{equation}
with $u = \frac{U}{\sqrt{y_1}}$ being the flavor Weyl-invariant CB parameter of \cite{Mitev:2014jza}.

\subsubsection{The Higgs Branch of the Rank-one \texorpdfstring{$E_3$}{E3} Theory}

To determine the Hilbert series for the Higgs branch of the (non-discretely-gauged) $E_3$ theory, we
first calculate the Coulomb branch Hilbert series of the unitary quiver in
equation \eqref{eq:unitaryE3}. The Hilbert
series expressed in terms of $\su(3)$ characters is
\begin{equation}\label{eq:Osu3unitaryrefin}
    \HS_{\mathcal{O}^{\su(3)}_\text{\tiny min}}(t,\mathbf{x}) = \PE\left[\chi^A_{[1,1]}(\mathbf{x})t^2-\left(1+\chi^A_{[1,1]}(\mathbf{x})\right)t^4+2\chi^A_{[1,1]}(\mathbf{x}) t^6+\mathcal{O}(t^8)\right]\coma
\end{equation}
or, unrefined,
\begin{equation}
        \HS_{\mathcal{O}^{\su(3)}_\text{\tiny min}}(t,1) = \frac{1+4t^2 + t^4}{(1-t^2)^4}
         = \PE\left[8t^2-9t^4+16t^6+\mathcal{O}(t^8)\right]\fstop
\end{equation}
The Hilbert series for $E_3$ is then given from the union as 
\begin{equation}\label{eq:E3HS-uni}
\begin{split}
    \HS_{E_3}(t,\mathbf{x},z) = &\, \HS_{\mathcal{O}^{\su(3)}_\text{\tiny min}}(t,\mathbf{x})+ \HS_{E_1}(t,z) -1 \\ 
    = &\,\scalebox{0.97}{$\displaystyle\PE\left[\left(\chi^A_{[1,1]}(\mathbf{x})+\chi^A_{[2]}(z)\right)t^2-\left(2+\left(1+\chi^A_{[2]}(z)\right)\chi^A_{[1,1]}(\mathbf{x})\right)t^4+\mathcal{O}(t^6)\right]\coma$}
\end{split}
\end{equation}
or unrefined as
\begin{equation}
    \begin{split}
        \HS_{E_3}(t,\mathbf{1},1) = \frac{1+7 t^2-6 t^4+5 t^6-t^8}{\left(1-t^2\right)^4} = \PE\left[11 t^2-34 t^4+159 t^6+\mathcal{O}(t^8)\right]\fstop
    \end{split}
\end{equation}

\subsubsection{The \texorpdfstring{$\ZZ_2^{(1)}$}{Z2(1)} Discrete Gauging of the Rank-one \texorpdfstring{$E_3$}{E3} Theory}\label{sec:E3wrZ21}

\paragraph{Wreathing the Magnetic Quiver:} In this section, we compute the $\ZZ_2^{(1)}$-wreathing of the magnetic quiver in equation \eqref{eq:unitaryE3} that cuts a bifundamental edge of the quiver in half. We will first show how the usual prescription to compute the wreathing would fail to give the correct answer, while by using the modification introduced in Section \ref{sec:Z2cuttingedges}, we will recover the expected $\so(3)$ symmetry associated to $\mathcal{O}^{\su(3)}_\mathrm{min}\wr \ZZ_2^{(1)}$. The wreathing we want to perform on equation \eqref{eq:unitaryE3} can be represented as
\begin{equation}\label{eq:unitaryE3-wreathedZ2}
    \begin{tikzpicture}[baseline=0,font=\footnotesize]
        \node (A0) at (0,0) {};
        \node[node, label=right:{$1$}] (A1) [above=4mm of A0] {};
        \node[node, label=below:{$1$}] (A2) [below left=6mm of A0] {};
        \node[node, label=below:{$1$}] (A3) [below right=6mm of A0] {};
        \draw (A1) -- (A2) -- (A3) -- (A1);
        \draw[Triangle-Triangle,red] ([yshift=-4mm]A2.south) to[bend right=50] node[below,pos=0.5] {$\wr \ZZ_2^{(1)}$} ([yshift=-4mm]A3.south);
        \node (Z2) [above right=5mm of A3] {$/\U(1)$};
        \draw[densely dashed, red,thin]  ([yshift=3mm]A1.north) -- ([yshift=-9mm]A0.south);
    \end{tikzpicture}\fstop
\end{equation}

The usual way of computing the wreathed Hilbert series proceeds as follows. This Hilbert series computation appeared first in \cite[(4.20)]{Grimminger:2024mks}, since it corresponds to the discrete gauging of charge conjugation of SQED with 3 flavors. 
The $\ZZ_2$-wreathing is obtained by summing over two contributions:
\begin{equation}
    \HS_{\mathcal{O}^{\su(3)}_\text{\tiny min}\wr \ZZ_2}(t,z) = \frac{1}{2} \left(\HS_{\mathcal{O}^{\su(3)}_\text{\tiny min}}^{\text{\tiny contr. 1}}(t,z)+\HS_{\mathcal{O}^{\su(3)}_\text{\tiny min}}^{\text{\tiny contr. 2}}(t,z)\right)\coma
\end{equation}
where
\begin{equation}\label{eq:E3wrZ2contr12}
    \begin{split}
        \HS_{\mathcal{O}^{\su(3)}_\text{\tiny min}}^{\text{\tiny contr. 1}}(t,z) & =\frac{1-t^2}{(1-t^2)^3}\sum_{m_1,m_2\geq -\infty}^\infty t^{2\Delta_1}z^{m_1+m_2}\coma\\
        \HS_{\mathcal{O}^{\su(3)}_\text{\tiny min}}^{\text{\tiny contr. 2}}(t,z) & =  \frac{1-t^2}{(1-t^2)(1-t^4)}\sum_{m_1\geq -\infty}^\infty t^{4\Delta_2}z^{2m_1}\coma
    \end{split}
\end{equation}
with 
\begin{equation}
        \Delta_1 = \frac{1}{2}\left(|m_1-m_2|+|m_2|+|m_1|\right)\coma \qquad
        \Delta_2 = \frac{1}{2}|m_1|\coma
\end{equation}
and we have decoupled the $\U(1)$ at the top by setting $m_3 = 0$. The resulting unrefined and refined Hilbert series are
\begin{equation}
\begin{aligned}
    \HS_{\mathcal{O}^{\su(3)}_\text{\tiny min}\wr \ZZ_2}(t,1) &= \frac{1+2t^2+2t^4+t^6}{\left(1-t^2\right)^3(1-t^4)} = \PE[5 t^2-t^6] = \frac{1-t^6}{\left(1-t^2\right)^5}\coma \\[0.5em]
     \HS_{\mathcal{O}^{\su(3)}_\text{\tiny min}\wr \ZZ_2}(t,z) &= \PE\left[t^2 \chi^A _{[4]}(z)-t^6 \right]\,,
\end{aligned}
\end{equation} 
where we have denoted by $\chi_{[4]}(w)$ the representation of $\su(2)$ associated with the Dynkin label $[4]$.
This Hilbert series is inconsistent with extended supersymmetry: the moment-maps \emph{must} transform in the adjoint representation of some reductive Lie algebra. We remark that this issue is not due to the wreathing method introduced in \cite{Grimminger:2024mks}; if instead one uses the definition of wreathing in the monopole formula as in \cite{Bourget:2020bxh,Arias-Tamargo:2021ppf}, one gets the same inconsistent result.

The reason this computation did not lead to the correct representation of the moment maps is that it is not correctly projecting over the invariant Coulomb branch moduli space under the involution, as explained in Section \ref{sec:Z2cuttingedges}. The way to project onto the correct moduli space is to modify the second contribution in equation \eqref{eq:E3wrZ2contr12} by adding the $(-1)^{m_1}$ phase:
\begin{equation}
    \HS_{\mathcal{O}^{\su(3)}_\text{\tiny min}}^{\text{\tiny contr. 2}}(t,z)  =  \frac{1-t^2}{(1-t^2)(1-t^4)}\sum_{m_1\geq -\infty}^\infty(-1)^{m_1}t^{4\Delta_2}z^{2m_1}\fstop
\end{equation}
In this way, by considering $z=y$, as by equation \eqref{eq:A2k_map}$_{k=1}$, we obtain
\begin{equation}\label{eq:E3wrZ2-HS}
        \HS_{\mathcal{O}^{\su(3)}_{\mathrm{min}}\wr \ZZ_2}(t,y) = \PE\Big[t^2 \chi^A_{[2]}(y)+t^4 \chi^A_{[8]}(y)-t^6 \chi^A_{[6]}(y)+\mathcal{O}\left(t^7\right)\Big]\coma
\end{equation}
or, unrefined,
\begin{equation}
\HS_{\mathcal{O}^{\su(3)}_{\mathrm{min}}\wr \ZZ_2}(t,1) = \frac{1+t^2+8 t^4+t^6+t^8}{\left(1-t^2\right)^2 \left(1-t^4\right)^2} = \PE[3 t^2+9 t^4-7 t^6+\mathcal{O}( t^8)] \coma
\end{equation}
that correctly has the dimension of the adjoint of $\su(2)$ as expected. The generalization of this computation for any affine $A$ quiver is explained in Appendix \ref{app:birch}.

We can then use equation \eqref{eq:E3wrZ2-HS} to correctly compute the Hilbert series for the Higgs branch of $E_3 / \mathbb{Z}_2^{(1)}$, obtaining
\begin{align}
        \HS_{E_3\wr \ZZ_2^{(1)}}(t) & =  \HS_{\mathcal{O}^{\su(3)}_\text{\tiny min}\wr \ZZ_2^{(1)}}(t)+ \HS_{E_1}(t) -1
        = \PE[6 t^2-t^4-13 t^6+\mathcal{O}\left(t^8\right)]\fstop\label{eq:O3Z2E1-HS}
\end{align}

\paragraph{Curve Analysis:}
In this case, we identify the action which correctly exchanges the legs of the $(p,q)$-web:
\begin{equation}
\label{eq:isoz2outere3}
    (x,y)  \to  (y,x) \,.
\end{equation}
Furthermore, we identify the action of the wreathing on the flavor algebra as the outer automorphism of its $A_2$ factor. This acts as 
\begin{equation}
\label{eq:outerausu3e3}
    (y_1,y_2,y_3,m) \quad \to (y_1^{-1},y_2^{-1},y_3^{-1},m)\,.
\end{equation}
The combination of equations \eqref{eq:isoz2outere3} and \eqref{eq:outerausu3e3} leaves the curve in equation \eqref{eq:e3curve_param} invariant only if we impose
 \begin{equation}
     y_1 = y_2^{-1}\coma  y_3 = 1 \,.
 \end{equation}
 This constraint determines the flavor algebra after the discrete gauging, which is isomorphic to $\mathfrak{so}(3)$, and thus in agreement with the wreathing result. 

\subsubsection{The \texorpdfstring{$\ZZ_2^{(2)}$}{Z2(2)} Discrete Gauging of the Rank-one \texorpdfstring{$E_3$}{E3} Theory}

\paragraph{Wreathing the Magnetic Quiver:}

The $\ZZ_2^{(2)}$ subgroup of $\operatorname{Aut}(\widetilde{\mathcal{D}}_{\mathfrak{e}_3})$ acts only on the affine $\mathfrak{su}(2)$ factor, which we discussed in the context of the $E_1$ theory in Section \ref{sec:E1th}. 
Therefore, the Coulomb branch Hilbert series for the $\mathbb{Z}_2^{(2)}$-wreathed theory is straightforwardly determined from the already-worked-out Hilbert series for $\mathcal{O}^{\mathfrak{su}(3)}_\text{min}$ and $(\mathbb{C}^2 / \mathbb{Z}_2)/\mathbb{Z}_2$:
\begin{equation} \label{eq:O3E1Z2-HS}
\begin{split}
      \HS_{E_3\wr \ZZ_2^{(2)}}(t) & =  \HS_{\mathcal{O}^{\su(3)}_\text{\tiny min}}(t)+ \HS_{E_1\wr \ZZ_2^{(2)}}(t) -1 \\
        & = \frac{1 + 6t^2 + 5t^4 - 3t^6 + 4t^8 - t^{10}}{(1-t^2)^3(1-t^4)}= \PE[9 t^2-15 t^4+37 t^6+\mathcal{O}\left(t^8\right)]\fstop
\end{split}
\end{equation}
Similarly, checking the refined index reveals the expected $\mathfrak{su}(3) \oplus \mathfrak{u}(1)$ Coulomb symmetry.

\paragraph{Curve Analysis:}
The action of this wreathing acts exclusively on the $A_1$ factor of the flavor symmetry. On the geometry, this corresponds to the central inversion of the coordinates:
\begin{equation}
(x,y) \to (x^{-1}, y^{-1}) \fstop
\end{equation}
To keep the curve $P_{E_3}(x,y)$ invariant, this geometric transformation must be accompanied solely by an action on the fugacity of the $\mathfrak{su}(2)$ factor, namely
\begin{equation}
m \to m^{-1} \coma
\end{equation}
leaving the $y_i$ parameters invariant. The rank of the flavor symmetry is not reduced, confirming the $\mathfrak{su}(3) \oplus \mathfrak{u}(1)$ algebra (rank $3$) predicted by the magnetic quiver analysis.

\subsubsection{The \texorpdfstring{$\ZZ_2^{DT}$}{Z2DT} Discrete Gauging of the Rank-one \texorpdfstring{$E_3$}{E3} Theory}

\paragraph{Wreathing the Magnetic Quiver:} In computing the wreathing of the $E_3$ theory, it is possible to consider the diagonal $\ZZ_2$ contained in $\ZZ_2^{(1)}$ and $\ZZ_2^{(2)}$. In order to compute this wreathing, instead of considering the two magnetic quivers separately, one should consider the Hilbert series of the whole $E_3$ theory at once, i.e., formally,
\begin{equation}\label{eq:E3Z2DT-formally}
    \HS_{E_3\wr\ZZ_2^{DT}}(t) = \frac{1}{2}\left(\HS_{E_3}^{\text{contr.} 1}(t)+\HS_{E_3}^{\text{contr.} 2}(t)\right)\fstop
\end{equation}
However, the moduli space of $E_3$ is the union of the $\mathcal{O}^{\su(3)}_\mathrm{min}$ and $E_1$ moduli spaces, with only the origin as their intersection. In fact, as we have done so far, the total Hilbert series is the sum of the two Hilbert series minus the identity element. Since each moduli space has only a single $\ZZ_2$ symmetry, equation \eqref{eq:E3Z2DT-formally} is effectively given by the sum of the $\ZZ_2$-wreathed Hilbert series of each moduli space:
\begin{equation} \label{eq:O3E1Z2DT-HS}
    \begin{split}
         \HS_{E_3\wr \ZZ_2^{DT}}
        & =  \frac{1}{2}\left(\HS_{E_3}^{\text{contr.} 1}(t)+\HS_{E_3}^{\text{contr.} 2}
        (t)\right)\\
        & = \frac{1}{2}\left(\HS_{\mathcal{O}^{\su(3)}_\text{\tiny min}}^{\text{contr.} 1}(t)+\HS_{E_1}^{\text{contr.} 1}(t)+\HS_{\mathcal{O}^{\su(3)}_\text{\tiny min}}^{\text{contr.} 2}
        (t)+\HS_{E_1}^{\text{contr.} 2}
        (t)\right) -1 \\
        & = \HS_{\mathcal{O}^{\su(3)}_\text{\tiny min}\wr \ZZ_2^{(1)}}(t)+ \HS_{E_1\wr \ZZ_2^{(2)}}(t) -1
        = \scalemath{1}{\PE[4 t^2+8 t^4-19 t^6+\mathcal{O}\left(t^8\right)]\fstop}
    \end{split}
\end{equation}
Again, the Hilbert series shows a $\su(2)\oplus\mathfrak{u}(1)$ flavor symmetry, as expected.

\paragraph{Curve Analysis:}
This subgroup is generated by the double transposition $st$. Geometrically, this corresponds to composing the reflection across the diagonal with the central inversion, yielding the action
\begin{equation}
(x,y) \to (y^{-1}, x^{-1}) \fstop
\end{equation}
On the parameters, this symmetry combines the outer automorphism of $\mathfrak{su}(3)$ with the inner automorphism of $\mathfrak{su}(2)$:
\begin{equation}
(y_1,y_2,y_3,m) \to (y_1^{-1}, y_2^{-1}, y_3^{-1}, m^{-1}) \fstop
\end{equation}
Requiring the curve to be invariant under this transformation imposes the exact same constraints on the $\mathfrak{su}(3)$ fugacities as generated by the $\mathbb{Z}_2^{(1)}$ action:
\begin{equation}
y_1 = y_2^{-1}\coma y_3 = 1 \coma
\end{equation}
while the parameter $m$ remains unconstrained. The total rank of the mass parameters drops from $3$ to $2$, consistent with the expected $\mathfrak{su}(2) \oplus \mathfrak{u}(1)$ residual flavor symmetry.

\subsubsection{The \texorpdfstring{$\ZZ_3^{}$}{Z3} Discrete Gauging of the Rank-one \texorpdfstring{$E_3$}{E3} Theory}

\paragraph{Wreathing the Magnetic Quiver:} The $\ZZ_3$-wreathing of $\mathcal{O}_\ttiny{min.}^{\su(3)}$ can be obtained following the prescription introduced in \cite{Giacomelli:2024sex}, i.e., by summing over the fluxes restricting $m_1<m_2\geq m_3$ and $m_1=m_2=m_3$. The resulting Hilbert series is given by
\begin{equation}\label{eq:Osu3minHBZ3}
    \HS_{\mathcal{O}^{\su(3)}_\text{\tiny min}\wr \ZZ_3}(t) = \frac{1-t^2+6 t^4-t^6+t^8}{\left(1-t^2\right)^3(1-t^6)} = \PE[2 t^2+6 t^4+6 t^6-15 t^8+\mathcal{O}\left(t^{10}\right)]\fstop
\end{equation}
Another way to obtain the same Hilbert series for $\mathcal{O}^{\su(3)}_\mathrm{min}\wr \ZZ_3$ is to consider the elements of the $\ZZ_3$ group over the three nodes, i.e., in cycle notation,
\begin{equation}
       \{\ID,(123),(132)\}\fstop
\end{equation}
Then, as explained in \cite{Grimminger:2024mks}, the wreathed Hilbert series can be seen as the sum of the Hilbert series associated to $\mathcal{O}^{\su(3)}_\text{\tiny min}$ and twice the Hilbert series associated to a single $\U(1)$ node where all fugacities have been elevated to the third power. This is the generalization of what we have discussed in Section \ref{sec:E3wrZ21}, and we will extensively use this prescription in the following sections. The resulting Hilbert series is then obtained as
\begin{equation}
    \HS_{\mathcal{O}^{\su(3)}_\text{\tiny min}\wr \ZZ_3}(t) = \frac{1}{3}\left(\frac{1+4t^2 + t^4}{(1-t^2)^4}+2\frac{(1-t^2)}{(1-t^6)}\right) = \PE[2 t^2+6 t^4+6 t^6-15 t^8+\mathcal{O}\left(t^{10}\right)]\fstop
\end{equation}
The Hilbert series of $E_3\wr \ZZ_3$ is given by the sum of the Hilbert series for $\mathcal{O}^{\su(3)}_\text{\tiny min.}\wr \ZZ_3$ and $E_1$:
\begin{equation}\label{eq:O3Z3E1-HS}
    \begin{split}
        \HS_{E_3\wr \ZZ_3}(t) =& \HS_{\mathcal{O}^{\su(3)}_\text{\tiny min.}\wr \ZZ_3}(t) + \HS_{E_1}(t) -1 
        = \PE[5 t^2-t^4-t^6+6 t^{10}+\mathcal{O}\left(t^{12}\right)]\fstop
    \end{split}
\end{equation}

\paragraph{Curve Analysis:}
The cyclic permutation of the three nodes of the magnetic quiver in equation \eqref{eq:unitaryE3} must be induced by the action of the isometry on the legs of the quiver. In this case, this implies that the isometry must be of the form
\begin{equation}
\label{eq:isometryz3e3aux}
    (x,y)  \to  (x',y') = \left(\frac{a_1}{xy}, a_2 x\right)\fstop
\end{equation}
On the fugacities, the action must be the same as the $\mathbb Z_3$ inner automorphism of $\mathfrak{su}(3)$
\begin{equation}
\label{eq:innerautz3e3}
    (y_1,y_2,y_3,m)  \to  (y_2,y_3,y_1,m)\fstop
\end{equation}
Requiring the joint action of equations \eqref{eq:isometryz3e3aux} and \eqref{eq:innerautz3e3} to act as 
\begin{equation}
    P_{E_3} \to \alpha P_{E_3}, \quad \alpha \in \mathbb C^*\coma
\end{equation}
fixes
\begin{equation}
    a_1 = -m^{-1}, \quad a_2 = 1, \quad \alpha = \sqrt{y_1y_3^{-1}}\fstop
\end{equation}
Summing up, the action is 
\begin{equation}
    (x,y)  \to  \left(-(mxy)^{-1}, x\right),\qquad  (y_1,y_2,y_3,m)  \to  (y_2,y_3,y_1,m)\coma
\end{equation}
and we see no rank reduction in agreement with the wreathing result.

\subsubsection{The \texorpdfstring{$\ZZ_2^{(1)} \times \ZZ_2^{(2)}$}{Z2(1)xZ2(2)} Discrete Gauging of the Rank-one \texorpdfstring{$E_3$}{E3} Theory}\label{sec:E3wrZ21Z22}

\paragraph{Wreathing the Magnetic Quiver:}
We have determined the Hilbert series for $\mathcal{O}_\text{min}^{\mathfrak{su}(3)} \wr \ZZ_2$ and $E_1 \wr \mathbb{Z}_2$ already; the Hilbert series for the wreathing of $\ZZ_2^{(1)} \times \ZZ_2^{(2)}$ follows directly:
\begin{equation}
    \begin{split}
        \HS_{E_3\wr (\ZZ_2^{(1)}\times \ZZ_2^{(2)})}(t) 
        & =  \HS_{\mathcal{O}^{\su(3)}_\text{\tiny min}\wr \ZZ_2^{(1)}}(t)+ \HS_{E_1\wr \ZZ_2^{(2)}}(t) -1 \\
        &= \PE[4 t^2+8 t^4-19 t^6+\mathcal{O}\left(t^8\right)]\fstop \label{eq:O3Z2E1Z2-HS}
    \end{split}
\end{equation}
This Hilbert series is the same as that for $E_3 \,/\, \mathbb{Z}_2^{DT}$ written in equation \eqref{eq:O3E1Z2DT-HS}. This is expected: since the Higgs branch is the union of $\mathcal{O}_\text{min}^{\mathfrak{su}(3)}$ and $\mathbb{C}^2 / \mathbb{Z}_2$, there are no chiral ring operators charged under both $\ZZ_2^{(1)}$ and $\ZZ_2^{(2)}$. Therefore, $\ZZ_2^{DT}$ projects out the same operators.

\paragraph{Curve Analysis:}
This subgroup is generated independently by the reflection $s$ and the inversion $t$. In order for the curve to be invariant under the entire group, the mass parameters must satisfy the intersection of the constraints from both generators. The action of $t$ ($m \to m^{-1}$) is an inner automorphism and does not fix $m$, whereas the action of $s$ is an outer automorphism and imposes $y_1 = y_2^{-1}$ and $y_3 = 1$. The combined conditions thus coincide with those of $s$ alone. The rank is reduced to $2$, confirming the $\mathfrak{su}(2) \oplus \mathfrak{u}(1)$ algebra.

\subsubsection{The \texorpdfstring{$\ZZ_3 \times \ZZ_2^{(2)}$}{Z3xZ2(2)} Discrete Gauging of the Rank-one \texorpdfstring{$E_3$}{E3} Theory}\label{sec:E3wrZ3Z22}

\paragraph{Wreathing the Magnetic Quiver:}

Again, this subgroup of $\operatorname{Aut}(\widetilde{\mathcal{D}}_{\mathfrak{e}_3})$ is simply the combination of the $\ZZ_3$ action on $\mathcal{O}^{\mathfrak{su}(3)}_\text{min}$ factor with the $\ZZ_2$ action on the $\mathbb{C}^2 / \mathbb{Z}_2$ factor, for which the wreathed Hilbert series were determined in equations \eqref{eq:Osu3minHBZ3} and \eqref{eq:E1wreathed}, respectively. The Hilbert series is therefore obtained from the union:
\begin{equation}\label{eq:O3Z3E1Z2-HS}
    \begin{split}
        \HS_{E_3\wr (\ZZ_3\times \ZZ_2^{(2)})}(t) =& \HS_{\mathcal{O}^{\su(3)}_\text{\tiny min}\wr \ZZ_3}(t) + \HS_{E_1\wr \ZZ_2^{(2)}}(t)-1 = \PE[3 t^2+6 t^4-3 t^6+O\left(t^{8}\right)]\coma
    \end{split}
\end{equation}
confirming the expected $\mathfrak{u}(1)^3$ flavor symmetry.

\paragraph{Curve Analysis:}
The action of this group is generated by the rotation $r$ and the inversion $t$. Both generators correspond exclusively to inner automorphisms of their respective subalgebras: $r$ applies a Weyl permutation to the fugacities $y_i$, and $t$ applies a Weyl reflection inverting $m$; the curve $P_{E_3}$ transforms covariantly under these actions, without the need to impose any rigid constraints (no parameters are forced to $1$). The rank of the curve remains $3$, consistent with a flavor symmetry of $\mathfrak{u}(1)^{\oplus 3}$.

\subsubsection{The \texorpdfstring{$S_3$}{S3} Discrete Gauging of the Rank-one \texorpdfstring{$E_3$}{E3} Theory}

\paragraph{Wreathing the Magnetic Quiver:} The $(S_3\simeq D_6)$-wreathing of $\mathcal{O}_\ttiny{min.}^{\su(3)}$ contains as contributions both the $\ZZ_3$- and the $\ZZ_2$-wreathings that we considered before, the latter of which required the modification of the wreathing procedure described in Section \ref{sec:Z2cuttingedges}. In fact, by numbering the nodes $1$, $2$, and $3$, the gauging of the $S_3$ automorphism of the quiver corresponds to summing over 
\begin{equation}
    \{\ID,(12),(23),(31),(123),(132)\}\coma
\end{equation}
meaning that the resulting Hilbert series is obtained by summing six contributions: one will be the Hilbert series of $\mathcal{O}_\ttiny{min.}^{\su(3)}$, three will be the Hilbert series of $\mathcal{O}_\ttiny{min.}^{\su(3)}\wr \ZZ_2$, and two will be the Hilbert series of $\mathcal{O}_\ttiny{min.}^{\su(3)}\wr \ZZ_3$ we computed before. The resulting Hilbert series is
\begin{equation}\label{eqn:mewtwo}
\begin{split}
    \HS_{\mathcal{O}^{\su(3)}_\text{\tiny min}\wr S_3}(t) &=\frac{1-t^2+4 t^4+4 t^6+4 t^8-t^{10}+t^{12}}{\left(1-t^2\right) \left(1-t^4\right)^2 \left(1-t^6\right)} \\
    &= \PE[6 t^4+9 t^6+2 t^8-21 t^{10}+\mathcal{O}\left(t^{11}\right)]\fstop
\end{split}
\end{equation}
The fact that the leading order is at $t^4$ shows that the $S_3$-wreathing leads to no residual flavor symmetry, as expected. Similarly to the $\mathbb{Z}_3$-wreathing, the Higgs branch of $E_3 \wr S_3$ is given by the sum of the Hilbert series for $\mathcal{O}^{\su(3)}_\text{\tiny min}\wr S_3$ and that of $E_1$. The final Hilbert series is
\begin{equation}\label{eq:O3S3E1-HS}
    \begin{split}
        \HS_{E_3\wr S_3}(t) &= \HS_{\mathcal{O}^{\su(3)}_\text{\tiny min}\wr S_3}(t) + \HS_{E_1}(t)-1\\
        &=  \PE[3 t^2+5 t^4-9 t^6-t^8+30 t^{10}+\mathcal{O}\left(t^{11}\right)]\coma
    \end{split}
\end{equation}
where the only moment maps come from the $\su(2)$ flavor symmetry of the $E_1$ theory.  

\paragraph{Curve Analysis:}
The $S_3$ group is generated by the rotation $r$ and the reflection $s$. The invariance of the curve must hold for both generators simultaneously. As discussed previously, the action of the outer automorphism $s$ imposes the constraints $y_1 = y_2^{-1}$ and $y_3 = 1$. On the other hand, the action of the inner automorphism $r$ requires the curve to be invariant under the cyclic permutation $(y_1,y_2,y_3) \to (y_2,y_3,y_1)$.\footnote{The perceptive reader will note that, with the $r$ and $s$ actions defined in this way on the mass parameters, they commute: $srs = r$. However, as mentioned in Section \ref{sec:swcurvediscrgauging}, the space of inequivalent mass parameters of the Seiberg--Witten curve is the Cartan torus modulo Weyl transformations, $T_{\mathbb{C}} / W$, and $s$ acts in the same way on $T_{\mathbb{C}} / W$ as $s'$ defined via $(y_1, y_2, y_3) \rightarrow (y_2^{-1}, y_1^{-1}, y_3^{-1})$. Then $r$ and $s'$ obey the $S_3$ relation: $s'rs' = r^{-1}$. This does not change the invariant locus of the mass-parameter space, and so we do not belabor the point.} Intersecting these conditions forces all the parameters of the $\mathfrak{su}(3)$ sector to collapse to the same value, namely:
\begin{equation}
y_1 = y_2 = y_3 = 1 \fstop
\end{equation}
The parameter $m$, being unaffected by either generator, remains completely unconstrained. The total rank of the flavor symmetry is therefore reduced from $3$ to $1$, reproducing the correct fugacity for the residual $\mathfrak{su}(2)$ algebra.

\subsubsection{The \texorpdfstring{$S_3 \times \ZZ_2^{(2)}$}{S3xZ2(2)} Discrete Gauging of the Rank-one \texorpdfstring{$E_3$}{E3} Theory}\label{sec:E3wrS3Z22}

\paragraph{Wreathing the Magnetic Quiver:}

We can also consider combining the $S_3$ action with the $\mathbb{Z}_2^{(2)}$, which acts on the $\mathbb{C}^2/\mathbb{Z}_2$ factor in the Higgs branch. The resulting Hilbert series again follow from the union:
\begin{equation}\label{eq:O3S3E1Z2-HS}
    \begin{split}
        \HS_{E_3\wr (S_3\times \ZZ_2)}(t) &= \HS_{\mathcal{O}^{\su(3)}_\text{\tiny min}\wr S_3}(t) + \HS_{E_1\wr \ZZ_2}(t) -1\\
        &=  \PE[t^2+8 t^4+3 t^6-20 t^8-14 t^{10}+\mathcal{O}\left(t^{11}\right)]\coma
    \end{split}
\end{equation}
with a $\mathfrak{u}(1)$ flavor symmetry arising from the $E_1\wr \ZZ_2^{(2)}$ sector of the theory.

\paragraph{Curve Analysis:}
This total symmetry combines the full $S_3$ group with the inversion of the $A_1$ branch, $t$. To guarantee the invariance of the curve, we must simultaneously apply the constraints originating from the $S_3$ sector and the $\mathbb{Z}_2^{(2)}$ sector. As calculated for $S_3$, the $\mathfrak{su}(3)$ portion of the algebra is completely broken, leaving only the rank-$1$ portion associated with $m$ intact. The final total rank is $1$, which is perfectly compatible with the residual $\mathfrak{u}(1)$ symmetry.

\subsubsection{The \texorpdfstring{$S_3^{DT}$}{S3DT} Discrete Gauging of the Rank-one \texorpdfstring{$E_3$}{E3} Theory}

\paragraph{Wreathing the Magnetic Quiver:} As discussed already for the $\ZZ_2^{DT}$ and $\ZZ_2^{(1)} \times \ZZ_2^{(2)}$ wreathings, the structure of the moduli space as a union of cones implies that the double transposition is identical to performing the elements of the double transposition separately on the two moduli spaces. Therefore, the Hilbert series of the $S_3^{DT}$-wreathing is the same as given for $S_3 \times \mathbb{Z}_2^{(2)}$ in equation \eqref{eq:O3S3E1Z2-HS}. This can also be confirmed by brute-force application of the wreathed monopole formula. 

\paragraph{Curve Analysis:}
This subgroup is generated by the rotation $r$ and the double transposition $st$. As in the previous cases, the invariance under $S_3$ fully breaks the $\mathfrak{su}(3)$ algebra and no constraint on $m$, leaving only one free parameter, compatibly with the $\mathfrak{u}(1)$ flavor read from the wreathing.

\subsection{The Rank-one \texorpdfstring{$E_4$}{E4} Theory}

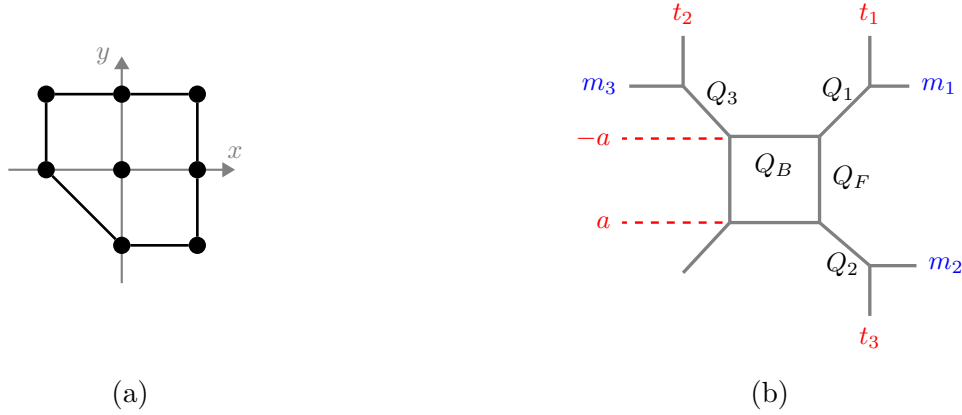
\begin{figure}[t]
\centering
\begin{subfigure}[b]{0.49\textwidth}
\centering
\begin{tikzpicture}[font=\footnotesize,baseline=0]
  \path (0, 2.5) -- (0, -2.5);
  \draw[thick,gray,-Triangle] (-1.5,0) -- node[above,pos=1] {$x$} (1.5,0);
  \draw[thick,gray,-Triangle] (0,-1.5) -- node[left,pos=1] {$y$} (0,1.5);

  \node[node, fill=black] (O2) at (0,0) {};
  \node[node, fill=black] (A2) at (-1,0) {};
  \node[node, fill=black] (B2) at (0,1) {};
  \node[node, fill=black] (C2) at (1,-1) {};
  \node[node, fill=black] (D2) at (-1,1) {};
  \node[node, fill=black] (E2) at (0,-1) {};
  \node[node, fill=black] (F2) at (1,0) {};
  \node[node, fill=black] (G2) at (1,1) {};

  \draw[line width=1pt] (B2)--(D2)--(A2)--(E2)--(C2)--(F2)--(G2)--(B2);
\end{tikzpicture}
\caption{}
\end{subfigure}\hfill
\begin{subfigure}[b]{0.49\textwidth}
\centering
\begin{tikzpicture}[
  scale=.95,
  font=\footnotesize,baseline=0,
  brane/.style={gray,line width=1.25pt},
  midarrow/.style={
    postaction={decorate},
    decoration={markings,
      mark=at position .55 with {\arrow[black]{Stealth[length=2.2mm,width=2.2mm]}}
    }
  },
  sublab/.style={font=\large,gray}
]
  \def\w{1.25}
  \def\hu{0.55}
  \def\hd{0.65}

  \path (0, 2.5) -- (0, -2.5);

  \coordinate (C) at (0, \hu);
  \coordinate (D) at (0, -\hd);
  \coordinate (B) at (-\w, \hu);
  \coordinate (E) at (-\w, -\hd);

  \coordinate (A) at ($(B) + (-0.65, 0.70)$);
  \coordinate (F) at ($(E) + (-0.65, -0.70)$);
  \coordinate (G) at ($(C) + (0.70, 0.70)$);
  \coordinate (H) at ($(D) + (0.70, -0.60)$);

  \draw[brane] (C) -- (B);
  \draw[brane] (E) -- (B);
  \draw[brane] (C) -- (D);
  \draw[brane] (E) -- (D);

  \draw[brane] (A) -- (B);
  \draw[brane] (C) -- (G);
  \draw[brane] (E) -- (F);
  \draw[brane] (H) -- (D);

  \draw[brane] (A) -- ++(-0.75, 0) node[left, blue] {$m_3$};
  \draw[brane] ($(G) + (0.55, 0)$) node[right, blue] {$m_1$} -- (G);
  \draw[brane] (H) -- ++(0.65, 0) node[right, blue] {$m_2$};

  \draw[brane] (G) -- ++(0, 0.70) node[above, red] {$t_1$};
  \draw[brane] (A) -- ++(0, 0.70) node[above, red] {$t_2$};
  \draw[brane] (H) -- ++(0, -0.70) node[below, red] {$t_3$};

  \draw[red,dashed,line width=.9pt] ($(B) + (-0.1, -0.03)$) -- ++(-1.40, 0) node[left] {$-a$};
  \draw[red,dashed,line width=.9pt] ($(E) + (-0.1, 0)$) -- ++(-1.40, 0) node[left] {$a$};

  \node at ($(A) + (0.55, -0.10)$) {$Q_3$};
  \node at ($(B) + (0.62, -0.41)$) {$Q_B$};
  \node at ($(D) + (0.45,  0.63)$) {$Q_F$};
  \node at ($(G) + (-0.44, -0.05)$) {$Q_1$};
  \node at ($(H) + (-0.38, 0.02)$) {$Q_2$};
\end{tikzpicture}
\caption{}
\end{subfigure}
\caption{Toric diagram and $(p,q)$-web (from \cite{Mitev:2014jza,Kim:2014nqa}) for the $E_4$ theory. The theory can be realized as $\mathbb P^1 \times \mathbb P^1$, blown up in three points, represented by the vertices $(-1,1)$, $(1,1)$, $(1,-1)$ of the toric diagram. The K\"ahler volumes become, in the $(p,q)$-web, the lengths of the various edges.}
\label{fig:e4pqwebtoric}
\end{figure}

We now turn to the rank-one $E_4$ theory, of which we have depicted the toric diagram and the $(p,q)$-web in Figure \ref{fig:e4pqwebtoric}. As usual, we first describe the mirror curve and the magnetic quiver that capture, respectively, the Coulomb branch and Higgs branch phases of the theory. 

\paragraph{Discrete Symmetries from Magnetic Quivers for the Higgs Branch:} For the $E_4$ theory, the Higgs branch is known to be isomorphic to the closure of the minimal nilpotent orbit of $\mathfrak{su}(5)$: $\mathcal{O}^{\su(5)}_\ttiny{min}$. A unitary magnetic quiver for the Higgs branch is thus
\begin{equation}\label{eq:unitaryE4}
    \begin{tikzpicture}[baseline=0,font=\footnotesize]
        \def\radius{1cm};
        \node[node, label=above:{$1$}] (A1) at (90:\radius) {}; 
        \node[node, label=above:{$1$}] (A2) at (162:\radius) {}; 
        \node[node, label=below:{$1$}] (A3) at (234:\radius) {};
        \node[node, label=below:{$1$}] (A4) at (306:\radius) {};
        \node[node, label=above:{$1$}] (A5) at (18:\radius) {};
        \draw (A1) -- (A2) -- (A3) -- (A4) -- (A5) -- (A1);
        \node (Z2) [right=5mm of A5] {$/\U(1)\fstop$};
    \end{tikzpicture} 
\end{equation}
As discussed, we propose that automorphisms of this magnetic quiver for the Higgs branch are discrete global symmetries of the $E_4$ SCFT. The group of quiver automorphisms of the quiver in equation \eqref{eq:unitaryE4} is the group of automorphisms of the affine $A_4$ Dynkin diagram, which is the dihedral group of order ten:
\begin{equation}
    \operatorname{Aut}\left(\widetilde{\mathcal{D}}_{\mathfrak{su}(5)}\right) = D_{10}\fstop
\end{equation}
This is generated by an order five generator, $r$, and an order two generator, $s$, which act on the quiver as follows:
\begin{equation}
\label{eq:symmetriesquivere4}
    r\,:\,\, \begin{tikzpicture}[baseline=0,font=\footnotesize]
        \def\radius{1cm};
        \node[node, label=right:{$1$}] (A1) at (90:\radius) {}; 
        \node[node, label=above:{$1$}] (A2) at (162:\radius) {}; 
        \node[node, label=below:{$1$}] (A3) at (234:\radius) {};
        \node[node, label=below:{$1$}] (A4) at (306:\radius) {};
        \node[node, label=above:{$1$}] (A5) at (18:\radius) {};
        \draw (A1) -- (A2) -- (A3) -- (A4) -- (A5) -- (A1);
        \draw[-Triangle, red] (A1) to[bend right=25] (A2);
        \draw[-Triangle, red] (A2) to[bend right=25] (A3);
        \draw[-Triangle, red] (A3) to[bend right=25] (A4);
        \draw[-Triangle, red] (A4) to[bend right=25] (A5);
        \draw[-Triangle, red] (A5) to[bend right=25] (A1);
    \end{tikzpicture}\qquad,\qquad s\,:\,\, \begin{tikzpicture}[baseline=0,font=\footnotesize]
        \def\radius{1cm};
        \node[node, label=right:{$1$}] (A1) at (90:\radius) {}; 
        \node[node, label=above:{$1$}] (A2) at (162:\radius) {}; 
        \node[node, label=below:{$1$}] (A3) at (234:\radius) {};
        \node[node, label=below:{$1$}] (A4) at (306:\radius) {};
        \node[node, label=above:{$1$}] (A5) at (18:\radius) {};
        \draw (A1) -- (A2) -- (A3) -- (A4) -- (A5) -- (A1);
        \draw[Triangle-Triangle,red] ([yshift=-5mm]A3.south) to[bend right=60] 
        ([yshift=-5mm]A4.south);
        \draw[densely dashed, red,thin]  ([yshift=3mm]A1.north) -- ([yshift=-20mm]A1.south);
    \end{tikzpicture}\fstop
\end{equation}
We have summarized the four conjugacy classes of subgroups of $D_{10}$, and the fixed-point subalgebras of $\mathfrak{e}_4 = \mathfrak{su}(5)$ for each such conjugacy class, utilizing the homomorphism as in equation \eqref{eqn:BIGHOMO}, in Table \ref{tbl:E4dg}. This is the prediction for the continuous flavor symmetry of the discretely-gauged SCFT, $E_4 / H$.

\begin{table}[t]
\centering
\begin{tabular}{c|c|c}
$H$ & Generators & $\mathfrak{g}^H$ \\\hhline{=|=|=}
$1$ & $1$ & $\mathfrak{su}(5)$ \\
$\mathbb{Z}_2$ & $s$ & $\mathfrak{so}(5)$ \\
$\mathbb{Z}_5$ & $r$ & $\mathfrak{u}(1)^{\oplus 4}$ \\
$D_{10}$ & $r, s$ & $\emptyset$
\end{tabular}
\caption{For each subgroup $H$ of $\operatorname{Aut}(\widetilde{\mathcal{D}}_{\mathfrak{su}(5)})$, we list the fixed-point subalgebra of $\mathfrak{su}(5)$, $\mathfrak{g}^H$, as defined in equation \eqref{eqn:gHdef}.}
\label{tbl:E4dg}
\end{table}

There exists an alternative magnetic quiver for the Higgs branch of the $E_4$ theory, which is orthosymplectic instead of unitary \cite{Bourget:2020xdz}. This magnetic quiver is:
\begin{equation}\label{eq:orthosymE4}
    \begin{tikzpicture}[baseline=0,font=\footnotesize]
        \node[node, label=below:{$2$},fill=blue] (A2) {};
        \node[node, label=below:{$2$},fill=red] (Al) [left=6mm of A2] {};
        \node[node, label=below:{$2$},fill=red] (Ar) [right =6mm of A2]{};
        \node[node, label=right:{$1$}] (A3) [above=4mm of A2] {};
        \node[flavor, label=right:{$1$}] (A7) [above=6mm of A3] {};
        \draw (A2.north) -- (A3.south);
        \draw[snake it] (A3.north) -- (A7.south);
        \draw (Al.east) -- (A2.west);
        \draw (A2.east) -- (Ar.west);
        \node (Z2) [right=6mm of Ar] {$/\ZZ_2$};
    \end{tikzpicture} \,.
\end{equation}
It is clear that this quiver admits a $\mathbb{Z}_2$ diagram automorphism by which we can wreath. 
Furthermore, we see that the quiver in equation \eqref{eq:orthosymE4} sees only a $\ZZ_2$ discrete symmetry of the $E_4$ theory, as opposed to the $D_{10}$ seen from the unitary magnetic quiver.

\paragraph{Mirror Curve:}
The curve for the $E_4$ theory is
\begin{align}\label{eq:SW_E4_first}
    P(x,y) = U + c_{(1,0)} x + c_{(0,1)} y +  c_{(-1,0)} \frac{1}{x} +  c_{(0,-1)} \frac{1}{y}
    + c_{(1,1)} xy + c_{(-1,1)} \frac{y}{x} + c_{(1,-1)} \frac{x}{y}\fstop 
\end{align}
Following \cite{Kim:2014nqa}, we use the rescaling freedoms to write this curve as
\begin{equation}\label{eq:curvefornowrandomlabel}
 U
+
xy-(t_1+t_2)y+t_1t_2\frac{y}{x}
-(m_1+m_2)x
+m_1m_2 \frac{x}{y}
-\frac{m_1m_2t_3}{y}
-\frac{m_3t_1t_2}{x}
 = 0 \fstop
\end{equation}
The curve has an explicit $\mathfrak{su}(2)\oplus \mathfrak{su}(2)$ flavor algebra, with Cartan subalgebra being 
\begin{equation}
    T \equiv \frac{t_1}{t_2}\coma  M \equiv \frac{m_1}{m_2}\fstop \end{equation}
The relation between the $(p,q)$-web parameters in Figure \ref{fig:e4pqwebtoric} and the parameters in the mirror curve in equation \eqref{eq:curvefornowrandomlabel} is
\begin{equation}
\label{eq:massesfromfiguree4}
\renewcommand{\arraystretch}{1.2}
\begin{array}{rclrclrcl}
  m_1 & = & \sqrt{Q_F} Q_1\coma  & m_2 & = & \dfrac{1}{Q_2\sqrt{Q_F}} \coma & m_3 & = & \sqrt{Q_F} Q_3\coma\\
  t_1 & = & \sqrt{Q_B} Q_1\coma  & t_2 & = & \dfrac{1}{Q_3\sqrt{Q_B}} \coma & t_3 & = & \sqrt{Q_B} Q_2\fstop
\end{array} 
\end{equation}
We can also express the curve parameters in terms of the Cartan of the UV flavor symmetry. To do so, let us introduce the $\mathfrak{su}(5)$ fugacities $y_1,\cdots,y_5$ such that
\begin{equation}
\label{eq:fugacitiessu5}
    y_1 y_2 y_3 y_4 y_5 = 1\fstop 
\end{equation}
To express the $m_i,t_i$ in terms of the $y_j$, one first gives the K\"ahler volumes in terms of the $y_j$ \cite{Kim:2014nqa}:
\begin{equation}
\label{eq:mitev338}
\renewcommand{\arraystretch}{1.2}
\begin{array}{c}
  Q_F = u^2 y_1\coma \qquad Q_B = u^2 y_2\coma \\
  \begin{array}{rclrclrcl}
    Q_1 & = & u^{-1} y_3 y_4\coma  & Q_2 & = & u^{-1} y_3 y_5\coma & Q_3 & = & u^{-1} y_4 y_5\fstop
  \end{array}
\end{array} 
\end{equation}
Then, plugging equation \eqref{eq:mitev338} into equation \eqref{eq:massesfromfiguree4}:
\begin{equation}
\label{eq:massestoy}
\renewcommand{\arraystretch}{1.2}
\begin{array}{rclrclrcl}
  m_1 & = & \sqrt{y_1}\,y_3y_4\coma  & m_2 & = & \dfrac{1}{\sqrt{y_1}\,y_3y_5} \coma & m_3 & = & \sqrt{y_1}\,y_4y_5\coma\\
  t_1 & = & \sqrt{y_2}\,y_3y_4\coma  & t_2 & = & \dfrac{1}{\sqrt{y_2}\,y_4y_5} \coma & t_3 & = & \sqrt{y_2}\,y_3y_5\fstop
\end{array} 
\end{equation}

This leads to the following expression for the curve in terms of the $y_i$; which will be useful for later purposes: 
\begin{equation}
\label{eq:curveyj}
\begin{split}
&\frac{\sqrt{y_3y_4}}{y_5}\,u+
xy-\left(\sqrt{y_2}\,y_3y_4+\frac{1}{\sqrt{y_2}\,y_4y_5}\right)y+\frac{y_3}{y_5}\frac{y}{x}
-\left(\sqrt{y_1}\,y_3y_4+\frac{1}{\sqrt{y_1}\,y_3y_5}\right)x
\\
&+\frac{y_4}{y_5}\frac{x}{y}
-\frac{\sqrt{y_2}\,y_3y_4}{y}
-\frac{\sqrt{y_1}\,y_3y_4}{x}
 = 0 \coma
\end{split}
\end{equation}
where we rescaled the CB coordinate as
\begin{equation}
    U = \frac{\sqrt{y_3y_4}}{y_5}\,u\fstop
\end{equation}

The toric diagram possesses just one evident symmetry that corresponds, in the mirror curve language, to the exchange of the vertical and horizontal axes. This corresponds, using the techniques described in Section \ref{sec:2}, to the $\mathbb Z_{2}$ symmetry $s$ in equation \eqref{eq:symmetriesquivere4}. On the other hand, the magnetic quiver also possesses a $\mathbb Z_5$ symmetry that has no counterpart in terms of a toric isometry readable from the toric diagram with the method of \cite{DeMarco:2025pza}. 

\subsubsection{The Higgs Branch of the Rank-one \texorpdfstring{$E_4$}{E4} Theory}

The Hilbert series for the unitary quiver in equation \eqref{eq:unitaryE4} has been computed in, for example, \cite{Ferlito:2017xdq}, and it can be obtained by parametrizing the lattice such that the magnetic fluxes are $m_1+m_2+m_3+m_4+m_5\in\{0,\cdots,4\}$. The resulting Hilbert series can then be expressed in terms of $\su(5)$ characters as
\begin{equation}\label{eq:E4-HS}
    \HS_{\mathcal{O}_\ttiny{min}^{\su(5)}}(t,\mathbf{x}) = \PE\left[\chi^A_{[1,0,0,1]}(\mathbf{x})t^2-\left(1+\chi^A_{[1,0,0,1]}(\mathbf{x})+\chi^A_{[0,1,1,0]}(\mathbf{x})\right)t^4+\mathcal{O}(t^6)\right]\fstop
\end{equation}
The unrefined Hilbert series is
\begin{equation}
    \HS_{\mathcal{O}_\ttiny{min}^{\su(5)}}(t,1) = \frac{1 + 16 t^2 + 36 t^4 + 16 t^6 + t^8}{(1-t^2)^8} = \PE\left[24 t^2-100 t^4+800 t^6+\mathcal{O}\left(t^8\right)\right]\fstop
\end{equation}

Similarly, the exact Hilbert series for the orthosymplectic quiver realization of $\mathcal{O}_\text{min}^{\mathfrak{su}(5)}$, in equation \eqref{eq:orthosymE4}, was computed in \cite{Bourget:2020xdz} and is
\begin{equation}\label{eq:HSE4-ortho}
\begin{split}
    \HS_{E_{4}}(t,\omega) &= \frac{\left(\begin{aligned}
        & 1+t^2 (8 \omega +12)+t^4 (48 \omega +58)+t^6 (136 \omega +124) \\
        &\quad +t^8 (176 \omega +170)+t^{10} (136 \omega +124)+\cdots+t^{16} 
    \end{aligned}\right)}{\left(1-t^2\right)^4 \left(1-t^4\right)^4} \\
    &= \PE\left[t^2 (8 \omega +16)-t^4 (48 \omega +52)+t^6 (416 \omega +384)
    +\mathcal{O}(t^{8})\right]\fstop
\end{split}
\end{equation}
By considering the decomposition $\su(5)\rightarrow \su(4)\oplus\U(1)$, we can see that $\omega$ is the $\ZZ_2^{[0]}$ fugacity that tracks the $\U(1)$ charge parity within this subalgebra embedding.

\subsubsection{The \texorpdfstring{$\ZZ_2$}{Z2} Discrete Gauging of the Rank-one \texorpdfstring{$E_4$}{E4} Theory}

In this section, we discuss the $\mathbb Z_2$ discrete gauging of the $E_4$ theory, associated with the symmetry that exchanges the $x$ and $y$ axes of the toric diagram in Figure \ref{fig:e4pqwebtoric}, or, equivalently, acts by the $s$ generator on the magnetic quiver for the Higgs branch as in equation \eqref{eq:symmetriesquivere4}. As we will see, this produces a putative new rank-one 5d SCFT with flavor algebra $\mathfrak{so}(5)$. 

\paragraph{Wreathing the Magnetic Quiver:}
The $\ZZ_2$-wreathing of the magnetic quiver for the $E_4$ theory is the generalization to five nodes of the $\ZZ_2$-wreathing we computed for the $\mathcal{O}^{\su(3)}_\mathrm{min}$ magnetic quiver and generalized to an arbitrary number of nodes in Appendix \ref{app:birch}.
By choosing to decouple the $\U(1)$ at the affine node of $A_4$ kept fixed by the $\ZZ_2$-wreathing, i.e., $m_5 = 0$, the Hilbert series is obtained by summing over two contributions:
\begin{equation}
    \HS_{\mathcal{O}^{\su(5)}_\text{\tiny min}\wr \ZZ_2}(t,\mathbf{z}) = \frac{1}{2} \left(\HS_{\mathcal{O}^{\su(5)}_\text{\tiny min}}^{\text{\tiny contr. 1}}(t,\mathbf{z})+\HS_{\mathcal{O}^{\su(5)}_\text{\tiny min}}^{\text{\tiny contr. 2}}(t,\mathbf{z})\right)\coma
\end{equation}
where
\begin{equation}\label{eqn:sparx}
    \begin{split}
        \HS_{\mathcal{O}^{\su(5)}_\text{\tiny min}}^{\text{\tiny contr. 1}}(t,\mathbf{z}) & =\frac{1-t^2}{(1-t^2)^5} \sum_{m_1, \dots, m_{4} \geq -\infty}^{\infty} t^{2 \Delta_1} z_1^{m_1+m_4} z_2^{m_2+m_3}\coma\\
        \HS_{\mathcal{O}^{\su(5)}_\text{\tiny min}}^{\text{\tiny contr. 2}}(t,\mathbf{z}) & =  \frac{1-t^2}{(1-t^2)(1-t^4)^2} \sum_{m_1, m_2 \geq -\infty}^\infty {\color{red}{(-1)^{m_2}}} \, t^{2 \Delta_2} z_1^{2m_1}z_2^{2m_2}\coma\\
    \end{split}
\end{equation}
and 
\begin{equation}
    \begin{split}
        \Delta_1 &= \frac{1}{2}\left(|m_1-m_2|+|m_2-m_3|+|m_3-m_4|+|m_4|+|m_1|\right)\coma\\
        \Delta_2 &= |m_1-m_2|+|m_1|\fstop
    \end{split}
\end{equation}
In the second contribution, we have written in {\color{red}red} the phase that we proposed, in Section \ref{sec:Z2cuttingedges}, must be included in the wreathed monopole formula for the Coulomb branch Hilbert series when the axis of wreathing bisects an edge of the quiver. Thus, by defining,
\begin{equation}\label{eq:A5_map}
    z_1 \rightarrow \frac{y_2^2}{y_{1}^2}\coma z_2 \rightarrow y_1\coma
\end{equation}
we obtain
\begin{equation}\label{eq:E4Z2-HS}
\begin{split}
    \HS_{\mathcal{O}^{\su(5)}_{\mathrm{min}}\wr \ZZ_2}(t,\mathbf{y}) = \,\PE&\Big[t^2 \chi^B_{[0,2]}(\mathbf{y})+t^4 \left(\chi^B_{[4,0]}(\mathbf{y})-\chi^B_{[1,0]}(\mathbf{y})\right) \\
    &+t^6 \left(\chi^B_{[1,0]}(\mathbf{y})-\chi^B_{[2,2]}(\mathbf{y})-\chi^B_{[3,2]}(\mathbf{y})\right)+\mathcal{O}\left(t^7\right)\Big]\coma
\end{split}
\end{equation}
or unrefined,
\begin{equation}
\begin{split}
    \HS_{\mathcal{O}^{\su(5)}_{\mathrm{min}}\wr \ZZ_2}(t,1) &= \frac{1+6 t^2+67 t^4+102 t^6+208 t^8+102 t^{10}+\cdots+t^{16}}{\left(1-t^2\right)^4 \left(1-t^4\right)^4} \\
    &=\PE\left[10 t^2+50 t^4-230 t^6+\mathcal{O}\left(t^7\right)\right]\fstop
\end{split}
\end{equation}
Therefore, the wreathed quiver appears to have a Coulomb branch global symmetry which is $\mathfrak{so}(5)$, exactly what we would expect from the discrete gauging on the Higgs branch of the $\mathbb{Z}_2$ symmetry associated with the outer-automorphism of the unwreathed magnetic quiver.

To emphasize the importance of the proposal in Section \ref{sec:Z2cuttingedges}, we now consider the Hilbert series obtained if we did \emph{not} include the phase written in {\color{red}red} in equation \eqref{eqn:sparx}. Appropriately redefining the fugacities, we would find
\begin{equation}\label{eq:A4Z2-HS}
    \HS_{\mathcal{O}^{\su(5)}_\text{\tiny min}\wr \ZZ_2}(t,\mathbf{y}) = \PE\left[\chi^B_{[2,0]}(\mathbf{y})t^2+\mathcal{O}(t^4)\right]\coma
\end{equation}
or, unrefined
\begin{equation}
    \HS_{\mathcal{O}^{\su(5)}_\text{\tiny min}\wr \ZZ_2}(t,1) = \frac{1+6 t^2+21 t^4+6 t^6+t^8}{\left(1-t^2\right)^8} = \PE\left[14 t^2-50 t^6+175 t^8+\mathcal{O}\left(t^9\right)\right]\fstop
\end{equation}
That is, we find moment maps transforming in the $[2,0]$ representation of $\mathfrak{so}(5)$. However, this representation does not coincide with any adjoint representation of a reductive Lie algebra. Thus, this expression cannot be the Coulomb branch Hilbert series of any 3d $\mathcal{N}=4$ theory, nor the Higgs branch Hilbert series of any eight-supercharge SCFT.

We can generalize this procedure to the $\ZZ_2$-wreathing of the magnetic quivers for all $\mathcal{O}^{\su(2n+1)}_\mathrm{min}$, for $n > 2$. If one computes the wreathing without introducing the phase of Section \ref{sec:Z2cuttingedges}, the leading order of the Hilbert series will have a $[2,0,\ldots,0]$ of $B_n$, as it was derived in \cite{Grimminger:2024mks}. However, the invariant moduli space is obtained by introducing the phase to the twisted contribution of the Hilbert series, and the resulting leading order will be the $[0,1,0\ldots,0]$ representation of $B_n$, leading to the correct flavor symmetry of the wreathed moduli space. We defer to Appendix \ref{app:birch} for more details of this generalization.

\paragraph{Curve Analysis:}
To realize the $\mathbb Z_2$ wreathing, we reflect with respect to the bisector of the first and third quadrants, that is, the only axis of symmetry in Figure \ref{fig:e4pqwebtoric}:
\begin{equation}
\label{eq:actionz2e4}
     (x,y)   \to  \left(\frac{y_5}{y_4}y \,,
\frac{y_5}{y_3}x\right) \,,
\end{equation}
together with $u \to u$.
The map exchanges $Q_F \leftrightarrow Q_B^{-1}, Q_2 \leftrightarrow Q_3^{-1}$ or, equivalently, via equation \eqref{eq:massestoy}, 
\begin{equation}
\label{eq:outeraut}
    (y_1,y_2,y_3,y_4,y_5) \quad \to  \quad (y_2^{-1},y_1^{-1},y_4^{-1},y_3^{-1},y_5^{-1})\,,
\end{equation}
that is the action of the outer automorphism.\footnote{A standard realization of the outer automorphism is $g \to -g^T, g \in \mathfrak{su}(5)$, which sends $y_i \to y_i^{-1}$. Here, we further compose with an inner automorphism, namely an element of the Weyl (permutation) group.}
On the $m_i,t_i$, equation \eqref{eq:outeraut} acts as 
\begin{equation}
(m_1,m_2,m_3,t_1,t_2,t_3) \to  (t_1^{-1},t_2^{-1},t_3^{-1},m_1^{-1},m_2^{-1},m_3^{-1})\fstop
\end{equation}
Applying equation \eqref{eq:outeraut} first and then equation \eqref{eq:actionz2e4} to \eqref{eq:curveyj} transforms the curve as
\begin{equation}
    P_{E_4}(x,y) \to  \frac{y_5^2}{y_3 y_4} P_{E_4}(x,y) \,,
\end{equation} 
hence, leaving invariant its zero locus, up to imposing
\begin{equation}
\label{eq:costre41}
   y_1  = y_4^{-1} y_5^{-3},\qquad  y_2  = y_3^{-1} y_5^{-3}\fstop
\end{equation}
Let us interpret equation \eqref{eq:costre41}. Plugging equation \eqref{eq:costre41} into equation \eqref{eq:fugacitiessu5}, we get: 
\begin{equation}
\label{eq:costre41second}
    y_5^5 = 1\coma
\end{equation}
and interpreting the fifth-roots of unity ambiguity as the central element of $\SU(5)$, we have that in $\text{PSU}(5)$ in equation \eqref{eq:costre41second} implies $y_5 = 1$. Inserting $y_5 = 1$ into equation \eqref{eq:costre41}, the constraints for the invariance of the curve are  
\begin{equation}
\label{eq:finalresulte4modz2}
    y_5 = 1, \quad y_1 = y_4^{-1}, \quad y_2 = y_3^{-1}\fstop  
\end{equation}
Equation \eqref{eq:finalresulte4modz2} is an invariant subalgebra with respect to the outer automorphism in equation \eqref{eq:outeraut}, in the sense that acting with equation \eqref{eq:outeraut} on a solution of equation \eqref{eq:finalresulte4modz2}, we still obtain a solution of equation \eqref{eq:finalresulte4modz2}.  Hence, by general Lie algebra theory, the subalgebra in equation \eqref{eq:finalresulte4modz2} $\mathfrak{so}(5) \cong \mathfrak{usp}(4)$ is a subalgebra of $\mathfrak{su}(5)$, compatible with the magnetic quiver computation.

To conclude, it is instructive to see the mechanism that allowed us to balance, in the mirror curve, the action on $x,y$ with the outer automorphism representative in equation \eqref{eq:outeraut}. Indeed, following \cite{Eguchi:2002fc,Eguchi:2002nx}, the Weierstrass form of the curve is determined by
    \begin{align}
    \label{eq:weiforme4}
g_2(u)
&=
-\left(
-\frac{1}{12}u^4
+\frac{2}{3}\chi_1 u^2
+2\chi_2 u
-\frac{4}{3}\chi_1^2
+4\chi_3
\right),
\\[4pt]
g_3(u)
&=
-\frac{1}{216}u^6
+\frac{1}{18}\chi_1 u^4
+\frac{1}{6}\chi_2 u^3
+\left(
-\frac{2}{9}\chi_1^2
+\frac{1}{3}\chi_3
\right)u^2
\nonumber\\
&\qquad
+\left(
-\frac{2}{3}\chi_1\chi_2
+4
\right)u
+\frac{8}{27}\chi_1^3
-\frac{4}{3}\chi_1\chi_3
-\chi_2^2
+4\chi_4 \,,
\end{align}
where $\chi_i$ denote the fundamental characters of the flavor algebra $\mathfrak{su}(5)$, labeled from left to right in the $A_4$ Dynkin diagram. The curve associated with equation \eqref{eq:weiform} is not invariant under equation \eqref{eq:outeraut}. Indeed, the characters are 
\begin{equation}
    \label{eq:charactersliste4}
    \chi_1 = \sum_{i} y_i, \quad \chi_2 = \sum_{i<j} y_iy_j, \quad \chi_3 = \sum_{i<j} (y_iy_j)^{-1}, \quad \chi_4 = \sum_i y_i^{-1} \,,
\end{equation}
and they are exchanged as $\chi_1 \leftrightarrow \chi_4, \chi_2 \leftrightarrow \chi_3$ by equation \eqref{eq:outeraut}.
However, if we \textit{first} impose equation \eqref{eq:finalresulte4modz2}, it is easy to see that all the characters become invariant under equation \eqref{eq:outeraut}. The reason is that the action in equation \eqref{eq:outeraut}  is conjugate to the action for which equation \eqref{eq:finalresulte4modz2} is the fixed-point subalgebra. Since the characters are invariant under inner automorphisms, checking the invariance under equation \eqref{eq:outeraut} is equivalent to checking the invariance under the representative that fixes equation \eqref{eq:finalresulte4modz2}. By definition, this latter representative does not act on the Cartan coordinates that remain after imposing equation \eqref{eq:finalresulte4modz2}, and hence also on the characters evaluated on them. The same argument can be used to justify the balancing of the $(x,y)$ transformation with outer automorphisms in all the analyzed cases. 

\subsubsection{The \texorpdfstring{$\ZZ_5$}{Z5} Discrete Gauging of the Rank-one \texorpdfstring{$E_4$}{E4} Theory}

Let us now discuss the $\mathbb Z_5$ wreathing associated with the cyclic permutation denoted by $r$ in equation \eqref{eq:symmetriesquivere4}. 
As we discussed, this symmetry of the magnetic quiver has no counterpart in terms of a symmetry of the toric diagram in Figure \ref{fig:e4pqwebtoric}, and hence the analysis of the Seiberg--Witten curve is absent.

\paragraph{Wreathing the Magnetic Quiver:}
We can consider the $\ZZ_5$-wreathing of the magnetic quiver for the Higgs branch of $E_4$ by performing a similar computation to that used in Section \ref{sec:E3th} when considering the $\ZZ_3$-wreathing of $\mathcal{O}^{\su(3)}_\mathrm{min}$. Specifically, we consider the elements of $\ZZ_5$, and we write down the contribution for each gauge node following the prescription of \cite{Grimminger:2024mks}. 
Effectively, there will be a contribution coming from the Hilbert series of $\mathcal{O}^{\su(5)}_\mathrm{min}$ and four contributions of a single $\U(1)$ node with all fugacities elevated to the fifth power. We then obtain the following Hilbert series:
\begin{equation}\label{eq:E4Z5-HS}
\begin{split}
 \HS_{E_4\wr \ZZ_5}(t) &  = \frac{1-3 t^2+33 t^4-31 t^6+70 t^8-31 t^{10}+33 t^{12}-3 t^{14}+t^{16}}{\left(1-t^2\right)^7(1-t^{10})}\\
    & = \PE\left[4 t^2+30 t^4+60 t^6-305 t^8-1500 t^{10}+\mathcal{O}\left(t^{11}\right)\right]\fstop
\end{split}
\end{equation}
The resulting flavor symmetry appears to be $\mathfrak{u}(1)^{\oplus 4}$, which matches the expectation from the discrete gauging, and it is the direct generalization of the $\mathbb{Z}_3$-wreathing of $\mathcal{O}^{\su(3)}_\mathrm{min}$.\footnote{Also, generalizing this case, it is possible to show that the flavor symmetry of the $\ZZ_{2n+1}$ wreathing of $\mathcal{O}^{\mathfrak{su}(2n+1)}_\ttiny{min}$ is $\mathfrak{u}(1)^{\oplus 2n}$.}

\subsubsection{The \texorpdfstring{$D_{10}$}{D10} Discrete Gauging of the Rank-one \texorpdfstring{$E_4$}{E4} Theory}

Combining the  $\mathbb Z_2$ and $\mathbb Z_5$ actions, we have a $D_{10}$ action on the magnetic quiver of the $E_4$ theory. Since the $\mathbb Z_5$ symmetry has no counterpart in Figure \ref{fig:e4pqwebtoric}, this also implies that such a $D_{10}$ group cannot be fully captured by the toric Calabi--Yau associated with the $E_4$ theory. Thus, we only study the wreathed magnetic quiver, and verify that the Coulomb branch Hilbert series reproduces the expected flavor symmetry after discrete gauging.

\paragraph{Wreathing the Magnetic Quiver:}
The dihedral group $D_{10}$ has ten elements that can be grouped into four rotations of the five $\mathfrak{u}(1)$ nodes, five reflections, and the identity element. The reflections must be dealt with according to the prescription for edge-cutting diagram automorphisms in Section \ref{sec:Z2cuttingedges}.  Putting  everything together, the wreathed Hilbert series is
\begin{equation}\label{eq:E4D5-HS}
\begin{split}
 \HS_{E_4\wr D_{10}}(t) &  = \scalebox{1}{$\displaystyle\frac{1-3 t^2+24 t^4+26 t^6+84 t^8+85 t^{10}+126 t^{12}+85 t^{14}+\cdots+t^{24}}{\left(1-t^2\right)^3 \left(1-t^4\right)^4 \left(1-t^{10}\right)}$}\\
    & = \PE\left[25 t^4+90 t^6+60 t^8-1180 t^{10}+\mathcal{O}\left(t^{11}\right)\right]\fstop
\end{split}
\end{equation}
Similarly to the $(S_3 \simeq D_6)$-wreathing of $\mathcal{O}_\ttiny{min}^{\su(3)}$ in equation \eqref{eqn:mewtwo}, there is no residual flavor symmetry after the wreathing, matching our expectation from $\mathfrak{su}(5)^{D_{10}}$.

\subsubsection{The (Orthosymplectic) \texorpdfstring{$\ZZ_{2}$}{Z2}-Wreathing of the Rank-one \texorpdfstring{$E_4$}{E4} Theory}

The orthosymplectic quiver admits a $\ZZ_2$-wreathing involving the two $\mathfrak{so}(2)$ groups, and the Hilbert series can be computed either by using the procedure introduced in \cite{Lawrie:2025exx} or by extending the results in \cite{Grimminger:2024mks} to orthosymplectic groups. 
Also in this case, this symmetry of the magnetic quiver has no counterpart in the toric diagram in Figure \ref{fig:e4pqwebtoric}, and we discuss only the wreathing of the magnetic quiver for the Higgs branch.

\paragraph{Wreathing the Magnetic Quiver:} The prescription to determine the Coulomb branch Hilbert series for wreathed orthosymplectic quivers was spelled out in \cite{Lawrie:2025exx}, and we simply state the final result here:
\begin{equation}\label{eq:HSE4wrZ2-ortho}
\begin{split}
    \HS_{E_{4}\wr \ZZ_2}(t,\omega) &=  \frac{\left(\begin{aligned}
        & 1+t^2 (6 \omega +6)+t^4 (24 \omega +34)+t^6 (70 \omega +54) \\
        &\quad +t^8 (80 \omega +90)+\cdots+t^{16} 
    \end{aligned}\right)}{\left(1-t^2\right)^4 \left(1-t^4\right)^4} \\
    &= \PE\left[t^2 (6 \omega +10)-t^4 (12 \omega +4)+t^6 (8 \omega -8)
    +\mathcal{O}\left(t^{8}\right)\right]\fstop
\end{split}
\end{equation}
We note that the resulting flavor symmetry is compatible with $\so(6)\oplus \mathfrak{u}(1)$. However, it is possible to interpret the flavor symmetry after the $\ZZ_2$-wreathing from the original $\su(5)$ flavor symmetry. Let us consider the following branching rule,
\begin{equation}
     \begin{split}
        \mathfrak{su}(5) &\rightarrow \mathfrak{so}(6) \oplus \mathfrak{u}(1)_A \\
        \mathbf{24} &\rightarrow \mathbf{15}_0 \oplus \mathbf{1}_0 \oplus \mathbf{4}_5 \oplus \overline{\mathbf{4}}_{-5} \fstop
    \end{split}
\end{equation}
That is, we can think of the unwreathed quiver as having an $\mathfrak{so}(6) \oplus \mathfrak{u}(1)$ global symmetry together with additional moment maps in the spinor and conjugate-spinor representations of $\mathfrak{so}(6)$. In fact, we can see that the $\mathbb{Z}_2^{[0]}$ zero-form symmetry which is parametrized by the fugacity $\omega$ is the $\mathbb{Z}_2$ inside of the $\mathfrak{u}(1)$ in this decomposition. On the Higgs branch, we expect to project out the $\mathfrak{su}(5)$ moment maps that are charged under the $\mathbb{Z}_2$ that we discretely-gauge. We can observe a $\mathbb{Z}_2$ compatible with the wreathed Coulomb branch Hilbert series as follows. Consider the further decomposition
\begin{equation}
     \begin{split}
        \mathfrak{so}(6) &\rightarrow \mathfrak{u}(3)\simeq \su(3)\oplus \mathfrak{u}(1)_B  \\
        \mathbf{15} &\rightarrow \mathbf{9}_0 \oplus \mathbf{3}_{4} \oplus \overline{\mathbf{3}}_{-4} \\
        \mathbf{4} &\rightarrow \mathbf{3}_1 \oplus \mathbf{1}_{-3} \\
        \bm{\overline{4}} &\rightarrow \bm{\overline{3}}_{-1} \oplus \mathbf{1}_{3}\,.
    \end{split}
\end{equation}
We then define our putative $\mathbb{Z}_2$ symmetry via
\begin{equation}\label{eqn:bianca}
    r = (-1)^{(q_A - 5q_B)/20} \,,
\end{equation}
where $q_A$ and $q_B$ are the charges under the two $\mathfrak{u}(1)$ factors in the decomposition. Since $\mathcal{O}_\text{min}^{\mathfrak{su}(5)}$ is generated by the moment map, the fact that the exponent in equation \eqref{eqn:bianca} is integer for all representations appearing in the branching rule of the $\bm{24}$ indicates the $\mathbb{Z}_2$ is well-defined on the Higgs branch. The representations of $\mathfrak{u}(3) \oplus \mathfrak{u}(1)$ coming from the moment map of $\mathfrak{su}(5)$ which are uncharged under $r$ are
\begin{equation}
    \bm{9}_{0,0} \oplus \bm{1}_{0,0} \oplus \bm{3}_{4,0} \oplus \bm{\overline{3}}_{-4,0}\coma
\end{equation}
where we have taken an illuminating linear combination of the two $\mathfrak{u}(1)$ factors. These recombine into the adjoint representation of an enhanced $\mathfrak{su}(4) \oplus \mathfrak{u}(1)$, which matches the wreathed Hilbert series in equation \eqref{eq:HSE4wrZ2-ortho}, including the $\mathbb{Z}_2^{[0]}$ zero-form symmetry.

\subsection{The Rank-one \texorpdfstring{$E_5$}{E5} Theory}
\label{sec:E5th}

\begin{figure}[t]
\centering

\begin{subfigure}[b]{0.49\textwidth}
\centering
\begin{tikzpicture}[font=\footnotesize,baseline=0]
  \path (0, 2.5) -- (0, -2.5);

  \draw[thick,gray,-Triangle]
    (-1.6,0) -- node[above,pos=1] {$x$} (1.6,0);
  \draw[thick,gray,-Triangle]
    (0,-1.6) -- node[left,pos=1] {$y$} (0,1.6);

  \node[node, fill=black] (C)  at (0,0) {};

  \node[node, fill=black] (TL) at (-1,1) {};
  \node[node, fill=black] (TM) at (0,1) {};
  \node[node, fill=black] (TR) at (1,1) {};

  \node[node, fill=black] (ML) at (-1,0) {};
  \node[node, fill=black] (MR) at (1,0) {};

  \node[node, fill=black] (BL) at (-1,-1) {};
  \node[node, fill=black] (BM) at (0,-1) {};
  \node[node, fill=black] (BR) at (1,-1) {};

  \draw[line width=1pt]
    (TL)--(TM)--(TR)--(MR)--(BR)--(BM)--(BL)--(ML)--(TL);

\end{tikzpicture}
\caption{}
\end{subfigure}
\begin{subfigure}[b]{0.49\textwidth}
\centering
\begin{tikzpicture}[
  scale=.95,
  font=\footnotesize,baseline=0,
  brane/.style={gray,line width=1.25pt},
  midarrow/.style={
    postaction={decorate},
    decoration={markings,
      mark=at position .55 with {\arrow[black]{Stealth[length=2.2mm,width=2.2mm]}}
    }
  },
  sublab/.style={font=\large,gray}
]
  \def\w{0.95}
  \def\h{0.58}
  \def\dx{0.78}
  \def\dy{0.78}

  \path (0, 2.6) -- (0, -2.6);

  \coordinate (UL) at (-\w,\h);
  \coordinate (UR) at (\w,\h);
  \coordinate (BR) at (\w,-\h);
  \coordinate (BL) at (-\w,-\h);

  \coordinate (TL) at ($(UL)+(-\dx,\dy)$);
  \coordinate (TR) at ($(UR)+(\dx,\dy)$);
  \coordinate (BRR) at ($(BR)+(\dx,-\dy)$);
  \coordinate (BLL) at ($(BL)+(-\dx,-\dy)$);

  \draw[brane] (UL) -- (UR);
  \draw[brane] (UR) -- (BR);
  \draw[brane] (BR) -- (BL);
  \draw[brane] (BL) -- (UL);

  \draw[brane] (UL) -- (TL);
  \draw[brane] (UR) -- (TR);
  \draw[brane] (BR) -- (BRR);
  \draw[brane] (BL) -- (BLL);

  \draw[brane] (TL)  -- ++(-0.82,0) node[left,blue]  {$m_4$};
  \draw[brane] (TR)  -- ++( 0.82,0) node[right,blue] {$m_1$};
  \draw[brane] (BRR) -- ++( 0.82,0) node[right,blue] {$m_2$};
  \draw[brane] (BLL) -- ++(-0.82,0) node[left,blue]  {$m_3$};

  \draw[brane] (TL)  -- ++(0, 0.82) node[above,red] {$t_4$};
  \draw[brane] (TR)  -- ++(0, 0.82) node[above,red] {$t_1$};
  \draw[brane] (BRR) -- ++(0,-0.82) node[below,red] {$t_2$};
  \draw[brane] (BLL) -- ++(0,-0.82) node[below,red] {$t_3$};

  \draw[red,dashed,line width=.9pt] ($(UL)+(-0.08,0)$) -- ++(-1.35,0) node[left] {$-a$};
  \draw[red,dashed,line width=.9pt] ($(BL)+(-0.08,0)$) -- ++(-1.35,0) node[left] {$a$};

  \node at ($(UL)!0.5!(UR)+(0,0.20)$) {$Q_B$};
  \node at ($(BL)!0.5!(BR)+(0,-0.20)$) {$Q_F$};

  \node at ($(TL)!0.52!(UL)+(0.2,0.14)$) {$Q_4$};
  \node at ($(UR)!0.52!(TR)+(-0.25,0.14)$) {$Q_1$};
  \node at ($(BR)!0.52!(BRR)+(-0.2,-0.14)$) {$Q_2$};
  \node at ($(BLL)!0.52!(BL)+(0.2,-0.14)$) {$Q_3$};
\end{tikzpicture}
\caption{}
\end{subfigure}
\caption{Toric diagram and original $(p,q)$-web (from \cite{Mitev:2014jza,Kim:2014nqa}) for the $E_5$ theory. The toric diagram represents the SCFT point, while the $(p,q)$-web depicts a Coulomb phase of the theory.}
\label{fig:e5pqwebtoric}
\end{figure}
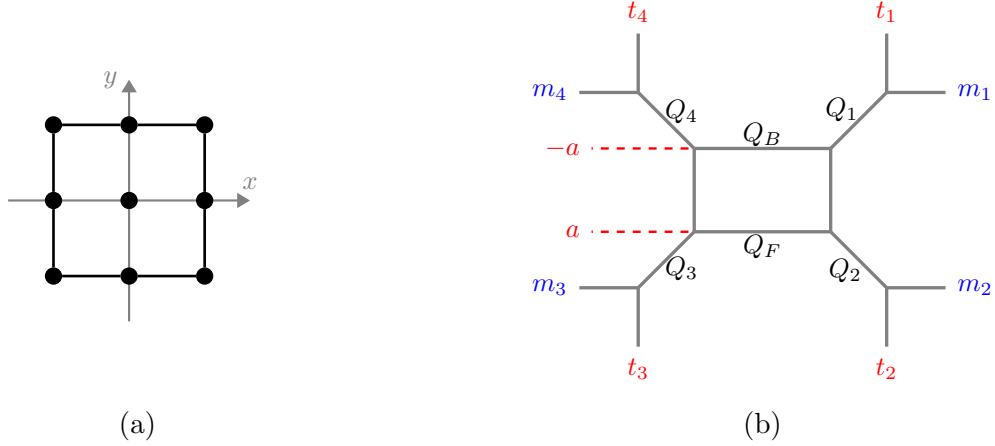

We now turn to the rank-one $E_5$ theory, which, next to the $E_3$ theory, has the most involved set of discrete symmetries that we can consider gauging through our methodology. We depicted the $(p,q)$-web and the toric diagram of the theory in Figure \ref{fig:e5pqwebtoric}. 

\paragraph{Discrete Symmetries from Magnetic Quivers for the Higgs Branch:} 
The unitary magnetic quiver for the Higgs branch of the rank-one $E_5$ SCFT takes the form of the affine $D_5$ Dynkin diagram, that is:
\begin{equation}\label{eq:unitaryE5}
    \begin{tikzpicture}[baseline=0,font=\footnotesize]
        \node[node, label=below:{$2$}] (A2) {};
        \node[node, label=below:{$2$}] (A3) [right=6mm of A2] {};
        \node[node, label=left:{$1$}] (Al1) [above left=6mm of A2] {};
        \node[node, label=left:{$1$}] (Al2) [below left =6mm of A2]{};
        \node[node, label=right:{$1$}] (Ar1) [above right=6mm of A3] {};
        \node[node, label=right:{$1$}] (Ar2) [below right =6mm of A3]{};
        \draw (Al1) -- (A2) -- (A3) -- (Ar1);
        \draw   (Al2) -- (A2) ;
        \draw   (Ar2) -- (A3) ;
        \node (Z2) [right=6mm of Ar2] {$/\U(1)\fstop$};
    \end{tikzpicture}
\end{equation}
This quiver has a diagram automorphism group $D_{8}$, the dihedral group of order eight. We label the nodes $1$ through $6$ where $1,2$ and $5,6$ are the nodes associated with the left (top and bottom) and right (top and bottom) $\U(1)$s, respectively, and $3,4$ are the nodes of the two $\U(2)$s. Then, in cycle notation, the $D_{8}$ automorphism group is generated by
\begin{equation}
    r \,: \,\, (1526)(34) \qquad\qquad s \, : \,\,  (12) \,.
\end{equation}
We can see that $r$ and $s$, respectively, generate $\mathbb{Z}_4$ and $\mathbb{Z}_2$ subgroups. There are eight conjugacy classes of subgroups of $D_{8}$ which we list together with their generators and the corresponding fixed-point subalgebras of $\mathfrak{e}_5 = \mathfrak{so}(10)$ in Table \ref{tbl:E5dg}.

\begin{table}[t]
    \centering
    \begin{tabular}{c|c|c}
        $H$ & Generators & $\mathfrak{g}^H$ \\\hhline{=|=|=}
        $1$ & $1$ & $\mathfrak{so}(10)$ \\
        $\mathbb{Z}_2^{(1)}$ & $s$ & $\mathfrak{so}(9)$ \\
        $\mathbb{Z}_2^{(2)}$ & $r^2s$ & $\mathfrak{so}(9)$ \\
        $\mathbb{Z}_2^{DT}$ & $r^2$ & $\mathfrak{so}(8) \oplus \mathfrak{u}(1)$ \\
        $\mathbb{Z}_2^{(3)}$ & $rs$ & $\mathfrak{so}(5) \oplus \mathfrak{so}(5)$ \\
        $\mathbb{Z}_4$ & $r$ & $\mathfrak{so}(4) \oplus \mathfrak{so}(4) \oplus \mathfrak{u}(1)$ \\
        $\mathbb{Z}_2^{(1)} \times \mathbb{Z}_2^{(2)}$ & $s,r^2 s$ & $\mathfrak{so}(8)$ \\
        $\mathbb{Z}_2^{DT} \times \mathbb{Z}_2^{(3)}$ & $r^2, rs$ & $\mathfrak{so}(4) \oplus \mathfrak{so}(4)$ \\
        $D_8$ & $r, s$ & $\mathfrak{so}(4) \oplus \mathfrak{so}(4)$
    \end{tabular}
    \caption{For each subgroup $H$ of $\operatorname{Aut}(\widetilde{\mathcal{D}}_{\mathfrak{so}(10)})$, we listed the fixed-point subalgebra of $\mathfrak{so}(10)$, $\mathfrak{g}^H$, as defined in equation \eqref{eqn:gHdef}. We have labeled the $\mathbb{Z}_2$ subgroups via superscripts. $\ZZ_2^{(1)}$ and $\ZZ_2^{(2)}$ are conjugate, but we include both in this table for future convenience.}
    \label{tbl:E5dg}
\end{table}

Similarly to the $E_4$ theory, there exists an alternative, but orthosymplectic, magnetic quiver for the Higgs branch of the $E_5$ theory which is
\begin{equation}\label{eq:orthosymE5}
   \begin{tikzpicture}[baseline=0,font=\footnotesize]
        \node[node, label=below:{$4$},fill=red] (A2) {};
        \node[node, label=below:{$2$},fill=blue] (Al1) [left=6mm of A2] {};
        \node[node, label=below:{$2$},fill=blue] (Ar1) [right =6mm of A2]{};
        \node[node, label=below:{$2$},fill=red] (Al2) [left=6mm of Al1] {};
        \node[node, label=below:{$2$},fill=red] (Ar2) [right =6mm of Ar1]{};
        \node[node, label=right:{$1$}] (A3) [above=4mm of A2] {};
        \draw (A2.north) -- (A3.south);
        \draw (Al2) -- (Al1) -- (A2) -- (Ar1) -- (Ar2);
        \node (Z2) [right=6mm of Ar2] {$/\ZZ_2$};
    \end{tikzpicture} \,.
\end{equation}
This orthosymplectic realization has a single $\ZZ_2$ diagram automorphism and thus sees only one $\ZZ_2$ discrete symmetry of the $E_5$ theory. We will see anon that this corresponds to a specific $\ZZ_2$ visible from the automorphism group of the unitary magnetic quiver.

\paragraph{Mirror Curve:} To parametrize the curve for this theory in a way that makes explicit the fugacities of the flavor group $D_5$, $y_i$ with $i=1,\dots,5$, we will follow \cite{Kim:2014nqa} and \cite{Mitev:2014jza} as in the previous $E_4$ analysis. On these parameters, the Weyl group of $D_5$ acts via permutations combined with an even number of reflections: $y_i \to y_i^{-1}$.  The curve can thus be written as
\begin{equation}
    \begin{split}
P(x,y) = &\,x \left(y + (-m_1 - m_2) + m_1 m_2 \frac{1}{y} \right)  \\ &+
\left((-t_1 - t_4) y + \sqrt{\frac{y_3}{y_4 y_5}} u + (-t_2 - t_3) m_1 m_2 \frac{1}{y}\right)  \\ &+ \frac{1}{x}\left(t_1 t_4 y + (-m_3 - m_4) t_1 t_4 + t_2 t_3 m_1 m_2 \frac{1}{y}\right) \, .
\end{split}
\end{equation}
Using the parametrization in terms of fugacities
\begin{equation}
\scalebox{0.94}{$\displaystyle
\renewcommand{\arraystretch}{1.7}
\begin{array}{rclrclrclrcl}
  m_1 & = & \sqrt{\dfrac{y_2 y_3}{y_4 y_5}} \coma & m_2 & = & \sqrt{\dfrac{y_3 y_4}{y_2 y_5}} \coma & m_3 & = & \sqrt{\dfrac{1}{y_2 y_3 y_4 y_5}} \coma & m_4 & = & \sqrt{\dfrac{y_2 y_4}{y_3 y_5}}\coma\\
  t_1 & = & \sqrt{\dfrac{y_1 y_3}{y_4 y_5}} \coma & t_2 & = & \sqrt{\dfrac{y_1 y_5}{y_3 y_4}} \coma & t_3 & = & \sqrt{\dfrac{1}{y_1 y_3 y_4 y_5}} \coma & t_4 & = & \sqrt{\dfrac{y_3 y_5}{y_1 y_4}}\coma
\end{array}$} 
\end{equation}
the curve becomes
\begin{equation}
    \begin{split}
        P(x,y) = &\,x \left(y - \left(\sqrt{\frac{y_2 y_3}{y_4 y_5}} + \sqrt{\frac{y_3 y_4}{y_2 y_5}}\right) + \frac{y_3}{y_5} \frac{1}{y} \right)  \\ &-
\left(\left(\sqrt{\frac{y_1 y_3}{y_4 y_5}} + \sqrt{\frac{y_3 y_5}{y_1 y_4}}\right) y - \frac{\sqrt{y_3} u}{\sqrt{y_4 y_5}} + \left(\sqrt{\frac{y_1 y_5}{y_3 y_4}} + \sqrt{\frac{1}{y_1 y_3 y_4 y_5}}\right) \frac{y_3}{y_5} \frac{1}{y}\right)  \\ &+ \frac{1}{x}\left(\frac{y_3}{y_4} y - \left(\sqrt{\frac{1}{y_2 y_3 y_4 y_5}} + \sqrt{\frac{y_2 y_4}{y_3 y_5}}\right) \frac{y_3}{y_4} + \frac{1}{y_4 y_5} \frac{1}{y}\right)\fstop
    \end{split}
\end{equation}

\subsubsection{The Higgs Branch of the Rank-one \texorpdfstring{$E_5$}{E5} Theory}

We first compute the refined Hilbert series for the unwreathed quiver. Define
$w_1$, $w_2$, $w_3$, and $w_4$ as the fugacities of the $\U(1)$s, and $w_5$, $w_6$ as the
fugacities for the $\U(2)$s. We decouple the $\U(1)$ by imposing $m_1\leq
m_2=0$ from, e.g., the left $\U(2)$ factor, and further impose
\begin{equation}
    w_6^2 = \frac{1}{w_1w_2w_3w_4w_5^2}\fstop
\end{equation}
In this formulation, there is a manifest $\so(10)$ flavor symmetry, observed by redefining the fugacities as follows:
\begin{equation}
    w_1\to \frac{x_2}{x_1^2}\coma w_2\to x_2\coma w_3\to \frac{x_3x_5}{x_2x_4}\coma w_4\to \frac{x_3x_4}{x_2x_5}\coma w_6\to \frac{x_2}{x_4x_5}\fstop
\end{equation}
Computing the refined Hilbert series, we see that it is
\begin{equation}\label{eq:E5-HS}
     \HS_{E_{5}}(t,\mathbf{x}) = \PE\left[\chi^D_{[0,1,0,0,0]}(\mathbf{x})t^2+\mathcal{O}(t^{4})\right]\fstop
\end{equation}
Furthermore, we can determine the unrefined Hilbert series, which is
\begin{equation}
\begin{split}
    \HS_{E_5}(t,1) = & \frac{1+31 t^2+231 t^4+595 t^6+595 t^8+231 t^{10}+31 t^{12}+t^{14}}{\left(1-t^2\right)^{14}} \\ 
    = &\PE[45 t^2-265 t^4+3354 t^6-53295 t^8+\mathcal{O}\left(t^{10}\right)]\fstop
\end{split}
\end{equation}

The exact Hilbert series for the orthosymplectic realization of the Higgs branch of the $E_5$ theory was given in \cite[Figure 11]{Bourget:2020xdz}; here we write only the expansion
\begin{equation}\label{eq:HSE5-ortho}
\begin{split}
    \HS_{E_{5}}(t,\omega) = & \PE\Big[t^2 (16 \omega +29)-t^4 (128 \omega +137)+t^6 (1712 \omega +1642)\\
    &-t^8 (26848 \omega +26447)+\mathcal{O}(t^{10})\Big]\,.
\end{split}
\end{equation}
We note that the fugacity $\omega$ distinguishes the parity of the operators under the $\U(1)$ appearing in the decomposition of the flavor: $\so(10)\rightarrow \so(8)\oplus \U(1)$.

\subsubsection{The \texorpdfstring{$\ZZ_2^{(1)}$}{Z2(1)} Discrete Gauging of the Rank-one \texorpdfstring{$E_5$}{E5} Theory}

\paragraph{Wreathing the Magnetic Quiver:}

First, we consider the $\mathbb{Z}_2$ diagram automorphism that swaps the two $\mathfrak{u}(1)$ nodes on the left (or, equivalently, on the right). It is well-known \cite{Hanany:2018dvd,Hanany:2018cgo} that wreathing a bouquet of $\mathfrak{u}(1)$s leads to a replacement by a single unitary node together with an adjoint-valued hypermultiplet. In this case, we get
\begin{equation}\label{eqn:swiper}
\begin{tikzpicture}[baseline=0,font=\footnotesize]
       \node[node, label=below:{$2$}] (A2) {};
        \node[node, label=below:{$2$}] (A3) [right=6mm of A2] {};
        \node[node, label=left:{$1$}] (Al1) [above left=6mm of A2] {};
        \node[node, label=left:{$1$}] (Al2) [below left =6mm of A2]{};
        \node[node, label=right:{$1$}] (Ar1) [above right=6mm of A3] {};
        \node[node, label=right:{$1$}] (Ar2) [below right =6mm of A3]{};
        \draw (Al1) -- (A2) -- (A3) -- (Ar1);
        \draw   (Al2) -- (A2) ;
        \draw   (Ar2) -- (A3) ;
        \node (Z2) [right=6mm of Ar2] {$/\U(1)\fstop$};
        \draw[Triangle-Triangle,red] ([xshift=-4mm]Al1.west) to[bend right=50] node[left,pos=0.5] {$\wr \ZZ_2^{(1)}$} ([xshift=-4mm]Al2.west);
\end{tikzpicture}
\end{equation}
We can proceed with the computation of the Hilbert series as before, where we now see a manifest $\mathfrak{su}(2) \oplus \mathfrak{su}(4)$ flavor symmetry, and we find:
\begin{equation}\label{eq:E5Z21-HS}
\begin{split}
     \HS_{E_{5}\wr \ZZ_2^{(1)}}(t,x,\mathbf{z}) = \PE&\left[\left(\chi^A_{[2]}(x)+\chi^A_{[1,0,1]}(\mathbf{z})+\chi^A_{[2]}(x)\chi^A_{[0,1,0]}(\mathbf{z})\right)t^2+\mathcal{O}(t^{4})\right]\fstop
\end{split}
\end{equation}
From this refined Hilbert series, we can read that the flavor symmetry corresponds to $\so(9)$, written in terms of the characters of $\su(2)\oplus\su(4)$ via the branching rule
\begin{equation}
    \begin{split}
        \mathfrak{so}(9) &\rightarrow \mathfrak{su}(2) \oplus \mathfrak{su}(4) \\
        \mathbf{36} &\rightarrow (\mathbf{3}, \mathbf{1}) \oplus (\mathbf{1}, \mathbf{15}) \oplus (\mathbf{3}, \mathbf{6})\fstop
    \end{split}
\end{equation}
Unrefining the flavor symmetry fugacities, we find the closed form for the Hilbert series:
\begin{equation}
\begin{split}
    \HS_{E_5\wr \ZZ_2^{(1)}}(t,1,1) = & \frac{1+22 t^2+126 t^4+280 t^6+280 t^8+126 t^{10}+22 t^{12}+t^{14}}{\left(1-t^2\right)^{14}} \\ 
    = &\PE[36 t^2-127 t^4+1050 t^6-11340 t^8+\mathcal{O}\left(t^{10}\right)]\fstop
\end{split}
\end{equation}
This is precisely the Hilbert series for the closure of the next-to-minimal nilpotent orbit of $\mathfrak{so}(9)$, $\mathcal{O}_{[3,1^6]}^{\mathfrak{so}(9)}$, as written in \cite[Table 18]{Cabrera:2017ucb}. The computation of the Hilbert series for the $\mathbb Z_2^{(2)}$ wreathing gives the same result.

\paragraph{Curve Analysis:}
The action on the geometry is 
\begin{align}
    (x,y) \to \left( \frac{1}{y_4 x},y \right)\coma
\end{align}
together with the action on the fugacities given by
\begin{align}
\label{eq:masstrtrial}
    (y_1,y_2,y_3,y_4,y_5) \to \left( y_1,\frac{1}{y_2},\frac{1}{y_3},y_4,\frac{1}{y_5} \right)\fstop
\end{align}
The curve is invariant (up to an overall rescaling that does not modify the zero-locus) only if we impose $y_5=1$. This effectively reduces the flavor rank by one, consistent with the wreathing analysis of the previous sections, where the flavor reduced to $\so(9)$.

A completely analogous discussion applies to the wreathing of the $\mathbb Z_2^{(2)}$ symmetry that flips the two nodes on the RHS of the magnetic quiver, instead of the LHS. 

\subsubsection{The \texorpdfstring{$\ZZ_2^{DT}$}{Z2DT} Discrete Gauging of the Rank-one \texorpdfstring{$E_5$}{E5} Theory}

\paragraph{Wreathing the Magnetic Quiver:}

We now consider the $\ZZ_2$ subgroup of the $\mathbb{Z}_4$ generated by $r$. We can see that this acts on the nodes of the quiver via the simultaneous exchange of the left and right $\mathfrak{u}(1)$ nodes; that is, it is a double transposition subgroup of $D_{8}$. We depict this wreathing as follows:
\begin{equation}\label{eq:E5quiverwrDT}
    \begin{tikzpicture}[baseline=0,font=\footnotesize]
        \node[node, label=below:{$2$}] (A2) {};
        \node[node, label=below:{$2$}] (A3) [right=6mm of A2] {};
        \node[node, label=left:{$1$}] (Al1) [above left=6mm of A2] {};
        \node[node, label=left:{$1$}] (Al2) [below left =6mm of A2]{};
        \node[node, label=right:{$1$}] (Ar1) [above right=6mm of A3] {};
        \node[node, label=right:{$1$}] (Ar2) [below right =6mm of A3]{};
        \draw (Al1) -- (A2) -- (A3) -- (Ar1);
        \draw   (Al2) -- (A2) ;
        \draw   (Ar2) -- (A3) ;
        \node (Z2) [right=8mm of Ar2] {$\qquad /\U(1)\fstop$};
        \draw[Triangle-Triangle,red] ([xshift=-4mm]Al1.west) to[bend right=50] node[left,pos=0.5] {$\wr \ZZ_2^{DT}$} ([xshift=-4mm]Al2.west);
        \draw[Triangle-Triangle,red] ([xshift=4mm]Ar1.east) to[bend left=50] node[right,pos=0.5] {$\wr \ZZ_2^{DT}$} ([xshift=+4mm]Ar2.east);
    \end{tikzpicture}
\end{equation}
The Hilbert series after wreathing this diagram automorphism reads
\begin{equation}\label{eq:E5DT-HS}
\begin{split}
     \HS_{E_{5}\wr\ZZ_2^{DT}}(t,x,\mathbf{z}) = \PE&\left[\left(1+\chi_{[2,0]}^{C}(\mathbf{z})+\chi^A_{[2]}(x)\left(1+\chi_{[0,1]}^{C}(\mathbf{z})\right)\right)t^2\right.\\
     &+\left(1+2\chi^A_{[4]}(x)+2\chi^A_{[2]}(x)\chi_{[0,1]}^C(\mathbf{z})+2\chi_{[0,2]}^C(\mathbf{z})\right.\\
     &\left.\left.-2\chi^A_{[2]}(x)\chi_{[2,0]}^C(\mathbf{z})-2\chi_{[0,1]}^C(\mathbf{z})\right)t^4+\mathcal{O}(t^{6})\right]\coma
\end{split}
\end{equation}
where the fugacities are for a manifest $\mathfrak{su}(2) \oplus \mathfrak{usp}(4)$ global symmetry. The unrefined Hilbert series, where we have suppressed explicit terms in the numerator since it is palindromic, is
\begin{equation}
\begin{split}
    \HS_{E_5\wr\ZZ_2^{DT}}(t,1,1) &= \frac{\left(\begin{aligned}
        & 1+22 t^2+245 t^4+1442 t^6+5355 t^8+12978 t^{10} \\
        &\quad +21919 t^{12}+25900 t^{14}+21919 t^{16}+\cdots +t^{28} 
    \end{aligned}\right)}{\left(1-t^2\right)^{7}\left(1-t^4\right)^{7}} \\
    &=\PE[29 t^2-t^4-406 t^6+3633 t^8+\mathcal{O}\left(t^{10}\right)]\fstop
\end{split}
\end{equation}
The representations at the leading order of the refined Hilbert series are compatible with a flavor symmetry of $\so(8) \oplus \mathfrak{u}(1)$ expressed in terms of characters of $\su(2)\oplus\usp(4)$ by the following branching rule:\footnote{The $S$ on the arrow indicates this is a special embedding.} 
\begin{equation}\label{eqn:ripto}
    \begin{split}
        \mathfrak{so}(8) &\stackrel{S}{\rightarrow} \mathfrak{su}(2) \oplus \mathfrak{usp}(4) \\
        \mathbf{28} &\rightarrow (\mathbf{3},\mathbf{1})\oplus (\mathbf{3}, \mathbf{5}) \oplus (\mathbf{1}, \mathbf{10})\fstop
    \end{split}
\end{equation}
It is interesting to note that the next order is also compatible with this decomposition, since
\begin{equation}\label{eq:so8tosu2usp4BR}
    \begin{split}
        \mathfrak{so}(8) &\stackrel{S}{\rightarrow} \mathfrak{su}(2) \oplus \mathfrak{usp}(4) \\
        \mathbf{35_s} &\rightarrow (\mathbf{1},\mathbf{1})\oplus (\mathbf{5}, \mathbf{1}) \oplus (\mathbf{3},\mathbf{5}) \oplus (\mathbf{1},\mathbf{14})\\
        \mathbf{35_v} &\rightarrow (\mathbf{1},\mathbf{5})\oplus (\mathbf{3}, \mathbf{10}) \\
        \mathbf{35_c} &\rightarrow (\mathbf{1},\mathbf{5})\oplus (\mathbf{3}, \mathbf{10})\coma
    \end{split}
\end{equation}
and thus, the order $t^4$ coefficient can be written as 
\begin{equation}
    2
    \chi_{\mathbf{35_s}}^{\so(8)} - \chi_{\mathbf{35_v}}^{\so(8)} - \chi_{\mathbf{35_c}}^{\so(8)}  -1\fstop
\end{equation}

\paragraph{Curve Analysis:}
We have the further involution with respect to the origin, given by
\begin{align}
    (x,y) \to \left( \frac{1}{x},\frac{1}{y} \right)\coma
\end{align}
as well as
\begin{align}
    (y_1,y_2,y_3,y_4,y_5) \to \left( y_1,y_2,y_3,\frac{1}{y_4},\frac{1}{y_5} \right)\coma
\end{align}
and rescaling the polynomial by 
\begin{align}
    P(x,y) \to P\left( \frac{1}{x},\frac{1}{y} \right)/(y_4 y_5)\fstop
\end{align}
leads to an invariant curve without imposing any further constraints.

\subsubsection{The \texorpdfstring{$\ZZ_2^{(3)}$}{Z2(3)} Discrete Gauging of the Rank-one \texorpdfstring{$E_5$}{E5} Theory}

\paragraph{Wreathing the Magnetic Quiver:}

We are now going to consider the $\ZZ_2$-wreathing of the quiver in equation \eqref{eq:unitaryE5} that reflects with respect to the vertical line passing through the bifundamentals in the $\U(2)\times\U(2)$ representation, i.e., 
\begin{equation}\label{eq:E5unitary-wrVL}
    \begin{tikzpicture}[baseline=0,font=\footnotesize]
        \node[node, label=below:{$2$}] (A2) {};
        \node[node, label=below:{$2$}] (A3) [right=6mm of A2] {};
        \node[node, label=left:{$1$}] (Al1) [above left=6mm of A2] {};
        \node[node, label=left:{$1$}] (Al2) [below left =6mm of A2]{};
        \node[node, label=right:{$1$}] (Ar1) [above right=6mm of A3] {};
        \node[node, label=right:{$1$}] (Ar2) [below right =6mm of A3]{};
        \draw (Al1) -- (A2) -- (A3) -- (Ar1);
        \draw   (Al2) -- (A2) ;
        \draw   (Ar2) -- (A3) ;
        \draw[Triangle-Triangle,red] ([yshift=3mm,xshift=3mm]Al1.north) to[bend left=50] node[above,pos=0.5] {$\wr \ZZ_2^{(3)}$} ([yshift=3mm,xshift=-3mm]Ar1.north);
        \node (Z2) [right=6mm of Ar2] {$/\U(1)\fstop$};
        \draw[densely dashed, red,thin] (0.42,1) -- (0.42,-1);
    \end{tikzpicture}
\end{equation}
The wreathed Hilbert series exhibits a manifest $\so(5)$ symmetry, and at the leading orders, it can be written as 
\begin{equation}\label{eq:E5Z23-HS}
\begin{aligned}
     \HS_{E_{5}\wr \ZZ_2^{(3)}}(t,\mathbf{x}) = \PE&\Bigg[2 t^2 \chi^B_{[0,2]}(\mathbf{x})+t^4 \Big(\chi^B_{[0,2]}(\mathbf{x})+\chi^B_{[0,4]}(\mathbf{x})-2 \chi^B_{[1,0]}(\mathbf{x})\\
     &+\chi^B_{[2,0]}(\mathbf{x})+\chi^B_{[2,2]}(\mathbf{x})+\chi^B_{[4,0]}(\mathbf{x})\Big)+ \mathcal{O}(t^{6})
     \Bigg]\fstop
\end{aligned}
\end{equation}
We emphasize that, since the axis around which we wreath bisects an edge of the quiver, we had to use the modified prescription for the wreathed monopole formula from Section \ref{sec:Z2cuttingedges}. We can see that the manifest $\mathfrak{so}(5)$ appears to be the diagonal of an $\mathfrak{so}(5) \oplus \mathfrak{so}(5)$ global symmetry, exactly as predicted by discrete gauging. 
Unrefined, the Hilbert series is
\begin{equation}
\begin{split}
    \HS_{E_{5}\wr \ZZ_2^{(3)}}(t,1,1) &= \frac{\left(\begin{aligned}
        & 1+13 t^2+269 t^4+1323 t^6+5691 t^8+12415 t^{10} \\
        &\quad +22775 t^{12}+24850 t^{14}+22775 t^{16}+\cdots+t^{28} 
    \end{aligned}\right)}{\left(1-t^2\right)^7 \left(1-t^4\right)^7} \\
    &=\PE\left[20 t^2+185 t^4-1446 t^6
     +\mathcal{O}\left(t^8\right)\right]\fstop
\end{split}
\end{equation}

\paragraph{Curve Analysis:}
The action on the geometry is 
\begin{align}
    (x,y) \to \left( y, x \right)\coma
\end{align}
together with the action on the fugacities given by
\begin{align}
    (y_1,y_2,y_3,y_4,y_5) \to \left( y_2, y_1, \frac{1}{y_3}, y_5, y_4 \right)\coma
\end{align}
 we find that the curve returns to itself only if we impose $y_3=1$. This effectively reduces the flavor rank by one, consistent with the wreathing analysis, where the flavor has been computed to reduce to $\so(5)\oplus \so(5)$.

\subsubsection{The \texorpdfstring{$\ZZ_4$}{Z4} Discrete Gauging of the Rank-one \texorpdfstring{$E_5$}{E5} Theory}

\paragraph{Wreathing the Magnetic Quiver:}

Another possible wreathing of the $E_5$ magnetic quiver is the $\ZZ_4$-wreathing from the diagram automorphism generated by $r$. Labeling the nodes of the quiver as before, the $\ZZ_4$ group has elements
\begin{equation}
    \{\ID, (12)(56), (1526)(34),(1625)(34)  \}\,.
\end{equation}
When considering the $(34)$ cycle, we need to introduce the phase of Section \ref{sec:Z2cuttingedges}, but otherwise, we can use the prescription of \cite{Grimminger:2024mks} to obtain the wreathed Hilbert series:
\begin{equation}\label{eq:E5Z4-HS}
\begin{aligned}
    \HS_{E_5\wr\ZZ_4}(t,z) = \PE&\Bigg[t^2 \left(4 \chi ^A_{[2]}(z)+1\right)+t^4 \Big(14 \chi ^A_{[2]}(z)+10 \chi ^A_{[4]}(z)\\
    &+3 \chi ^A_{[6]}(z)+\chi ^A_{[8]}(z)+7\Big)-t^6 \Big(26 \chi ^A_{[2]}(z)+10 \chi ^A_{[4]}(z)\\
    &-2 \chi ^A_{[6]}(z)-2 \chi ^A_{[8]}(z)+12\Big)+\mathcal{O}\left(t^8\right)\Bigg]\coma
\end{aligned}
\end{equation}
in terms of a manifest $\mathfrak{su}(2)$ flavor symmetry. This is compatible with an enhanced $\so(4)\oplus \so(4)\oplus \mathfrak{u}(1)$ flavor symmetry, where the $\mathfrak{su}(2)$ is the diagonal of the four $\mathfrak{su}(2)$ factors; this is as expected from the discrete gauging arguments that generated Table \ref{tbl:E5dg}. We can also determine the unrefined, resummed, Hilbert series, which is
\begin{equation}
\begin{split}
    \HS_{E_5\wr\ZZ_4}(t,1) &= \frac{\left(\begin{aligned}
        & 1+6 t^2+146 t^4+698 t^6+3139 t^8+8576 t^{10} \\
        &\quad +19460 t^{12}+34536 t^{14}+51932 t^{16}+65752 t^{18} \\
        &\quad +70804 t^{20}+65752 t^{22}+\cdots+t^{40} 
    \end{aligned}\right)}{\left(1-t^2\right)^7 \left(1-t^4\right)^4 \left(1-t^8\right)^3} \\
    &= \PE\left[13 t^2+129 t^4-108 t^6+
\mathcal{O}\left(t^8\right)\right] \fstop
\end{split}
\end{equation}

\paragraph{Curve Analysis:}
The action on the geometry is 
\begin{align}
    (x,y) \to \left( \frac{1}{y}, x \right)\coma
\end{align}
together with the action on the fugacities given by
\begin{align}
    (y_1,y_2,y_3,y_4,y_5) \to \left( y_2, y_1, \frac{1}{y_3}, \frac{1}{y_5}, y_4 \right)\coma
\end{align}
and the rescaling of the curve
\begin{align}
    P(x,y) \to \frac{y_3}{y_5} P\left( \frac{1}{y}, x\right)\fstop
\end{align}
Under these transformations, the curve is invariant. Since no extra identification is required, the rank of the curve does not drop.

\subsubsection{The \texorpdfstring{$\ZZ_2^{(1)}\times\ZZ_2^{(2)}$}{Z2(1)xZ2(2)} Discrete Gauging of the Rank-one \texorpdfstring{$E_5$}{E5} Theory}

\paragraph{Wreathing the Magnetic Quiver:}

There is a $\mathbb{Z}_2 \times \mathbb{Z}_2$ subgroup of $D_8$ which is generated by the single transposition ($s$) and the double transposition ($r^2$). The resulting group can alternatively be generated by the transpositions of both the left and the right pairs of $\mathfrak{u}(1)$s in the quiver, corresponding to $s$ and $r^2 s$. We depict this group action on the quiver as
\begin{equation}\label{eq:E5wreathedleftright}
\begin{tikzpicture}[baseline=0,font=\footnotesize]
    \node (A) at (0,0) {\begin{tikzpicture}[baseline=0,font=\footnotesize]
       \node[node, label=below:{$2$}] (A2) {};
        \node[node, label=below:{$2$}] (A3) [right=6mm of A2] {};
        \node[node, label=left:{$1$}] (Al1) [above left=6mm of A2] {};
        \node[node, label=left:{$1$}] (Al2) [below left =6mm of A2]{};
        \node[node, label=right:{$1$}] (Ar1) [above right=6mm of A3] {};
        \node[node, label=right:{$1$}] (Ar2) [below right =6mm of A3]{};
        \draw (Al1) -- (A2) -- (A3) -- (Ar1);
        \draw   (Al2) -- (A2) ;
        \draw   (Ar2) -- (A3) ;
        \node (Z2) [right=6mm of Ar2] {$\qquad/\U(1)$};
        \draw[Triangle-Triangle,red] ([xshift=-4mm]Al1.west) to[bend right=50] node[left,pos=0.5] {$\wr \ZZ_2^{(1)}$} ([xshift=-4mm]Al2.west);
        \draw[Triangle-Triangle,red] ([xshift=4mm]Ar1.east) to[bend left=50] node[right,pos=0.5] {$\wr \ZZ_2^{(2)}$} ([xshift=+4mm]Ar2.east);
    \end{tikzpicture}};
\end{tikzpicture}
\end{equation}
which also has a standard Lagrangian description for the wreathing of a bouquet of $\mathfrak{u}(1)$s \cite{Hanany:2018dvd,Hanany:2018cgo}.  Since this wreathed quiver has an alternative description as an unwreathed quiver, the computation of the Coulomb branch Hilbert series is textbook, and we obtain:
\begin{equation}
\label{eq:E5wrZ2U2adjoints}
\begin{aligned}
     \HS_{E_{5}\wr \left(\ZZ_2^{(1)}\times\ZZ_2^{(2)}\right)}&(t,x,\mathbf{z}) = \PE\Bigg[\left(\chi_{[2,0]}^{C}(\mathbf{z})+\chi^A_{[2]}(x)(1+\chi_{[0,1]}^{C}(\mathbf{z}))\right)t^2\\
     &+(1+\chi^A_{[4]}(x)+\chi^A_{[2]}(x)\chi_{[0,1]}^C(\mathbf{z})+\chi_{[0,2]}^C(\mathbf{z})\\
     &-2\chi^A_{[2]}(x)\chi_{[2,0]}^C(\mathbf{z})-2\chi_{[0,1]}^C(\mathbf{z}))t^4+\mathcal{O}(t^{6})\Bigg]\fstop
\end{aligned}
\end{equation}
The wreathed quiver has a manifest $\mathfrak{su}(2) \oplus \mathfrak{usp}(4)$ flavor symmetry, and, as for the $\mathbb{Z}_2^{DT}$-wreathing, we can see that the moment maps recombine, following the branching rule in equation \eqref{eqn:ripto}, to an enhanced $\mathfrak{so}(8)$ symmetry. Furthermore, using the branching rules in equation \eqref{eq:so8tosu2usp4BR}, we can see that the $t^4$ coefficient in the wreathed Hilbert series is
\begin{equation}
    \chi_{\mathbf{35_s}}^{\so(8)} - \chi_{\mathbf{35_v}}^{\so(8)} - \chi_{\mathbf{35_c}}^{\so(8)}\fstop
\end{equation}
We can also resum the unrefined Hilbert series, and we find the palindromic expression:
\begin{equation}
\begin{split}
    \HS_{E_5\wr \left(\ZZ_2^{(1)}\times\ZZ_2^{(2)}\right)}(t,1,1) &= \frac{\left(\begin{aligned}
        & 1+21 t^2+189 t^4+931 t^6+3003 t^8+6615 t^{10} \\
        &\quad +10567 t^{12}+12258 t^{14}+10567 t^{16}+\cdots+t^{28} 
    \end{aligned}\right)}{\left(1-t^2\right)^{7}\left(1-t^4\right)^{7}} \\
    &=\PE[28 t^2-35 t^4+42 t^6+336 t^8+\mathcal{O}\left(t^{10}\right)]\fstop
\end{split}
\end{equation}

\paragraph{Curve Analysis:} 
The action on the geometry is 
\begin{align}
    (x,y) \to \left( \frac{1}{y_4 x}, \frac{1}{y} \right)\coma
\end{align}
together with the action on the fugacities given by
\begin{align}
    (y_1,y_2,y_3,y_4,y_5) \to \left( \frac{1}{y_1}, \frac{1}{y_2}, y_3, y_4, y_5 \right)\coma
\end{align}
preserve the zero-locus of the curve only if we impose $y_5 = 1$. 
This effectively reduces the flavor rank by one, consistent with the wreathing analysis.

\subsubsection{The \texorpdfstring{$\ZZ_2^{DT}\times\ZZ_2^{(3)}$}{Z2DTxZ2(3)} Discrete Gauging of the Rank-one \texorpdfstring{$E_5$}{E5} Theory}

\paragraph{Wreathing the Magnetic Quiver:}

Penultimately, we can consider the $\mathbb{Z}_2 \times \ZZ_2$ subgroup of $D_8$ formed by combining the generators of the $\mathbb{Z}_2^{DT}$ and the $\mathbb{Z}_2^{(3)}$. Schematically, this wreathing action on the quiver can be depicted as
\begin{equation}\label{eq:E5unitary-wrVL-3}
    \begin{tikzpicture}[baseline=0,font=\footnotesize]
         \node[node, label=below:{$2$}] (A2) {};
        \node[node, label=below:{$2$}] (A3) [right=6mm of A2] {};
        \node[node, label=left:{$1$}] (Al1) [above left=6mm of A2] {};
        \node[node, label=left:{$1$}] (Al2) [below left =6mm of A2]{};
        \node[node, label=right:{$1$}] (Ar1) [above right=6mm of A3] {};
        \node[node, label=right:{$1$}] (Ar2) [below right =6mm of A3]{};
        \draw (Al1) -- (A2) -- (A3) -- (Ar1);
        \draw   (Al2) -- (A2) ;
        \draw   (Ar2) -- (A3) ;
       \draw[Triangle-Triangle,red] ([xshift=-4mm]Al1.west) to[bend right=50] node[left,pos=0.5] {$\wr \ZZ_2^{DT}$} ([xshift=-4mm]Al2.west);
        \draw[Triangle-Triangle,red] ([xshift=4mm]Ar1.east) to[bend left=50] node[right,pos=0.5] {$\wr \ZZ_2^{DT}$} ([xshift=+4mm]Ar2.east);
        \draw[Triangle-Triangle,red] ([yshift=3mm]Al1.north) to[bend left=50] node[above,pos=0.5] {$\wr \ZZ_2^{(3)}$} ([yshift=3mm]Ar1.north);
        \node (Z2) [right=8mm of Ar2] {$\qquad/\U(1)\fstop$};
        \draw[densely dashed, red,thin] (0.42,1) -- (0.42,-1);
    \end{tikzpicture}
\end{equation}
This particular finite subgroup of $D_8$ contains four elements which are given by 
\begin{equation}
    \{\ID, (12)(56), (16)(25)(34),(15)(26)(34)  \}\coma
\end{equation}
in the standard labeling of the nodes of the quiver. The resulting Hilbert series needs to be computed using the new prescription in Section \ref{sec:Z2cuttingedges}, due to the $\ZZ_2^{(3)}$ action, and the final expression has only a single fugacity for an $\su(2)$ symmetry, $z$. The final expression is
\begin{equation}\label{eq:E5DtZ23-HS}
\begin{aligned}
    \HS_{E_5\wr\left(\ZZ_2^{DT}\times\ZZ_2^{(3)}\right)}(t,z) = \PE&\Bigg[4 t^2 \chi^A_{[2]}(z)+t^4 \Big(16 \chi^A_{[2]}(z)+12 \chi^A_{[4]}(z)+3 \chi^A_{[6]}(z)\\
    &+\chi^A_{[8]}(z)+11\Big)-t^6 \Big(29 \chi^A_{[2]}(z)+11 \chi^A_{[4]}(z)+\chi^A_{[6]}(z)\\
    &-\chi^A_{[8]}(z)+10\Big)+\mathcal{O}\left(t^8\right)\Bigg]\coma
\end{aligned}
\end{equation}
or, unrefined: 
\begin{equation}
\begin{split}
    \HS_{E_5\wr \left(\ZZ_2^{DT}\times\ZZ_2^{(3)}\right)}(t,1) &= \frac{\left(\begin{aligned}
        & 1+5 t^2+157 t^4+595 t^6+3003 t^8+5863 t^{10} \\
        &\quad +11879 t^{12}+11906 t^{14}+11879 t^{16}+\cdots+t^{28} 
    \end{aligned}\right)}{\left(1-t^2\right)^7 \left(1-t^4\right)^7} \\
    &= \PE\left[12 t^2+149 t^4-150 t^6+
\mathcal{O}\left(t^8\right)\right] \fstop
\end{split}
\end{equation}
The Coulomb branch Hilbert series of the wreathed quiver is consistent with an $\mathfrak{su}(2)^{\oplus 4} \cong \mathfrak{so}(4)^{\oplus 2}$ flavor symmetry, as expected from looking at the fixed-point subalgebra in Table \ref{tbl:E5dg}.

\paragraph{Curve Analysis:}
The action on the geometry is obtained by composing the $\ZZ_2^{DT}$ and $\ZZ_2^{(3)}$ actions:
\begin{align}
    (x,y) \to \left( \frac{1}{y},\frac{1}{x} \right)\coma
\end{align}
together with the combined action on the fugacities given by
\begin{align}
    (y_1,y_2,y_3,y_4,y_5) \to \left( y_2, y_1, \frac{1}{y_3}, \frac{1}{y_5}, \frac{1}{y_4} \right)\coma
\end{align}
and the rescaling of the curve
\begin{align}
    P(x,y) \to \frac{y_3}{y_4 y_5} P\left( \frac{1}{y},\frac{1}{x} \right)\fstop
\end{align}

Under these transformations, we get the curve to return to itself only if we impose $y_3=1$. This effectively reduces the flavor rank by one, consistent with the wreathing analysis of the previous sections, where the flavor has been computed to reduce to $\so(4)\oplus \so(4)$ (the inner automorphism part associated with $\ZZ_2^{DT}$ does not drop the rank any further).

\subsubsection{The \texorpdfstring{$D_8$}{D8} Discrete Gauging of the Rank-one \texorpdfstring{$E_5$}{E5} Theory}

\paragraph{Wreathing the Magnetic Quiver:}
The remaining wreathing is the full $D_8$-wreathing of the quiver. In the usual cycle notation, the $D_8$ group contains the following elements 
\begin{equation}
 \scalebox{0.93}{$\displaystyle \{\ID, (12), (56), (12)(56), (15)(26)(34), (16)(25)(34), (1526)(34), (1625)(34) \}\fstop $}
\end{equation}
Conveniently, we have determined the contributions to the wreathed Hilbert series from each of these elements in the previous sections. Thus, summing them appropriately, we find that the Coulomb branch Hilbert series of the $D_8$-wreathed magnetic quiver is
\begin{equation}\label{eq:E5D8-HS}
\begin{aligned}
    \HS_{E_5\wr D_8}(t,z) &= \PE\Bigg[4 t^2 \chi ^A_{[2]}(z)+t^4 \Big(11 \chi ^A_{[2]}(z)+9 \chi ^A_{[4]}(z)+3 \chi ^A_{[6]}(z)+\chi ^A_{[8]}(z)+6\Big) \\
    &\qquad\qquad\qquad -t^6 \Big(25 \chi ^A_{[2]}(z)+15 \chi ^A_{[4]}(z)+4 \chi ^A_{[6]}(z)+10\Big)+\mathcal{O}\left(t^8\right)\Bigg]\,.
\end{aligned}
\end{equation}
Similarly to the previous case, there is a manifest $\mathfrak{su}(2)$ Coulomb branch symmetry in the wreathed quiver, and we can see from the Hilbert series that this appears to enhance to an $\mathfrak{so}(4) \oplus \mathfrak{so}(4)$ global symmetry. This is precisely the symmetry expected from the $D_8$ discrete gauging in Table \ref{tbl:E5dg}. We can also determine the, resummed, unrefined Hilbert series:
\begin{equation}
\begin{aligned}
    \HS_{E_5\wr D_8}(t,1) &= \frac{\left(\begin{aligned}
        & 1+5 t^2+125 t^4+397 t^6+2048 t^8+4170 t^{10} \\
        &\quad +11122 t^{12} +15778 t^{14} +28047 t^{16}+29314 t^{18} \\
        &\quad +37634 t^{20}+29314 t^{22}+\cdots+t^{40}
    \end{aligned}\right)}{\left(1-t^2\right)^7 \left(1-t^4\right)^4 \left(1-t^8\right)^3} \\
    &= \PE\left[12 t^2+114 t^4-188 t^6+
\mathcal{O}\left(t^8\right)\right] \,.
\end{aligned}
\end{equation}

\paragraph{Curve Analysis:}
The full $D_8$ discrete symmetry is generated by the $\ZZ_4$ rotation and the $\ZZ_2^{(3)}$ reflection. The action on the geometry for the rotation is 
\begin{align}
    (x,y) \to \left( \frac{1}{y}, x \right)\coma
\end{align}
together with the action on the fugacities given by
\begin{align}
    (y_1,y_2,y_3,y_4,y_5) \to \left( y_2, y_1, \frac{1}{y_3}, \frac{1}{y_5}, y_4 \right)\coma
\end{align}
and the rescaling of the curve
\begin{align}
    P(x,y) \to \frac{y_3}{y_5} P\left( \frac{1}{y}, x \right)\fstop
\end{align}
Conversely, the action on the geometry of the reflection is 
\begin{align}
    (x,y) \to \left( y, x \right)\coma
\end{align}
together with the action on the fugacities given by
\begin{align}
    (y_1,y_2,y_3,y_4,y_5) \to \left( y_2, y_1, \frac{1}{y_3}, y_5, y_4 \right)\coma
\end{align}
and the associated rescaling of the curve
\begin{align}
    P(x,y) \to \frac{1}{y_3}P\left( y , x \right)\fstop
\end{align}
Under these combined transformations,\footnote{Note that if we call $R$ the combined action on the geometric coordinates and the mass parameters for the rotation, and $S$ similarly for the reflection, then $R^4 = S^2 = 1$ and $SRS = R^{-1}$, as expected from $D_8$.} requiring the curve to return to itself under the entire $D_8$ group dictates that we must satisfy the constraints of both generators simultaneously. This translates to imposing $y_3=1$. This effectively reduces the flavor rank by one, which is fully consistent with the wreathing analysis where the continuous flavor algebra reduces from $\so(10)$ to $\so(4)\oplus \so(4)$.

\subsubsection{The (Orthosymplectic) \texorpdfstring{$\ZZ_2$}{Z2}-Wreathing of the Rank-one \texorpdfstring{$E_5$}{E5} Theory}\label{sec:orthoE5}

We can also consider the wreathing of the $\mathbb{Z}_2$ diagram automorphism of the orthosymplectic magnetic quiver for $E_5$ given in equation \eqref{eq:orthosymE5}. The resulting wreathed Coulomb branch Hilbert series is
\begin{equation}\label{eq:HSE5wrZ2-ortho}
\begin{split}
    \HS_{E_{5}\wr \ZZ_2}(t,\omega) = & \PE\Big[t^2 (12 \omega +17)+t^4 (15-16 \omega)-t^6 (188 \omega +218)\\
    &\qquad+t^8 (2072 \omega +1561)+\mathcal{O}(t^{10})\Big]\fstop
\end{split}
\end{equation}
The Hilbert series evinces a flavor symmetry with a $29$-dimensional adjoint representation. This is consistent with the same $\mathfrak{so}(8) \oplus \mathfrak{u}(1)$ flavor obtained from the discrete gauging of the $\mathbb{Z}_2^{DT}$ symmetry, and, indeed, the Hilbert series agrees to the orders computed here. It therefore seems natural to identify the $\mathbb{Z}_2$ manifest in the orthosymplectic quiver with the $\mathbb{Z}_2^{DT}$ subgroup of the $D_8$ automorphism group visible from the unitary quiver.

\subsection{The Rank-one \texorpdfstring{$E_6$}{E6} Theory -- Continued}
\label{sec:E56th_more}

We studied the discrete gauging of a particular $\ZZ_2$ symmetry of the rank-one $E_6$ theory in the warm-up in Sections \ref{sec:E6th} and \ref{sec:warmupcurve}. In this section, we extend that analysis to the discrete gauging of the other symmetries of the $E_6$ theory that we observe.

In Section \ref{sec:E6th}, we explained that the unitary magnetic quiver for the Higgs branch of the $E_6$ SCFT (given in equation \eqref{eq:unitaryE6}) has an $S_3$ diagram automorphism group. The fixed-point algebras associated to the automorphisms of $\mathfrak{e}_6$ induced by the homomorphism in equation \eqref{eqn:BIGHOMO} for each nontrivial subgroup of $S_3$ are
\begin{equation}
    \mathfrak{e}_6^{\ZZ_2} = \mathfrak{f}_4 \,, \qquad \mathfrak{e}_6^{\ZZ_3} = \mathfrak{so}(8) \oplus \mathfrak{u}(1)^{\oplus 2} \,, \qquad \mathfrak{e}_6^{S_3} = \mathfrak{so}(8) \,,
\end{equation}
as already discussed in Section \ref{sec:dg}. We now demonstrate that $\ZZ_3$ and $S_3$ wreathings reproduce these fixed-point algebras as flavor algebras, and show that this is also consistent with the discrete action on the Seiberg--Witten curve in the case of $\ZZ_3$.

The rank-one $E_6$ has an alternative, orthosymplectic, magnetic quiver for the Higgs branch, which also evinces a $\mathbb{Z}_2$ diagram automorphism which can be wreathed. We analyze this discrete symmetry and its gauging in Section \ref{sec:orthoE6}.

\subsubsection{The \texorpdfstring{$\ZZ_3$}{Z3} Discrete Gauging of the Rank-one \texorpdfstring{$E_6$}{E6} Theory}

\paragraph{Wreathing the Magnetic Quiver:}
 
Similarly to the $\ZZ_2$-wreathing, we can perform the $\ZZ_3$-wreathing of equation \eqref{eq:unitaryE6}, which acts on all three legs of the quiver, depicted via
\begin{equation}\label{eq:E6wrZ3-quiver}
    \begin{tikzpicture}[baseline=0,font=\footnotesize]
        \node[node, label=below:{$3$}] (A2) {};
        \node[node, label=below:{$2$}] (Al1) [left=6mm of A2] {};
        \node[node, label=below:{$2$}] (Ar1) [right =6mm of A2]{};
        \node[node, label=below:{$1$}] (Al2) [left=6mm of Al1] {};
        \node[node, label=below:{$1$}] (Ar2) [right =6mm of Ar1]{};
        \node[node, label=right:{$2$}] (A3) [above=4mm of A2] {};
        \node[node, label=right:{$1$}] (A3a) [above=4mm of A3] {};
        \draw (A2) -- (A3) -- (A3a);
        \draw   (Al2) -- (Al1) -- (A2) -- (Ar1) -- (Ar2) ;
        \node (Z2) [right=6mm of Ar2] {$/\U(1)\fstop$};
        \draw[-Triangle,red] ([shift=(15:1.1cm)]A2) arc (15:75:1.1cm);
        \draw[-Triangle,red] ([shift=(105:1.1cm)]A2) arc (105:165:1.1cm);
        \draw[-Triangle,red] ([shift=(195:1.1cm)]A2) arc (195:345:1.1cm);
    \end{tikzpicture}
\end{equation}
We can decompose the computation of the Hilbert series in terms of the three $T_{[1,1,1]}[\SU(3)]$ tails. The $\ZZ_3$-wreathing first identifies the fugacities of the three tails, i.e.,
\begin{equation}\label{eq:E6-Z3wrfugiden}
    w_5=w_3 = w_1 \coma w_6=w_4=w_2\fstop
\end{equation}
Then, we consider the elements of $\ZZ_3$, i.e.,
\begin{equation}
\{\ID, (123), (132)\}\coma
\end{equation}
that is, there are two elements whose cycle has length three. The Hilbert series is obtained by summing the contributions from each element of $\ZZ_3$:
\begin{equation}
\begin{split}
     \text{HS}_{E_6\wr \ZZ_3}(t,\mathbf{w}) = &\,(1-t^2)\sum_{-\infty\leq m_1\leq m_2\leq m_3 = 0}^\infty P_{\U(3)}(t,\mathbf{m}) w_7^{m_1+m_2+m_3}t^{2\Delta_\text{res.}}\times \\
     &\frac{1}{3}\left(\text{HS}_{T_{[1,1,1]}[\SU(3)]}(t,w_1,w_2,\mathbf{m})^3+2\text{HS}_{T_{[1,1,1]}[\SU(3)]}(t^3,w_1^3,w_2^3,\mathbf{m})\right)\fstop
\end{split}
\end{equation}
The result, after an appropriate redefinition of the fugacities, is given in terms of the fugacity of a single $\mathfrak{su}(3)$ flavor symmetry: 
\begin{equation}\label{eq:E6Z3-HS}
    \HS_{E_{6}\wr \ZZ_3}(t,\mathbf{x}) = \PE\left[\left(2+\chi^A_{[3,0]}(\mathbf{x})+\chi^A_{[0,3]}(\mathbf{x})+\chi^A_{[1,1]}(\mathbf{x})\right)t^2 +\mathcal{O}(t^{4})\right]\fstop
\end{equation}
We can unrefine the manifest $\mathfrak{su}(3)$ flavor symmetry and determine the fully-resummed Hilbert series, which is
\begin{equation}
\begin{split}
    \HS_{E_{6}\wr \ZZ_3}(t,1) &= \frac{\left(\begin{aligned}
        & 1+15 t^2+483 t^4+4920 t^6 +34284 t^8+160449 t^{10}+569592 t^{12} \\
        &\quad +1587058 t^{14}+3594105 t^{16}+6781096 t^{18} \\
        &\quad +10770924 t^{20}+14614305 t^{22}+16982046 t^{24} \\
        &\quad +16982046 t^{26}+14614305 t^{28}+\cdots+t^{50} 
    \end{aligned}\right)}{\left(1-t^2\right)^{15} \left(1-t^6\right)^7} \\
    &= \PE\left[30t^2 +363t^4-1198 t^6+\mathcal{O}(t^{8})\right]\coma
\end{split}
\end{equation}
where the $\cdots$, as usual, denotes palindromicity. The flavor symmetry of the theory is enhanced from the manifest $\mathfrak{su}(3)$, and we find that it is $\so(8)\oplus \mathfrak{u}(1)\oplus \mathfrak{u}(1)$ where the $\mathfrak{su}(3)$ is a subalgebra of $\mathfrak{so}(8)$ under the following special embedding:
\begin{equation}
    \begin{split}\label{eq:so8tosu3S}
        \so(8) &\stackrel{S}{\rightarrow} \su(3)\\
        \mathbf{28} &\rightarrow \mathbf{10}\oplus \overline{\mathbf{10}} \oplus \mathbf{8}\fstop
    \end{split}
\end{equation}
This is precisely the global symmetry we expected from the fixed-point subalgebra $\mathfrak{e}_6^{\ZZ_3}$.

\paragraph{Curve Analysis:}

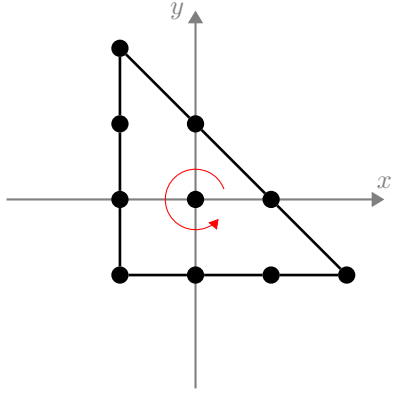
\begin{figure}[t]
\centering
\begin{tikzpicture}[scale=1,font=\footnotesize,baseline=0]
    \draw[thick,gray,-Triangle] (-2.5,0) -- node[above,pos=1] {$x$} (2.5,0);
    \draw[thick,gray,-Triangle] (0,-2.5) -- node[left,pos=1] {$y$} (0,2.5);
    
    \node[node, fill=black] (O) at (0,0) {};
    \node[node, fill=black] (A) at (-1,-1) {};
    \node[node, fill=black] (B) at (0,-1) {};
    \node[node, fill=black] (C) at (1,-1) {};
    \node[node, fill=black] (D) at (2,-1) {};
    \node[node, fill=black] (E) at (1,0) {};
    \node[node, fill=black] (F) at (0,1) {};
    \node[node, fill=black] (G) at (-1,2) {};
    \node[node, fill=black] (H) at (-1,1) {};
    \node[node, fill=black] (I) at (-1,0) {};

    \draw[line width=1pt] (A) -- (B) -- (C) -- (D) -- (E) -- (F) -- (G) -- (H) -- (I) -- (A);

    \draw[-Triangle, red]
        ($(0,0)+(20:0.4)$) arc[start angle=20,end angle=320,radius=0.4];
\end{tikzpicture}
\caption{Isometry associated with the $\mathbb Z_3$ inner automorphism action.}
\end{figure}

The curve for the $E_6$ theory prior to the discrete gauging was given in equation \eqref{eq:e6curvekimyagi}. The isometry action which exchanges the points of the toric diagram as required is 
\begin{equation}\label{eqn:gnastygnorc2}
    (x,y)  \to  \left(-\frac{y}{x},\frac{1}{x}\right)\,.
\end{equation}
It can be composed with the following action on the mass parameters,
\begin{equation}
\label{eq:z3innere6}
    (M_i,N_i,L_i)  \to  (N_i^{-1},L_i,M_i^{-1})\,.
\end{equation}
Again, it is possible to check that equation \eqref{eq:z3innere6} is an inner automorphism and that equations \eqref{eqn:gnastygnorc2} and \eqref{eq:z3innere6} preserve equation \eqref{eq:e6curvekimyagi}, without imposing any conditions on the mass parameters. Hence, the rank of the flavor group after the discrete gauging of the associated zero-form symmetry is still six, in agreement with the wreathing result. 

\subsubsection{The \texorpdfstring{$S_3$}{S3} Discrete Gauging of the Rank-one \texorpdfstring{$E_6$}{E6} Theory}

\paragraph{Wreathing the Magnetic Quiver:}

Finally, we can take the $S_3$-wreathing of the theory. First, as for the $\ZZ_3$-wreathing, we identify the fugacities of the three tails as in equation \eqref{eq:E6-Z3wrfugiden}. Then, we consider the group elements of $S_3$:
\begin{equation}
    \{\ID,(12),(23),(31),(123),(132)\}\fstop
\end{equation}
This means that to compute the $S_3$-wreathing of $E_6$, we have the contributions that we considered for both the $\ZZ_2$ and $\ZZ_3$ wreathings, i.e.,
\begin{equation}
\begin{split}
     \text{HS}_{E_6\wr S_3}(t,\mathbf{w}) = &\,(1-t^2)\sum_{-\infty\leq m_1\leq m_2\leq m_3 = 0}^\infty P_{\U(3)}(t,\mathbf{m})w_7^{m_1+m_2+m_3}t^{2\Delta_\text{res.}} \times \\
     &\frac{1}{6}\left(\text{HS}_{T_{[1,1,1]}[\SU(3)]}(t,w_1,w_2,\mathbf{m})^3+2\text{HS}_{T_{[1,1,1]}[\SU(3)]}(t^3,w_1^3,w_2^3,\mathbf{m})\right.\\
     &\left.+3\text{HS}_{T_{[1,1,1]}[\SU(3)]}(t^2,w_1^2,w_2^2,\mathbf{m})\text{HS}_{T_{[1,1,1]}[\SU(3)]}(t,w_1,w_2,\mathbf{m})\right)\fstop
\end{split}
\end{equation}
The refined Hilbert series can be expressed in terms of the characters of $\mathfrak{su}(3)$ as we did for the $\ZZ_3$-wreathing, obtaining
\begin{equation}\label{eq:E6S3-HS}
    \HS_{E_{6}\wr S_3}(t,\mathbf{x}) = \PE\left[\left(\chi^A_{[3,0]}(\mathbf{x})+\chi^A_{[0,3]}(\mathbf{x})+\chi^A_{[1,1]}(\mathbf{x})\right)t^2 +\mathcal{O}(t^{4})\right]\fstop
\end{equation}
Unrefined, we can see that the higher-orders are given by:
\begin{equation}
\begin{split}
    \HS_{E_{6}\wr S_3}(t,1) &= \frac{\left(\begin{aligned}
        & 1+13 t^2+261 t^4+2490 t^6 +17142 t^8+80127 t^{10}+284611 t^{12} \\
        &\quad +793414 t^{14}+1797228 t^{16}+3391000 t^{18} \\
        &\quad +5385844 t^{20}+7307055 t^{22}+8490453 t^{24} \\
        &\quad +8490453 t^{26}+7307055 t^{28}+\cdots+t^{50} 
    \end{aligned}\right)}{\left(1-t^2\right)^{15} \left(1-t^6\right)^7} \\
    &= \PE\left[28t^2 +170t^4-168 t^6+\mathcal{O}(t^{8})\right]\fstop
\end{split}
\end{equation}
The nontrivial $\mathfrak{su}(3)$-characters are the same as for the $\ZZ_3$-wreathing of the theory, leading to an enhanced $\so(8)$ flavor symmetry by the special embedding in equation \eqref{eq:so8tosu3S}, and again reproducing the expected fixed-point subalgebra from discrete gauging.

\paragraph{Curve Analysis:} As in the $D_8$-discrete gauging of $E_5$, the $S_3$-discrete gauging of $E_6$ involves the combination of the actions of the $\ZZ_2$ and $\ZZ_3$ that we have already discussed. Since $\ZZ_3$ is a cyclic inner automorphism of $\mathfrak{e}_6$, the combined rank reduction is the same as for just the $\ZZ_2$, and hence we see a rank-four flavor algebra after discrete gauging. This is compatible with the $\mathfrak{so}(8)$ predicted from the $S_3$-wreathing.

\subsubsection{The (Orthosymplectic) \texorpdfstring{$\ZZ_2$}{Z2}-Wreathing of the Rank-one \texorpdfstring{$E_6$}{E6} Theory}\label{sec:orthoE6}

We can also consider the orthosymplectic realization of $E_6$, given by
\begin{equation}
\begin{tikzpicture}[baseline=(OR.base)]
   \node (OR) {or};
   \node (A) [left=1mm of OR]
   {\begin{tikzpicture}[baseline=0,font=\footnotesize]
        \node[node, label=below:{$4$},fill=blue] (A2) {};
        \node[node, label=below:{$4$},fill=red] (Al1) [left=6mm of A2] {};
        \node[node, label=below:{$4$},fill=red] (Ar1) [right =6mm of A2]{};
        \node[node, label=below:{$2$},fill=blue] (Al2) [left=6mm of Al1] {};
        \node[node, label=below:{$2$},fill=blue] (Ar2) [right =6mm of Ar1]{};
        \node[node, label=below:{$2$},fill=red] (Al3) [left=6mm of Al2] {};
        \node[node, label=below:{$2$},fill=red] (Ar3) [right =6mm of Ar2]{};
        \node[node, label=right:{$2$},fill=red] (A3) [above=4mm of A2] {};
        \draw (A2.north) -- (A3.south);
        \draw (Al3) -- (Al2) -- (Al1) -- (A2) -- (Ar1) -- (Ar2) -- (Ar3);
        \node (Z2) [right=2mm of Ar3] {$/\ZZ_2$};
    \end{tikzpicture}};
    \node (B) [right=1mm of OR] { \begin{tikzpicture}[baseline=0,font=\footnotesize]
        \node[node, label=below:{$4$},fill=blue] (A2) {};
        \node[node, label=below:{$4$},fill=red] (Al1) [left=6mm of A2] {};
        \node[node, label=below:{$4$},fill=red] (Ar1) [right =6mm of A2]{};
        \node[node, label=below:{$2$},fill=blue] (Al2) [left=6mm of Al1] {};
        \node[node, label=below:{$2$},fill=blue] (Ar2) [right =6mm of Ar1]{};
        \node[node, label=below:{$2$},fill=red] (Al3) [left=6mm of Al2] {};
        \node[node, label=below:{$2$},fill=red] (Ar3) [right =6mm of Ar2]{};
        \node[node, label=right:{$1$}] (A3) [above=4mm of A2] {};
        \draw (A2.north) -- (A3.south);
        \draw (Al3) -- (Al2) -- (Al1) -- (A2) -- (Ar1) -- (Ar2) -- (Ar3);
        \node (Z2) [right=2mm of Ar3] {$/\ZZ_2$};
    \end{tikzpicture}};
\end{tikzpicture} \,.
\end{equation}
The exact Hilbert series for this quiver is in \cite{Bourget:2020xdz} and is given by the expansion
\begin{equation}\label{eq:HSE6-ortho}
\begin{split}
    \HS_{E_{6}}(t,\omega) = & \PE\Big[t^2 (32 \omega +46)-t^4 (320 \omega +331)+t^6 (6240 \omega +6136)\\
    &-t^8 (148768 \omega +148178)+\mathcal{O}(t^{10})\Big]\fstop
\end{split}
\end{equation}
The fugacity $\omega$ distinguishes the parity of the operators under the $\U(1)$ of the $\so(10)\oplus \U(1)$ decomposition of $\mathfrak{e}_6$. This quiver has a $\ZZ_2$ diagram automorphism.

Interestingly, the $\ZZ_2$-wreathing of the orthosymplectic quiver does not lead to the same result we obtained for the $\mathbb{Z}_2$-wreathing of the unitary quiver. After the wreathing, we obtain the following Hilbert series:
\begin{equation}\label{eq:HSE6wrZ2-ortho}
\begin{split}
    \HS_{E_{6}\wr \ZZ_2}(t,\omega) &= \frac{\left(\begin{aligned}
        & 1+t^2 (20 \omega +15)+t^4 (380 \omega +436)+t^6 (5232 \omega +4944) \\
        &\quad +t^8 (39452 \omega +40540)+t^{10} (215052 \omega +211473) \\
        &\quad +t^{12} (819124 \omega +828489)+t^{14} (2398192 \omega +2378492) \\
        &\quad +t^{16} (5331044 \omega +5366246)+t^{18} (9428704 \omega +9375232) \\
        &\quad +t^{20} (13105232 \omega +13173184)+t^{22} (14711104 \omega +14637864) \\
        &\quad +t^{24} (13105232 \omega +13173184)+\cdots+t^{44} 
    \end{aligned}\right)}{\left(1-t^2\right)^{11} \left(1-t^4\right)^{11}} \\
    &= \PE\Big[t^2 (20 \omega +26)+t^4 (80 \omega+117)-t^6 (2028 \omega +2076) \\
    &\quad\quad\quad +t^8 (8012 \omega +7234)+\mathcal{O}(t^{10})\Big]\fstop
\end{split}
\end{equation}
We can see that the proposed discretely-gauged SCFT has a continuous flavor symmetry with a $46$-dimensional adjoint representation. 

We can interpret the wreathing of the Coulomb branch of the magnetic quiver directly on the Higgs branch side as follows. Let us consider the branching rules 
\begin{equation}\label{eqn:gnastygnorc}
     \begin{split}
        \mathfrak{e}_6 &\rightarrow \mathfrak{so}(10) \oplus \mathfrak{u}(1)_A \\
        \mathbf{78} &\rightarrow \mathbf{45}_0 \oplus \mathbf{1}_0 \oplus \mathbf{16}_{3} \oplus \overline{\mathbf{16}}_{-3} \fstop
    \end{split} \qquad \text{ and } \qquad \begin{split}
        \mathfrak{so}(10) &\rightarrow \mathfrak{u}(5)\simeq \su(5)\oplus \mathfrak{u}(1)_B  \\
        \mathbf{45} &\rightarrow \mathbf{24}_0 \oplus \mathbf{1}_0 \oplus \mathbf{10}_4 \oplus \overline{\mathbf{10}}_{-4} \\
        \bm{16} &\rightarrow \bm{\overline{5}}_{-3} \oplus \bm{10}_1 \oplus \bm{1}_5 \\
        \bm{\overline{16}} &\rightarrow \bm{5}_{3} \oplus \bm{\overline{10}}_{-1} \oplus \bm{1}_{-5} \coma
    \end{split} 
\end{equation}
where we have labeled the two $\mathfrak{u}(1)$ factors for convenience. We can then define a $\mathbb{Z}_2$ symmetry by
\begin{equation}
    r = (-1)^{(3q_B - q_A)/12} \,,
\end{equation}
where $q_A$, $q_B$ are the charges under the respective $\mathfrak{u}(1)$s. We then propose that $r$ is the generator of the $\mathbb{Z}_2$ symmetry that we are discrete gauging.\footnote{We emphasize that this is not a derivation; we merely argue that the discrete gauging of this particular symmetry on the Higgs branch  reproduces the first orders of the $\ZZ_2$-wreathed Coulomb branch Hilbert series.} Under the decomposition to $\mathfrak{u}(5) \oplus \mathfrak{u}(1)$, the $\bm{78}$ representation of $\mathfrak{e}_6$ branches to
\begin{equation}
  \begin{aligned}
    &\left(\bm{24}_{0,0} \oplus \bm{1}_{0,0} \oplus \bm{1}_{0,0}  \oplus \bm{10}_{-4,0} \oplus \bm{\overline{10}}_{4,0}\right)_+ \\ &\qquad\qquad \oplus \left(\bm{10}_{-4,-12} \oplus \bm{\overline{10}}_{4,12} \oplus \bm{5}_{0,-12} \oplus \bm{\overline{5}}_{0,12} \oplus \bm{1}_{-8,-12} \oplus \bm{1}_{8,12}\right)_- \coma
  \end{aligned}
\end{equation}
where the $\pm$ subscript denotes the parity under $r$, and we have taken linear combinations of $\mathfrak{u}(1)_A$ and $\mathfrak{u}(1)_B$ to put the charges in a suggestive form. We can see that the preserved moment maps, those charged $+$ under $r$, transform in the adjoint representation of an enhanced $\mathfrak{so}(10) \oplus \mathfrak{u}(1)$ global symmetry. We note that this is a \emph{different} $\mathfrak{so}(10) \oplus \mathfrak{u}(1)$ symmetry than the one appearing in the decomposition in equation \eqref{eqn:gnastygnorc}.

We can use the method of Kac diagrams discussed in Section \ref{sec:dg} to classify the (conjugacy classes of) homomorphisms from $\mathbb{Z}_2$ into the automorphism group of $\mathfrak{e}_6$. There are precisely four such nontrivial conjugacy classes, and their fixed-point subalgebras are
\begin{equation}
    \mathfrak{f}_4 \,, \qquad \mathfrak{usp}(8) \,, \qquad \mathfrak{su}(6) \oplus \mathfrak{su}(2) \,, \qquad \mathfrak{so}(10) \oplus \mathfrak{u}(1) \,,
\end{equation}
where the latter two involve a homomorphism whose image is contained in the group of inner automorphisms. Therefore, we see that the wreathed quiver is consistent with a discrete gauging operation; it remains, however, to check that the quiver automorphism indeed induces precisely the automorphism of $\mathfrak{e}_6$ that preserves $\mathfrak{so}(10) \oplus \mathfrak{u}(1)$. 

\subsection{The Rank-one \texorpdfstring{$E_7$}{E7} Theory}
\label{sec:E7th}

Next, we consider the rank-one $E_7$ theory, which can be obtained by considering the GTP written in Table \ref{tab:magquiverspqwebs}. The theory has a nontrivial Higgs branch, which is the closure of the minimal nilpotent orbit of $\mathfrak{e}_7$. As usual, we first analyze the known magnetic quivers for this Higgs branch, summarize their quiver automorphisms, and then review the mirror curve.

\paragraph{Discrete Symmetries from Magnetic Quivers for the Higgs Branch:} As discussed, the unitary magnetic quiver for the Higgs branch is known, and it is:
\begin{equation}\label{eq:unitaryE7}
    \begin{tikzpicture}[baseline=0,font=\footnotesize]
        \node[node, label=below:{$4$}] (A2) {};
        \node[node, label=below:{$3$}] (Al1) [left=6mm of A2] {};
        \node[node, label=below:{$3$}] (Ar1) [right =6mm of A2]{};
        \node[node, label=below:{$2$}] (Al2) [left=6mm of Al1] {};
        \node[node, label=below:{$2$}] (Ar2) [right =6mm of Ar1]{};
        \node[node, label=below:{$1$}] (Al3) [left=6mm of Al2] {};
        \node[node, label=below:{$1$}] (Ar3) [right =6mm of Ar2]{};
        \node[node, label=right:{$2$}] (A3) [above=4mm of A2] {};
        \draw (A2) -- (A3);
        \draw  (Al3) -- (Al2) -- (Al1) -- (A2) -- (Ar1) -- (Ar2) -- (Ar3) ;
        \node (Z2) [right=6mm of Ar3] {$/\U(1)\fstop$};
    \end{tikzpicture}
\end{equation}
The automorphism group of this quiver is simply the $\mathbb{Z}_2$ which swaps the two equal length legs, and, under the homomorphism in equation \eqref{eqn:BIGHOMO}, this produces a specific $\mathbb{Z}_2$ subgroup of the inner automorphism group of $\mathfrak{e}_7$. The fixed-point subalgebra is
\begin{equation}\label{eq:glob_sym_E7}
    \mathfrak{e}_7^{\mathbb{Z}_2} = \mathfrak{e}_6 \oplus \mathfrak{u}(1)\fstop
\end{equation}
This is the continuous flavor symmetry of the putative $\mathbb{Z}_2$-gauged SCFT, $E_7 / \mathbb{Z}_2$.

Furthermore, the closure of the minimal nilpotent orbit of $\mathfrak{e}_7$ also has a realization as the Coulomb branch of an orthosymplectic 3d $\mathcal{N}=4$ quiver. This takes the form
\begin{equation}\label{eq:orthosymE7}
    \begin{tikzpicture}[baseline=0,font=\footnotesize]
        \node[node, label=below:{$6$},fill=red] (A2) {};
        \node[node, label=below:{$4$},fill=blue] (Al1) [left=6mm of A2] {};
        \node[node, label=below:{$4$},fill=blue] (Ar1) [right =6mm of A2]{};
        \node[node, label=below:{$4$},fill=red] (Al2) [left=6mm of Al1] {};
        \node[node, label=below:{$4$},fill=red] (Ar2) [right =6mm of Ar1]{};
        \node[node, label=below:{$2$},fill=blue] (Al3) [left=6mm of Al2] {};
        \node[node, label=below:{$2$},fill=blue] (Ar3) [right =6mm of Ar2]{};
        \node[node, label=below:{$2$},fill=red] (Al4) [left=6mm of Al3] {};
        \node[node, label=below:{$2$},fill=red] (Ar4) [right =6mm of Ar3]{};
        \node[node, label=right:{$2$},fill=blue] (A3) [above=4mm of A2] {};
        \node[node, label=right:{$1$}] (A3a) [above=4mm of A3] {};
        \draw (A2) -- (A3) -- (A3a);
        \draw (Al4) -- (Al3) -- (Al2) -- (Al1) -- (A2) -- (Ar1) -- (Ar2) -- (Ar3) -- (Ar4);
        \node (Z2) [right=6mm of Ar4] {$/\ZZ_2 \,.$};
    \end{tikzpicture}
\end{equation}
This quiver also admits a $\mathbb{Z}_2$ diagram automorphism that exchanges the two horizontal legs.

\paragraph{Mirror Curve:}

The (generalized) toric diagram is in Table \ref{tab:magquiverspqwebs}. The curve is
\begin{align}
    P(x,y)&= u + c_{(-1,-1)}\frac{1}{x y}+c_{(0,-1)}\frac{1}{y}+c_{(1,-1)}\frac{x}{y}+c_{(2,-1)}\frac{x^2}{y}+c_{(3,-1)}\frac{x^3}{y}\nonumber\\
    &+c_{(-1,0)}\frac{1}{x}+c_{(1,0)}x+c_{(2,0)}x^2+c_{(-1,1)}\frac{y}{x}+c_{(0,1)}y+c_{(1,1)}xy\nonumber\\
    &+c_{(-1,2)}\frac{y^2}{x}+c_{(0,2)}y^2+c_{(-1,3)}\frac{y^3}{x} \,.
\end{align}
The presence of the white dots on the diagonal side of the toric diagram implies that not all of the coefficients $c_{(a,b)}$ are free, even before taking into account the various redundancies in the description, in order to correctly reproduce the fact that the two legs of the brane web must end on the same seven branes \cite{Kim:2014nqa,Arias-Tamargo:2024fjt}. Following \cite{Kim:2014nqa}, we can impose these conditions and use the rescaling freedom to fix the curve as follows:
\begin{equation}
\begin{split}
    P(x,y)&=u+\frac{1}{xy}+\chi_1(N)\frac{1}{y}+\chi_2(N)\frac{x}{y}+\chi_3(N)\frac{x^2}{y}+\frac{x^3}{y}\\
   & +\chi_1(M)\frac{1}{x}+\chi_2(M)\frac{y}{x}+[\chi_3(N)+\chi_1(L)\chi_3(M)]x\\
   &+[\chi_3(M)+\chi_1(L)\chi_3(N)]y+[(\chi_1(L)-1)^2+1]xy+\chi_3(M)\frac{y^2}{x}+\frac{y^3}{x}\coma
\end{split}
\end{equation}
where
\begin{align}
    \chi_1(A)=\sum_{i} A_i\,,\quad \chi_2(A)=\sum_{i<j} A_i A_j\,,\quad \chi_3(A)=\sum_{i<j<k} A_i A_j A_k\coma
\end{align}
and
\begin{align}
    M_1 M_2 M_3 M_4 = N_1 N_2 N_3 N_4 = L_1 L_2 = 1\coma
\end{align}
so that the coefficients in the curve are characters of the subalgebra $\mathfrak{su}(4)\oplus\mathfrak{su}(4)\oplus \mathfrak{su}(2)\subset \mathfrak{e}_7$.

\subsubsection{The Higgs Branch of the Rank-one \texorpdfstring{$E_7$}{E7} Theory}

We first compute the refined Hilbert series for the unitary quiver in equation \eqref{eq:unitaryE7}, that is, for the unwreathed theory. We introduce eight fugacities, $w_i$, where $w_1,w_4$ are associated to the two $\U(1)$s, $w_2,w_5,w_7$ are associated to three $\U(2)$s, $w_3,w_6$ are the fugacities for the two $\U(3)$s, and finally $w_8$ is for the $\U(4)$. For simplicity, we decouple the overall $\U(1)$ by summing over fluxes of $\U(4)$ such that $m_1\leq m_2\leq m_3\leq m_4=0$, and impose
\begin{equation}
    w_8^4 = \frac{1}{w_1w_2^2w_3^3w_4w_5^2w_6^3w_7^2}\fstop
\end{equation}
With this prescription, the Hilbert series has a manifest $\mathfrak{e}_7$ flavor symmetry, observed by redefining the fugacities as
\begin{equation}
\renewcommand*{\arraystretch}{1.5}
\begin{array}{c}
  w_1\to \dfrac{z_1 z_6}{z_4}\coma w_2\to \dfrac{z_3}{z_6 z_7}\coma w_3\to \dfrac{z_4 z_7}{z_3 z_5}\coma w_4\to \dfrac{z_2^2}{z_1 z_3}\coma\\
  w_5\to \dfrac{z_1 z_4 z_6}{z_2 z_5}\coma  w_6\to \dfrac{z_5}{z_6 z_7}\coma w_7\to \dfrac{z_1 z_5 z_7}{z_3 z_6}\fstop
\end{array}
\end{equation}
The refined Hilbert series is
\begin{equation}\label{eq:E7-HS}
    \HS_{E_{7}}(t,\mathbf{z}) = \PE\left[\chi^{\mathfrak{e}_7}_{[1,0,0,0,0,0,0]}(\mathbf{z})t^2-(1 + \chi^{\mathfrak{e}_7}_{[0,0,0,0,1,0,0]}(\mathbf{z}))t^4+\mathcal{O}(t^{6})\right]\coma
\end{equation}
and unrefined reads
\begin{equation}\label{eq:HSE7-unrefined}
    \HS_{E_{7}}(t, 1) = \PE\left[133t^2 -1540t^4+42427t^6 +\mathcal{O}(t^{8})\right]\fstop
\end{equation}

To aid in the following discussion, we note that the same fugacities can be redefined to make an $\so(12)\oplus \su(2)$ symmetry explicit instead. Specifically, redefining the fugacities as
\begin{equation}
  \begin{gathered}
    w_1\to \frac{x_1^2}{x_2}\coma w_2\to \frac{x_2}{x_1x_3}\coma w_3\to x_2\coma w_4\to \frac{x_5^2}{x_4}\coma \\[-0.3em] w_5\to \frac{x_3}{x_5x_6}\coma  w_6\to \frac{x_6^2}{x_4}\coma w_7\to y_1^2\coma
  \end{gathered}
\end{equation}
then the refined Hilbert series, in terms of this manifest $\mathfrak{so}(12)_{\mathbf{x}} \oplus \mathfrak{su}(2)_y$ global symmetry, is
\begin{equation}\label{eq:E7-HS-so12su2}
    \HS_{E_{7}}(t,\mathbf{x},y) = \PE\left[\left(\chi^A_{[2]}(y)+\chi^D_{[0,0,0,0,0,1]}(\mathbf{x})\chi^A_{[1]}(y)+\chi^D_{[0,1,0,0,0,0]}(\mathbf{x})\right)t^2+\mathcal{O}(t^{4})\right]\fstop
\end{equation}
This could equally have been obtained directly from equation \eqref{eq:E7-HS} by considering the decomposition and branching rule:
\begin{equation}\label{eqn:oak}
    \begin{split}
      \mathfrak{e}_7 & \rightarrow \so(12)\oplus \su(2)\coma\\
      \mathbf{133} & \rightarrow (\mathbf{1}, \mathbf{3})\oplus(\overline{\mathbf{32}}, \mathbf{2}) \oplus (\mathbf{66}, \mathbf{1})\coma
    \end{split}
\end{equation}
which precisely matches the $t^2$-coefficients in equations \eqref{eq:E7-HS} and \eqref{eq:E7-HS-so12su2}.

The exact expression for the Hilbert series of the orthosymplectic magnetic quiver for the Higgs branch in equation \eqref{eq:orthosymE7} was determined in \cite{Bourget:2020xdz}. We report here only that the series expansion is:
\begin{equation}\label{eq:HSE7-ortho}
\begin{split}
    \HS_{E_{7}}(t,\omega) = & \PE\Big[t^2 (64 \omega +69)-t^4 (768 \omega +772)+t^6 (21248 \omega +21179)+\mathcal{O}(t^{8})\Big]\fstop
\end{split}
\end{equation}
The $\ZZ_2^{[0]}$ fugacity $\omega$ distinguishes between the even/odd charged operators under the $\U(1)$ algebra if we decompose $\mathfrak{e}_7$ into $\so(12)\oplus \U(1)$. That is, the $\ZZ_2^{[0]}$ charge is the charge under the center of the $\mathfrak{su}(2)$ in equation \eqref{eqn:oak}.

\subsubsection{The \texorpdfstring{$\ZZ_2$}{Z2} Discrete Gauging of the Rank-one \texorpdfstring{$E_7$}{E7} Theory}

\paragraph{Wreathing the Magnetic Quiver:} The affine $\mathfrak{e}_7$ quiver in equation \eqref{eq:unitaryE7} admits a single diagram automorphism, which is the $\mathbb{Z}_2$ that exchanges the two identical legs:
\begin{equation}\label{eq:E7-wrZ2}
    \begin{tikzpicture}[baseline=0,font=\footnotesize]
        \node[node, label=below right:{$4$}] (A2) {};
        \node[node, label=below:{$3$}] (Al1) [left=6mm of A2] {};
        \node[node, label=below:{$3$}] (Ar1) [right =6mm of A2]{};
        \node[node, label=below:{$2$}] (Al2) [left=6mm of Al1] {};
        \node[node, label=below:{$2$}] (Ar2) [right =6mm of Ar1]{};
        \node[node, label=below:{$1$}] (Al3) [left=6mm of Al2] {};
        \node[node, label=below:{$1$}] (Ar3) [right =6mm of Ar2]{};
        \node[node, label=right:{$2$}] (A3) [above=4mm of A2] {};
        \draw (A2) -- (A3);
        \draw  (Al3) -- (Al2) -- (Al1) -- (A2) -- (Ar1) -- (Ar2) -- (Ar3) ;
        \draw[Triangle-Triangle,red] ([yshift=-5mm]Al2.south) to[bend right=50] node[below,pos=0.5] {$\wr \ZZ_2$} ([yshift=-5mm]Ar2.south);
        \node (Z2) [right=6mm of Ar3] {$/\U(1)\fstop$};
        \draw[densely dashed, red, thin] ([yshift=3mm]A3.north) -- ([yshift=-3mm]A2.south);
    \end{tikzpicture}
\end{equation}
We can therefore consider wreathing by this $\mathbb{Z}_2$, in an analogous way as we did in the previous sections, e.g., Section \ref{sec:E6th}. The way in which the wreathing of equation \eqref{eq:E7-wrZ2} can be computed is by noting that the quiver is given by the gauging of the $\su(4)$ of two copies of $T_{[1,1,1,1]}[\SU(4)]$ tails and a $T_{[2,2]}[\SU(4)]$ tail, decoupling the center-of-mass $\U(1)$ on the central node. The wreathing acts on the two $T_{[1,1,1,1]}[\SU(4)]$ tails. In fact, the unwreathed Hilbert series in equation \eqref{eq:E7-HS} can be obtained by considering the Hilbert series
\begin{equation}
\begin{split}
    \text{HS}_{E_7}(t,\mathbf{w}) = &\,(1-t^2)\sum_{m_1\leq \cdots\leq m_4 = 0} P_{\U(4)}(t,\mathbf{m})w_8^{m_1+m_2+m_3+m_4}t^{2\Delta_\text{res.}} \times \\
    &\text{HS}_{T_{[1,1,1,1]}[\SU(4)]}(t,w_1,w_2,w_3,\mathbf{m})\text{HS}_{T_{[1,1,1,1]}[\SU(4)]}(t,w_4,w_5,w_6,\mathbf{m})\times \\
    &\text{HS}_{T_{[2,2]}[\SU(4)]}(t,w_7,\mathbf{m})\coma    
\end{split}
\end{equation}
where $\Delta_\text{res.}$ is the contribution associated to the gauging of the $\U(4)$ gauge group. The action of the wreathing is to first identify the fugacities of the two $T_{[1,1,1,1]}[\SU(4)]$ tails, i.e.,
\begin{equation}
    w_4 = w_1 \coma w_5 = w_2 \coma w_6 = w_3\fstop
\end{equation}
Then, following \cite[Appendix A]{Grimminger:2024mks}, the wreathed Hilbert series is given by 
\begin{equation}
\scalebox{0.92}{$\displaystyle
\begin{aligned}
     \text{HS}_{E_7\wr \ZZ_2}(t,\mathbf{w}) = \,&(1-t^2)\sum_{m_1\leq \cdots\leq m_4 = 0} P_{\U(4)}(t,\mathbf{m}) \text{HS}_{T_{[2,2]}[\SU(4)]}(t,w_7,\mathbf{m})w_8^{m_1+m_2+m_3+m_4}t^{2\Delta_\text{res.}}\times \\
     &\frac{1}{2}\left(\text{HS}_{T_{[1,1,1,1]}[\SU(4)]}(t,w_1,w_2,w_3,\mathbf{m})^2+\text{HS}_{T_{[1,1,1,1]}[\SU(4)]}(t^2,w_1^2,w_2^2,w_3^2,\mathbf{m})\right)\fstop
\end{aligned}$}
\end{equation}
The refined wreathed Hilbert series can be determined while keeping a manifest $\mathfrak{su}(4)_{\mathbf{x}} \oplus \mathfrak{su}(2)_{y}$ global symmetry; define
\begin{equation}
    w_1 \rightarrow \frac{x_1^2}{x_2}\coma w_2 \rightarrow \frac{1}{x_1x_3}\coma w_3 \rightarrow \frac{x_3^2}{x_2}\coma w_7 \rightarrow y^2\fstop
\end{equation}
We then find
\begin{equation}\label{eq:E7Z2-HS}
  \begin{split}
    \HS_{E_{7}\wr \ZZ_2}(t,\bm{x},y) = \PE&\left[\left(1 + \chi_{[1,0,1]}^A(\bm{x}) + \chi^A_{[2]}(y)\right.\right.\\
    &\left.\left.+\left(\chi_{[2,0,0]}^A(\bm{x}) + \chi_{[0,0,2]}^A(\bm{x})\right)\chi^A_{[1]}(y)+\chi_{[0,2,0]}^A(\bm{x}) \right)t^2\right.\\
    &+\left.\left(1+2\left(\chi^A_{[0,2,0]}(\bm{x})+\chi^A_{[1,0,1]}(\bm{x})+ \chi^A_{[2,0,2]}(\bm{x})\right.\right.\right.\\
    &+\left.\left.\left.
    \left(\chi^A_{[0,0,2]}(\bm{x})+ \chi^A_{[1,1,1]}(\bm{x})+ \chi^A_{[2,0,0]}(\bm{x})\right)\chi^A_{[1]}(y)\right.\right.\right.\\
    &\left.\left.\left.+
    \left(1+\chi^A_{[0,2,0]}(\bm{x})\right)\chi^A_{[2]}(y)\right)\right)t^4+\mathcal{O}(t^{6})\right]\fstop
  \end{split}
\end{equation}
Studying the $t^2$ term, we find that this Hilbert series realizes an $\mathfrak{e}_6 \oplus \mathfrak{u}(1)$ global symmetry via the composition of the maximal embeddings associated with the following branching rules:
\begin{equation}
    \begin{split}
        \mathfrak{e}_6 &\rightarrow \mathfrak{su}(2) \oplus \mathfrak{su}(6) \\
        \mathbf{78} &\rightarrow (\mathbf{3}, \mathbf{1}) \oplus (\mathbf{1}, \mathbf{35}) \oplus (\mathbf{2}, \mathbf{20})
    \end{split} \qquad \text{ and } \qquad \begin{split}
        \mathfrak{su}(6) &\stackrel{S}{\rightarrow} \mathfrak{su}(4) \\
        \mathbf{35} &\rightarrow \mathbf{15}\oplus \mathbf{20}' \\
        \mathbf{20} &\rightarrow \mathbf{10}\oplus \overline{\mathbf{10}} \fstop
    \end{split}
\end{equation}
In fact, we can use the further branching rules
\begin{equation}
    \begin{split}
        \mathfrak{e}_6 &\rightarrow \mathfrak{su}(2) \oplus \mathfrak{su}(6) \\
        \mathbf{351}' &\rightarrow (\mathbf{1},\overline{\mathbf{15}}) \oplus (\mathbf{3},\overline{\mathbf{21}}) \oplus (\mathbf{2},\mathbf{84}) \oplus (\mathbf{1},\overline{\mathbf{105}}') 
    \end{split} \quad \text{ and } \quad \begin{split}
        \mathfrak{su}(6) &\stackrel{S}{\rightarrow} \mathfrak{su}(4) \\
        \mathbf{15} &\rightarrow \mathbf{15} \\
        \mathbf{21} &\rightarrow \mathbf{1}\oplus \mathbf{20}' \\
        \mathbf{84} &\rightarrow \mathbf{10} \oplus \overline{\mathbf{10}}\oplus \mathbf{64}\\
        \mathbf{105}' &\rightarrow \mathbf{1}\oplus \mathbf{20}'\oplus \mathbf{84}\coma
    \end{split}
\end{equation}
to see that the $t^4$ term can be rewritten in terms of characters as $\chi^{\mathfrak{e}_6}_{\mathbf{351}'}(\bm{x}) + \chi^{\mathfrak{e}_6}_{\overline{\mathbf{351}'}}(\bm{x}) - 1$ under this putative enhanced global symmetry. Therefore, the Hilbert series of the $\mathbb{Z}_2$-wreathed quiver can be written as
\begin{equation}\label{eqn:bluey}
    \begin{split}
    \HS_{E_{7}\wr \ZZ_2}&(t,\bm{x}) = \PE\left[\left(1 + \chi_{\bm{78}}^{\mathfrak{e}_6}(\bm{x}) \right)t^2 + \left( \chi^{\mathfrak{e}_6}_{\mathbf{351}'}(\bm{x}) + \chi^{\mathfrak{e}_6}_{\overline{\mathbf{351}'}}(\bm{x}) - 1\right) t^4 + \mathcal{O}(t^6)\right]\fstop
  \end{split}
\end{equation}

\paragraph{Curve Analysis:}
The unitary magnetic quiver for the $E_7$ SCFT only admits one wreathing, which corresponds to exchanging two external legs of the corresponding brane web. This amounts to the action
\begin{align}\label{eq:E7innercoord-trans}
    (x,y)\to (y,x) \,,
\end{align}
on the curve coordinates. We can compose this with a transformation of the characters
\begin{align}\label{eq:E7innerMass-trans}
    (M_i,N_i,L_i)\to (N_i,M_i,L_i^{-1})\fstop
\end{align}
This naturally corresponds to an inner automorphism of the flavor algebra (there are no outer automorphisms of $\mathfrak{e}_7$); and accordingly, we observe that the curve is invariant under these transformations without the need to impose any further conditions on the $M, N, L$. This is in agreement with the expectation in equation \eqref{eq:glob_sym_E7} and the wreathing result in equation \eqref{eqn:bluey}, where there is no rank reduction of the flavor symmetry.\footnote{The transformations in equations \eqref{eq:E7innercoord-trans} and \eqref{eq:E7innerMass-trans} are those that, if applied to the $E_6$ curve, would also realize the $\ZZ_2$ wreathing associated to the orthosymplectic magnetic quiver for the $\mathfrak{e}_6$ theory we discussed in Section \ref{sec:orthoE6}. It would be interesting to understand the origin of such automorphisms from the curve analysis.}

\subsubsection{The (Orthosymplectic) \texorpdfstring{$\ZZ_2$}{Z2}-wreathing of the Rank-one \texorpdfstring{$E_7$}{E7} Theory}

A similar conclusion, but with different branching rules, can be obtained by considering the $\ZZ_2$-wreathing of the orthosymplectic quiver in equation \eqref{eq:orthosymE7}. The $\mathbb{Z}_2$-wreathed Hilbert series can be determined analogously to the unitary version of the quiver, and we find
\begin{equation}\label{eq:HSE7wrZ2-ortho}
    \HS_{E_{7}\wr \ZZ_2}(t,\omega) = \PE\left[t^2 (40 \omega +39)+t^4 (336 \omega +365)-t^6 (9888 \omega +9833)+\mathcal{O}(t^{8})\right]\fstop
\end{equation}
The Coulomb branch flavor symmetry after wreathing appears to again be $\mathfrak{e}_6 \oplus \mathfrak{u}(1)$, indicating that the two $\mathbb{Z}_2$ manifest in the unitary and orthosymplectic quivers are, in fact, corresponding to the same discrete symmetry of the $E_7$ SCFT.

To determine the expected flavor symmetry of the discretely-gauged SCFT, we follow the same pattern that was observed in six dimensions in \cite{Lawrie:2025exx}. We look at the manifest $\mathfrak{so}(12)$ Coulomb branch symmetry in the orthosymplectic quiver, and we decompose it as
\begin{equation}\label{eqn:potato}
    \mathfrak{so}(12) \rightarrow \mathfrak{u}(6)\fstop
\end{equation}
If we start with an $\mathfrak{e}_7$ representation generated from the adjoint, and decompose it first under equation \eqref{eqn:oak} followed by equation \eqref{eqn:potato}, then all $\mathfrak{u}(1)$ charges are even. We can then naturally define a $\mathbb{Z}_2$ inside this $\mathfrak{u}(1)$ such that, after scaling the $\U(1)$ generator by $1/2$, even/odd $\mathfrak{u}(1)$ charges are trivial/nontrivial under the $\mathbb{Z}_2$. We propose that this $\mathbb{Z}_2$ is the $\mathbb{Z}_2$ that we are discretely gauging.
The branching rules for the relevant representations are
\begin{equation}\label{eq:so12tou6}
     \begin{split}
        \mathfrak{so}(12) &\rightarrow \mathfrak{u}(6)\simeq \su(6)\oplus \mathfrak{u}(1)  \\
        \mathbf{66} &\rightarrow \mathbf{35}_0 \oplus\mathbf{1}_0 \oplus \mathbf{15}_{2} \oplus \overline{\mathbf{15}}_{-2} \coma\\
        \mathbf{32} &\rightarrow \mathbf{20}_0 \oplus \mathbf{6}_2 \oplus \overline{\mathbf{6}}_{-2}\fstop
    \end{split}
\end{equation}
As we can see, the anti-symmetric (and conjugate) representations of the adjoint representation are projected out, and the fundamental (and conjugate) representations are projected out from the spinor of the $\mathfrak{so}(12)$. Thus, we have moment maps in the representations
\begin{equation}
    (\bm{35}, \bm{1}) \oplus (\bm{1}, \bm{3}) \oplus (\bm{1}, \bm{1}) \oplus (\mathbf{20}, \mathbf{2})\coma
\end{equation}
under the naive $\mathfrak{u}(6) \oplus \mathfrak{su}(2)$ global symmetry after wreathing. We can see that these are simply the representations that appear in the branching rule of the adjoint representation of an enhanced $\mathfrak{e}_6 \oplus \mathfrak{u}(1)$ global symmetry, and it agrees with the wreathed Hilbert series given in equation \eqref{eq:HSE7wrZ2-ortho}, including the refinement by the $\mathbb{Z}_2^{[0]}$ fugacity.

Now that we have provided evidence that the $\mathbb{Z}_2$ inner automorphism of $\mathfrak{e}_7$ which is discretely gauged in this section is the $\mathbb{Z}_2$ inside the $\mathfrak{u}(1)$ factor in equation \eqref{eqn:potato}, it is possible to track which operators of the original theory are projected out by the discrete gauging, directly. This $\mathfrak{u}(1)$ is \emph{not} part of the enhanced $\mathfrak{e}_6$ symmetry after discrete gauging, so we can just look at the branching rules under the decomposition $\mathfrak{e}_7 \rightarrow \mathfrak{e}_6 \oplus \mathfrak{u}(1)$. For example, expanding the plethystic exponential in equation \eqref{eq:HSE7-ortho}, we see that the coefficient of $t^4$ is the character of the $\bm{7371}$ representation of $\mathfrak{e}_7$. Under the decomposition we have 
\begin{equation}
    \begin{split}
        \mathfrak{e}_7 &\rightarrow \mathfrak{e}_6\oplus\mathfrak{u}(1)\\
        \mathbf{7371} &\rightarrow {\color{red}{\mathbf{1}_{(0)}}} \oplus {\color{red}{\mathbf{78}_{(0)}}} \oplus {\color{red}{\mathbf{650}_{(0)}}} \oplus {\color{red}{\mathbf{2430}_{(0)}}} \oplus {\color{red}{\mathbf{351}'_{(-4)}}} \oplus {\color{red}{\overline{\mathbf{351}'}_{(4)}}} \\
    &\quad \oplus \mathbf{1728}_{(2)} \oplus \mathbf{27}_{(2)} \oplus \overline{\mathbf{1728}}_{(-2)} \oplus \overline{\mathbf{27}}_{(-2)} \coma
    \end{split}
\end{equation}
where we have highlighted in {\color{red}{red}} the representations which are uncharged under the $\mathbb{Z}_2$, and thus survive the discrete gauging. This reproduces the $t^4$ term in the refined wreathed Hilbert series in equation \eqref{eqn:bluey}; we can incorporate the refinement by $\omega$ in equation \eqref{eq:HSE7wrZ2-ortho} by checking the charges under the center of the latter factor in the decomposition $\mathfrak{e}_6 \rightarrow \su(6)\oplus \su(2)$. 

\subsection{The Rank-one \texorpdfstring{$E_8$}{E8} Theory}
\label{sec:E8th}

Finally, we consider the rank-one $E_8$ theory, for which we present the results compactly. The unitary magnetic quiver capturing the Higgs branch takes the following form:
\begin{equation}\label{eq:unitaryE8}
    \begin{tikzpicture}[baseline=0,font=\footnotesize]
        \node[node, label=below:{$6$}] (A2) {};
        \node[node, label=below:{$5$}] (Al1) [left=6mm of A2] {};
        \node[node, label=below:{$4$}] (Ar1) [right =6mm of A2]{};
        \node[node, label=below:{$4$}] (Al2) [left=6mm of Al1] {};
        \node[node, label=below:{$2$}] (Ar2) [right =6mm of Ar1]{};
        \node[node, label=below:{$3$}] (Al3) [left=6mm of Al2] {};
        \node[node, label=below:{$2$}] (Al4) [left =6mm of Al3]{};
        \node[node, label=below:{$1$}] (Al5) [left =6mm of Al4]{};
        \node[node, label=right:{$3$}] (A3) [above=4mm of A2] {};
        \draw (A2) -- (A3);
        \draw  (Al5) -- (Al4) -- (Al3) -- (Al2) -- (Al1) -- (A2) -- (Ar1) -- (Ar2);
        \node (Z2) [right=6mm of Ar2] {$/\U(1)\fstop$};
    \end{tikzpicture}
\end{equation}
The Hilbert series of this quiver has been computed in the literature (see, for example, \cite{Ferlito:2017xdq}), revealing a manifest $\mathfrak{e}_8$ flavor symmetry. One way to decouple the overall $\U(1)$ is by setting the fluxes of the $\U(1)$ node to zero and
\begin{equation}
    z_1 = \frac{1}{z_2^2 z_3^3 z_4^4 z_5^5 z_6^6 z_7^4 z_8^2 z_9^3}\fstop
\end{equation}
A manifest $\mathfrak{e}_8$ flavor symmetry is found by the substitution of the remaining fugacities:
\begin{equation}
\renewcommand*{\arraystretch}{1.5}
\begin{array}{rclrclrclrcl}
  z_2 &\to & \dfrac{x_4 x_7 x_8}{x_3 x_5}\coma & z_3 &\to & \dfrac{x_3}{x_8^2}\coma & z_4& \to & \dfrac{x_1 x_5 x_8}{x_2 x_4}\coma & z_5 & \to & \dfrac{x_2}{x_1 x_6}\coma\\
z_6 & \to& \dfrac{x_4}{x_2}\coma & z_7 & \to &\dfrac{x_2 x_6}{x_1 x_4}\coma & z_8 & \to & \dfrac{x_1 x_7}{x_6}\coma & z_9 & \to & \dfrac{x_1 x_6}{x_5 x_7}\fstop
\end{array}
\end{equation}
In this way, the Coulomb branch Hilbert series at leading order is the expected
\begin{equation}\label{eq:E8HS-unitary}
\begin{split}
    \HS_{E_{8}}(t,\mathbf{x}) = \PE&\left[\chi^{\mathfrak{e}_8}_{[0,\ldots,0,1,0]}(\mathbf{x})t^2+\mathcal{O}(t^{4})\right]\fstop
\end{split}
\end{equation}

The unitary realization of the Higgs branch of the $E_8$ theory has no diagram automorphism since the affine $\mathfrak{e}_8$ Dynkin diagram has no automorphisms, and thus there is no wreathing to perform. Instead, we can consider the orthosymplectic magnetic quiver for the Higgs branch of the $E_8$ theory, which is
\begin{equation}\label{eq:orthosymE8}
    \begin{tikzpicture}[baseline=0,font=\footnotesize]
        \node[node, label=below:{$8$},fill=red] (A2) {};
        \node[node, label=below:{$6$},fill=blue] (Al1) [left=6mm of A2] {};
        \node[node, label=below:{$6$},fill=blue] (Ar1) [right =6mm of A2]{};
        \node[node, label=below:{$6$},fill=red] (Al2) [left=6mm of Al1] {};
        \node[node, label=below:{$6$},fill=red] (Ar2) [right =6mm of Ar1]{};
        \node[node, label=below:{$4$},fill=blue] (Al3) [left=6mm of Al2] {};
        \node[node, label=below:{$4$},fill=blue] (Ar3) [right =6mm of Ar2]{};
        \node[node, label=below:{$4$},fill=red] (Al4) [left=6mm of Al3] {};
        \node[node, label=below:{$4$},fill=red] (Ar4) [right =6mm of Ar3]{};
        \node[node, label=below:{$2$},fill=blue] (Al5) [left=6mm of Al4] {};
        \node[node, label=below:{$2$},fill=blue] (Ar5) [right =6mm of Ar4]{};
        \node[node, label=below:{$2$},fill=red] (Al6) [left=6mm of Al5] {};
        \node[node, label=below:{$2$},fill=red] (Ar6) [right =6mm of Ar5]{};
        \node[node, label=right:{$2$},fill=blue] (A3) [above=4mm of A2] {};
        \draw (A2) -- (A3);
        \draw (Al6) -- (Al5) -- (Al4) -- (Al3) -- (Al2) -- (Al1) -- (A2) -- (Ar1) -- (Ar2) -- (Ar3) -- (Ar4) -- (Ar5) -- (Ar6);
       \node (Z2) [right=6mm of Ar6] {$/\ZZ_2$};
    \end{tikzpicture}\fstop
\end{equation}
We can see that this quiver has a $\mathbb{Z}_2$ diagram automorphism that exchanges the two identical legs; we can consider wreathing by this $\mathbb{Z}_2$.

The Coulomb branch Hilbert series for the quiver in equation \eqref{eq:orthosymE8} was found in \cite{Bourget:2020xdz}, and we provide the expansion here:
\begin{equation}\label{eq:HSE8-ortho}
    \HS_{E_{8}}(t,\omega) = \PE\left[t^2 (128 \omega +120)+\mathcal{O}(t^{4})\right]\fstop
\end{equation}
We can interpret this Hilbert series by considering the quiver as having a manifest $\mathfrak{so}(16)$ global symmetry together with additional moment maps in the Weyl spinor representation.\footnote{Whether the positive or negative chirality spinor representation is not particularly germane.} We can identify the $\mathbb{Z}_2^{[0]}$ zero-form symmetry, with fugacity $\omega$, as the diagonal of the $\mathbb{Z}_2 \times \mathbb{Z}_2$ center of $\operatorname{Spin}(16)$; in particular, the adjoint has charge zero, and the spinor has charge one.

\subsubsection{The (Orthosymplectic) \texorpdfstring{$\ZZ_2$}{Z2}-Wreathing of the Rank-one \texorpdfstring{$E_8$}{E8} Theory}

Using the techniques we have employed in great detail throughout this paper, we can see that
the $\ZZ_2$-wreathing of the quiver in equation \eqref{eq:orthosymE8} gives the following Hilbert series:
\begin{equation}\label{eq:HSE8wrZ2-ortho}
    \HS_{E_{8}\wr \ZZ_2}(t,\omega) = \PE\left[t^2 (72 \omega +64)+\mathcal{O}(t^{4})\right]\fstop
\end{equation}
We can understand the wreathed flavor symmetry as follows. Consider the $\mathfrak{u}(8)$ subalgebra of the manifest $\mathfrak{so}(16)$, and suppose that the automorphism of $\mathfrak{e}_8$ induced by the quiver automorphism is identified with the $\mathbb{Z}_2$ inside of the $\mathfrak{u}(1)$ factor; the generator acts on a state of charge $q$ by  $(-1)^{q/2}$.\footnote{Note that under the decomposition $\mathfrak{e}_8 \rightarrow \mathfrak{so}(16) \rightarrow \mathfrak{u}(8)$ all the $\mathfrak{u}(1)$ charges are necessarily even.} Then we can write
\begin{equation}\label{eq:so16tou8}
    \begin{split}
        \so(16) &\rightarrow \mathfrak{u}(8)\simeq \su(8)\oplus \mathfrak{u}(1)\\
        \mathbf{120} &\rightarrow \mathbf{1}_0 \oplus \mathbf{28}_2 \oplus \overline{\mathbf{28}}_{-2} \oplus \mathbf{63}_0\\
        \mathbf{128} &\rightarrow \mathbf{1}_{4} \oplus \mathbf{1}_{-4} \oplus \mathbf{28}_{-2} \oplus \overline{\mathbf{28}}_{2} \oplus \mathbf{70}_0\fstop
    \end{split}
\end{equation}
Since they are charged under the $\mathbb{Z}_2$, the $\mathbf{28} \oplus \overline{\mathbf{28}}$ from both the adjoint representation and the spinor representation are projected out. Therefore, we see that there are moment maps remaining which transform under the following representations of $\mathfrak{u}(8)$:
\begin{equation}
    \bm{63}_0 \oplus \bm{70}_0 \oplus \bm{1}_{2} \oplus \bm{1}_{0} \oplus\bm{1}_{-2}\coma
\end{equation}
where we have scaled the $\mathfrak{u}(1)$ charges by a factor of $1/2$. It is easy to see that these representations recombine into the adjoint representation of 
\begin{equation}
    \mathfrak{e}_7 \oplus \mathfrak{su}(2)\coma
\end{equation}
which we propose is then the enhanced global symmetry of the discretely-gauged rank-one $E_8$ theory, $E_8 / \ZZ_2$. The resulting decomposition is compatible with the Hilbert series in equation \eqref{eq:HSE8wrZ2-ortho}, where the $\omega$ distinguishes the representations charged or uncharged under the $\ZZ_2^{[0]}$ zero-form symmetry. Since the unitary quiver of $E_8$ did not admit any wreathing, this flavor symmetry is only observed from the orthosymplectic formulation of the $E_8$ Higgs branch.

\section{Higher Rank \texorpdfstring{\boldmath{$E_n$}}{En} SCFTs}\label{sec:higherrank}

The discrete gaugings discussed for the rank-one $E_n$ SCFTs in this paper extend, essentially mutatis mutandis, to the higher-rank $E_n$ SCFTs. We refer to these rank-$N$ SCFTs as $E_n^{(N)}$, and we briefly discuss their discrete gauging here.

The magnetic quivers for the Higgs branches of these rank-$N$ $E_n$ SCFTs are the same as those given in Table \ref{tbl:Endata}, except the $\U(c)$ gauge algebra at each node is replaced with $\U(Nc)$. That is, the Higgs branches are the reduced moduli spaces of $N$ $E_n$-instantons on $\CC^2$; for $N \geq 2$ the flavor symmetry is
\begin{equation}
    \mathfrak{f} \left( \, E_n^{(N)} \, \right) = \mathfrak{e}_n \oplus \mathfrak{su}(2) \,.
\end{equation}
It is apparent from this description that the group of quiver automorphisms is independent of $N$, and thus each automorphism listed in Table \ref{tbl:Uwreathing}, and studied for the rank-one case, is equally an automorphism, acting in the same way on the flavor symmetry, for arbitrary rank. Similarly, the pure and mixed-gravitational 't Hooft anomalies of these discrete symmetries, generically, are again forced to vanish, following Section \ref{sec:tHooftanomalies}. 

While the rank-one chiral ring, for $n \geq 4$, is completely defined by the single generator in the adjoint representation of $\mathfrak{e}_n$ and the ideal generated by the Joseph relations, for $N \geq 2$, the Higgs branch chiral ring is rather more complicated for higher ranks. In \cite{Cremonesi:2014xha}, it was proposed that there are generators in the following representations of the flavor algebra
\begin{equation}\label{eqn:GinstCONJGEN}
    \bigoplus_{d=2}^{N}\,\, (\bm{1},\, \bm{d+1}\,)_{t^d} \,\,\oplus\,\, \bigoplus_{d=1}^{N}\,\, (\mathbf{Adj},\, \bm{d}\,)_{t^{d+1}} \,,
\end{equation}
where the subscript denotes the order in $t$ that they first enter in the Hilbert series. A priori, we have established how the quiver automorphism $H$ acts on the $\mathfrak{e}_n$ flavor symmetry, but we have not specified the action on the other generators. For example, in principle it could act nontrivially on the $\mathfrak{su}(2)$ factor. We conjecture that the $H$-action on the SCFT acts only on the exceptional flavor algebra, via the automorphism induced by the quiver symmetry. Hence, for all $N \geq 2$ and $H \subseteq \operatorname{Aut}(\widetilde{\mathcal{D}}_{\mathfrak{e}_n})$, we propose the existence of a rank-$N$ discretely-gauged theory $E_n^{(N)} \, / \, H$ with flavor symmetry
\begin{equation}
    \mathfrak{f} \left( \, E_n^{(N)} \, / \, H \, \right) = \mathfrak{e}_n^H \oplus \mathfrak{su}(2) \,.
\end{equation}

Let us now exemplify this discussion via an explicit example. We will consider the rank $N$ $E_6$ SCFT; the generalized toric polygon which engineers this theory has been constructed in \cite{Benini:2009gi}. The GTP is a triangle of side length $3N$, with each edge divided into three blocks of lattice length $N$. The endpoints of these blocks are black, while the $N-1$ points between each adjacent pair of black dots are white. Upon choosing an $s$-rule-compatible tessellation, the boundary $s$-rule propagates into the interior and generates further white dots; these depend on the chosen tessellation, and are related to alternative tessellations via flop transitions. For each $N$, there exists an internal tessellation such that the geometry retains the $S_3$ symmetry as in the rank-one theory, indicating an honest discrete symmetry of the SCFT. The magnetic quiver for the Higgs branch is:
\begin{equation}\label{eq:unitaryNE6}
    \begin{tikzpicture}[baseline=0,font=\footnotesize]
        \node[node, label=below:{$3N$}] (A2) {};
        \node[node, label=below:{$2N$}] (Al1) [left=6mm of A2] {};
        \node[node, label=below:{$2N$}] (Ar1) [right =6mm of A2]{};
        \node[node, label=below:{$N$}] (Al2) [left=6mm of Al1] {};
        \node[node, label=below:{$N$}] (Ar2) [right =6mm of Ar1]{};
        \node[node, label=right:{$2N$}] (A3) [above=4mm of A2] {};
        \node[node, label=right:{$N$}] (A3a) [above=4mm of A3] {};
        \draw (A2) -- (A3) -- (A3a);
        \draw   (Al2) -- (Al1) -- (A2) -- (Ar1) -- (Ar2) ;
        \node (Z2) [right=6mm of Ar2] {$/\U(1)\fstop$};
    \end{tikzpicture}
\end{equation}

For $N > 1$, the monopole formula cannot be applied directly to the affine quiver in equation \eqref{eq:unitaryNE6}, since non-central magnetic charges along the affine null direction have vanishing conformal dimension. Following \cite{Cremonesi:2014xha}, we therefore introduce the auxiliary overextended quiver
\begin{equation}\label{eq:unitaryNE6over}
    \begin{tikzpicture}[baseline=0,font=\footnotesize]
        \node[node, label=below:{$3N$}] (A2) {};
        \node[node, label=below:{$2N$}] (Al1) [left=6mm of A2] {};
        \node[node, label=below:{$2N$}] (Ar1) [right =6mm of A2]{};
        \node[node, label=below:{$N$}] (Al2) [left=6mm of Al1] {};
        \node[node, label=below:{$N$}] (Ar2) [right =6mm of Ar1]{};
        \node[node, label=right:{$2N$}] (A3) [above=4mm of A2] {};
        \node[node, label=right:{$N$}] (A3a) [above=4mm of A3] {};
        \node[node, label=right:{$1$}] (Ar3) [above =4mm of A3a]{};
        \draw (A2) -- (A3) -- (A3a) -- (Ar3);
        \draw   (Al2) -- (Al1) -- (A2) -- (Ar1) -- (Ar2);
        \node (Z2) [right=6mm of Ar2] {$/\U(1)\coma$};
    \end{tikzpicture}
\end{equation}
for which the Coulomb branch is the full $N$-instanton moduli space. By considering the overextended Dynkin diagram, we break the manifest $S_3$ automorphism group of the quiver down to a $\mathbb{Z}_2$ subgroup. We emphasize however that this is not a physical statement, the Higgs branch is the Coulomb branch of the $S_3$-manifest quiver in equation \eqref{eq:unitaryNE6}; the overextended diagram is a tool that allows us to use the monopole formula to calculate the Coulomb branch Hilbert series of equation \eqref{eq:unitaryNE6}. 

Let us now further specialize to the rank-two $E_6$ theory. The (unwreathed) Coulomb branch Hilbert series of the magnetic quiver for the Higgs branch is \cite{Hanany:2012dm}
\begin{equation}\label{eqn:2e6HS}
  \begin{aligned}
    \HS_{E_6^{(2)}}(t, \bm{x}, y) = \PE\big[ &\left(\chi_{\bm{78}}(\bm{x}) + \chi_{\bm{3}}(y)\right)t^2 + \chi_{\bm{2}}(y)\chi_{\bm{78}}(\bm{x})t^3 \\ &\quad - t^4 - \left(  \chi_{\bm{2}}(y)(1 + \chi_{\bm{78}}(\bm{x}) + \chi_{\bm{650}}(\bm{x}) )\right)t^5 + \cdots \big] \,,
  \end{aligned}
\end{equation}
where $\bm{x}$ and $y$ are the fugacities of the $\mathfrak{e}_6$ and $\mathfrak{su}(2)$, respectively. We see precisely the conjectured chiral ring generators from equation \eqref{eqn:GinstCONJGEN}. 

As discussed in Section \ref{sec:dg}, the adjoint representation of $\mathfrak{e}_6$ carries commuting actions of
$\mathfrak{so}(8)$ and $S_3$. Expanding the plethystic exponential in equation \eqref{eqn:2e6HS}, the nontrivial $\mathfrak{e}_6$ representations that appear in the Hilbert series up to order $t^5$ are the
\begin{equation}
    \bm{78} \,, \quad \bm{650} \,, \quad \bm{2430} \,, \quad \bm{2925} \,.
\end{equation}
We can determine the branching rules under the commuting actions for these representations as follows. The adjoint branching rule was given in
equation \eqref{eq:branchinge6sec2}, and for the $\bm{650}$ it is:
\begin{equation}
\label{eq:branchinge6sec4}
  \begin{aligned}
    \bm{650} &\rightarrow 3(\bm{1}, \bm{1}) \oplus 3(\bm{1}, \bm{2}) \oplus 2(\bm{28}, \bm{1_\mathrm{sgn}}) \\ &\qquad\quad \oplus \bigoplus_{\alpha = v,s,c} \bigg( (\bm{35_\alpha}, \bm{1}) \oplus (\bm{8_\alpha}, \bm{1}) \oplus (\bm{8_\alpha}, \bm{1_\mathrm{sgn}}) \oplus 2(\bm{8_\alpha}, \bm{2}) \oplus (\bm{56_\alpha}, \bm{2}) \bigg)\fstop
  \end{aligned}
\end{equation}
For the remaining two representations, we can use the fact that
\begin{equation}
    \bm{2430} = \operatorname{Sym}^2 \bm{78} - \bm{1} - \bm{650} \qquad \text{and} \qquad \bm{2925} = \wedge^2 \bm{78} - \bm{78} \,,
\end{equation}
to derive the branching rules from the branching rules for the $\bm{78}$ and $\bm{650}$. Then, if we are correct in our assumption that the $S_3$ only acts on the Higgs branch of the SCFT via the induced automorphism of the $\mathfrak{e}_6$ factor of the global symmetry, we can determine the Hilbert series of the proposed discretely-gauged theories. Projecting the expanded Hilbert series coefficient-by-coefficient and then taking the plethystic log, we find
\begin{align}\label{eqn:rank2prop}
    \nonumber
    \HS_{E_6^{(2)} /\, \ZZ_2}(t, \bm{x}, y) &= \begin{aligned}[t]
         \PE\big[ &\left(\chi_{\bm{52}}(\bm{x}) + \chi_{\bm{3}}(y)\right)t^2 + \chi_{\bm{2}}(y)\chi_{\bm{52}}(\bm{x})t^3 \\ & +  \left(\chi_{\bm{26}}(\bm{x}) +  \chi_{\bm{324}}(\bm{x})\right)t^4  + \chi_{\bm{2}}(y)\big(\chi_{\bm{273}}(\bm{x})  - 1\big)t^5 + \cdots \big] \,.
    \end{aligned} \\[0.6em]
    \HS_{E_6^{(2)} /\, \ZZ_3}(t, \bm{x}, y) &= \begin{aligned}[t]
         \PE\big[ &\left(2 + \chi_{\bm{28}}(\bm{x}) + \chi_{\bm{3}}(y)\right)t^2 + \chi_{\bm{2}}(y)(2 + \chi_{\bm{28}}(\bm{x}))t^3 \\& +  \left( 2 + 3\chi_{\bm{28}}(\bm{x}) + 2X_{\bm{8}}(\bm{x}) + X_{\bm{35}}(\bm{x}) + 2X_{\bm{56}}(\bm{x})\right)t^4 \\ & + \chi_{\bm{2}}(y)\big(3\chi_{\bm{28}}(\bm{x}) + 2X_{\bm{8}}(\bm{x}) + X_{\bm{35}}(\bm{x}) + 4X_{\bm{56}}(\bm{x})\big)t^5 + \cdots \big] \,.
    \end{aligned} \\[0.6em]
    \HS_{E_6^{(2)} /\, S_3}(t, \bm{x}, y) &= \begin{aligned}[t]
         \PE\big[ &\left(\chi_{\bm{28}}(\bm{x}) + \chi_{\bm{3}}(y)\right)t^2 + \chi_{\bm{2}}(y)\chi_{\bm{28}}(\bm{x})t^3 \\ &  + \left( 5 + X_{\bm{8}}(\bm{x}) + X_{\bm{35}}(\bm{x}) + X_{\bm{56}}(\bm{x})\right)t^4 \\& +  \chi_{\bm{2}}(y)\big(3 + 2\chi_{\bm{28}}(\bm{x}) + X_{\bm{8}}(\bm{x}) + 2X_{\bm{56}}(\bm{x})\big)t^5 + \cdots \big]\,.\nonumber
    \end{aligned}
\end{align}
Here, $\bm{x}$ is the fugacity of the semisimple part of the fixed-point subalgebra, either $\mathfrak{f}_4$ or $\mathfrak{so}(8)$, and we have defined
\begin{equation}
    X_{\bm{d}}(\bm{x}) = \sum_{\alpha = v,s,c} \chi_{\bm{d}_\alpha}(\bm{x}) \,.
\end{equation}

To verify the first of these putative Higgs branch Hilbert series, we can compute the Coulomb branch Hilbert series of the $\mathbb{Z}_2$-wreathing of the quiver in equation \eqref{eq:unitaryNE6}, via the artifice of the $\ZZ_2$-wreathed monopole formula applied to the quiver in equation \eqref{eq:unitaryNE6over}. This is possible since the overextension does not spoil the manifest $\ZZ_2$ diagram automorphism. Altogether, the final result is
\begin{equation}\label{eqn:rank2wr}
    \HS_{E_6^{(2)}\, \wr\, \ZZ_2}(t,1,1) = \PE\left[55 t^2+104 t^3+350 t^4+544 t^5+\mathcal{O}(t^6)\right] \,.
\end{equation}
While we have only shown the unrefined Hilbert series here, we can confirm from the refined Hilbert series that the $55$ moment maps correspond to a Coulomb branch flavor symmetry of
\begin{equation}
    \mathfrak{f}_4 \oplus \mathfrak{su}(2) \,.
\end{equation}
Furthermore, we can see that the numerical coefficients appearing in the Hilbert series in equation \eqref{eqn:rank2wr} are precisely those obtained from equation \eqref{eqn:rank2prop} after setting $\bm{x} = y = 1$.

\section{Discussion}\label{sec:disc}

Taken altogether, and subject to caveats in specific cases, the geometric, anomaly, Higgs branch, and Seiberg--Witten analyses presented above provide strong and mutually consistent evidence that the discrete gaugings considered here define genuine rank-one five-dimensional SCFTs, $E_n / H$. For the unitary magnetic quivers, the relevant diagram automorphisms can often be lifted to automorphisms of the generalized toric polygons and brane webs, thereby establishing corresponding symmetries of the full engineering geometry; our anomaly analysis shows that the gaugings of these symmetries are generically unobstructed. The resulting theories remain rank one:, while the mass deformations are restricted to the invariant subspace. Their Higgs branch and flavor algebra are
\begin{equation}
\mathcal H(E_n/H)
= \mathcal H(E_n)/H
\cong \mathcal C\left(\mathcal{T}_M \,\wr\, H\right),
\qquad
\mathfrak f(E_n/H)
= \mathfrak e_n^{\rho(H)} \,,
\end{equation}
where $\mathcal{T}_M$ is a magnetic quiver for the Higgs branch of $E_n$ which realizes $H$ as a diagram automorphism, and $\rho$ defines the induced action of $H$ on $\mathfrak{e}_n$.

The Hilbert series of the wreathed magnetic quivers reproduce these fixed-point flavor algebras in every case, while the Seiberg--Witten geometries, whenever available, provide an independent check through the number of invariant mass parameters. These quotients furnish a broad class of rank-one SCFTs whose Higgs branches are generally not closures of nilpotent orbits.\footnote{Any (refined) Hilbert series considered in this paper cannot be interpreted as the Hilbert series of the closure of a nilpotent orbit of some Lie algebra if there is a character in the coefficient of $t^4$ in the plethystic log with a positive sign. This implies the existence of an additional generator beyond the moment-map, contradicting the assumption of it being a nilpotent orbit closure.} We moreover conjecture that the same construction extends to the rank $N$ $E_n$ theories, with
\begin{equation}
\mathfrak f\left(E_n^{(N)}/H\right)
= \mathfrak e_n^{\rho(H)}\oplus\mathfrak{su}(2),
\qquad N\geq 2 \,,
\end{equation}
and the rank-two $E_6/\mathbb Z_2$ computation provides a nontrivial check of this proposal.

We now briefly highlight a few natural questions and future directions that follow from our work here.

\paragraph{Geometric Engineering:} Our construction provides strong bottom-up evidence for the existence of the discretely gauged theories $E_n/H$, but does not generally furnish a direct string-theoretic engineering of them. Even when the action of $H$ lifts to an automorphism of the generalized toric polygon or brane web of the parent $E_n$ theory, it is not automatic to use such data to produce a geometric engineering of $E_n/H$. Besides providing an independent construction of these SCFTs, this would clarify how they fit into or extend existing top-down geometric classifications of 5d SCFTs. It is, however, tempting to mention two facts. First, since we have obtained the SW geometries for the discretely-gauged theories, one possible strategy would be to apply mirror symmetry to them again. This would imply that the Type IIA/M-theory auxiliary geometry capturing the BPS spectra of the 4d KK theories presented in this paper is the same Calabi--Yau engineering the $E_n$ theories, with some parameters frozen, possibly due to the presence of torsional fluxes \cite{deBoer:2001wca}. However, as pointed out in Section \ref{sec:swcurvediscrgauging}, the curve written in terms of the $x,y$ coordinates has to be thought of as a gauge-covariant object after the gauging, preventing an immediate application of the previous construction and an interpretation in terms of Type IIA geometric engineering. 

On the other hand, e.g., for the $E_{6}/\mathbb Z_2$ theory presented in Section \ref{sec:2}, we would like to mention that there exists a possible candidate Calabi--Yau manifold. Indeed, the Weierstrass model in equation (5.55) of \cite{Anderson:2023wkr}
    displays a non-compact curve of $E_6$ singularities, which, due to standard monodromy arguments \cite{Aspinwall:2000kf}, is associated with an $\mathfrak{f}_4$ symmetry of the SCFT localized at its maximal singular point. Such a maximal singular point has vanishing orders (4,6,12) for the $g_2, g_3, \Delta$ coefficients of its Weierstrass model, suggesting that, in analogy with the $E$-string, it can be resolved by a single compact divisor. Taken together, these two features would suggest that the corresponding theory produced by M-theory has rank one and an $\mathfrak f_{4}$ flavor symmetry. 
\paragraph{Superconformal Index:} Here, we have given the Higgs branch Hilbert series for our proposed discretely-gauged 5d SCFTs. It would be interesting to determine further properties of these theories, such as the 5d superconformal index \cite{Bhattacharya:2008zy}.
    The indices of the parent rank-one $E_n$ theories have been determined using localization and instanton counting \cite{Kim:2012gu,Hwang:2014uwa}. Since $S^4$ is simply connected, gauging the finite zero-form symmetry $H$ should project the Hilbert space on $S^4$ onto its $H$-invariant subspace. The resulting index probes a substantially broader protected operator spectrum than the Higgs branch Hilbert series and would therefore provide a stronger consistency test of the proposed $E_n/H$ theories. It would likewise be interesting to construct the corresponding $H$-discretely-gauged versions of the Nekrasov partition function \cite{Nekrasov:2002qd,Nekrasov:2003rj}, which was recently used in \cite{Harding:2026gzj} precisely to study global symmetries and 't Hooft anomalies for 5d SCFTs.
    Whenever the action of $H$ admits a suitable geometric realization on the brane web, these quantities may be accessible by combining the refined topological vertex formalism \cite{Aganagic:2003db,Iqbal:2007ii,Taki:2007dh,Hayashi:2013qwa} with existing quotient and S-fold constructions \cite{Acharya:2021jsp,Kim:2021fxx}. The $\mathbb{Z}_2$ orbifold/orientifold prescription of \cite{Kim:2024ufq} provides a related concrete example in which quotient data can be incorporated into a topological vertex computation.
\paragraph{Discrete Gauging and Affine Schemes:} In recent work \cite{Kang:2025zub,Kang:2026nge,NONPOINT,SUPERPOINT}, it has been proposed that the Macdonald index, and thus the Schur index and Higgs branch Hilbert series,\footnote{Two different limits of the Macdonald index recover the Schur index and the Hall--Littlewood index \cite{Gadde:2011uv}. In many examples, the Hall--Littlewood index is identical to the Higgs branch Hilbert series \cite{Kang:2022zsl}.} of a broad collection of 4d $\mathcal{N}=2$ SCFTs is encoded in a (syzygy-maximized) bifiltered affine scheme, which reduces as a variety to the Higgs branch. Now that we have developed the technology for discrete gauging on the Higgs branch in this paper, it is natural to ask if the magnetic quiver diagram symmetry can be uplifted to the scheme-ification of the Higgs branch. Via analogous geometric and anomaly arguments as those presented in this paper, we expect that these discrete gaugings lead to novel 4d $\mathcal{N}=2$ SCFTs.
\paragraph{6d SCFT Origins:} 
    There is by now significant evidence that compactifications of 6d $(1,0)$ SCFTs provide an organizational principle for large classes of known 4d $\mathcal{N}=2$ SCFTs \cite{Heckman:2022suy,Giacomelli:2025zqn,Giacomelli:2024ycb,Giacomelli:2024dbd}. In particular, 6d $(1,0)$ A-type orbi-instanton theories \cite{Aspinwall:1997ye,DelZotto:2014hpa,Heckman:2015bfa,Mekareeya:2017jgc}, i.e., 6d $\mathcal{N}=(1,0)$ theories realized on M5-branes, probing an M9-brane \cite{Horava:1996ma}, on a $\CC^2/\ZZ_k$ singularity, were recently shown to provide the origin of class $\mathcal{S}$ \cite{Gaiotto:2009we,Gaiotto:2009gz} theories of type A with both regular and irregular untwisted punctures \cite{Giacomelli:2024ycb,Giacomelli:2025zqn}. Do the discrete gaugings of 5d SCFTs discussed here, or more generally 4d $\mathcal{N}=2$ SCFTs, descend by compactification from the same 6d SCFTs as their non-discretely-gauged cousins, or are there discretely-gauged SCFTs also in 6d (perhaps along the lines of \cite{Lawrie:2025exx}) that function as the parent theories for the discretely-gauged theories discussed herein? A natural place to explore such questions is when the lower dimensional theory has multiple origins in 6d, such as studied in, for example, \cite{Baume:2021qho,Distler:2022kjb}. We intend to return to some of these questions in future work.

\subsection*{Acknowledgements}

We thank Riccardo Comi, Michele Del Zotto, Julius Grimminger, William Harding, Jonathan Heckman, Max Hubner, Chris Hull, Max LVC Hutt, Noppadol Mekareeya, Matteo Sacchi, and Andrea Sangiovanni for essential discussions. 
This work was performed in part while A.~M.~was at the Aspen Center for Physics, which is supported by National Science Foundation grant PHY-2210452. C.~L., S.~N.~M., and A.~M. thank the Simons Center for Geometry and Physics for hospitality during the ``23rd Simons Physics Summer Workshop: Theory, Experiment and the Emerging New Physics''. 
G.~A.~T.~is supported by the STFC Consolidated Grant ST/X000575/1.
The research of M.~D.~M.~is funded through an ARC advanced project and is further supported by IISN-Belgium (convention 4.4503.15). 
C.~L.~acknowledges the Deutsche Forschungsgemeinschaft under Germany's Excellence Strategy - EXC 2121 ``Quantum Universe" - 390833306 and the Collaborative Research Center - SFB 1624 ``Higher Structures, Moduli Spaces, and Integrability" - 506632645. The work of S.~N.~M.~is supported by DOE (HEP) Award DE-SC0013528, BSF grant 2022100, and a University Research Foundation grant at the University of Pennsylvania.
A.~M.~is supported in part by the DOE (HEP) Awards DE-SC0017647 and DE-SC0023719.

\appendix

\addtocontents{toc}{\protect\setcounter{tocdepth}{1}}

\section{Affine \texorpdfstring{$A$}{A} Theories}\label{app:birch}

The (unwreathed) Coulomb branch global symmetry of the affine $A_{2k}$ quiver is $\mathfrak{g} = \mathfrak{su}(2k+1)$. The corresponding untwisted affine Dynkin diagram, $\widetilde{\mathcal{D}}_{\mathfrak{g}} = \widetilde{A}_{2k}$, consists of $2k+1$ nodes arranged in a single cycle. Its automorphism group is the dihedral group
\begin{equation}
    \operatorname{Aut}(\widetilde{A}_{2k}) \cong D_{2(2k+1)}\fstop
\end{equation}
We consider the discrete gauging of a subgroup $H \cong \mathbb{Z}_2$ generated by a reflection $s \in \operatorname{Aut}(\widetilde{A}_{2k})$. The diagram $\widetilde{A}_{2k}$ has an odd number of nodes, and thus every reflection axis in $D_{2(2k+1)}$ passes through exactly one node and bisects one opposite link.

In the magnetic quiver, the overall $\mathfrak{u}(1)$ center of mass decouples because physical observables depend exclusively on relative flux differences $n_i = m_i - m_{i+1}$. We parameterize the group-element-twisted summation directly over these independent edges. An overall shift $m_i \to m_i + C$ leaves the relative differences invariant. To evaluate the invariant lattice $\mathbb{Z}^M/\mathbb{Z}$, we express all fluxes $m_i$ in terms of the differences $n_i$ relative to a reference node. We eliminate the overall $\mathfrak{u}(1)$ by imposing $z_0 = \prod z_i^{-1}$, with $z_0$ the fugacity associated with the reference node.

Using the Kac diagram formalism introduced in Section \ref{sec:dg}, we should consider labelings of the twisted affine $\mathfrak{su}(2k+1)$ Dynkin diagram, which is
\begin{equation}
    \begin{tikzpicture}[baseline=0,font=\footnotesize]
        \node[node, label=below:{$1$}, label=above:{$0$}] (A0) at (0,0) {}; 
        \node[node, label=below:{$2$}, label=above:{$1$}] (A1) at (1.5,0) {}; 
        \node (Adots) at (3,0) {$\cdots$};
        \node[node, label=below:{$2$}, label=above:{$k-1$}] (Ak1) at (4.5,0) {};
        \node[node, label=below:{$2$}, label=above:{$k$}] (Ak) at (6,0) {};
        
        \draw[-implies, double distance=3pt] (A0) -- (A1);
        \draw (A1) -- (Adots);
        \draw (Adots) -- (Ak1);
        \draw[-implies, double distance=3pt] (Ak1) -- (Ak);
    \end{tikzpicture} \,.
\end{equation}
Here we have written the affine marks below the node and an index labeling the node above.
The single $\mathbb{Z}_2$ embedding visible from such Kac diagrams (and thus the only embedding which involves the outer automorphism subgroup of the automorphism group of $\mathfrak{su}(2k+1)$) is specified by $s_0 = 1$. The fixed-point subalgebra is then the special orthogonal algebra:
\begin{equation}
    \mathfrak{su}(2k+1)^{\rho(\mathbb{Z}_2)} = \mathfrak{so}(2k+1)\fstop
\end{equation}
This proves that the flavor algebra for this particular $\ZZ_2$ wreathing of $\widetilde{A}_{2k}$ is expected to be $\mathfrak{so}(2k+1)$. This can also be shown by computing the Hilbert series using \cite[Appendix A]{Grimminger:2024mks} and the prescription for cutting the edge as in Section \ref{sec:Z2cuttingedges}. The refined wreathed Hilbert series is given by:
\begin{equation}
    H_{\widetilde{A}_{2k}\,\wr\, \ZZ_2}(t, \mathbf{z}) = \frac{1}{2} \big( H_{\text{free}}(t, \mathbf{z}) + H_{\mathbb{Z}_2}(t, \mathbf{z}) \big) \fstop
\end{equation}
For any $M$-node affine $A$ quiver, $H_{\text{free}}(t, \mathbf{z})$ is computed over the full lattice $\mathbb{Z}^M/\mathbb{Z}$. The Hilbert series is 
\begin{equation}
    H_{\text{free}}(t, \mathbf{z}) = \frac{1-t^2}{(1-t^2)^M} \sum_{m_1, \dots, m_{M-1} \geq -\infty}^{\infty} t^{2 \Delta(m)} \prod_{i=1}^{M-1} z_i^{m_i} \coma
\end{equation}
where 
\begin{equation}
\Delta(m) = \frac{1}{2} \sum_{i=0}^{M-1} |m_i - m_{i+1}|\coma
\end{equation}
and we decouple the overall $\mathfrak{u}(1)$ by setting $m_0 = m_M = 0$. The sector $H_{\mathbb{Z}_2}(t,\mathbf{z})$ is computed as usual by looking at each nontrivial group element contribution under the $\ZZ_2$-wreathing. This is given by $m_{2k+1-i} = m_i$, requiring $m_k = m_{k+1}$, with a corresponding modification of the prefactor as explained in \cite{Lawrie:2025exx}. The result is
\begin{equation}
    H_{\mathbb{Z}_2}^{A_{2k}}(t, \mathbf{z}) = \frac{1-t^2}{(1-t^2)(1-t^4)^k} \sum_{m_1, \dots, m_k \geq -\infty}^\infty (-1)^{m_k} \, t^{2 \Delta(\mathbf{m})} \prod_{i=1}^k z_i^{2m_i} \coma
\end{equation}
with 
\begin{equation}
    \Delta(\mathbf{m}) = \sum_{i=0}^{k-1} |m_i - m_{i+1}|\fstop
\end{equation}
By summing the two contributions, we obtain $H_{\widetilde{A}_{2k}\,\wr \,\ZZ_2}(t, \mathbf{z})$ which has a leading order corresponding to the adjoint of $\so(2k+1)$, obtained by defining (for $k > 2$)
\begin{equation}\label{eq:A2k_map}
    z_1 \mapsto \frac{y_{k-2}y_k^2}{y_{k-1}^2}\coma \quad z_i \mapsto \frac{y_{k-i-1}y_{k-i+1}}{y_{k-i}^2} \quad (1 < i < k-1)\coma \quad z_{k-1} \mapsto \frac{y_2}{y_1^2}\coma \quad z_k \mapsto y_1\fstop
\end{equation}

For completeness, we can now compute the $\ZZ_2$-wreathing of the affine $A_{2k-1}$ quiver that has an even number of nodes. The dihedral group contains two conjugacy classes of reflections. One axis passes through two opposite nodes (fixing two nodes). The other axis bisects two opposite edges (fixing zero nodes). We can understand these two types of reflections using the results of Kac, as utilized in Section \ref{sec:dg}. There are two conjugacy classes of homomorphisms
\begin{equation}
    \mathbb{Z}_2 \rightarrow \operatorname{Aut}(\mathfrak{su}(2k)) \,,
\end{equation}
such that the image is nontrivial inside of the outer automorphism group of $\mathfrak{su}(2k)$. These are captured via a Kac labeling of the twisted affine $\mathfrak{su}(2k)$ Dynkin diagram, depicted here:
\begin{equation}
    \begin{tikzpicture}[baseline=0,font=\footnotesize]
        \node[node, label=left:{$1$}, label=above:{$0$}] (A0) at (0, 0.5) {}; 
        \node[node, label=left:{$1$}, label=below:{$1$}] (A1) at (0, -0.5) {}; 
        \node[node, label=below:{$2$}, label=above:{$2$}] (A2) at (1, 0) {}; 
        \node (Adots) at (2.5, 0) {$\cdots$};
        \node[node, label=below:{$2$}, label=above:{$k-1$}] (Ak1) at (4, 0) {};
        \node[node, label=below:{$1$}, label=above:{$k$}] (Ak) at (5.5, 0) {};
        
        \draw (A0) -- (A2);
        \draw (A1) -- (A2);
        \draw (A2) -- (Adots);
        \draw (Adots) -- (Ak1);
        \draw[implies-, double distance=3pt] (Ak1) -- (Ak);
    \end{tikzpicture} \,.
\end{equation}
The two choices (up to symmetry of the diagram) of Kac labels capturing the nontrivial $\mathbb{Z}_2$ embeddings are
\begin{equation}
    s_0 = 1 \qquad \text{ and } \qquad s_k = 1 \,.
\end{equation}
The fixed-point subalgebras of these two embedded $\ZZ_2$ are easily seen, using the diagram formed by the zero-labels of the Kac diagram, to be
\begin{equation}
    \mathfrak{usp}(2k) \qquad \text{ and } \qquad \mathfrak{so}(2k) \,,
\end{equation}
respectively. It is direct to determine which of the two $\mathbb{Z}_2$ affine Dynkin diagram automorphisms map to which of the two Kac labelings under the homomorphism in equation \eqref{eqn:BIGHOMO}: $s_0 = 1$ fixes two nodes and $s_k = 1$ fixes zero nodes.

We can confirm these symmetries after the $\ZZ_2$ quotient by turning again to the wreathing analysis. First, we consider the reflection passing through two opposite nodes, like,\footnote{We draw the quivers for $k=3$ as a visual aid. The extension to arbitrary $k$ is obvious.}
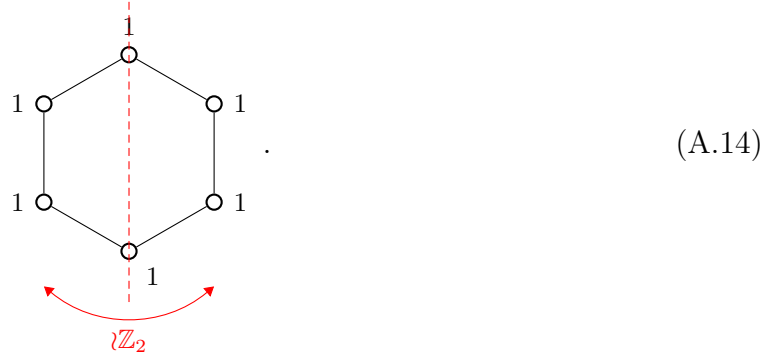
\begin{equation}\label{eq:A2km1_node_quiver}
    \begin{tikzpicture}[baseline=0,font=\footnotesize]
        \def\radius{1.3cm};
        \node[node, label=above:{$1$}] (A0) at (90:\radius) {}; 
        \node[node, label=left:{$1$}] (A1) at (150:\radius) {}; 
        \node[node, label=left:{$1$}] (A2) at (210:\radius) {};
        \node[node, label=below right:{$1$}] (A3) at (270:\radius) {};
        \node[node, label=right:{$1$}] (A4) at (330:\radius) {};
        \node[node, label=right:{$1$}] (A5) at (30:\radius) {};
        \draw (A0) -- (A1) -- (A2) -- (A3) -- (A4) -- (A5) -- (A0);
        \draw[densely dashed, red, thin] (90:2.0cm) -- (270:2.0cm);
        \draw[Triangle-Triangle,red] ([yshift=-10mm]A2.south) to[bend right=40] node[below,pos=0.5] {$\wr \ZZ_2$} ([yshift=-10mm]A4.south);
    \end{tikzpicture} \,.
\end{equation}
One of the two fixed nodes can be identified with the affine node. We can then compute the Hilbert series, and specifically, only the $\ZZ_2$ sector must be modified from the discussion above. Setting the reference node $m_0 = 0$, the invariant locus is $m_{2k-i} = m_i$. This yields
\begin{equation}\label{eq:A2km1_node}
    H_{\mathbb{Z}_2}(t, \mathbf{z}) = \frac{1-t^2}{(1-t^2)^{2}(1-t^4)^{k-1}} \sum_{m_1, \dots, m_k \geq -\infty}^\infty t^{2\Delta(\mathbf{m})} z_k^{m_k} \prod_{i=1}^{k-1} z_i^{2m_i} \coma
\end{equation}
with 
\begin{equation}
    \Delta(\mathbf{m}) =  \sum_{i=0}^{k-1} |m_i - m_{i+1}|\fstop
\end{equation}
The fugacities $z_i$ map directly to those of $\mathfrak{usp}(2k)$ by the following redefinition:
\begin{equation}\label{eq:A2km1_node_map}
    z_1 \mapsto \frac{y_1^2}{y_2}\coma \quad z_i \mapsto \frac{y_i^2}{y_{i-1}y_{i+1}} \quad (1 < i < k)\coma \quad z_k \mapsto \frac{y_k^2}{y_{k-1}^2}\fstop
\end{equation}

Alternatively, if we choose the axis of reflection passing through two edges,   
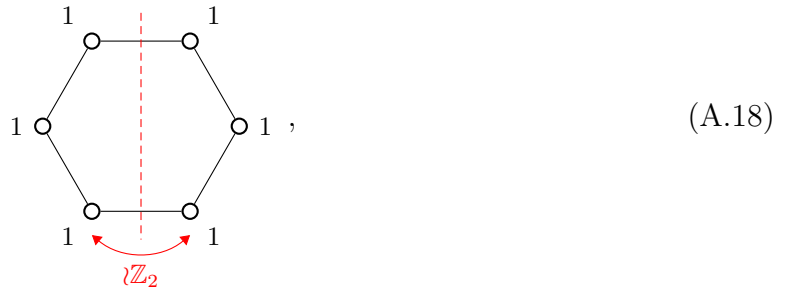
\begin{equation}\label{eq:A2km1_edge_quiver}
    \begin{tikzpicture}[baseline=0,font=\footnotesize]
        \def\radius{1.3cm};
        \node[node, label=above right:{$1$}] (A0) at (60:\radius) {}; 
        \node[node, label=above left:{$1$}] (A1) at (120:\radius) {}; 
        \node[node, label=left:{$1$}] (A2) at (180:\radius) {};
        \node[node, label=below left:{$1$}] (A3) at (240:\radius) {};
        \node[node, label=below right:{$1$}] (A4) at (300:\radius) {};
        \node[node, label=right:{$1$}] (A5) at (360:\radius) {};
        \draw (A0) -- (A1) -- (A2) -- (A3) -- (A4) -- (A5) -- (A0);
        \draw[densely dashed, red, thin] (90:1.5cm) -- (270:1.5cm);
        \draw[Triangle-Triangle,red] ([yshift=-2mm]A3.south) to[bend right=40] node[below,pos=0.5] {$\wr \ZZ_2$} ([yshift=-2mm]A4.south);
    \end{tikzpicture} \,,
\end{equation}
then there are no fixed nodes. When considering the $\ZZ_2$-wreathing of the quiver, we find that the $\ZZ_2$ sector contributing to the Hilbert series is now
\begin{equation}\label{eq:A2km1_edge}
    H_{\mathbb{Z}_2}(t, \mathbf{z}) = \frac{1-t^2}{(1-t^4)^k} \sum_{m_1, \dots, m_{k-1} \geq -\infty}^\infty (-1)^{m_{k-1}} \, t^{2 \Delta(\mathbf{m})} \prod_{i=1}^{k-1} z_i^{2m_i} \coma
\end{equation}
with
\begin{equation}
   \Delta(\mathbf{m})= \sum_{i=0}^{k-2} |m_i - m_{i+1}|\fstop
\end{equation}
The reason why there is only one $(-1)$-factor is that we have chosen the axis to bisect the edges $(2k-1, 0)$ and $(k-1, k)$, and we have set $m_0 = 0$.  This quiver has only $k-1$ fugacities, while the $\mathfrak{so}(2k)$ adjoint representation requires rank $k$. However, the resulting character is the character of the adjoint of $\mathfrak{so}(2k)$ expressed in terms of its $\mathfrak{so}(2k-1)$ subalgebra. This can be shown by considering the character of $\mathfrak{so}(2k)$ and setting, e.g., $y_k=1$. The remaining fugacities are related to those in the Hilbert series by:
\begin{equation}
    z_i = \frac{y_i}{y_{i+1}} \quad (1 \leq i \leq k-2) \coma \quad z_{k-1} = y_{k-1} \fstop
\end{equation}

\bibliographystyle{sortedbutpretty}
\bibliography{mybib}

\end{document}